\documentclass[a4paper,12pt,times,numbered,print,index]{Classes/PhDThesisPSnPDF}

\ifsetCustomMargin
  \RequirePackage[left=37mm,right=30mm,top=35mm,bottom=30mm]{geometry}
  \setFancyHdr % To apply fancy header after geometry package is loaded
\fi

\ifsetCustomFont
\fi

\usepackage{amsmath,graphicx,float,bm,amssymb,mathtools,calrsfs}
\usepackage{hyperref}
\usepackage{MnSymbol}
\usepackage{tikz,subcaption}
\usepackage[export]{adjustbox}
\usepackage[font=small]{caption}
\usepackage{notoccite}
\newcommand{\eqsp}[1]{\hspace{0.5 cm} \text{#1} \hspace{0.5cm}}

\DeclareSymbolFont{newfont}{OML}{cmm}{m}{it}% Computer Modern math font
\DeclareMathSymbol{\Epsilon}{3}{newfont}{15}% Symbol 

\newenvironment{mytitle}
{
\clearpage
\setsinglecolumn 
\thispagestyle{empty}
\centering 
}

\newenvironment{myabstract}{
\cleardoublepage
\setsinglecolumn
\chapter*{\vspace{-4.1 cm}\centering \LARGE Abstract \vskip -3 mm}
\thispagestyle{empty}
}

\RequirePackage[labelsep=space,tableposition=top]{caption}
\usepackage{subcaption}

\usepackage{booktabs} % For professional looking tables
\usepackage{multirow}

\usepackage{siunitx} % use this package module for SI units

\ifuseCustomBib
   \RequirePackage[square, sort, numbers, authoryear]{natbib} % CustomBib

\fi

\ifdefineAbstract
\fi

\ifdefineChapter
\fi

\begin{document}

\frontmatter

%\maketitle

\begin{mytitle}

\includegraphics[height = 7 cm]{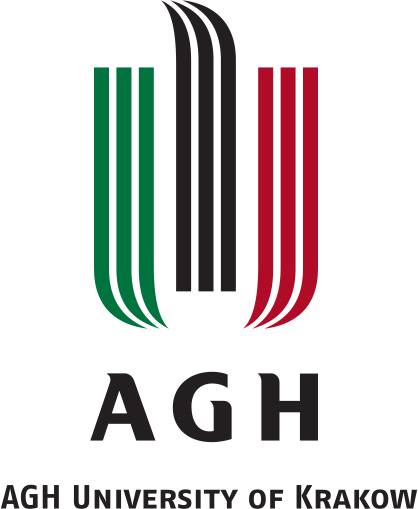}

\vskip 1.2 cm
\textbf{\Large FIELD OF SCIENCE : \  NATURAL SCIENCES}
\vskip 3 mm
{\large SCIENTIFIC DISCIPLINE : \  PHYSICAL SCIENCES }

\vskip 1.3 cm
\textbf{\huge DOCTORAL DISSERTATION }

\vskip 1.5 cm
%\textbf{\Large Correlation and fluctuation of collective observables in ultrarelativistic heavy-ion collision (tentative)}

\textit{\textbf{\Large Study of the hottest droplet of fluid through correlations and fluctuations of collective variables}}

\vskip 1.5 cm
{\large Author : Rupam Samanta \\ \vskip 3 mm
Supervisor: Prof. dr. hab. Piotr Bo\.zek}

\vskip 1.2 cm
{\large AGH University of Krakow \\
Faculty of Physics and Applied Computer Science\\
}
\vskip  1.4 cm

{\large Krak\'ow, 2024}

\end{mytitle}

% ******************************* Thesis Dedidcation ********************************

\begin{dedication} 
\vskip 4 cm
{\it I would like to dedicate this thesis to my beloved parents and my brother Dr. Chandan Samanta. } 

\end{dedication}
%\include{Declaration/declaration}
%\include{Acknowledgement/acknowledgement}
% ************************** Thesis Abstract *****************************
% Use `abstract' as an option in the document class to print only the titlepage and the abstract.
\begin{myabstract}
Collisions of two heavy nuclei at relativistic speeds at the Relativistic Heavy Ion collider (RHIC) at BNL and the Large Hadron Collider (LHC) at CERN, create a state of matter which has a temperature $10^5$ times that of Sun's core, a size of the order of nuclear radius (femtometer) and which behaves like a perfect fluid with minimal viscosity. This matter under extreme condition, is a medium where the quarks and gluons, normally existing as bound states in hadrons, travel freely with color degrees of freedom, with their interactions governed by Quantum Chromodynamics (QCD). This hot, dense, fluid-like droplet of deconfined state of quarks and gluons is known as the Quark Gluon Plasma (QGP). The QGP medium, surviving for a very short time ($~10^{-22}$ s) with its evolution dynamics described by the relativistic viscous hydrodynamics, creates thousands of particles hitting the detectors at the end. One of the most remarkable features is the collective flow of these particles, serving as a key phenomenon for probing the QGP medium in high energy nuclear collisions. The most peculiar and intriguing characteristics of the collective anisotropic flow, quantified in terms of flow harmonics, is the importance of event-by-event fluctuations, stemming mostly from event-by-event fluctuations in the initial state. In this thesis, we focus on fluctuations and correlations between the collective observables such as mean transverse momentum per particle ($[p_T]$) and harmonic flow coefficients ($v_n$) etc. Specifically, we show that the fluctuations of harmonic flow can be probed by the factorization-breaking coefficients between flow vectors in different $p_T$-bins. Experimental difficulty can be reduced by taking one of the flow vectors momentum averaged. Fluctuations cause a decorrelation between the flow vectors, which can be attributed to equal contributions from the flow magnitude and flow angle decorrelation. We study fluctuations of mean transverse momentum per particle ($[p_T]$) in ultra-central collisions and show that our model can explain the steep fall of its variance observed by the ATLAS collaboration. We also present robust predictions for the skewness and kurtosis, and highlight the role of impact parameter fluctuations in ultracentral collisions. We study the Pearson correlation coefficients between $[p_T]$ and $v_n^2$, which can map the initial state correlations between the shape and size of the fireball. We show that higher order normalized and symmetric cumulants between these observables can be constructed, which put useful additional constraints on the initial state properties. Furthermore, we study the momentum dependent Pearson correlation between $[p_T]$ and the transverse momentum dependent flow. It shows sensitivity to the Gaussian width of the nucleon at the initial state. Finally, we show that such correlations and fluctuations of collective observables can be used to study nuclear deformation and put robust constraints on their deformation parameters through high energy nuclear collisions. The research presented in this thesis has significantly contributed to the advancement of the field leaving ample opportunities for further developments in future, which remain beyond its current scope.      
\end{myabstract}

%\include{Abstract/abstractPolish}

% *********************** Adding TOC and List of Figures ***********************

\tableofcontents

%\listoffigures

%\listoftables

% \printnomenclature[space] space can be set as 2em between symbol and description
%\printnomenclature[3em]

%\printnomenclature

\chapter{ Notations}

\begin{itemize}
    \item We use natural units : $\hbar=c=k_B =1$ .
    \item Metric : $g^{\mu \nu} = diag(1, -1, -1 ,-1)$ .
    \item Position four vector : $x^\mu=(t,x,y,z)$ .
    \item Fluid velocity : $u^\mu = (1,\vec{v})$ . 
    \item The projector : $\Delta^{\mu\nu} = g^{\mu\nu}-u^\mu u^\nu  $ .
    \item Rank 4 tensor : $ \Delta^{\mu \nu}_{\alpha \beta} =  \frac{1}{2} (\Delta^\mu_\alpha \Delta^\nu_\beta+\Delta^\mu_\beta \Delta^\nu_\alpha -\frac{2}{3} \Delta^{\mu \nu} \Delta^{\alpha \beta} ) \ $ .
    \item The angular brackets and the parenthesis used : 
    \begin{equation}
    \begin{aligned}
    X^{\langle \mu \rangle} = \Delta^\mu_\nu X^\nu ,\eqsp{} Y^{\langle \mu \nu \rangle} = \Delta^{\mu \nu}_{\alpha \beta} Y^{\alpha \beta} \eqsp{and} Z^{(\mu \nu)} = \frac{1}{2} (Z^{\mu \nu}+Z^{\nu \mu}) \ \nonumber.
    \end{aligned}
    \end{equation}
\end{itemize}

\subsection*{Transverse Momentum}

\begin{itemize}
    \item General notation for transverse momentum : $p_T$ .
    \item Mean transverse momentum per particle in an event : $[p_T]$ .
    \item Average transverse momentum ( averaged over events) : $\langle p_T \rangle $ or $\overline{p_T}$ $\equiv \langle [p_T] \rangle$ .
    \item We use $\delta p_T= p_T - \langle p_T \rangle $ or $ p_T - \overline{p_T}$  .
    \item Other notations used for transverse momentum to avoid confusions whenever necessary : $p$ (Factorization-breaking coefficients) or $q$ (momentum dependent Pearson correlation coefficients) .
\end{itemize}

\subsection*{Harmonic flow}

\begin{itemize}
    \item Harmonic flow vector in an event : $V_n = v_n e^{i n \Psi_n }$ .
    where, flow magnitude : $v_n = V_n  V_n^\star$ and flow angle : $\Psi_n$. \item $v_n$ denotes the integrated flow in an event .
    \item Two-particle cumulant for flow harmonics, which is an event averaged quantity : $v_n\{2\} = \sqrt{\langle V_n  V_n^\star \rangle}$,  the integrated flow, averaged over events .
    \item Transverse momentum dependent or differential harmonic flow in an  event : $V_n(p_T)$, other notations : $V_n(p)$ or $V_n(q)$, with $V_n(p) = v_n(p) e^{i n \Psi_n(p) }$ .
    \item Event averaged $p_T$-differential flow : 
     $$v_n\{2\}(p) =\frac{\langle V_nV_n^*(p) \rangle}{\sqrt{\langle V_n V_n^* \rangle}} \eqsp{and} v_n[2](p) =\sqrt{\langle V_n(p)V_n^*(p) \rangle} = \sqrt{\langle v_n(p)^2 \rangle} \ .$$
   
\end{itemize}

% ******************************** Main Matter *********************************
\mainmatter

%*******************************************************************************
%*********************************** First Chapter (Introduction) *****************************
%*******************************************************************************

\chapter{Introduction}  %Title of the First Chapter

\ifpdf
    \graphicspath{{Chapter1/Figs/Raster/}{Chapter1/Figs/PDF/}{Chapter1/Figs/}}
\else
    \graphicspath{{Chapter1/Figs/Vector/}{Chapter1/Figs/}}
\fi

The field of high-energy physics which can be broadly classified into nuclear and particle physics, deals with understanding the fundamental forces of nature governing the interactions between its microscopic constituents. The physics of the interaction between fundamental particles that are found in nature is largely governed by the {\it Standard Model} (SM) of particle physics\footnote{The Standard Model does not describe the gravitational force which is one of the four fundamental forces in nature.}~\cite{Gaillard:1998ui,Thomson_2013} and accessed by colliding particles using high energy accelerators~\cite{Kobayashi}. 

Initial years of nuclear physics focused on the properties and interactions of atomic nuclei at low energies. However, with the advancement of accelerator and collider technology, it became possible to explore nuclear collisions at much higher energies. In these colliders, by colliding two heavy nuclei such as gold (Au) or lead (Pb) at a speed very close to the speed of light $c$, it is possible to create a state of matter which existed just microseconds after the Big Bang~\cite{Heinz:2013wva}. This is known as {\it ultrarelativistic heavy-ion collision} which lies at the frontier of high-energy physics facilitating the study of fundamental properties of matter under extreme conditions~\cite{Voloshin:2008dg,florkowski2010phenomenology,Heinz:2013th,Busza:2018rrf}. The study of high-energy nuclear collisions is important for advancing our understandings of {\it Quantum Chromodynamics} (QCD), the theory governing the strong interaction~\cite{Fritzsch:1973pi,Weinberg:1973un}, one of the four fundamental forces in nature. 

The state of matter created in ultrarelativistic heavy-ion collision is an extremely hot, dense, tiny medium of quarks and gluons, the fundamental constituents of hadrons, known as the {\it Quark-Gluon-Plasma} (QGP), which behaves like an almost perfect fluid. This QGP medium survives for a very short time (few fm/c) and what is ultimately seen in the detectors of such heavy-ion experiments, is a large number of hadrons produced in the collisions. A single head-on collision of Pb+Pb at 5.02 TeV centre of mass energy, creates around $\sim 35000$ hadrons which are mostly pions\footnote{If we consider only the identified charged hadrons, a large fraction of them are pions ($\sim 80 \%$) while the rest are kaons ($\sim 17 \%$) and protons ($\sim 3 \% $) contributing to significantly smaller fractions.}. The study of the properties and the dynamics of the QGP medium accessed through those detected particles can unravel many interesting aspects of the matter under exotic conditions. The field of high-energy heavy-ion physics has evolved considerably over the past three decades answering some of the most fundamental questions of physics.       

\section{The Quark-Gluon-Plasma : hottest fluid ever known} 
\label{the QGP}
The Quark-Gluon-Plasma created at the collision of two heavy nuclei moving at relativistic speed has an effective temperature of 212 MeV ($\equiv 10^{12}$K)~\cite{Gardim:2019xjs}, a size in the scale of femtometer (nuclear radius) and it is a nearly perfect fluid with a small viscosity~\cite{Bernhard:2019bmu}. The core of the Sun has a temperature $~10^7$ K, the QGP medium's temperature is another five orders of magnitude larger than that. Therefore, the QGP medium turns out to be the hottest, most dense and tiniest droplet of fluid that is ever known to the mankind and which can be crated at the laboratory~\cite{Gyulassy:2004zy}. 
\begin{figure}[ht!]
    \includegraphics[height=8 cm]{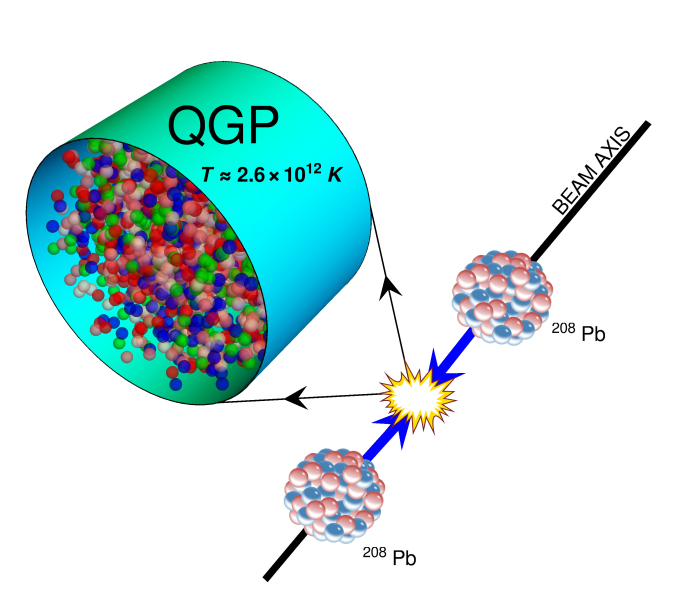}  
    \centering
    \caption{Pictorial representation of Pb+Pb collision at the LHC and formation of the deconfined state of the Quark Gluon Plasma (QGP) medium, the hottest fluid (T$\sim 10^{12}$ K) that can be created in a laboratory. Figure taken from~\cite{Gardim:2019xjs}.}
   \label{fig: heavy-ion collision}
\end{figure}
Normally the quarks and gluons are never seen as free particles, rather they are always found as bound states such as proton, neutron or hadrons in general.  The interaction between the quarks and gluons is governed by  QCD involving their {\it color charges }. The quarks and gluons, called {\it partons} in general, individually carry color charges (or color quantum number) but the observed hadrons which are bound states of them are color neutral. This is known as {\it color confinement}. 

However, in QCD theory the interaction strength between the quarks and gluons is stronger at larger distance and become weaker or asymptotically zero at very short distance, a phenomena known as {\it the asymptotic freedom}~\cite{Gross:1973ju,Gross:1974cs}. In ultrarelativistic heavy-ion collision, which occur at very high energy, these quarks and gluons become deconfined creating a quasi-free state : the Quark-Gluon-Plasma, making the color degrees of freedom deconfined. Thus the study of this little droplet of fluid would be very interesting, and could deepen our understanding of the strong nuclear force, QCD and basic building blocks of nature.  

\section{Experiments in heavy-ion collision} 
\label{HI experiemnts}
Almost fifty years ago, the first relativistic heavy-ion collision experiments were performed at the Bevatron in Berkeley at energies $\sim$ 1-2 GeV. Since then heavier ions were collided at higher energies at facilities such as the Alternating Gradient Synchrotron (AGS) at the Brookhaven National Laboratory (BNL) and the Super Proton Synchrotron (SPS) at CERN providing early evidence of QGP formation.
\begin{figure}[ht!]
    \includegraphics[height=6 cm]{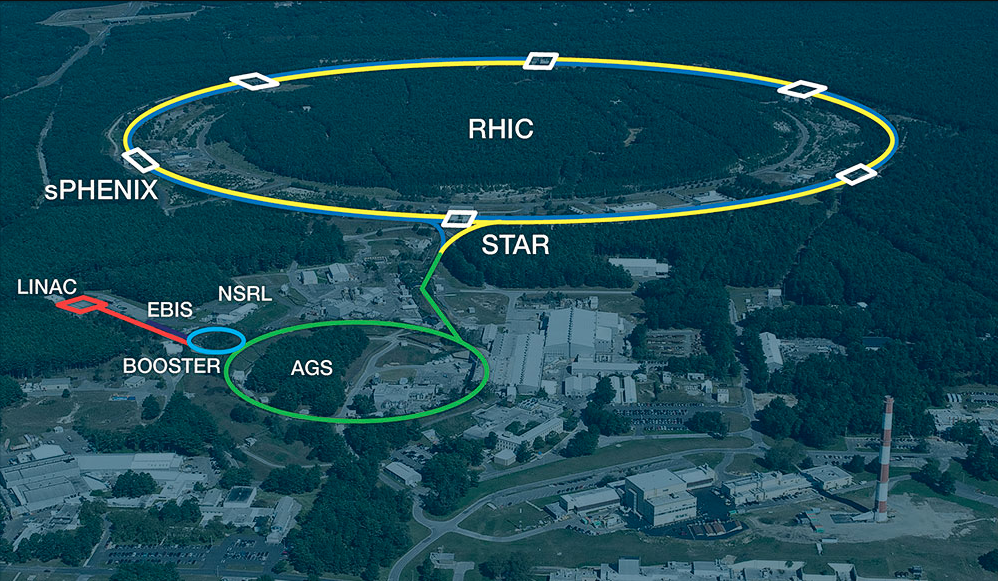}  
    \centering
    \caption{The aerial view of the Relativistic Heavy Ion Collider (RHIC), located at the Brookhaven National Laboratory (BNL) in the United States, with its key experimental collaborations (e.g. STAR). Source : BNL .}
   \label{fig: rhic}
\end{figure}

However, with the rapid progress in accelerator and collider technology, soon there were dedicated heavy-ion colliders such as the Relativistic Heavy Ion Collider (RHIC) (Fig.~\ref{fig: rhic}) at BNL which started colliding heavy-ions since 2000 and the Large Hadron Collider (LHC) at CERN colliding heavy-ions and hadrons since 2010.   
\begin{figure}[ht!]
    \includegraphics[height=6 cm, width= 10 cm]{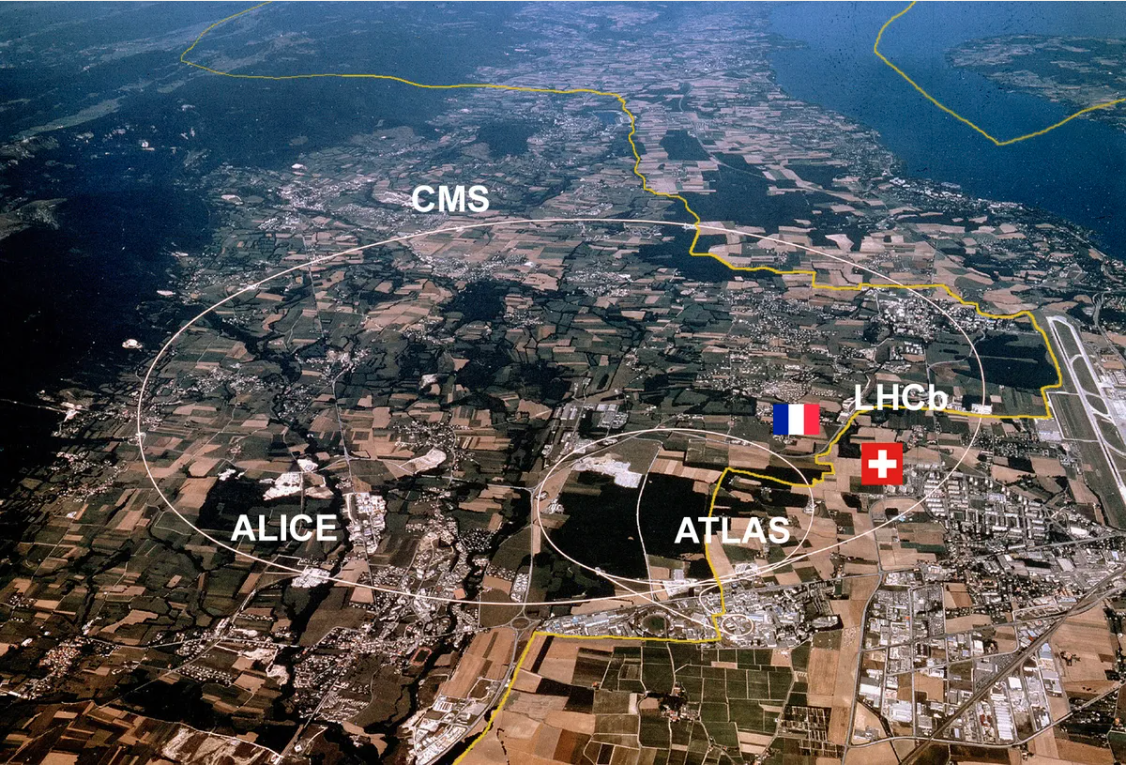}  
    \centering
    \caption{The aerial view of the Large Hadron Collider (LHC), located at CERN near the France-Switzerland border, along with its key experimental collaborations: ALICE, ATLAS CMS and LHCb. Source : CERN .}
   \label{fig: lhc}
\end{figure}
At RHIC, located at BNL in the United States, gold (Au) ions are collided at up to 200 GeV center of mass energy per nucleon pair creating the QGP medium~\cite{Gyulassy:2004zy}. On the other hand heavy-ion collision entered its first TeV collision energy scale at the LHC (Fig.~\ref{fig: lhc}), located at CERN near France-Switzerland border, where two lead (Pb) ions are collided at a center of mass energy 2.76 TeV or 5.02 TeV, reinforcing the extreme conditions necessary for the QGP formation~\cite{Muller:2012zq}. 

These collider facilities utilize sophisticated detectors to capture and analyze the plethora of particles produced in each collision. In particular, the key detectors or experimental collaborations at RHIC are STAR, PHENIX and very recently sPHENIX. At the LHC, dedicated measurements on heavy-ion collision are performed by the ALICE collaboration in addition to the ATLAS and CMS collaborations. These detectors use modern technology and intricate methods to track thousands of particles created in a collision providing detailed information about their momenta, energies and interaction patterns.   

\section{Evolution of the QGP : different stages of HI collision} 
\label{QGP evolution and stages of HI collision}
The QGP medium created in a heavy-ion collision lasts for a very short time $\sim$ 10 fm/c. However, within such a short time the QGP fireball evolves and leaves some of its most exclusive signatures on the dynamics of the final state particles. The QGP medium has a very small\footnote{The shear viscosity to entropy ratio $\eta/s\sim 0.1$} viscosity, making it the most perfect fluid and its evolution can be effectively described by relativistic viscous hydrodynamics, where the macroscopic physics can be applied at the femtoscale~\cite{Huovinen:2003fa,Kolb:2003dz,Huovinen:2006jp,Ollitrault:2007du,Dusling:2007gi,Hirano:2008hy,Luzum:2008cw,Heinz:2009xj,Jaiswal:2016hex,Romatschke:2017ejr}.   

Specifically, an event of a heavy-ion collision, from the time of collision until the detection of particles, passes through a number of successive stages~\cite{Jaiswal:2016hex,Moreland:2018gsh,JETSCAPE:2020mzn}. At the time of the collision, the colliding nuclei deposit energy or entropy at the overlap region, which serves as the {\it initial state} or {initial condition} of the collision. Then there exist a very short {\it pre-equilibrium phase} before the created fireball achieves {\it local  thermal equilibrium }.  After that, the thermalized QGP medium undergoes a relativistic viscous hydrodynamic evolution, where it {\it collectively expands} until it cools down to a certain temperature ($T_c$), where phase-transition occurs and the quarks and gluons again confine into hadroinc bound states. This is called the {\it QCD phase transition}~\cite{Stephanov:1998dy,Fodor:2004nz,Aoki:2006we,Stephanov:2008qz,Bazavov:2011nk,Borsanyi:2020fev} where a {\it smooth cross-over transition} from the QGP phase to {Hadron Resonance Gas} (HRG) phase occurs. Then the hadrons undergo elastic and in elastic collisions until a certain state is reached when such processes cease to occur, called the {\it freezeout}. After freezeout the hadrons stream towards the detectors. 
%where they might undergo secondary interactions during their flight before hitting the detectors. 

It should be noted that relativistic hydrodynamics is one of the theoretical approaches to study the QGP medium. The properties and dynamics of the fireball created in heavy-ion collision can be studied in the light of other theoretical frameworks~\cite{Gelis:2021zmx} such as relativistic kinetic theory approach or a transport model with string melting~\cite{Zhang:1999bd,Lin:2004en}. All these theories can reasonably describe the properties of the final state particles across different momentum range. However, in this thesis we will be limited to the soft processes at low {\it transverse momentum}, where relativistic hydrodynamical picture is more applicable. High transverse momentum particles are useful for studying the {\it jet physics}\cite{Mehtar-Tani:2013pia,Blaizot:2015lma}.    

\section{Collective flow and fluctuations}
\label{collectivity and fluctuations}
One of the most interesting and unique features of ultrarelativistic heavy-ion collisions is the {\it collective flow} of the final state particles~\cite{Ollitrault:1992bk,Voloshin:1994mz,Sorge:1996pc,Poskanzer:1998yz,Teaney:2000cw,Bleicher:2000sx,Huovinen:2001cy,Bozek:2004dt,Alver:2010gr,Schenke:2010rr,Bozek:2010bi,Luzum:2010sp,PHENIX:2002hqx,STAR:2004jwm,ALICE:2010suc,ATLAS:2012at,CMS:2012zex}. This collective dynamics originate as an effect of the initial state of the collision and of the collective expansion of the fireball. The most notable is the {\it anisotropic flow}~\cite{Ollitrault:1992bk,Gardim:2012yp,Ollitrault:2023wjk,ALICE:2016ccg,ATLAS:2018ezv} which originates due to the spatial anisotropy of the density distribution at the initial state of the collision. This spatial azimuthal anistoropy of the entropy or energy density distribution, through the evolution and collective expansion, translates into the azimuthal anisotropy of the momentum distribution of final state particles. This azimuthal anisotropy can be understood through different order of flow harmonics ($v_n$) such as the {\it elliptic flow}, {\it triangular flow} etc~\cite{Schenke:2011bn,Qiu:2011hf,STAR:2000ekf,PHENIX:2003qra,ALICE:2010suc}. 
\begin{figure}[ht!]
    \includegraphics[height=10 cm]{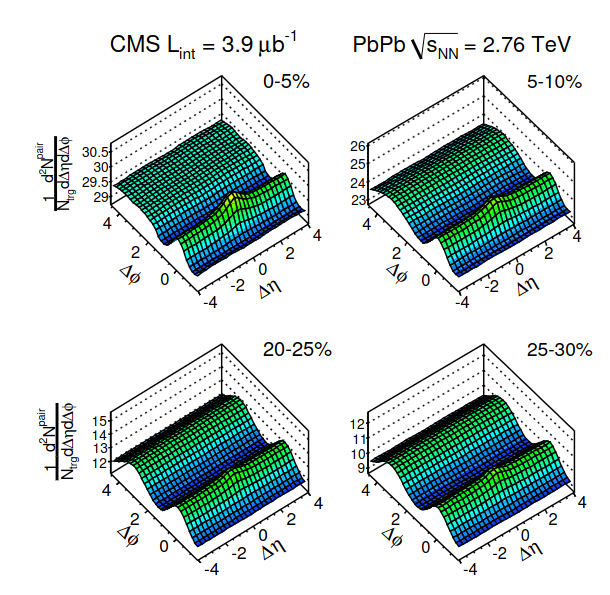}  
    \centering
    \caption{Collective flow in ultrarelativistic heavy-ion collision seen from the {\it ridge}-like structures of the pair distribution of particles on the ($\eta,\phi$) plane for different collision centrality as measured by the CMS collaboration. Figure taken from~\cite{CMS:2012xss}.}
   \label{fig: collective flow cms}
\end{figure}

The anisotropic flow depends on the {\it centrality} of the collision, defined by the transverse distance between the centers of colliding nuclei or the impact parameter ($b$) of the collision~\cite{Broniowski:2001ei,Das:2017ned} and determined in experiments through charged particle multiplicity $N_{ch}$ or other centrality estimators~\cite{PHOBOS:2000wxz, STAR:2001eyo,PHENIX:2004vdg,ALICE:2013hur,ATLAS:2011ah,CMS:2011aqh}. As $b$ become smaller, the collision is more central. The experimental evidence of the anispotropic flow, which caries an evidence of the formation of the tiny droplet of fluid, can be understood from the ridge like structure of the particle-pair distribution $d^2N^{pair}/d\eta d\phi$ on the ($\eta$,$\phi$) plane, as measured by the CMS collaboration and shown in Fig.~\ref{fig: collective flow cms}, where $\phi$ is the azimuthal angle on the transverse plane and $\eta$ is the {\it pseudorapidity} of the particles. Moreover, the collective nature of the particle spectra can be realized through the {\it mean transverse momentum per particle}, which can be calculated from the transverse momentum ($p_T$) distribution of the particles in each event. These mean transverse momentum and harmonic flow coefficients constitute the basic collective observables in a heavy-ion collision event. Collective flow can be used to constrain the medium properties of the QGP such as the effective temperature, the viscosity and to gain broad insights of the dynamics of its evolution.     

Another most peculiar characteristics of heavy-ion collisions are {\it event-by-event} fluctuations of collective flow~\cite{Aguiar:2001ac,Ollitrault:2009ie,Qiu:2011iv,Bzdak:2012tp,Heinz:2013bua,Jia:2014vja,Bhalerao:2014mua,Bozek:2017qir,PHOBOS:2010ekr,CMS:2015xmx,ALICE:2016kpq,ALICE:2017lyf,ATLAS:2017rij,ALICE:2022dtx}, which originate due to event-by-event fluctuations of the initial state~\cite{Broniowski:2007ft,Alver:2010gr,Qin:2010pf,Teaney:2010vd,Holopainen:2010gz,Gardim:2011xv,Bhalerao:2011yg,Schenke:2012wb} as well as from dynamical fluctuations during evolution. These fluctuations lead to many interesting effects which could be studied by constructing suitable observables and can carry observable signature of fundamental properties of the QGP medium such as {\it thermalization} at the initial stage. 

Additionally, one can construct correlation coefficients between the collective observables which contain important information and can be used to probe the initial state of collision providing useful constraints on the parameters~\cite{Bozek:2016yoj,Schenke:2020uqq, Giacalone:2020dln,Giacalone:2021clp,Mordasini:2019hut,ATLAS:2019pvn,ALICE:2021gxt,ATLAS:2022dov}. Moreover, such correlation coefficients can be used to study nuclear structure and deformation in high energy nuclear collisions by colliding nuclei with different shapes and sizes (e.g. U+U, Xe+Xe, Ru+Ru and  Zr+Zr etc.)~\cite{Filip:2009zz,Giacalone:2019pca,Li:2019kkh,Giacalone:2020awm,Giacalone:2021udy,Jia:2021wbq,Jia:2021tzt,Bally:2021qys,Jia:2021qyu,Xu:2021uar,Zhang:2021kxj,Jia:2021oyt,Jia:2022qgl,Jia:2022qrq,Bally:2023dxi,Ryssens:2023fkv,STAR:2015mki,ALICE:2018lao,CMS:2019cyz,ATLAS:2019dct,ATLAS:2020sgl,ALICE:2021gxt,ATLAS:2022dov,STAR:2024eky}. Such studies impose robust constraints on nuclear deformation parameters.

The main goal of this thesis is to look into different aspects of these event-by-event fluctuations of harmonic flow and mean transverse momentum, to study the correlations between the collective observables, proposing new constructions that can be measured in experiments and can potentially impose new robust constraints on the initial state, and to study nuclear structure (deformation) through similar observables by colliding deformed nuclei. Such studies will be helpful to broaden our knowledge  about the initial state, dynamics and properties of the QGP medium and unravel nuclear structure at the ultrarelativistic energy scale. 

\section{Outline of the thesis}  
\label{thesis outline}

This thesis represents the culmination of the research conducted during my PhD study. In particular, the thesis is based on the following publications :

\begin{itemize}
    \item Piotr Bozek, Rupam Samanta, `{\it Higher order cumulants of transverse momentum and harmonic flow in relativistic heavy ion collisions}', PRC 104 (2021) 1, 014905. \href{https://arxiv.org/pdf/2103.15338}{arXiv: 2103.15338 [nucl-th]}.  
    \item Piotr Bozek, Rupam Samanta, `{\it Factorization breaking for higher moments of harmonic flow}', PRC 105 (2022) 3, 034904. \href{https://arxiv.org/pdf/2109.07781}{arXiv: 2109.07781 [nucl-th]}. 
    \item Rupam Samanta, Piotr Bozek, `{\it Momentum-dependent flow correlations in deformed nuclei at collision energies available at the BNL Relativistic Heavy Ion Collider}', PRC 107 (2023) 3, 054916. \href{https://arxiv.org/pdf/2301.10659}{arXiv: 2301.10659 [nucl-th]}.
    \item Rupam Samanta, Joao Paulo Picchetti, Matthew Luzum, Jean-Yves Ollitrault, `{\it Non-Gaussian transverse momentum fluctuations from impact parameter fluctuations}', PRC 108 (2023)  2, 024908. \href{https://arxiv.org/pdf/2306.09294}{arXiv: 2306.09294 [nucl-th]}.
    
    \item Rupam Samanta, Somadutta Bhatta, Jiangyong Jia, Matthew Luzum, Jean-Yves Ollitrault, `{\it Thermalization at the femtoscale seen in high-energy Pb+Pb collisions}', PRC 109 (2024) 5, L051902. \href{https://arxiv.org/pdf/2303.15323}{arXiv: 2303.15323 [nucl-th]}.

    \item Rupam Samanta, Pitor Bozek, `{\it  Momentum dependent measures of correlations between mean transverse momentum and harmonic flow in heavy ion collisions }', Phys.Rev.C 109 (2024) 6, 064910. \href{https://arxiv.org/pdf/2308.11565}{arXiv: 2308.11565 [nucl-th]}. 
\end{itemize}

The thesis aims to provide an overview of the ultrarelativistic heavy-ion collision and the physics of the Quark-Gluon-Plasma with a special focus on the collective dynamics of the final state particles realized through correlations and fluctuations of the collective observables. The thesis is structured as follows :

\begin{itemize}
    \item In Chapter-2, we provide an introductory overview of ultrarelativistic heavy-ion collision including its basic elements such as kinematics, collision geometry, collision centrality etc. We discuss Glauber modelling of nucleus-nucleus collision and its Monte-Carlo implementation. Next we briefly discuss the theory of relativistic ideal and viscous hydrodynamics which form the fundamental basis for the evolution of the QGP medium. We also briefly discuss Quantum Chromodynamics (QCD), the theory of strong interactions and how deconfinement occurs at the QGP state. Finally, we provide thorough discussions on the different stages of hydrodynamic framework of heavy-ion collisions, providing the details of the simulation set-up used in our analyses. Chapter-2 presents a short review on ultrarelativistic heavy-ion collision, based on published literature. 

    \item In Chapter-3, we discuss the most distinctive feature of the heavy-ion collision, the collective flow and its basic phenomenological properties. We provide an overview of the anisotropic flow, its origin, the theoretical and experimental methods of flow analysis. Next we discuss the most peculiar characteristics of the collective flow: its event-by-event fluctuations. We show how event-by-event fluctuations of harmonic flow can be probed by constructing factorization-breaking coefficients in different transverse momentum bins. We discuss how the experimental limitation due to low statistics can be removed while measuring such observables. We extend the study to momentum dependent mixed-flow correlations as a probe of non-linearity present in the system. Chapter-3 is partly a brief review on collective flow based on published literature and partly presents original results published in~\cite{Bozek:2021mov}. 

    \item In Chapter-4, we discuss the transverse momentum fluctuations in ultracentral Pb+Pb collisions. In the first part, we show how the sudden fall of the variance in ATLAS data can be explained by a simple model of correlated Gaussian distribution between multiplicity and mean transverse momentum per particle. We show separately different contributions to the variance in our model and highlight the remarkable effect of impact parameter fluctuations. We perform a model fit to the ATLAS data and based on our fit results we provide crucial physical argument that could be responsible behind such phenomena. We point out important features of the ATLAS measurements in terms of the effect of $p_T$-cut and different centrality estimator. In the second part, we study the non-Gaussian characteristics of $[p_T]$-fluctuation, namely the skewness and kurtosis. Based on a similar model, we present predictions for ultracentral behavior of those cumulants and compare with the existing experimental data. Unless otherwise stated, Chapter-4 presents results published in~\cite{Samanta:2023amp} and \cite{Samanta:2023kfk}.

    \item In Chapter-5, we discuss the correlation between mean transverse momentum $[p_T]$ and harmonic flow square $v_n^2$. We study the Pearson correlation coefficient $\rho([p_T],v_n^2)$ which can be used to probe correlation present in the initial state. We also present a linear predictor to map the final state to the initial state of the collision. We propose new higher order normalized and scaled symmetric cumulants which could potentially put useful additional constraints and measure genuine higher order correlations. In the second part of the chapter, we study correlation coefficient between $[p_T]$ and momentum dependent harmonic flow, which is independent of the specific $p_T$-cut and could be sensitive to particular properties of the initial state such as nucleon width $w$. We also propose experimentally feasible alternate definitions and direct measure of such correlations through normalized covariance. Unless otherwise specified, Chapter-5 presents results published in~\cite{Bozek:2021zim} and \cite{Samanta:2023rbn}.

    \item In Chapter-6, we discuss how nuclear deformation can be accessed through high energy nuclear collision experiments. In particular, we show how the deformation parameter $\beta$ of $^{238}$U nucleus can be constrained by studying similar correlations (e.g. $\rho([p_T],v_n^2)$ and symmetric cumulants) and the factorization-breaking coefficients probing fluctuations in central U+U collisions. Unless otherwise stated, Chapter-6 presents results published in~\cite{Bozek:2021zim} and \cite{Samanta:2023qem}.

    \item In Chapter-7, we summarize our main findings and draw conclusions based on our study. We briefly discuss scopes for further developments on the topics that we studied. We also present our plans with prospective research projects that we would be interested to pursue in future.  
\end{itemize}

%*******************************************************************************
%*********************************** Second Chapter *****************************
%*******************************************************************************

\chapter{ Ultrarelativistic heavy-ion collision}  %Title of the First Chapter

\ifpdf
    \graphicspath{{Chapter2/Figs/Raster/}{Chapter2/Figs/PDF/}{Chapter2/Figs/}}
\else
    \graphicspath{{Chapter2/Figs/Vector/}{Chapter1/Figs/}}
\fi

In ultrarelativistic heavy-ion collision experiments performed at the LHC and RHIC, two heavy nuclei collide at a speed close to the speed of light, producing the hot dense QGP matter at the point of collision. This QGP fireball expands and cools down in a very short time before producing thousands of particles, a fraction of which are detected at the detector~\cite{florkowski2010phenomenology,Heinz:2013th,Busza:2018rrf}. Therefore, in order to understand the properties and the evolution dynamics of the QGP medium, it is indispensable to discuss the kinematics of the collision, its different stages starting from the time of first hard collisions to the detection of the particles and the underlying theoretical frameworks that can be used to model the collision dynamics. In this chapter, we discuss these components of ultrarelativistic heavy-ion collisions with appropriate details wherever necessary. For the theoretical framework, we restrict ourselves to relativistic hydrodynamics~\cite{Huovinen:2003fa,Kolb:2003dz,Huovinen:2006jp,Ollitrault:2007du,Dusling:2007gi,Hirano:2008hy,Luzum:2008cw,Heinz:2009xj,Jaiswal:2016hex,Romatschke:2017ejr} with brief discussions on QCD and the deconfinement. Towards the end of this chapter, we briefly describe the simulation set-up used in our analysis for producing the results presented in the subsequent chapters.

\section{Kinematics and invariants} 
\label{kinematics}
\begin{figure}[ht!]
\includegraphics[height= 7 cm]{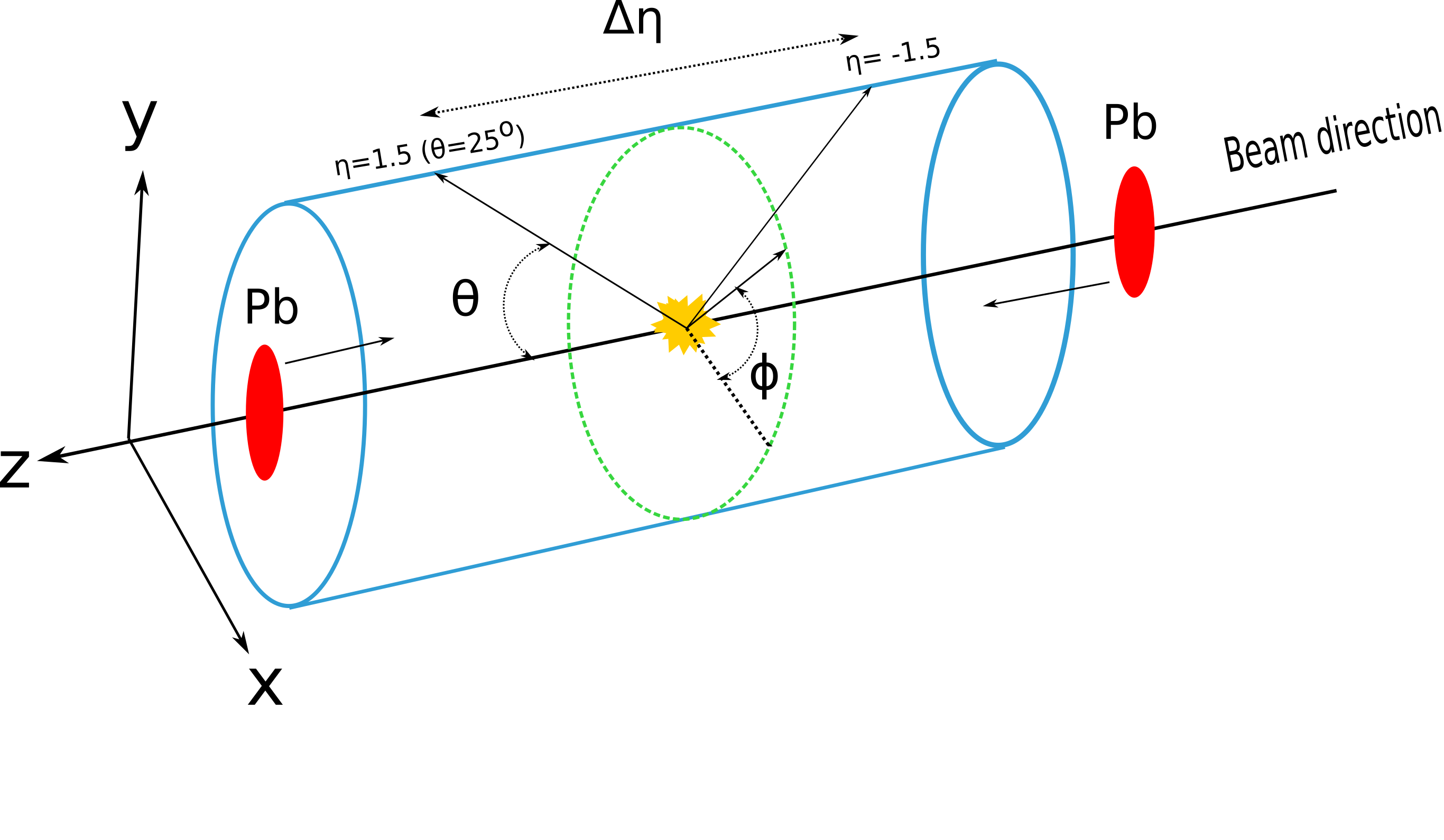}
\centering
\caption{Schematic representation of the geometry of ultrarelativistic heavy-ion collision (Pb+Pb) experiments. The collision axis is along the z-axis, and the detectors with a cylindrical form around this axis cover the full solid angle. The azimuthal angle $\phi$ is the angle on the transverse plane (x,y) perpendicular to the beam axis. The polar angle $\theta$ is associated with the pseudorapidity $\eta$, accounting for the longitudinal boost along the z-axis. The figure is motivated from~\cite{Giacalone:2020ymy}.}
\label{fig: colgeometry}
\end{figure}

The properties and dynamics of heavy-ion collisions are studied through experimental measurements of physical observables, which broadly depend on the kinematics of the measurements. Fig. \ref{fig: colgeometry} shows the geometry of the heavy-ion collision experiments. Two heavy nuclei (Pb in this case) move at a relativistic speed along the z-axis and collide at z=0. After each collision a large number of particles are produced, denoted by the black arrows, and hit the detector's wall. The plane (x, y), transverse to the direction of the two nuclei (beam axis), is called the {\it transverse plane}. In the laboratory frame, each particle carries momentum, and in the experiments particles are detected and identified by their momentum coordinates. 

In a Cartesian frame, the momentum space coordinates of a particle are denoted by the four vector, $p^\mu=\{p^0, p^i\}=\{ E, p_x, p_y, p_z\}$. The four-momentum $p^\mu$ is often represented in the context of collider experiments as, $p^\mu=\{ E, \vec{\bm p_T}, p_z\}$, where $\vec{\bm p_T}$ is called the {\it transverse momentum} of the particle and is defined as 
\begin{equation}
 \begin{aligned}
   \vec{\bm p_T}=\{ p_x, p_y \} \hskip 3em  \text{where} \hskip 3em p_T= \sqrt{p_x^2+p_y^2} \ .
  \end{aligned}
\label{eq: transverse momentum}
\end{equation}
The vector $\vec{\bm p_T}$ lies in the plane (x,y). In a collision event, particles are emitted in random directions on the transverse plane. The momentum conservation requires that the total transverse momentum $\sum \vec{\bm p_T} = 0$ in each event. More completely, one can write the
four momentum $p^\mu$ in terms of spherical coordinates as, $p^\mu=\{E, p sin \theta \cos \phi, p \sin \theta \sin \phi, p \cos \theta \}$ where $\phi$ is the azimuthal angle on the transverse plane with respect to the x axis and the angle $\theta$ is the polar angle on the (y,z) plane with respect to the z axis (Fig. \ref{fig: colgeometry}).

\begin{figure}[ht!]
\includegraphics[height= 9 cm]{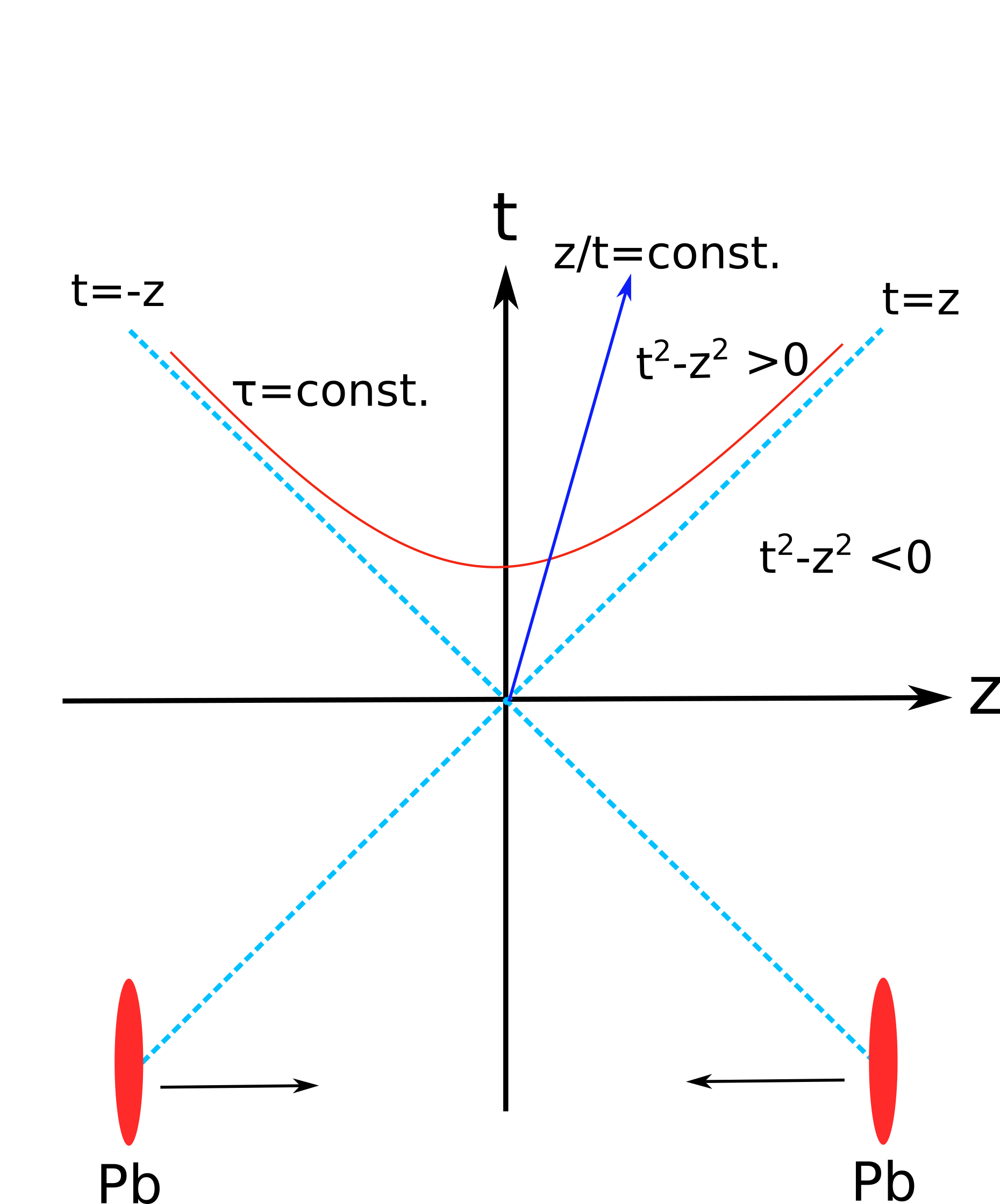}
\centering
\caption{Space-time representation of the collision in (t,z) plane. The vertical axis represents time and the horizontal axis denotes the beam direction. The lower and upper part of the plot represent before and after the collision respectively. The collision occurs at t=z=0. The diagonal lines t= $\pm$ z form the light cone. The region on the plot with $t^2 -z^2 > 0$ is called time-like region and $t^2 -z^2 < 0$ is called space-like region. Particle production occurs within the time-like region only. The figure is adapted from ~\cite{Ollitrault:2007du}}
\label{fig: t-z diagram}
\end{figure}

As the two nuclei move with an ultrarelativistic speed, in laboratory frame they appear like flat pancakes~\cite{Ollitrault:2023wjk}, when viewed vertically from x-axis. This occurs because the nuclei are Lorentz-contracted due to the boost along z-axis.  In general, the Lorentz transformation between the laboratory frame and the rest frame of the nuclei is given by, 
\begin{equation}
 \begin{aligned}
    X'^\mu = \Lambda^\mu_\nu \ X_\nu \ ,
  \end{aligned}
\label{eq: Lorentz transformation}
\end{equation}
where $X'^\mu = \{t, x, y, z \} = \{ t, \vec{x_T}, z \} $ is the position four vector, where $x_T=\sqrt{x^2+y^2}$ is the transverse distance on (x,y) plane. For Lorentz boost along the z-axis, 

\begin{equation}
 \begin{aligned}
    \Lambda^\mu_\nu  =  
    \begin{pmatrix}
     \gamma & 0 & 0 & -\beta \gamma \\
     0 & 1 & 0 & 0 \\
     0 & 0 & 1 & 0 \\
     -\beta \gamma & 0 & 0 & \gamma \\
    
    \end{pmatrix} \ ,
  \end{aligned}
\label{eq: Lorentz transformation matrix}
\end{equation}
 where $\gamma = \sqrt{1-\beta^2}$ is the Lorentz factor and $\beta = v$ is the velocity of the nuclei. [ Note: We use natural units, $\hslash=c=k_B=1$ and metric $g^{\mu \nu} = diag(1,-1,-1,-1)$ ]. Fig.~\ref{fig: t-z diagram} shows the space-time picture of a collision event in (t,z) plane.  The vertical and horizontal lines denote time and beam axis respectively. The lower part and the upper part of z-axis represent the scenarios before and after the collision respectively. The two nuclei collide at $t=z=0$. As this is a relativistic collision, it is more appropriate to use {\it proper time coordinates} ({\it Milne coordinates}) defined as ,
\begin{equation}
 \begin{aligned}
    \text{proper time,} \ \tau &= \sqrt{t^2-z^2} \\
    \text{ and space-time rapidity,} \ \eta_s &= \frac{1}{2} \ \text{ln} \ \frac{t+z}{t-z}\ .
  \end{aligned}
\label{eq: proper coordinates}
\end{equation}
The two diagonal lines defined as $t=\pm z$ define the {\it light-cone}; along these lines $t^2-z^2=0 ( \tau = 0) $. The region within the light-cone with $t^2-z^2 > 0 \ (\tau>0)$ is called {\it time-like} region and region outside with  $t^2-z^2 < 0 \ (\tau<0)$ is called {\it space-like} region. The space-time rapidity $\eta_s$ which spans from $-\infty$ to $+\infty$ is properly defined only in the time-like region and the particle production occurs on the upper-half part within this region. 

Next we introduce two important variables in heavy-ion collision, namely {\it rapidity} and {\it pseudorapidity}. The rapidity of a particle is defined as , 

\begin{equation}
 \begin{aligned}
    y = \frac{1}{2} \ \text{ln} \ \frac{E+p_z}{E-p_z}   \ ,
  \end{aligned}
\label{eq: rapidity}
\end{equation}
which is expressed in terms of longitudinal velocity of the particle, $\beta_z=p_z/E$ as, 
\begin{equation}
 \begin{aligned}
    y = \frac{1}{2} \ \text{ln} \ \frac{1+p_z/E}{1-p_z/E} = \tanh^{-1} \big(\frac{p_z}{E}\big) = \tanh^{-1} (\beta_z)  \ .
  \end{aligned}
\label{eq: rapidity-longitudinal velocity}
\end{equation}
An advantage of using rapidity over longitudinal velocity is that rapidity is additive under longitudinal boosts, therefore it provides a measure of Lorentz boost along z-axis. Rapidity can be understood as the relativistic analog of non-relativistic ($\beta_z<<1$) velocity. From Eq.~(\ref{eq: rapidity-longitudinal velocity}), 
\begin{equation}
 \begin{aligned}
    y = \frac{1}{2} \ \text{ln} \ \frac{1+\beta_z}{1-\beta_z} = \frac{1}{2} [ \text{ln}(1+\beta_z) - \text{ln}(1-\beta_z) ]  \simeq \beta_z \equiv v_z \ .
  \end{aligned}
\label{eq: rapidity-velocity analog}
\end{equation}
Similar to transverse momentum, one can also define the {\it transverse mass} of a particle, 
\begin{equation}
 \begin{aligned}
   m_T^2 = m^2 + p_T^2 = E^2 -p_z^2 \ ,
  \end{aligned}
\label{eq: transverse mass}
\end{equation}
which is invariant under the Lorentz boost along the z direction. $E$ and $p_z$ can be expressed in terms of rapidity and transverse mass, 
\begin{equation}
 \begin{aligned}
   E = m_T \cosh y \hskip 5 mm \text{and} \hskip 5mm p_z= m_T \sinh y \ .
  \end{aligned}
\label{eq: energy and long momentum in terms of trans mass}
\end{equation}
Therefore, the four momentum of a particle can be written as, $p^\mu=\{m_T\cosh y, p_x, p_y, m_T \sinh y \}$. On the other hand pseudorapidity accounts for the polar angle $\theta$ of the particle and is defined as, 

\begin{equation}
 \begin{aligned}
    \eta = - \text{ln} \tan (\theta/2) \ ,
  \end{aligned}
\label{eq: pseudorapidity}
\end{equation}

which is also expressed as , 

\begin{equation}
 \begin{aligned}
    \eta =  \frac{1}{2} \text{ln} \ \frac{| \bm p | + p_z}{| \bm p |-p_z} = \tanh^{-1} \big(\frac{p_z}{| \bm p |}\big) \ .
  \end{aligned}
\label{eq: pseudorapidity alternate}
\end{equation}

At ultrarelativistic energy, $p\gg m$ and the rapidity,

\begin{equation}
 \begin{aligned}
   y= \frac{1}{2}\text{ln} \ \frac{\sqrt{p^2+m^2} + p_z}{\sqrt{p^2+m^2} - p_z} \simeq  \frac{1}{2}\text{ln} \ \frac{p + p \cos \theta}{p - p \cos \theta} = -\text{ln} \tan \theta/2 = \eta \ .
  \end{aligned}
\label{eq: rapidity equals pseudorapidity}
\end{equation}
Therefore, for very high energy ultrarelativistic particles, rapidity is equal to pseudorapidity. Pseudorapidity is more conveniently used by experimentalists as it can be used for the unidentified particles (charged particles) because $\eta$ does not depend on the mass of the particles whereas $y$ does. 

The momentum distribution of particles is expressed in terms of particle spectrum,
\begin{equation}
 \begin{aligned}
   E\frac{d^3N}{d^3p} = E\frac{d^3N}{dp_zd^2p_T} =  \frac{d^3N}{dyd^2p_T} \ ,
  \end{aligned}
\label{eq: momentum distribution}
\end{equation}
which is Lorentz invariant because the element $d^3p/E$ (or $dp_z/E$) is Lorentz invariant. The transformation between rapidity and pseudorapidity distribution is given by, 
\begin{equation}
 \begin{aligned}
   \frac{d^3N}{d\eta d^2p_T} =\sqrt{1-\frac{m^2}{m_T^2 \cosh^2y}} \ \frac{d^3N}{dy d^2p_T}  \ .
 \end{aligned}
\label{eq: rapidity dist- pseudorapidity dist}
\end{equation}

\section{Collision geometry} 
\label{collision geometry}
In a heavy-ion collision, the geometry of the overlap region is important to determine the volume of the QGP medium and its evolution time. In particular, the initial state of the collision depends on the collision geometry, which in turn affect the final state observables. The main features of the collision geometry are described by the following quantities :

\subsection{Nuclear density distribution }
\label{nuclear distribution}
First, we need to know the density distribution of each of the colliding nuclei. One considers two kinds of density distribution: nuclear charge density distribution and nuclear matter (mass) density distribution. The charge density distribution is related to the distribution of protons, which is obtained from the electron scattering experiments. The matter density distribution considers both neutrons and protons.
\begin{figure}[ht!]
\includegraphics[height=7 cm]{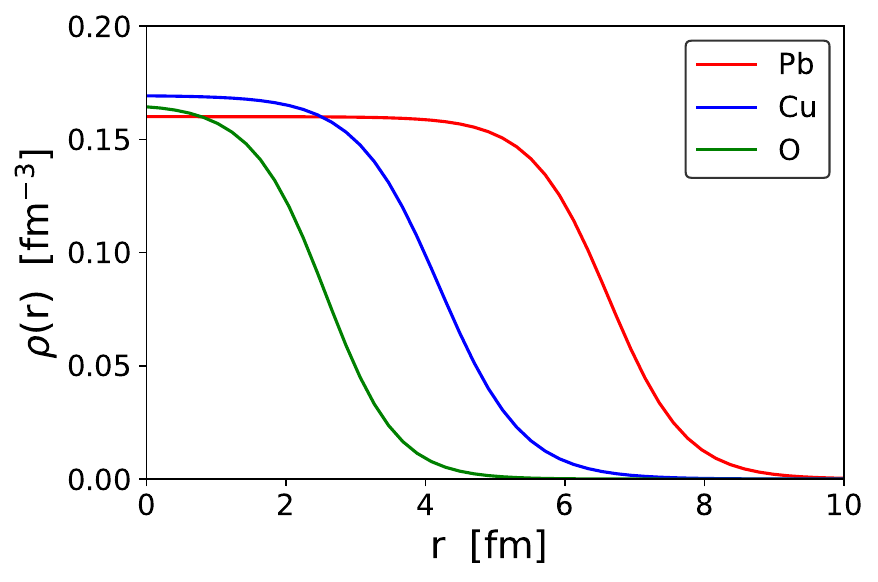}
\centering
\caption{ Woods-Saxon density distribution for lead (red) , copper (blue) and oxygen (green) nucleus.}
\label{fig: wood-saxon}
\end{figure}
It is usually a good assumption that the nuclear matter distribution is proportional to the charge distribution~\cite{Krane:1987ky} . The nuclear density distribution $\rho(r)$ at a distance r from the center, is often represented by a three parameter Fermi-distribution function~\cite{Alver:2008aq,Loizides:2014vua}: 
\begin{equation}
 \begin{aligned}
   \rho(r)=\rho_0\frac{1+w(\frac{r}{R})^2}{1+\exp(\frac{r-R}{a})}  \ , 
 \end{aligned}
\label{eq: wood-saxon general}
\end{equation}
which is known as {\it Woods-Saxon parametrization} of the nuclear density distribution. $\rho_0$ is the nucleon density, $R$ is the nuclear radius, $a$ is the {\it skin-depth} or nuclear diffusivity and the term with $w$ is a correction for small nuclei. The overall normalization factor $\rho_0$ can be obtained from
\begin{equation}
 \begin{aligned}
   \int d^3 r \ \rho(r) = A  \ ,
 \end{aligned}
\label{eq: wood-saxon rho0}
\end{equation}
where $A$ is the total number of nucleons. The values of the other parameters for few nuclei are listed in Table \ref{tab: Wood-Saxon params}. Fig. \ref{fig: wood-saxon} shows the Woods-Saxon nuclear density distribution for different nuclei. 
\begin{table}
\centering
\begin{tabular}{ c c c c }
   \toprule 
   Nucleus  & R (fm)  &  a (fm) & w \\ 
    \midrule
    $^{197}$Au & 6.38  & 0.535  & 0  \\
     
    $^{208}$Pb & 6.624 & 0.549  & 0  \\ 
    
    $^{16}$O & 2.608  & 0.513  & -0.051  \\
     
    $^{63}$Cu & 4.2 & 0.596  & 0  \\
     \bottomrule
\end{tabular}
\vskip 2mm
\caption{Values of Woods-Saxon parameters for nuclear density distribution for different nuclei~\cite{Alver:2008aq,Loizides:2014vua} }
\label{tab: Wood-Saxon params}
\end{table}

\subsection{Impact parameter} 
\label{impact param}
In an untrarelativistic nucleus-nucleus collision, impact parameter, $b$ is a crucial quantity and it is defined as the spatial distance between the centers of the two nuclei at the time of collision ($t=0$). The transverse shape and the size of the QGP medium is largely determined by the collision impact parameter. Impact parameter is a semi-classical quantity and cannot be measured experimentally. In a collision experiment, the direction of the impact parameter is random. However, for theoretical calculation  $\vec{\bf b}$ (which is a two dimensional vector) is commonly represented along the x axis. The plane formed by the impact parameter and z-axis is called the {\it reaction plane}, so that the (x,z) plane is the reaction plane~\cite{Ollitrault:1993ba,Voloshin:2008dg}. In theoretical simulations, impact parameter is generated with a probability distribution $d\sigma/db$ which is proportional to $b$ for $b\leq 2R$, where $R$ is the radius of the nuclei, in a symmetric collision. 
\begin{figure}[ht!]
\includegraphics[height=7 cm]{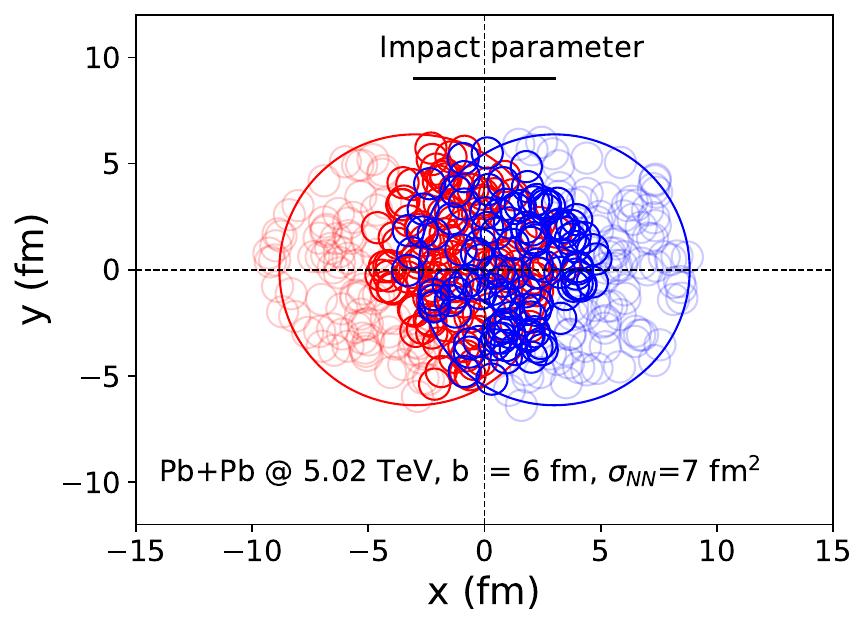}
\centering
\caption{ Schematic representation of Pb+Pb collision on the transverse plane at $\sqrt{s_{NN}}=5.02$ TeV corresponding to the nucleon-nucleon inelastic cross section, $\sigma_{NN}=7.2$ fm$^2$. The impact parameter of the collision is $b=6$ fm, shown by the black horizontal line on the top. The nucleons from each parent nuclei are shown by corresponding colored (red and blue) circles. The clear solid circles denote the participant nucleons and the blurry circles represent the spectator nucleons. The figure is motivated from~\cite{Alver:2008aq} and prepared using a MC-Glauber calculation.}
\label{fig: impact parameter}
\end{figure}
Fig. \ref{fig: impact parameter} shows the projection of Pb+Pb collision on the transverse (x,y) plane , where the center of mass energy of the collision is, $\sqrt{s_{NN}}$ = 5.02 TeV and impact parameter, $b=6$ fm. In the overlap region of the two nuclei, the nucleons from each nuclei interact. The nucleons which encounter at least one interaction with the nucleon from other nucleus, are called the {\it participant nucleons} or {\it wounded nucleons}~\cite{Bialas:1976ed} and each of such nucleon-nucleon interactions are labeled as {\it binary collision}. The rest of the nucleons which are not involved in any interactions and just pass by are called {\it spectator nucleons}.

\subsection{Centrality of the collision}
\label{centrality}
The medium produced in a nucleus-nucleus collision is expected to be a QGP medium and this is more likely if the interaction region during the collision is large. Thus if the collision is closer to {\it head-on} or in other word more {\it central}, the overlap area is larger and there is a higher chance of creating a QGP medium. On the other hand, if the collision is more glancing or {\it peripheral}, it is less likely that the system will achieve the condition for the QGP formation. 

Theoretically, the centrality of a collision is defined according to the impact parameter $b$. However, as a direct measurement of $b$ is not possible, in experiments the centrality of a collision can be estimated according to the total number of charged particles produced in the collision, which is known as {\it charged particle multiplicity} $N_{ch}$~\cite{PHOBOS:2000wxz, STAR:2001eyo,PHENIX:2004vdg,ALICE:2013hur} or the transverse energy deposition $E_T$ in the forward calorimeter~\cite{ATLAS:2011ah,CMS:2011aqh}. The relation between theory and experiment can be understood as the following: if the impact parameter is small (central collision), the volume of the QGP medium is large and the number of participant nucleons is large. On the other hand, $N_{ch}$ and $E_T$ both are increasing with the number of participant nucleons $N_{part}(b)$~\cite{Bialas:1976ed} (and also the number of binary collisions $N_{coll}(b)$~\cite{Kharzeev:2000ph}), which depends on the collision impact parameter~\cite{Broniowski:2001ei}. The opposite occurs if the impact parameter is large (peripheral). The dependence of $N_{part}$ (or $N_{coll}$) on $b$ can be calculated with great precision based on some theoretical models such as the two component {\it Glauber model}~\cite{Miller:2007ri}. Thus, although $b$ cannot be directly measured, it can be estimated from the experiments based on the mapping described above. 
\begin{figure}[ht!]
\includegraphics[height=11 cm]{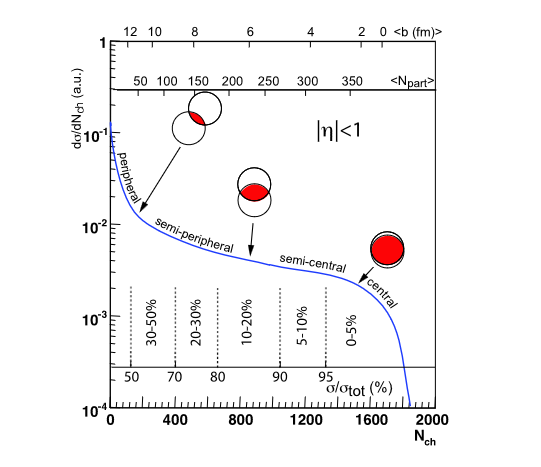}
\centering
\caption{Final state multiplicity distribution ($N_{ch}$) and centrality classification in experiments. The cartoons of the nucleus-nucleus collision represent corresponding impact parameter in a given centrality class. Figure is taken from~\cite{Miller:2007ri} }
\label{fig: multiplicity-centrality}
\end{figure}
The centrality classified according to the impact parameter is denoted by $c_b$ (sometimes called {\it b-centrality}) and given by~\cite{Broniowski:2001ei,Das:2017ned}, 
\begin{equation}
 \begin{aligned}
    c_b=\frac{1}{\sigma_{in}^{AB}} \int_0^b P_{in}(b') 2 \pi b' db' \ ,
 \end{aligned}
\label{eq: b-centrality}
\end{equation}
where $\sigma_{in}^{AB}$ is the inelastic nucleus-nucleus cross-section and $P_{in}(b)$ represents the probability that the two nuclei A and B collide at impact parameter $b$. The probability distribution of $c_b$ is uniform, $P(c_b)=1$ for $0<c_b<1$. On the other hand, in heavy-ion experiments, the centrality is defined as a cumulative distribution of $N_{ch}$ or $E_T$,
\begin{equation}
 \begin{aligned}
    c =\int_n^\infty P(n')dn' \hspace{1 cm} (n\equiv N_{ch} \ \text{or} \ E_T) \ .
 \end{aligned}
\label{eq: exp centrality}
\end{equation}
Again, the probability distribution of $c$ is uniform, $P(c)=1$ in $0<c<1$. Fig.~\ref{fig: multiplicity-centrality} shows the experimental classification of centrality based on the charged particle multiplicity ($N_{ch}$) and its corresponding mapping to collision impact parameter.

\section{Glauber modeling in nucleus-nucleus collision} 
\label{glauber model}
The Glauber model~\cite{Miller:2007ri,Glauber:1955qq,Glauber:lecture} serves as a fine theoretical tool to study high-energy nucleus-nucleus collision and it can provide an indirect mapping between the theoretical and experimental classification of collision centrality. In particular, with the help of Glauber model, one can calculate the number of participants ($N_{part}(b)$) and binary collisions ($N_{coll}(b)$) as a function of $b$ in a nucleus-nucleus collision. The model calculation can be based on a classical view of the quantum mechanical framework involving full multiple scattering integrals ~\cite{Bialas:1976ed,Bialas:1977pd,Glauber:2006gd}, known as the {\it optical limit} of Glauber model. However, it can be also calculated based on numerical Monte Carlo simulation~\cite{Wang:1991hta,Broniowski:2007nz}, which is known as {\it Monte Carlo Glauber or MC-Glauber} approach.    

\subsection{Optical Glauber model}
\label{optical Glauber}
The Galuber Model treats the nucleus-nucleus collision in terms of independent interactions of the constituent nucleons~\cite{Miller:2007ri} and assumes that at sufficiently high energies, the nucleons have momentum large enough that their trajectories are essentially undeflected, while travelling on a straight line independent of other nucleons. The individual nucleon-nucleon cross section is obtained from the phase shift using the {\it optical theorem}~\cite{Sakurai:1167961,Newton:1976} and the overall phase shift of the incoming wave associated to each nuclei is assumed to be the sum over all possible two-nucleon phase shifts. The model also assumes that the size of the nucleus is larger than the range of nucleon-nucleon interaction. Such assumptions make possible a simple calculation of the nucleus-nucleus cross section, or of the number of participant nucleons and binary collisions in terms of the nucleon-nucleon cross section.

\begin{figure}[ht!]
\begin{subfigure}{0.6\textwidth}
\centering
\includegraphics[height=6 cm]{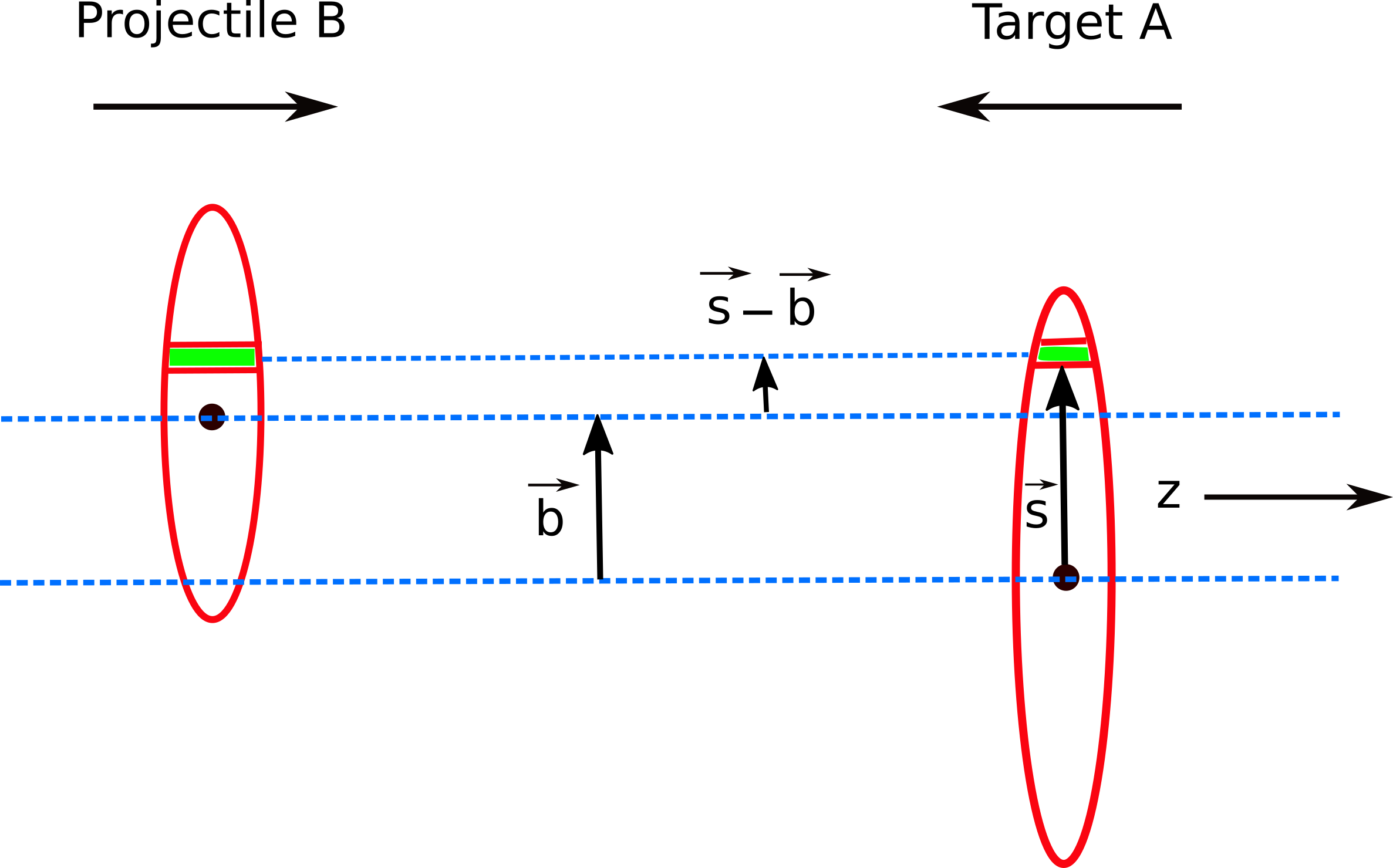}
\caption{Side view}
\centering
\end{subfigure}~
\begin{subfigure}{0.4\textwidth}
\centering
\includegraphics[height=5 cm]{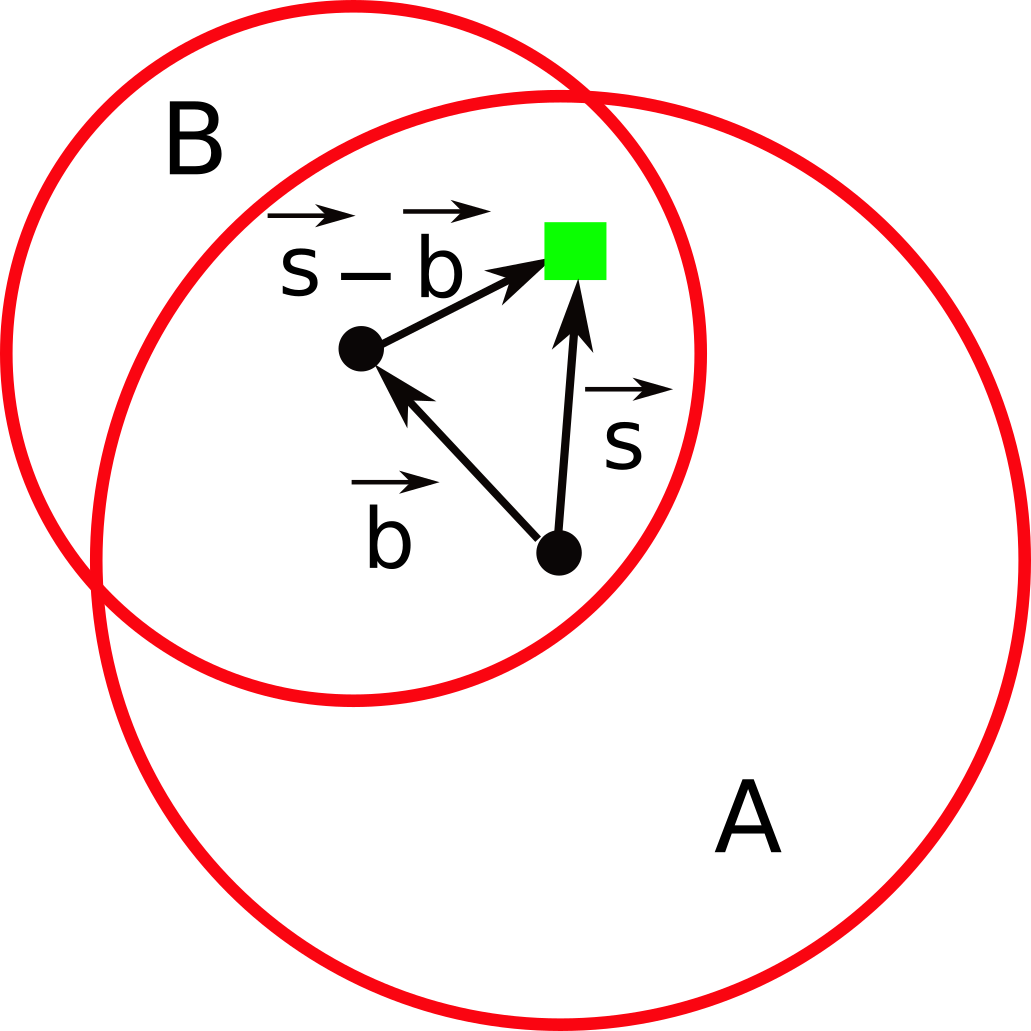}  
\caption{Beam-line view}
\end{subfigure}
\centering
\caption{Schematic representation of the geometry of optical Glauber model with longitudinal (b) and transverse (a) views. Two nuclei denoted by red curves collide along z axis at an impact parameter $\vec{\bf b}$ and the green colored region represents the overlap area during the collision. The figure is a modification from~\cite{Miller:2007ri}.}
\label{fig: optical glauber}
\end{figure}

Fig.~\ref{fig: optical glauber} shows the collision of two heavy nuclei A (say {\it projectile}) and B ({\it target}) along the z axis, colliding with a relativistic speed at an impact parameter $\vec{\bf b}$. Let us consider a flux-tube located at a displacement $\vec{\bf s}$ from the centre of the target nucleus and at a distance  $\vec{\bf s}-\vec{\bf b}$ from the centre of projectile. The vector $\vec{\bf s}$ represents the transverse position vector $\vec{\bf s}(x,y)$. At the time of collision, these two tubes overlap on each other. The probability of a given nucleon of being located within the target flux tube per transverse area is given by the {\it thickness function}, 

\begin{equation}
 \begin{aligned}
    T_A(\vec{\bf s}) = \int \rho_A(\vec{\bf s},z_A) \ dz_A \ ,
 \end{aligned}
\label{eq: target thickness}
\end{equation}
where $\rho_A(\vec{\bf s},z)$ is the nuclear density distribution at $\vec{\bf r}=(x,y,z)$ given by Eq.~(\ref{eq: wood-saxon general}). Similarly, the probability of finding a nucleon in the projectile flux tube per transverse area, 
\begin{equation}
 \begin{aligned}
    T_B(\vec{\bf s}-\vec{\bf b}) = \int \rho_B(\vec{\bf s}-\vec{\bf b},z_B) \ \ dz_B \ .
 \end{aligned}
\label{eq: projectile thickness}
\end{equation}
The product of the two, $T_A(\vec{\bf s})T_B(\vec{\bf s}-\vec{\bf b})$ represents the joint probability density per unit area of finding a nucleon from A and a nucleon from B at the overlap region with the collision impact parameter $\vec{\bf b}$. One can then define the {\it overlap function} $T(\vec{\bf b})$ by integrating over all $\vec{\bf s}$,
\begin{equation}
 \begin{aligned}
    T(\vec{\bf b}) = \int T_A(\vec{\bf s}) \ T_B(\vec{\bf s}-\vec{\bf b})\ \ d^2s \ .
 \end{aligned}
\label{eq:  thickness fucntion}
\end{equation}

Please note, $T(\vec{\bf b})$ has the unit of inverse area. The thickness function represents the effective overlap area between a particular nucleon from A and a given nucleon from B at the interaction region. The probability of such an interaction is given by $T(\vec{\bf b})  \sigma_{in}^{NN}$, where $\sigma_{in}^{NN}$ is the nucleon-nucleon inelastic cross section\footnote{Note that the elastic nucleon-nucleon collisions encounter very little energy loss and hence do not contribute to the Glauber model calculation}. The probability of $n$ such interactions between the nucleus A (let us say has $A$ nucleons) and the nucleus B (having $B$ nucleons) is given by the binomial distribution,
\begin{equation}
 \begin{aligned}
    P(n,\vec{\bf b}) = 
    \begin{pmatrix}
    AB\\
    n
    \end{pmatrix}
    [T_{AB}(\vec{\bf b})\sigma_{in}^{NN}]^n \ [1-T_{AB}(\vec{\bf b})\sigma_{in}^{NN}]^{AB-n} \ ,
 \end{aligned}
\label{eq: binomial-dist coll-prob}
\end{equation}
where in the above expression, the first term denotes the number of combinations in which $n$ nucleon-nucleon collisions out of $AB$ collisions occur, the second term represents the probability of exactly $n$ collisions, and the last term gives the probability that $AB-n$ collisions do not occur. The total probability of an interaction between nuclei A and B is,
\begin{equation}
 \begin{aligned}
    P_{in}^{AB}(\vec{\bf b})=\frac{d^2\sigma_{in}^{AB}}{db^2}=\sum_{n=1}^{AB} P(n,\vec{\bf b}) = 1-[1-T_{AB}(\vec{\bf b})\sigma_{in}^{NN}]^{AB} \ .
 \end{aligned}
\label{eq: total prob of AB collision}
\end{equation}
 The total inelastic cross section of A-B collision can be found by integrating the above equation over impact parameter\footnote{The scalar distance can be used instead of vector impact parameter if the colliding nuclei are not polarized.},

\begin{equation}
 \begin{aligned}
    \sigma_{in}^{AB}=\int_0^\infty [{1-[1-T_{AB}(b)\sigma_{in}^{NN}]^{AB}}] \ 2\pi b db \ .
 \end{aligned}
\label{eq: sigmaAB}
\end{equation}
The mean number of binary nucleon-nucleon collision is, 
\begin{equation}
 \begin{aligned}
    N_{coll}(b)=\sum_{n=1}^{AB} nP(n,b)=ABT_{AB}(b)\sigma_{in}^{NN}
 \end{aligned}
\label{eq: NcollAB}
\end{equation}
and the number of participants at impact parameter b is given by, 
\begin{equation}
 \begin{aligned}
    N_{part}(b)&=A \int T_A(\vec{\bf s}) P_{in}^{NB}(\vec{\bf s}-\vec{\bf b})\  d^2s + B \int T_B(\vec{\bf s}-\vec{\bf b}) P_{in}^{NA}(\vec{\bf s}) \ d^2s \\
     &= A\int T_A(\vec{\bf s}) \bigg\{ 1-[1-T_B(\vec{\bf s}-\vec{\bf b})\sigma_{in}^{NN}]^B \bigg\} \ d^2s\\
     &+ B\int T_B(\vec{\bf s}-\vec{\bf b}) \bigg\{ 1-[1-T_A(\vec{\bf s})\sigma_{in}^{NN}]^A \bigg\} \ d^2s \ ,
 \end{aligned}
\label{eq: NpartAB}
\end{equation}
where $P_{in}^{NA(B)}$ represents the probability of a nucleon-nucleus collision. Note that the total cross section of a nucleon-nucleus collision can be written as,
\begin{equation}
 \begin{aligned}
  \sigma_{in}^{NA(B)}=\int \bigg\{ 1-[1-T_{A(B)}(\vec{\bf s})\sigma_{in}^{NN}]^{A(B)} \bigg\} \ d^2s \ .
 \end{aligned}
\label{eq: sigmaNA}
\end{equation}

 As the Glauber model calculation of $\sigma_{in}^{AB}$, $N_{part}$ and $N_{coll}$ depends on the nucleon-nucleon cross section, it is important to mention the measurement of $\sigma_{in}^{NN}$ in different experiments and collision energies. From the energy dependence of $\sigma_{in}^{NN}$ on $\sqrt{s_{NN}}$ \cite{ParticleDataGroup:2006fqo}, it can be estimated that at the SPS energy ($\sqrt{s_{NN}}$ = 20 GeV), $\sigma_{in}^{NN}\simeq$ 3 fm$^2$, at the top RHIC energy ($\sqrt{s_{NN}} = $ 200 GeV), $\sigma_{in}^{NN}\simeq$ 4.2 fm$^2$~\cite{PHENIX:2015tbb} and at the LHC energies, $\sigma_{in}^{NN}\simeq$ 6.4 fm$^2$~\cite{ATLAS:2011ag} ($\sqrt{s_{NN}} $ = 2.76 TeV) and 7.0 fm$^2$~\cite{ALICE:2012xs} ($\sqrt{s_{NN}}$ = 5.02 TeV).

\subsection{Monte Carlo Glauber model}
\label{mc Glauber}
The optical Glauber model deals with the continuous distribution of nucleon density and does not really locate the nucleons at locations with specific spatial coordinates inside the nuclei. This is exactly the case in Monte Carlo approach of the Glauber model. The main feature of this method is that the quantities like $N_{part}$, $N_{coll}$ are calculated by colliding two nuclei event-by-event through computer simulation. Both models give very close results for the averaged quantities, $\langle N_{part}(b) \rangle$ and $\langle N_{coll}(b) \rangle$, where $\langle \dots \rangle$ denotes average over events. However, for the quantities where event-by-event fluctuations (as we will discuss later) are important, the two models give different results. Moreover, in the Monte Carlo Glauber approach, one can simulate the observables event-by-event (in a minimum bias way), which are measured in the experiment, such as multiplicity $N_{ch}$ and its distribution can be compared with the experimental distribution to classify similar centrality cuts as in the data. In the Glauber Monte Carlo approach \cite{Miller:2007ri}, first the nucleons are distributed in the two colliding nuclei A and B, in 3D coordinate space, according to the nucleon density distribution in Eq.~(\ref{eq: wood-saxon general}). Next, a random impact parameter is generated according to the impact parameter distribution: $d\sigma/db=2\pi b$  (where, $0 \leq b \leq R_max$, with $R_{max}>R_A+R_B$). Then, the collision between the two nuclei is assumed as a sequence of independent nucleon-nucleon collisions. It is assumed that the nucleons traverse through a straight line and in each collision, nucleon-nucleon cross section is independent of the collision history of the colliding nucleons. Next, in the simplest scenario, two nucleons are treated to have collided if they satisfy the following condition: 
\begin{equation}
 \begin{aligned}
   d_{NN} \leq \sqrt{\sigma_{in}^{NN}/\pi} \ ,
 \end{aligned}
\label{eq: hard-sphere wounding condition}
\end{equation}
where $d_{NN}$ is the transverse distance (on the plane perpendicular to beam-axis) between the two nucleons at the time of collision. This criterion is known as {\it hard-sphere wounding} and could be described for symmetric collisions, by the collision probability: 
\begin{equation}
 \begin{aligned}
   p(d_{NN}) = \Theta(R-d_{NN}) \ \ \eqsp{with} \ \ \sigma^{NN}_{in} = \pi R^2 \ .
 \end{aligned}
\label{eq: hard-sphere wounding probabilityalso}
\end{equation}
A {\it Gaussian-wounding} probability is also possible~\cite{Rybczynski:2011wv}(i.e. the nucleons are wounded with respect to a Gaussian probability distribution ): 
\begin{equation}
 \begin{aligned}
   p(d_{NN}) = C \ e^{\frac{-\pi \ C d_{NN}^2}{\sigma^{NN}_{in} }} \ ,
 \end{aligned}
\label{eq: Gaussian wounding probability}
\end{equation}
where $C$ is a constant fitted to data. In Fig. \ref{fig: impact parameter}, the solid colored circles denote the wounded nucleons selected according to the hard-sphere condition in Eq.~(\ref{eq: hard-sphere wounding condition}), for one Pb+Pb collision.

\section{Relativistic hydrodynamics} 
\label{hydrodynamics}
The relativistic hydrodynamic description of the evolution and the dynamics of the QGP medium produced in ultrarelativistic heavy-ion collision is a useful way to model the collision dynamics~\cite{Huovinen:2003fa,Kolb:2003dz,Huovinen:2006jp,Ollitrault:2007du,Dusling:2007gi,Hirano:2008hy,Luzum:2008cw,Heinz:2009xj,Romatschke:2017ejr}. The hot dense droplet of fluid-like QGP medium consists of strongly interacting quarks and gluons ({\it partons}). A formal physical description of such a system should be based on some rigorous microscopic theory. However, a microscopic description of a system consisting of many interacting particles (many degrees of freedom) is not a trivial task. In such situations, one looks for an effective description of the dynamics of the system, based on some macroscopic theory which considers only those degrees of freedom which are relevant at the larger length and time scales. This works because most of the microscopic variables encounter rapid fluctuations on a small length and time scale, which does not have a significant effect on the macroscopic observables measured in the experiments, only the average quantities on the macroscopic scale matter. On the contrary, the conserved quantities which vary slowly during the dynamical evolution of the system play a crucial role in the effective theory. {\it Hydrodynamics} (or {\it fluid dynamics} ) is a perfect example of such a macroscopic theory which can provide an effective description of the dynamics of the QGP medium.

Physically, a {\it fluid} can be described  as a continuous system of infinitesimal volume elements, called {\it fluid elements}. In hydrodynamics, it is assumed that each of these fluid elements is in local thermodynamic equilibrium during its space-time evolution i.e. they are homogeneous without any spatial gradients and can be characterized by thermodynamic relations. This means that the fluid cells are large enough in comparison to the microscopic scale so that a local thermodynamic equilibrium can be defined and at the same time they are small enough in comparison to the macroscopic scale that the continuum (infinitesimal volume)  limit is satisfied. We will assume the above criteria for the description of the QGP medium as a nearly fluid-like medium are fulfilled. The following discussions presented in this section is a general short review of some published literature on the theory of relativistic hydrodynamics in heavy-ion collision~\cite{Jaiswal:2016hex,Israel:1976tn,Israel:1979wp,Grad:1949zza,Denicol:2010xn,Jaiswal:2013npa,landau1959fluid,Eckart:1940zz,Pu:2009fj,Jaiswal:2013fc,Denicol:2008ha,Denicol:2012cn,Betz:2010cx,Baier:2006um}.

\subsection{Thermodynamics}
\label{thermodynamics}
Thermodynamics is a macroscopic theory describing the bulk properties and state of matter. It only deals with the average properties of the microscopic constituents (which are quite large in number) of a system and its fundamental basis lies in statistical mechanics which could be used to derive its basic laws. Here, we will briefly go through the thermodynamic identities and relations which are important and often used in a hydrodynamic model.  

A thermodynamic system is usually characterized by a set of few extensive variables such as volume ($V$), pressure ($P$), total energy ($E$), entropy ($S$) and the number of particles ($N$) in the system. The differential change in the total internal energy is given by,
\begin{equation}
 \begin{aligned}
   dE= dQ - PdV +\mu dN \ ,
 \end{aligned}
\label{eq: differntial energy}
\end{equation}
which is basically the {\it first law of thermodynamics}. On the right hand side, $dQ$ is the heat exchange, the second term is the work done and the third term is rather difficult to interpret. In a situation where particle is exchanged between two systems at same pressure and temperature but different number density, it represents the thermodynamic potential. However, it is important to note that $N$ is usually the number of particles if the system is non-relativistic. In a relativistic system, $N$ is no longer the number of particles as it is not conserved anymore. In a relativistic system, particles can be produced at the expense of energy or can be destroyed to produce energy. In such systems, $N$ represents some conserved quantities e.g. {\it baryon number}. If there are more than one conserved quantity, the term $\mu dN$ is replaced by $\sum_i \mu_i dN_i$, where $\mu_i$ is the chemical potential corresponding to the conserved quantity $N_i$. However, here we will use $N$ to denote baryon number only.  

For a reversible thermodynamic process, Eq.~(\ref{eq: differntial energy}) takes the form,
\begin{equation}
 \begin{aligned}
   dE= TdS - PdV +\mu dN \ ,
 \end{aligned}
\label{eq: first law of theromdynamics differential form}
\end{equation}
where the heat transfer is related to the change in entropy of the system. Using the above equation one can find the extensive variables,
\begin{equation}
 \begin{aligned}
   \frac{\partial E}{\partial S} \bigg\vert_{V,N} = T, \hspace{0.5cm} \frac{\partial E}{\partial V} \bigg\vert_{S,N} = -P \eqsp{and} \frac{\partial E}{\partial N} \bigg\vert_{S,V} = \mu \ ,
 \end{aligned}
\label{eq: extensive varriables as partial derrivatives}
\end{equation}
which lead to the relation between the thermodynamic variables, 
\begin{equation}
 \begin{aligned}
   E = -PV+TS+\mu N \ ,
 \end{aligned}
\label{eq: total energy }
\end{equation}
known as the {\it Euler's equation}. Differentiating the above equation one gets the {\it Gibbs-Duhem relation},
\begin{equation}
 \begin{aligned}
   VdP=SdT+Nd\mu \ .
 \end{aligned}
\label{eq: gibbs-duhem equation}
\end{equation}
In the context of hydrodynamics, these thermodynamic quantities  are expressed in terms of densities i.e. the energy density $\epsilon=E/V$, the entropy density $s=S/V$ and the baryon number density $n=N/V$, which are intensive quantities. In terms of densities, Eqs.~(\ref{eq: total energy }) and (\ref{eq: gibbs-duhem equation}) can be written as, 
\begin{equation}
 \begin{aligned}
   \epsilon = -P+Ts+\mu n
 \end{aligned}
\label{eq: energy density}
\end{equation}
and 
\begin{equation}
 \begin{aligned}
   dP=sdT+nd\mu \ .
 \end{aligned}
\label{eq: gibbs-duhem in termns of densities}
\end{equation}

\textbf{\textit {The thermodynamic equilibrium : }} It is the  thermodynamic state of a system where the system is in a stationary state and where the extensive and intensive variables defining the stationary state do not change with time. For example, the entropy of a system is known to either increase or remain constant from the {\it second law of thermodynamics}. If a system is in equilibrium, its entropy remains constant in time. But, if the system is in an out of equilibrium state, its entropy increases. 

\subsection{Relativistic ideal hydrodynamics}
We start with the dynamics of an ideal relativistic fluid~\cite{landau1959fluid,Israel:1979wp,Jaiswal:2016hex}. For an ideal fluid, the system is assumed to be in local thermal equilibrium i.e all of its fluid elements maintain a thermodynamic equilibrium state with each other. This implies that at each of the space-time coordinates $x \equiv x^\mu$ of the fluid, one can characterize the local temperature $T(x)$, chemical potential $\mu(x)$ and the {\it collective four-velocity} $u^\mu(x)$ defined as, 
\begin{equation}
 \begin{aligned}
   u^\mu(x)=\frac{dx^\mu}{d\tau} = \gamma (1,\vec{v}) \ ,
 \end{aligned}
\label{eq: fluid four-velocity}
\end{equation}
where $\tau$ is the proper time, $ \gamma = 1/\sqrt{1-\vec{v}^2}$ and $\vec{v}=\frac{d\vec{x}}{dt}$ is the spatial velocity of the fluid. 

The state of a relativistic fluid is described with the local energy-momentum tensor $T^{\mu \nu}$ and particle number current $N^\mu$. In the {\it local rest frame (LRF)} of the fluid, $\vec{v}=0$, making $u^\mu_{LRF}=(1,0)$ and the energy-momentum tensor takes the form,
\begin{equation}
 \begin{aligned}
  T^{\mu \nu}_{LRF}=
  \begin{pmatrix}
      \epsilon & 0 & 0 & 0 \\
          0  & P & 0 & 0 \\
          0 & 0 & P & 0 \\
          0 & 0 & 0 & P \\
  \end{pmatrix} \ ,
 \end{aligned}
\label{eq: Tmunu matrix LRF}
\end{equation}
where $\epsilon$ is the energy density and $P$ is the pressure of the fluid. In the laboratory frame, the energy momentum tensor is given by,
\begin{equation}
 \begin{aligned}
 T^{\mu \nu}_{(0)} =(\epsilon + P)u^\mu u^\nu -g^{\mu \nu} P \ ,
 \end{aligned}
\label{eq:  Tmunu labframe}
\end{equation}
for an ideal ( denoted by $``(0)"$) relativistic fluid. Similarly, the net particle number current and the entropy current are given by,
\begin{equation}
 \begin{aligned}
 N^\mu_{(0)} = n u^\mu \eqsp{and} S^\mu_{(0)} = s u^\mu \ .
 \end{aligned}
\label{eq:  Nmu and Smu}
\end{equation}

\subsubsection{Hydrodynamic equations of motion}
\label{ideal-hydro equations}
The equations governing the motion of an ideal fluid, encoding its dynamical description are the hydrodynamic equations of motion, which arise from the conservation laws during the time evolution. The energy-momentum conservation and the net baryon number conservation are given by, 
\begin{equation}
 \begin{aligned}
 \partial_\mu T^{\mu \nu} = 0 \eqsp{and} \partial_\mu N^\mu = 0 \ ,
 \end{aligned}
\label{eq: conservation equations }
\end{equation}
where $\partial_\mu \equiv \frac{\partial}{\partial x^\mu}$ and $N^\mu=(n,\vec{j})$, with $\vec{j}$ as the particle current vector. 

In the laboratory frame, the hydrodynamic equations of motion are obtained by projecting the conservation equations in Eq.~(\ref{eq: conservation equations }) along and orthogonal to the fluid velocity $u^\mu$. Projecting the energy-momentum conservation equation along $u^\mu$ gives the first equation of ideal hydrodynamics :
\begin{equation}
 \begin{aligned}
\frac{d \epsilon}{d \tau} = -(\epsilon+P) \theta \eqsp{or} D\epsilon +(\epsilon +P)\theta =0 \ ,
 \end{aligned}
\label{eq: first hydro equation }
\end{equation}
where
\begin{equation}
 \begin{aligned}
 D \equiv u^\mu \partial_\mu = \frac{d}{d\tau} \eqsp{and} \theta \equiv \partial_\mu u^\mu \ .
 \end{aligned}
\label{eq: projection of delmu along umu }
\end{equation}
$D$ is known as {\it hydro-derivative } or {\it convective derivative} which represents the projection of $\partial_\mu$ along $u^\mu$, and $\theta$ is called the {\it expansion scalar}. Similarly, taking the projection orthogonal to $u^\mu$, one obtains another three hydro-equations :
\begin{equation}
 \begin{aligned}
  a_\lambda = \frac{1}{\epsilon+P} \nabla_\lambda P \eqsp{or} (\epsilon +P)Du_\lambda - \nabla_\lambda P =0 \ ,
 \end{aligned}
\label{eq: Euler's equation hydro (three equations) }
\end{equation}
where
\begin{equation}
 \begin{aligned}
  a_\mu = Du_\mu = \frac{d u_\mu}{d\tau},  \eqsp{} \nabla_\mu \equiv \Delta^\nu_\mu \partial_\nu \eqsp{and}  \Delta^{\mu\nu} = g^{\mu\nu}-u^\mu u^\nu \ . 
 \end{aligned}
\label{eq: four acceleration and transverse gradient }
\end{equation}
$a_\mu$ is called the {\it four-acceleration}, $\nabla_\mu$ is the {\it transverse gradient} and $\Delta^{\mu \nu}$ is known as the {\it projector}. $\nabla_\mu$ denotes the projection of $\partial_\mu $ orthogonal to $u^\mu$.

Finally, from the conservation equation of particle number current  $\partial_\mu N^\mu_{(0)} =0$, we have the fifth equation for ideal hydrodynamics,
\begin{equation}
 \begin{aligned}
 \frac{dn}{d\tau}=-n\theta \eqsp{or} Dn+n\theta=0 \ .
 \end{aligned}
\label{eq: fifth hydro-equation }
\end{equation}
It is important to note that Eqs.~(\ref{eq: first hydro equation }),(\ref{eq: Euler's equation hydro (three equations) }) and (\ref{eq: fifth hydro-equation }) provide five equations of motion for an ideal fluid but there are six degrees of freedom: $\epsilon$, $P$, $n$ and $u_\mu$. The sixth equation comes from the thermodynamic equation of state $P=P(\epsilon,n)$ which relates the pressure to energy or number density. 

In the present context, it is necessary to mention that ideal fluid hydrodynamics is also {\it isentropic}, which means that the entropy also remain conserved during the hydrodynamic evolution of the fluid. Similar to the particle number conservation, the entropy conservation equation is given by $ \partial_\mu S^\mu_{(0)} = 0$ which leads to,
\begin{equation}
 \begin{aligned}
  Ds+s\theta = 0 \ ,
 \end{aligned}
\label{eq: entropy conservation }
\end{equation}
representing the equation of motion for the entropy current.

\subsection{Relativistic dissipative hydrodynamics}
\label{viscous hydro}

The formulation of ideal hydrodynamics relies on Lorentz covariance, conservation equations and the crucial assumption that the fluid is in local thermodynamic equilibrium. While the first two principles are quite robust, the assumption of a perfect thermal equilibrium is significantly crude and far for reality. In practice, the fluid elements are never in exact thermodynamic equilibrium due to the dissipative effects originating due to irreversible thermodynamic processes, frictions between the fluid elements during their motion etc. To properly describe the dynamics of a real fluid, these dissipative effects must be taken into account.   

The covariant formulation of relativistic dissipative hydrodynamics, known as the {\it first order theory}, was first given by Eckart~\cite{Eckart:1940zz} and later Landau and Lifshitz~\cite{landau1959fluid}. These theories are covariant generalization of the {\it Navier-Stokes theory}. However, the first order theory has a serious problem in its formulation. The relativistic Navier-Stokes theory is intrinsically unstable~\cite{Hiscock:1983zz,Hiscock:1985zz} because it violates the fundamental causality condition of the relativistic theory~\cite{Denicol:2008ha,Pu:2009fj}. In this theory, signals can travel at an infinite speed or instantaneously, which is not allowed by the principle of causality. Therefore, one needs to resort to the {\it second order theory} which takes into account the causality principle in its formulation. Among many second order dissipative hydrodynamic theories which addresses the acausal behavior~\cite{Israel:1979wp,Israel:1976tn,Grad:1949zza,carter1991cv,Grmela:1997zz}, the {\it Israel-Stewart theory}~\cite{Israel:1979wp,Israel:1976tn} is the most popular and widely used to describe the hydrodynamic formulation of the QGP. 

In relativistic dissipative hydrodynamics, the basic conservation laws for the energy-momentum and the particle number current remains unchanged ( Eq.~(\ref{eq: conservation equations })). However, for a dissipative fluid, the energy-momentum tensor $T^{\mu \nu}$ and the particle current $N^\mu$ assume additional terms $\tau^{\mu \nu}$ and dissipative current $n^\mu$ respectively,
\begin{equation}
 \begin{aligned}
  T^{\mu \nu} &= T^{\mu \nu}_{(0)} + \tau^{\mu \nu} = (\epsilon+P)u^\mu u^\nu - g^{\mu \nu}P + \tau^{\mu \nu} = \epsilon u^\mu u^\nu -P\Delta^{\mu \nu} +\tau^{\mu \nu}  \ , \\
  N^\mu &= N^\mu_{(0)}+n^\mu= n u^\mu +n^\mu \ ,
 \end{aligned}
\label{eq: Tmunu and Nmu for dissipative fluid }
\end{equation}
where $\tau^{\mu \nu}$ is a symmetric tensor. 

\subsubsection{Matching conditions}
\label{matching conditions}
The additional terms in the energy-momentum tensor and number current in Eq.~(\ref{eq: Tmunu and Nmu for dissipative fluid }), disrupt the local thermal equilibrium of the fluid, making the definitions of the thermodynamic variables ambiguous. In order to  address this situation, we need to define an equivalent thermodynamic equilibrium  so that the definitions of the thermodynamic variables and their relations remain consistent. The total energy density $\epsilon$ and the number density $n$ in the local rest frame of the fluid is defined with respect to the {\it matching conditions}~\cite{landau1959fluid},
\begin{equation}
 \begin{aligned}
\epsilon \equiv u_\mu u_\nu T^{\mu \nu} \eqsp{and} n\equiv u_\mu N^\mu \ .
 \end{aligned}
\label{eq: matching condition }
\end{equation}

\subsubsection{Irreducible decomposition of dissipative components}
\label{tensor decomposition}
 In general, any tensor can be decomposed into its irreducible components that can be a scalar or a four-vector or any other tensor of ranking lesser or equal to the original tensor rank. The particle number current $N^\mu$ which is a rank-1 tensor can be decomposed in two components,
\begin{equation}
 \begin{aligned}
 N^\mu = N_\nu g^{\mu \nu} = N_\nu (\Delta^{\mu \nu}+u^\mu u^\nu) = n u^\mu + n^\mu \ ,
 \end{aligned}
\label{eq: Tensor decomposition of Nmu }
\end{equation}
where we have used the relation $N_\mu u^\mu =n$ (matching condition in Eq.~(\ref{eq: matching condition })) and $n^\mu = \Delta^{\mu \nu}N_\nu$ which satisfies $n^\mu u_\mu =0$. Similarly, the energy momentum tensor $T^{\mu \nu}$ which is a second rank tensor can be decomposed as~\cite{Jaiswal:2016hex}, 
\begin{equation}
 \begin{aligned}
 T^{\mu \nu} =\epsilon u^\mu u^\nu - P \Delta^{\mu \nu} -\Pi \Delta^{\mu \nu} + 2 h^{(\mu} u^{\nu)} +  \pi^{\mu \nu} \ ,
 \end{aligned}
\label{eq: energy-momentum tensor with bulk and shear }
\end{equation}
where, 
\begin{equation}
 \begin{aligned}
 \epsilon = T_{\alpha \beta} u^\alpha u^\beta, \eqsp{} \Pi = - P - \frac{1}{3} T_{\alpha \beta}\Delta^{\alpha \beta}, \eqsp{} h^\mu = \Delta^{\alpha \mu} u^\beta T_{\alpha \beta} \ . 
 \end{aligned}
\label{eq: irr components of Tmunu }
\end{equation}
Please note that $\epsilon$ has same form as in Eq.~(\ref{eq: matching condition }). $\Pi$ is a scalar, known as the {\it bulk viscous pressure} , the field $h^\mu$ is the {\it energy diffusion four-current}. The last two terms in Eq.~(\ref{eq: energy-momentum tensor with bulk and shear }) are given by, 
\begin{equation}
 \begin{aligned}
  h^{(\mu} u^{\nu)} = \frac{1}{2} (h^\mu u^\nu+h^\nu u^\mu)  \eqsp{and} \pi^{\mu \nu} = T^{\alpha \beta} \Delta^{\mu \nu}_{\alpha \beta}    \ ,
 \end{aligned}
\label{eq:  shear and bulk in terms of projectors }
\end{equation}
where $\pi^{\mu \nu}$ is the {\it shear-stress tensor} which is is traceless and $\Delta^{\mu \nu}_{\alpha \beta}$ is a rank-4 projection operator which is doubly symmetric, traceless and orthogonal to $u^\mu$. $\Delta^{\mu \nu}_{\alpha \beta}$ is expressed in terms of the projectors as,
\begin{equation}
 \begin{aligned}
 \Delta^{\mu \nu}_{\alpha \beta} =  \frac{1}{2} (\Delta^\mu_\alpha \Delta^\nu_\beta+\Delta^\mu_\beta \Delta^\nu_\alpha -\frac{2}{3} \Delta^{\mu \nu} \Delta^{\alpha \beta} ) \ .
 \end{aligned}
\label{eq: rank-4 projection operator }
\end{equation} 
Comparing with the energy-momentum tensor for the dissipative fluid in Eq.~(\ref{eq: Tmunu and Nmu for dissipative fluid }), we have the dissipative term $\tau^{\mu \nu}$ in terms of irreducible dissipative components,
\begin{equation}
 \begin{aligned}
\tau^{\mu \nu} = -\Pi \Delta^{\mu \nu} + 2 h^{(\mu} u^{\nu)} +  \pi^{\mu \nu} \ .
 \end{aligned}
\label{eq:  components of taumunu}
\end{equation}
 
The symmetric tensor $T^{\mu \nu}$ has ten independent components and $N^\mu$ has four, which make a total fourteen independent components. The fields $n^\mu$ and $h^\mu$ have three independent components each ( both are orthogonal to $u^\mu$). The stress-tensor $\tau^{\mu \nu}$ being symmetric, traceless and orthogonal to $u^\mu$, has only five independent components. Remaining $\epsilon$, $\Pi$, $n$ and $u^\mu$ constitutes another six ( $\epsilon$ and $P$ are related by equation of state ) independent components. Therefore, in total the irreducible components of $T^{\mu \nu}$ and $N^\mu$ have seventeen independent components. There are extra three components, coming from the velocity field $u^\mu$ which is not well-defined. The fluid velocity needs to be defined properly by choosing appropriate frame.  

\subsubsection{ Definition of velocity field}
\label{def. velocity field}
For dissipative fluid, there is energy and particle diffusion in the $LRF$ of the fluid, which need to be taken into account while defining the velocity. In most of the cases, two possible choices are used:
\vskip 2mm
\textbf{\textit{I. Eckart frame :}} In this frame~\cite{Eckart:1940zz}, it is assumed that there is no net flow of particles or no particle diffusion  and that defines the fluid velocity. Using Eq.~(\ref{eq: Tensor decomposition of Nmu }), 
\begin{equation}
 \begin{aligned}
   n^\mu =0 \eqsp{} \Rightarrow \eqsp{} N^\mu = n u^\mu \ .
 \end{aligned}
\label{eq:  eckart frame}
\end{equation}

\textbf{\textit{II. Landau frame :}} In this frame~\cite{landau1959fluid}, the velocity is defined in a way such that there is no energy diffusion or no net flow of total energy. Using Eq.~(\ref{eq: irr components of Tmunu }), 
\begin{equation}
 \begin{aligned}
   h^\mu =0 \eqsp{} \Rightarrow \eqsp{} u_\nu T^{\mu \nu } = \epsilon u^\mu \ . 
 \end{aligned}
\label{eq:  landau frame}
\end{equation}
It should be noted  that either of the choices reduces three independent components. In the following discussions, we will use the Landau frame to define the fluid velocity, under which the conserved currents become, 
\begin{equation}
 \begin{aligned}
   T^{\mu \nu}=\epsilon u^\mu u^\nu - (P+\Pi) \Delta^{\mu \nu} + \pi^{\mu \nu} \eqsp{and} N^\mu = n u^\mu +n^\mu \ .
 \end{aligned}
\label{eq:  conserved currents under landau frame}
\end{equation}

\subsubsection{Hydrodynamic equations for dissipative fluid}
\label{dissp-hydro equations}
Similar to ideal fluids, we need to project the conservation equations (Eq.~(\ref{eq: conservation equations }) ) along and orthogonal to the velocity field $u^\mu$, in order to find the equations of motion. We use the expressions in Eq.~(\ref{eq:  conserved currents under landau frame}) and using similar contractions we get~\cite{Jaiswal:2016hex}: 
\begin{equation}
 \begin{aligned}
 u_\nu \partial_\mu T^{\mu \nu} =0 \eqsp{} &\Rightarrow \eqsp{} \dot{\epsilon}+(\epsilon + P+\Pi)\theta - \pi^{\mu \nu} \sigma_{\mu \nu} = 0 \ ,\\
\Delta^\alpha_\nu \partial_\nu T^{\mu \nu} = 0 \eqsp{} &\Rightarrow \eqsp{} (\epsilon + P+\Pi) \dot{u}^\alpha - \nabla^\alpha (P+\Pi) + \nabla^\alpha_\nu  \partial_\mu \pi^{\mu \nu} = 0 \ ,\\
\partial_\mu N^\mu = 0 \eqsp{} &\Rightarrow \eqsp{} \dot{n} +n \theta +\partial_\mu n^\mu =0 \ ,
 \end{aligned}
\label{eq:  first hydro equations for dissipative fluids}
\end{equation}
where $\dot{a} = Da=u^\mu \partial_\mu a$ and $\sigma^{\mu \nu}$ is called {\it shear tensor}, defined as, $\sigma^{\mu \nu} \equiv \nabla ^{\langle \mu} u^{\nu \rangle} = \Delta^{\mu \nu}_{\alpha \beta} \nabla^\alpha u^\beta$. Eq.~(\ref{eq:  first hydro equations for dissipative fluids}) gives us five equations for the relativistic viscous hydrodynamics, whereas $T^{\mu \nu}$ and $N^\mu$ have fourteen independent components. Therefore, we need nine more equations for a complete set, which are obtained from the dissipative terms i.e. the diffusion current $n^\mu$, the shear term $\pi^{\mu \nu}$ and the bulk term $\Pi$. The evolution equations for the dissipative terms are obtained from the second order Israel-Stewart theory which we discuss below.

\subsubsection{Israel-Stewart theory for viscous hydrodynamics}
\label{Israel-Stewart}
The Israel-Stewart theory~\cite{Israel:1979wp,Israel:1976tn,Grad:1949zza} of relativistic dissipative fluid is a second order theory which does not violate causality. In this theory, the entropy four-current for the fluid at non-equilibrium, is assumed to be a function of the dissipative currents in addition to the primary fluid-dynamic variables~\cite{Jaiswal:2016hex},
\begin{equation}
 \begin{aligned}
S^\mu = P \beta^\mu +\beta_\nu T^{\mu \nu} -\alpha N^\mu - Q^\mu(\delta N^\mu, \delta T^{\mu \nu}) \ ,
 \end{aligned}
\label{eq: entropy current israel-stewart}
\end{equation}
where $\beta^\mu = u^\mu / T$, $\beta = 1/T$, $\alpha = \mu/T$ with $\mu$ as the chemical potential and $Q^\mu$ is a function of the deviation of dissipative currents from equilibrium: $\delta N^\mu = N^\mu - N^\mu_{(0)}$ and $\delta T^{\mu \nu} = T^{\mu \nu} - T^{\mu \nu}_{(0)}$. Expanding $Q^\mu$ in Taylor's series up to the second order in dissipative flux (second order in $\delta$), we have~\cite{Jaiswal:2013fc,Jaiswal:2016hex},
\begin{equation}
 \begin{aligned}
S^\mu = s u^\mu -\alpha n^\mu &-\frac{\beta}{2}(\beta_0 \Pi^2-\beta_1 n_\nu n^\nu +\beta_2 \pi_{\rho \sigma} \pi^{\rho \sigma}) u^\mu \\
&-\beta (\alpha_0 \Pi \Delta^{\mu \nu} + \alpha_1 \pi^{\mu \nu})n_\nu+\mathcal{O}(\delta^3)) \ ,
 \end{aligned}
\label{eq: entropy current israel-stewart taylor expansion}
\end{equation}
where the coefficients $\beta_i$ and $\alpha_i$ are the thermodynamics coefficients of Taylor's series expansion, which depend on temperature and chemical potential. The next step would be to generate entropy by taking divergence $\partial_\mu S^\mu$ and then one applies the second law of thermodynamics for each fluid element i.e the entropy production always remains positive: $\partial_\mu S^\mu \geq 0$. This results in dynamical equations for the dissipative currents, which are of relaxation-type~\cite{Jaiswal:2016hex}:
\begin{equation}
 \begin{aligned}
\dot{\Pi} + \frac{\Pi}{\tau_\Pi} = -\frac{1}{\beta_0} [\theta+\beta_{\Pi \Pi}\Pi\theta+\psi\alpha_{n \Pi}n_\mu \dot{u}^\mu+\alpha_0 \nabla_\mu n^\mu +\psi \alpha_{\Pi n}n_\mu \nabla^\mu \alpha] \ , \\
\dot{n}^{\langle \mu \rangle} + \frac{n^\mu}{\tau_n} = -\frac{1}{\beta_1}  [T \nabla_\mu \alpha - \beta_{n n} n_\mu \theta + \alpha_0 \nabla_\mu \Pi + \alpha_1 \nabla_\nu \pi^\nu_\mu + \tilde{\psi}\alpha_{n \Pi}\Pi\dot{u}_\mu+ \\
\tilde{\psi}\alpha_{\Pi n}\Pi \nabla_\mu \alpha + \tilde{\chi}\alpha_{\pi n}\pi^\nu_\mu \nabla_\nu \alpha + \tilde{\chi}\alpha_{n \pi}\pi^\nu_\mu \dot{u}_\nu] \ , \\
\dot{\pi}^{\mu \nu} + \frac{\pi^{\mu \nu}}{\tau_\pi} = -\frac{1}{\beta_2}[\sigma_{\mu \nu} -\beta_{\pi \pi} \theta \pi_{\mu \nu} - \alpha_1 \nabla_{\langle \mu}n_{\nu \rangle}-\chi \alpha_{\pi n} n_{\langle \mu} \nabla_{\nu \rangle} \alpha - \\
\chi \alpha_{n \pi} n_{\langle \mu}\dot{u}_{\nu \rangle}] \ ,
 \end{aligned}
\label{eq: relaxation-type equation israel-stewart}
\end{equation}
where $\lambda \equiv \kappa / T$ and the coefficients $\zeta$, $\kappa$ and $\eta$ are called bulk viscosity, particle diffusion and shear viscosity of the fluid, respectively. The parameters 
\begin{equation}
 \begin{aligned}
\tau_\Pi \equiv \zeta \beta_0, \eqsp{} \tau_n \equiv \lambda \beta_1 = \kappa \beta_1/T \eqsp{and} \tau_\pi \equiv 2 \eta \beta_2 \ 
 \end{aligned}
\label{eq: relaxation-times israel-stewart}
\end{equation}
are positive and can be interpreted as relaxation times. Therefore, the coefficients $\beta_0$, $\beta_1$, and $\beta_2$ must also be positive. 

The presence of the relaxation time indicates that the hydrodynamic response to the dissipative currents occur within a time scale given by $\tau$, instead of an instantaneous effect i.e. the theory satisfies the causality requirement of the relativistic theory. However, the theory introduces five new parameters: $\beta_0$, $\beta_1$, $\beta_2$, $\alpha_0$ and $\alpha_1$, which can be obtained from more fundamental microscopic theory such as {\it relativistic kinetic theory} using the framework of Boltzman transport.

\subsection{Relativistic kinetic theory}
\label{rel. kinetic theory}
In relativistic kinetic theory, the macroscopic properties of a system can be expressed in the framework of statistical mechanics using a {\it single particle phase-space distribution function} $f(x,p)$. At each space-time point $x^\mu$, the quantity $f(x,p)\Delta^3x \Delta^3p$ denotes the average number of particles within the volume $\Delta^3x$ and having momenta between $\vec{p}$ and $\vec{p}+\Delta\vec{p}$. In terms of the distribution function $f(x,p)$, the particle number four-current or the {\it four-flow} is given by, 
\begin{equation}
 \begin{aligned}
 N^\mu(x) = \int \frac{d^3p}{p^0} \ p^\mu \ f(x,p) \ ,
 \end{aligned}
\label{eq: four-flow}
\end{equation}
and the energy-momentum tensor as,
\begin{equation}
 \begin{aligned}
 T^{\mu \nu}(x) = \int \frac{d^3p}{p^0} \ p^\mu \ p^\nu \ f(x,p) \ ,
 \end{aligned}
\label{eq: energy-momentum tensor kinetic theory}
\end{equation}
which is symmetric and involves the second moment of the distribution function. Furthermore, using Boltzmann's H-theorem, the general form of the entropy four current (or {\it entropy four-flow}) can be written as, 
\begin{equation}
 \begin{aligned}
 S^\mu(x) = - \int \frac{d^3p}{p^0} \ p^\mu \ [f(x,p)\ln f(x,p) + r \tilde{f}(x,p)\ln \tilde{f}(x,p) ] \ ,
 \end{aligned}
\label{eq: local entropy four-flow generalized form }
\end{equation}
where $\tilde{f}(x,p)=1-rf(x,p)$ and $r = 0$ (Maxwell-Boltzmann statistics), $+1$ (Fermi-Dirac statistics) or $-1$ (Bose-Einstein statistics). 

If the system is in equilibrium, the distribution function is given by, 
\begin{equation}
 \begin{aligned}
f(x,p) \equiv f_0(x,p) = \frac{1}{\exp(\beta p^\mu u_\mu-\alpha)+r} \ .
 \end{aligned}
\label{eq: equilibrium distribution function }
\end{equation}
But if the system is in a non-equilibrium state, the distribution function can be written as a small deviation from the equilibrium distribution function: $f=f_0+\delta f$, where $\delta f$ is the non-equilibrium correction. Furthermore, using the non-equilibrium distribution function in Eq.~(\ref{eq:  conserved currents under landau frame}), one can find the expressions for the dissipative quantities i.e. the bulk pressure, the particle diffusion current and the shear-stress tensor, in terms of the correction $\delta f$ as~\cite{Jaiswal:2016hex}, 
\begin{equation}
 \begin{aligned}
  \Pi &= -\frac{1}{3} \Delta_{\mu \nu} \int \frac{d^3p}{p^0} \ p^\mu \ p^\nu \ \delta f \ , \\ 
  n^\mu &= \Delta^{\mu \nu} \int \frac{d^3p}{p^0} \ p_\nu \ \delta f \ , \\
   \pi^{\mu \nu} &= \Delta^{\mu \nu}_{\alpha \beta} \int \frac{d^3p}{p^0} \ p^\alpha \ p^\beta \ \delta f \ .
 \end{aligned}
\label{eq: dissipative currents from kinetic theory}
\end{equation}

\subsubsection{Dissipative hydrodynamics from relativistic kinetic theory}
\label{dissipative hydro from rel. kin. theory}
Israel-Stewart theory for the relativistic viscous hydrodynamics introduced five new parameters:  $\beta_0$, $\beta_1$, $\beta_2$, $\alpha_0$ and $\alpha_1$, in the evolution equation for the dissipative currents (Eq.~(\ref{eq: relaxation-type equation israel-stewart})). These parameters can now be derived from the evolution equations obtained by solving the relativistic Boltzmann equation for the distribution function,
\begin{equation}
 \begin{aligned}
  p^\mu \partial_\mu f = C[f] \ 
 \end{aligned}
\label{eq: relativistic Boltzmann equation}
\end{equation}
 under the {\it relaxation time approximation} (RTA),
 \begin{equation}
 \begin{aligned}
  C[f] = - u_\mu p^\mu \frac{\delta f}{\tau_R} \ ,
 \end{aligned}
\label{eq: collision integral in RTA}
\end{equation}
 where $C[f]$ is called {\it collision functional} and $\tau_R$ is the relaxation time.
 However, for this one needs to assume some approximated form for $\delta f$. One of the most popular method is the {\it Grad's 14-moment approximation method}. 
\vskip 2mm
\textbf{\textit{ Grad's 14-moment approximation method :}} In this method, originally proposed by Grad~\cite{Grad:1949zza}, the correction $\delta f$ is obtained by expanding the distribution in Taylor's series around its local equilibrium in the power of momenta (truncating in second moment of momenta)~\cite{Israel:1979wp, Denicol:2010xn, Jaiswal:2012qm, Jaiswal:2013fc,Denicol:2012cn,Baier:2006um,Betz:2010cx},
\begin{equation}
 \begin{aligned}
  \delta f = f_0 \tilde{f_0} [\epsilon(x)+\epsilon_a(x)p^\alpha+\epsilon_{\alpha \beta}(x) p^\alpha p^\beta] + \mathcal{O}(p^3) \ ,
 \end{aligned}
\label{eq: Grad's 14-moment approximation}
\end{equation}
which involves 14-unknowns to be determined in order to obtain the distribution, hence called 14-moment approximation. While Israel-Stewart used second moment of Boltzmann equation~\cite{Israel:1979wp}, a more consistent approach was proposed by Denicol-Koide-Rischke (DKR) in~\cite{Denicol:2010xn} to obtain the evolution equation for the dissipative quantities~\cite{Denicol:2010xn,Jaiswal:2012qm},
\begin{equation}
 \begin{aligned}
 \dot{\Pi} &= -\frac{\Pi}{\tau_\Pi} - \beta_\Pi \theta - \ell_{\Pi n} \partial \cdot n - \tau_{\Pi n} n \cdot \dot{u} - \delta_{\Pi \Pi} \Pi \theta - \lambda_{\Pi n} n \cdot \nabla \alpha_0 + \lambda_{\Pi \pi} \pi^{\mu \nu} \sigma_{\mu \nu} \ , \\
 \dot{n}^{\langle \mu \rangle} &= - \frac{n^\mu}{\tau_n} + \beta_n \nabla^\mu \alpha_0 - n_\nu w^{\nu \mu} - \delta_{n n} n^\mu \theta - \ell_{n \Pi} \nabla^\mu \Pi + \ell_{n \pi} \Delta^{\mu \nu} \partial_\lambda \pi^\lambda_\nu  \\
 &+ \tau_{n\Pi} \Pi \dot{u}^\mu - \tau_{n\pi} \pi^\mu_\nu \dot{u}^\nu -  
 \lambda_{n n} n^\nu \sigma^\mu_\nu + \lambda_{n \Pi} \Pi \nabla^\mu \alpha_0 - \lambda_{n\pi} \pi^{\mu \nu} \nabla_\nu \alpha_0 \ , \\
 \dot{\pi}^{\langle \mu \nu \rangle} &= -\frac{\pi^{\mu \nu}}{\tau_\pi} + 2\beta_\pi \sigma^{\mu \nu} + 2 \pi_\alpha^{\langle \mu} w^{\nu \rangle \alpha} - \tau_{\pi n}n^{\langle \mu}\dot{u}^{\nu \rangle} + \ell_{\pi n}\nabla^{\langle \mu}n^{\nu \rangle} - \delta_{\pi \pi} \pi^{\mu \nu} \theta \\
 &- \tau_{\pi \pi} \pi_\alpha^{\langle \mu}\sigma^{\nu \rangle \alpha} + \lambda_{\pi n} n^{\langle \mu}\nabla^{\nu \rangle} \alpha_0 + \lambda_{\pi \Pi} \Pi \sigma^{\mu \nu} \ ,
 \end{aligned}
\label{eq: dissipative equation grad's method}
\end{equation}
where $w^{\mu \nu} = (\nabla^\mu u^\nu - \nabla^\nu u^\mu)/2$ is the vorticity tensor. The above derived equations contain 25 transport coefficients in general, among which three coefficients are shown below~\cite{Denicol:2010xn}, 
\begin{equation}
 \begin{aligned}
    \beta_\Pi &= (\frac{1}{3}-c_s^2)(\epsilon + P)- \frac{2}{9}(\epsilon-3P)-\frac{m^4}{9}\langle (u\cdot p)^{-2}\rangle_0 \ ,\\
    \beta_n &= - \frac{n^2}{\beta(\epsilon + P)}+ \frac{2\langle 1 \rangle_0}{3\beta}+\frac{m^2}{3\beta}\langle (u\cdot p)^{-2}\rangle_0 \ ,\\
    \beta_\pi &= \frac{4P}{5} + \frac{\epsilon - 3P}{15} - \frac{m^4}{15}\langle (u\cdot p)^{-2}\rangle_0 \ ,
 \end{aligned}
\label{eq: expression for three transport coefficients (beta) Grad's method}
\end{equation}
where $\langle \dots \rangle_0 = \int dp (\dots) f_0$ and $c_s^2=(dP/d\epsilon)_{s/n}$ is the speed of sound squared within the medium, which we discuss later.  

Besides, there exist another commonly used method to obtain the equations for the dissipative quantities, namely the {\it Chapman-Enskog expansion}, where the particle distribution function is expanded in powers of the space-time gradient around its equilibrium value in order to find an approximate expression for $\delta f$. For details see~\cite{Jaiswal:2013npa}.

\section{Quantum Chromodynamics (QCD) : theory of strong interaction} 
\label{hydrotheory}
The QGP medium created in the heavy-ion collision is a strongly interacting medium of quarks and gluons which are expected to be weekly coupled in the QGP state at asymptotically large temperature (only). The fundamental theory governing this strong interaction is Quantum Chromodynamics or QCD which deals with the quarks and gluons field and their interactions. The theory explains two fundamental phenomena : i) quarks and gluons cannot exist as free particles in nature. They are always found as bound states (hadrons), this phenomena is known as the {\it color confinement} ii) the interaction between quarks and gluons becomes weaker at short distance and stronger at longer distance, a phenomenon known as {\it asymptotic freedom}. We first discuss one of the most interesting feature of this theory; the existence of {\it color degrees of freedom or color charges} of quarks.

\textbf{\textit{ Color degrees of freedom :}} The first hint of color charges came when people encountered problem to explain the wave function of strange baryons e.g. $\Delta^{++}$, $\Omega^-$ etc.  For example, $\Omega^-$ hyperon consists of three $s$-quarks and has spin $3/2$. As a result, the spin and flavour wave function of  $\Omega$ is symmetric with respect to the change of identical valance $s$-quarks. According to the Pauli's exclusion principle, the full wave function of a particle containing three identical quarks must be antisymmetric. Therefore, it demands that the spatial wave-function of $\Omega^-$ has to be antisymetric. However, $\Omega^-$ is a stable particle and the ground state of a three $s-$quark system; its total wave function has to be symmetric. To solve this puzzle, people came up with a new quantum number associated with the $\Omega^-$ hyperon, which should have at least three different values corresponding to the three quarks within $\Omega^-$. This new quantum number is known as {\it color}, which is an additional degree of freedom of the quarks and gluons, the fundamental basis of QCD theory.  

\subsection{The QCD Lagrangian}
\label{QCD Lagrangian}
Quantum chromodynamics is a  SU(3) gauge theory, known as Yang-Mills gauge theory~\cite{Yang:1954ek}. The QCD Langrangian density is given by~\cite{Fritzsch:1973pi,Gross:1973ju,Gross:1974cs,Weinberg:1973un},
\begin{equation}
 \begin{aligned}
   \mathcal{L}_{QCD}= \bar{\psi^i_q}(x) \ [i\gamma^\mu D_\mu - m_q ]_{ij} \psi^j_q(x) - \frac{1}{4} F^a_{\mu \nu}F^{a \mu \nu} \ ,
 \end{aligned}
\label{eq: The QCD Lagrangian}
\end{equation}
where, $\psi^i_q(x)$ and $\bar{\psi^i_q}(x)$ are the spin-1/2 Dirac fields for quark and antiquark respectively, having color $i$, flavor $q$, and mass $m_q$, with $\bar{\psi}=\psi^\dagger \gamma^0$.  The gluons are represented by a field $A_\mu^a$, which has spin equal to $1$, zero mass and color index $a$ corresponding to the adjoint representation in SU(3) gauge group. In Eq.~(\ref{eq: The QCD Lagrangian}), it is assumed that there is a summation over repeated color and Lorentz indices. The indices  $i,j=1,2,3$ correspond to three possible colors for quarks whereas $a=1,\dots, 8$ represent the colors for eight gluon fields corresponding to the $8$ generators of SU(3) group. The covariant derivative $D_\mu$ is defined as,
\begin{equation}
 \begin{aligned}
   D_\mu = \partial_\mu - i g A_\mu = \partial_\mu - i g t^a A^a_\mu \ ,
 \end{aligned}
\label{eq: covariant derivative QCD Lagrangian}
\end{equation}
where $t^a$ are the $8$ generators of SU(3), given by $t^a=\lambda^a/2$, where $\lambda^a$ are the Gell-Mann matrices~\cite{Fritzsch:1973pi}. $F^a_{\mu \nu}$ are the non-abelian gluon field strength tensor, defined by,
\begin{equation}
 \begin{aligned}
   F^a_{\mu \nu}=\partial_\mu A^a_\nu - \partial_\nu A^a_\mu +g f^{abc} A^b_\mu A^c_\nu   \eqsp{or} F_{\mu \nu} = t^aF^a_{\mu \nu} = \frac{i}{g} [D_\mu, D_\nu] \ ,
 \end{aligned}
\label{eq: gluon field-strength tensor}
\end{equation}
where $f^{abc}$ are the structure constants of color SU(3) gauge group. The underlying gauge symmetry means that the Lagrangian is invariant under the local gauge transformation,
\begin{equation}
 \begin{aligned}
    \psi(x) \rightarrow \chi(x) \psi(x) , \eqsp{} A_\mu(x) \rightarrow \chi(x) A_\mu(x) \chi^{-1}(x) - \frac{i}{g} [\partial_\mu \chi(x)] \chi^{-1}(x) \ ,
 \end{aligned}
\label{eq: local gauge transformation}
\end{equation}
where $\chi(x) = e^{i \alpha^a(x)t^a}$. The Lagrangian in Eq.~(\ref{eq: The QCD Lagrangian}) is based on two fundamental assumptions which are confirmed by experimental observations : all hadrons consist of quarks and quarks are not observed as free particles. These two assumptions demand that there exists a particle which mediates the interaction between the quarks to form the bound state. This interaction has to be attractive i.e. it should depend on the quark colors which requires the mediating particle to be a vector boson of spin 1. In addition to that, the fact that quarks cannot exist as free particles, requires the force of attraction to be stronger at larger distance, which essentially needs the mediating particle to be massless. Thus the particle responsible for strong interaction is a non-Abelian massless vector boson, a gluon.

\subsection{Asymptotic freedom and confinement}
\label{asymtotic freedom}
Another very interesting and remarkable feature of QCD is the {\it asymptotic freedom} which tells us that the QCD coupling strength gets weaker at shorter distances i.e corresponding to larger values of four-momentum squared $q^2=-Q^2$, where $Q$ is a real number. The running coupling constant of QCD governing the interactions is given by~\cite{Gross:1973ju,Politzer:1974fr},
\begin{equation}
 \begin{aligned}
    \alpha_s(Q^2) = \frac{\alpha_s(\mu^2)}{1+\alpha_s(\mu^2) \ \beta_2 \ \ln(Q^2/\mu^2)}
 \end{aligned}
\label{eq: QCD running coupling}
\end{equation}
where,
\begin{equation}
 \begin{aligned}
    \beta_2 = \frac{11N_c-2N_q}{12\pi} \eqsp{and} \beta_{QCD}(\alpha) = -\beta_2 \alpha^2 + \mathcal{O}(\alpha^3) \ .
 \end{aligned}
\label{eq: QCD beta function}
\end{equation}
$\beta_{QCD}$ is known as QCD beta function, $N_c$ is the number of colors and $N_q$ is the number of quark flavors. While $N_q = 6$ in the Standard Model, the effective numbers of flavor depend on the momentum scale $Q$ and in principle could be smaller than six. Eq.~(\ref{eq: QCD running coupling}) clearly suggests that as $Q \rightarrow \infty$, the coupling constant $\alpha_s(Q^2)\rightarrow 0$. This means that the strong interaction between the quarks and gluons become smaller at larger momentum or asymptotically short distance, called the asymptotic freedom. Such a behavior is in striking contrast to other categories of interactions e.g. Quantum Electrodynamics or QED. In QED, $\beta_2$ is negative and as a result the QED coupling constant becomes stronger at the shorter distance or larger $Q^2$.  
\begin{figure}[ht!]
\includegraphics[height=9 cm]{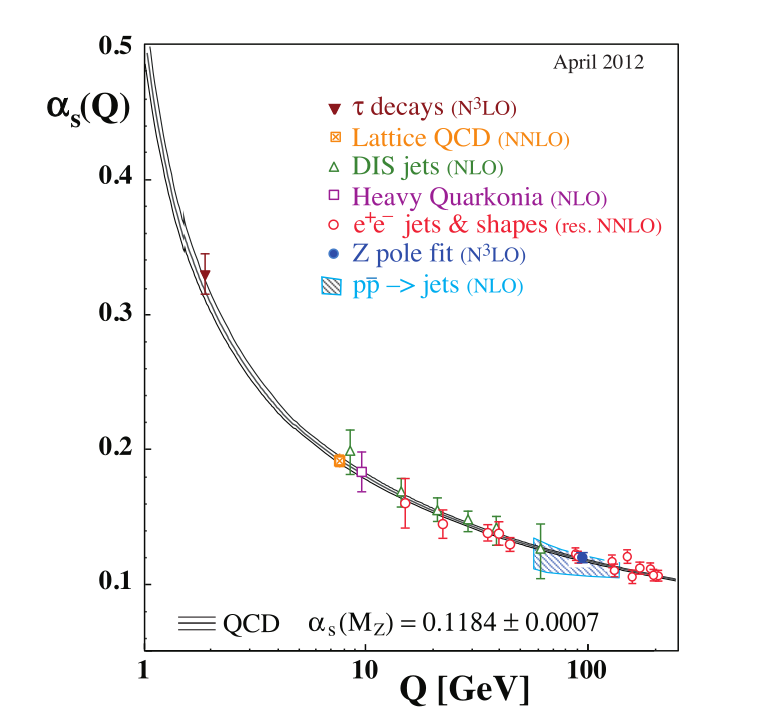}
\centering
\caption{ The running coupling of QCD, as a function of energy (momentum) scale $Q$, illustrating the asymptotic freedom. The figure shows a summary of different measurements of $\alpha_s$ at the respective energy scale. Figure is taken from~\cite{particlereview2012} }
\label{fig: QCD symptotic freedom}
\end{figure}
The quantity $\mu$ in Eq.~(\ref{eq: QCD running coupling}) represents an arbitrary scale, known as the {\it renormalization point}. Eq.~(\ref{eq: QCD running coupling}) can be rewritten as, 
\begin{equation}
 \begin{aligned}
    \alpha_s(Q^2) = \frac{1}{  \beta_2 \ \ln(Q^2/\Lambda^2_{QCD})} \ ,
 \end{aligned}
\label{eq: QCD running coupling  in terms of Lambda}
\end{equation}
where $\Lambda_{QCD}$ is the fundamental scale of QCD, having value $\simeq 200 - 300$ MeV. The exact value of $\Lambda_{QCD}$ depends on the method of renormalization. The strong coupling constant $\alpha_s(Q^2)$ become large when $Q \sim \Lambda_{QCD}$, which makes the interaction between quarks and gluons stronger leading to the confinement of  the quarks and gluons inside hadrons. Fig.~\ref{fig: QCD symptotic freedom} shows the dependence of the running coupling constant $\alpha_s(Q^2)$ on the energy scale ($Q$), depicting the asymptotic freedom.
 
\subsubsection{Deconfinement and formation of QGP}
\label{deconfinement}
The asymptotic freedom discussed above suggests that when the momentum-transfer squared is very large (($Q^2 \gg 1$)) i.e. when the collision energy is very high, the coupling strength of the strong interaction between the constituent partons become very week ($\alpha_s(Q^2) \rightarrow 0$). At asymptotically large temperature, the interaction strength between the partons become so week that within the hot dense matter, the quarks and gluons become asymptotically free and a new degrees of freedom (color) enters into the picture. This phenomena is called {\it color deconfinement } which is responsible for the formation of the Quark-Gluon-Plasma (QGP). The experimental program of heavy-ion collision aims at creating this hot and dense droplet of matter where this particular phenomenon of deconfinement transition takes place.

In the present context, it is important to mention another useful description of QCD theory, known as {\it Lattice QCD} (LQCD) which is used to calculate the QCD Equation of State (EoS) for the QGP medium at high temperature as well as to obtain the QCD phase diagram. Below we briefly discuss this.  

\subsection{QCD Equation of State and Lattice QCD}  
\label{EOS}
The Equation of State (EoS) of a system is defined as the relationship between the thermodynamic variables (state variables) of the system, i.e. pressure ($P$) or number density ($n$) to the energy density ($\epsilon$). EoS is extremely important because it describes the equilibrium properties of the QCD matter. To close the hydrodynamic set of equations, one needs an additional equation which is provided by the equation of state of the system, hence it is an important input for hydrodynamics.

If we treat the QGP medium as an ideal gas of massless quarks and gluons having zero net chemical potential ($\mu_B = 0 $), then thermodynamic quantities of the system can be calculated from the partition function $Z(T,V)$ of the system, 
\begin{equation}
 \begin{aligned}
 \text{energy density,} \eqsp{} \epsilon_{QGP} &= \nu_{QGP}  \frac{\pi^2}{30} T^4 \ , \\
  \text{pressure,} \eqsp{} P_{QGP} &= \nu_{QGP}  \frac{\pi^2}{90} T^4 \ ,\\
  \text{number density,} \eqsp{} n_{QGP} &\approx \nu_{QGP} \frac{1}{\pi^2} T^4 \ ,
 \end{aligned}
\label{eq: thermodynamic quantities of QGP in ideal gas limit}
\end{equation}
where $\nu_{QGP}$ is the total number of degrees of freedom of the system, given by,
\begin{equation}
 \begin{aligned}
 \nu_{QGP} = 16 + \frac{21}{2} N_f \ ,
 \end{aligned}
\label{eq: total degrees of freedom of QGP }
\end{equation}
where $N_f$ is the number of quark flavors. Then in the massless ideal gas limit, one has the EoS of the system: $P=\frac{1}{3} \epsilon$, the EoS for the ideal gas. 
\vskip 1 mm
\textbf{\textit{Speed of sound in QGP :}} In this context, it is important to discuss the speed of sound within the QGP medium. Sound is defined as a small disturbance that propagates through a uniform fluid at rest. For a QGP medium with pressure $P$ and energy density $\epsilon$, the speed of sound squared is defined as,
\begin{equation}
 \begin{aligned}
 c_s^2 = \frac{\partial P}{\partial \epsilon} \ . 
 \end{aligned}
\label{eq: sound velocity in QGP }
\end{equation}
Therefore, for a massless gas of quarks and gluons, $c_s^2 = \frac{1}{3}$. In the case of a baryon less QGP medium ($\mu_B = 0$), using Eqs.~(\ref{eq: gibbs-duhem in termns of densities}) and the relation between the densities, the speed of sound within the medium can be also written in terms of temperature and entropy density as~\cite{Ollitrault:2007du, Gardim:2019brr},
\begin{equation}
 \begin{aligned}
 c_s^2 = \frac{d \ln T}{d \ln s} \ . 
 \end{aligned}
\label{eq: sound velocity for baryonless QGP }
\end{equation}

However, in reality the quarks have mass and interact strongly. Matter created after the heavy-ion collision exhibit two phases (as we will describe later) during the hydrodynamic evolution : the QGP phase at high temperature and as the temperature drops, it gradually enters into the {\it hadron resonance gas} (HRG) phase. As a result, in the heavy-ion community it is a common practice to use Lattice QCD (LQCD) calculation at high temperature and HRG model at low temperature for the EoS with vanishing baryon chemical potential ($\mu_B = 0$)~\cite{Huovinen:2009yb}. 

At zero baryon chemical potential ($\mu_B=0$), there exist many methods, a combination of which could give us a good understanding of EoS at all temperatures. At extremely high temperature ($\alpha_s \ll 1$) i.e. in the pure QGP phase, EoS can be obtained by the perturbative QCD calculations~\cite{Braaten:1995ju,Blaizot:2003iq,Komoltsev:2021jzg}. On the other hand, at the low temperature ($\alpha_s$ large), i.e. in the hadron gas phase EoS is calculated from HRG model~\cite{Huovinen:2009yb}. Lattice QCD (LQCD) bridges the gap between the two calculations and captures the transition between the QGP and the hadron gas phase~\cite{Borsanyi:2013bia}. In the non-perturbative region, where $\alpha_s$ is not so small, LQCD calculation serves as the major tool to investigate equilibrium properties of QCD.  In lattice QCD calculations, a discretized 3+1D lattice space is created and through Monte Carlo approach, the partition function ($Z$) is evaluated on the lattice through path integral method. Once the partition function is defined, all the thermodynamic quantities can be calculated  at $\mu_B=0$ and for any temperature $T$, eventually providing the QCD EoS~\cite{Borsanyi:2010cj,Borsanyi:2013bia,HotQCD:2014kol} at $\mu_B=0$. In the case of finite baryon chemical potential ($\mu_B \neq 0$), the EoS is usually obtained through some approximation using Taylor series expansion around $\mu_B = 0$~\cite{Bazavov:2017dus,Borsanyi:2018grb,Bellwied:2019pxh,Bazavov:2020bjn}. LQCD calculation is used to obtain the equation of state at high temperatures and HRG model is used at low temperatures, with an interpolation procedure in between~\cite{Huovinen:2009yb,Jaiswal:2016hex}.

\section{Different stages of HI collision : hydrodynamic framework}
\label{stages HI collision}
An ultrarelativistic nucleus-nucleus collision at the LHC or RHIC produces hundreds or thousands of particles. From the time of the collision to the time of the detection of the produced particles, the time span is very short (few fm/c). However, within that little span of time, there exist different stages that the system undergoes before ending up as particles at the detectors~\cite{Müller_2013,Jaiswal:2016hex,Moreland:2018gsh,JETSCAPE:2020mzn}. Fig~\ref{fig: Different stages of HI collision} shows the schematic representation and timeline of different stages of a heavy-ion collision event, which we briefly describe below :
\begin{figure}[ht!]
\includegraphics[height=5.5 cm]{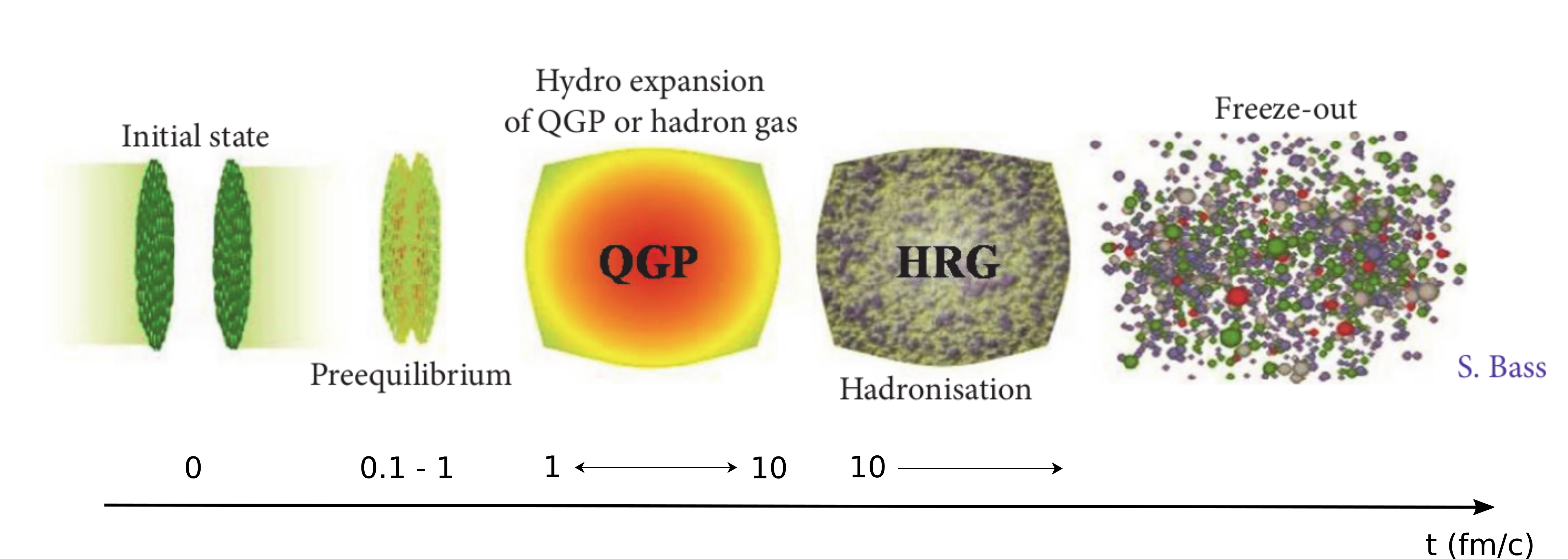}
\centering
\caption{ Schematic representation of different stages in heavy-ion collision along with the timeline for the stages. Figure is taken from~\cite{Jaiswal:2016hex}. }
\label{fig: Different stages of HI collision}
\end{figure}
\begin{itemize}
    \item At $t= 0$ fm/c, two incoming nuclei collide and at the overlap area of collision they deposit energy (or entropy), which serves as the {\it initial state} of the collision.
    \item The fireball created at the collision does not achieve thermal equilibrium immediately after the collision, rather it takes some time for its constituents to interact between each other so that the system gradually approach the equilibrium. This is known as the {\it pre-equilibrium state} which survives for $t \simeq 0.1-1$ fm/c. 
    \item At around $t \lesssim 1$ fm/c, the system is usually considered to have partially achieved thermodynamic equilibrium. Next the system of deconfined quarks and gluons starts to expand collectively staying close to the local thermal equilibrium and simultaneously cools down. At this stage, the QGP medium can be treated as a fluid medium so that it can be evolved through {\it hydrodynamics}. This phase lasts approximately until $t \sim 10$ fm/c.
    \begin{figure}[ht!]
    \includegraphics[height= 8cm]{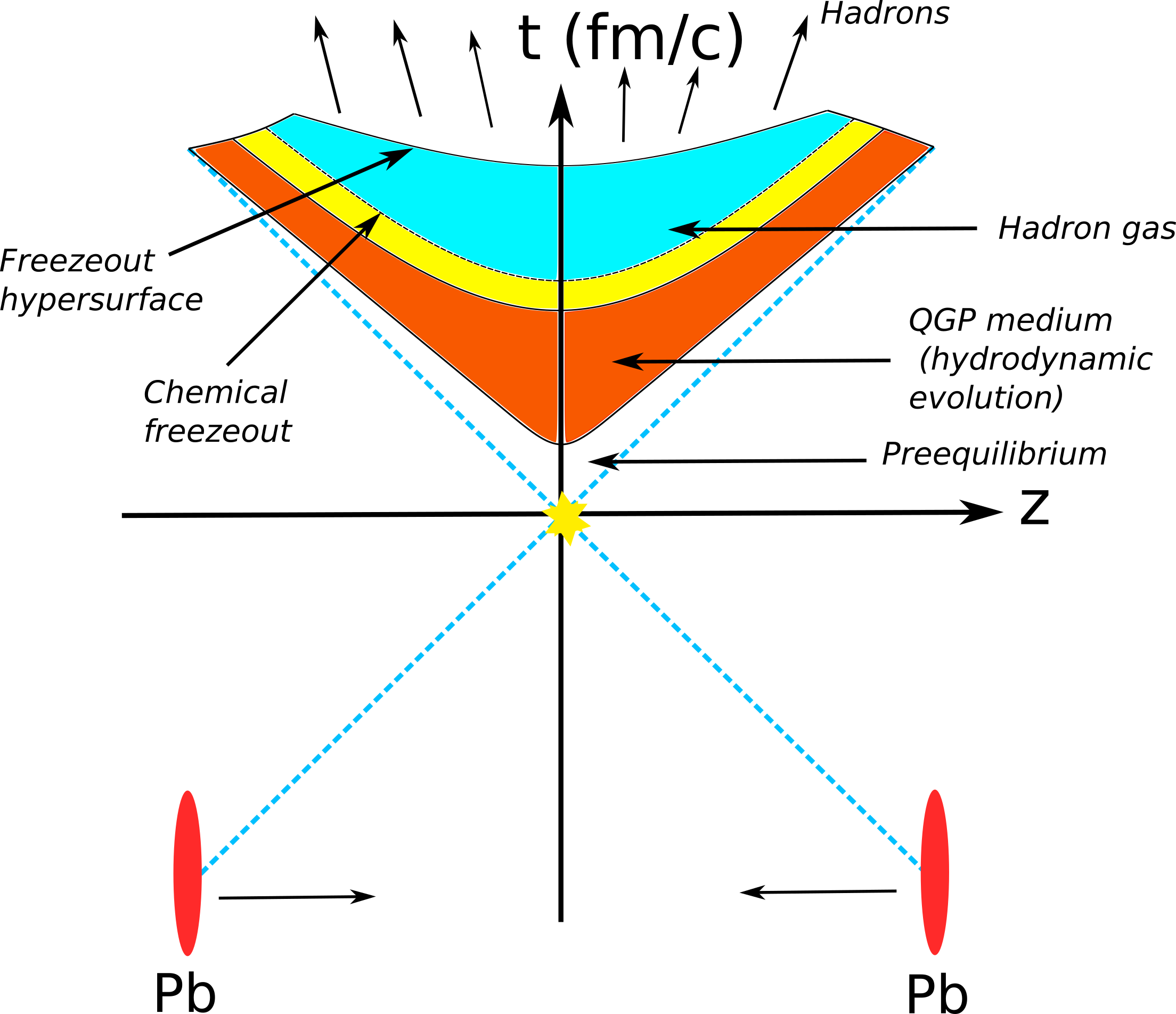}
    \centering
    \caption{ Space-time diagram of different stages of heavy-ion collision. There exist a preequilibrium phase denoted by the white space. Next, there is a QGP phase, where hydrodynamic evolution occurs, denoted by the orange color. The hadron gas phase is denoted by the combination of yellow and blue colors, where at the end of the yellow region chemical freezeout occurs.}
    \label{fig: t-z digram diff stages HI}
    \end{figure}
    \item After the hydrodynamic evolution of the QGP medium ($t \sim 10$ fm/c), when the system temperature drops below the phase-transition temperature ($T_c$) which is around $150$ MeV, the system encounters a phase transition, where the medium constituents or the partons combine with each other to produce hadrons. This is called {\it hadronization}\footnote{not to be confused with {\it particlization} which occur at later time} in which many hadrons and resonance particles are produced forming a system of hadrons, {\it the hadron-resonance gas} (HRG). At this stage, the produced hadrons undergo resonance decay and elastic or inelastic interactions with each other until they reach a certain point where the system ceases to produce new particles, known as the {\it chemical freezeout}. After chemical freezeout, the produced stable hadrons still undergo elastic interactions with each other until the system reach a certain state where the elastic interactions between the hadrons also stop, called  the {\it kinetic freezeout}\footnote{Please note until this stage the system is still evolving through hydrodynamics as the fluid picture was still valid. In reality, the precise time for the hydrodynamic evolution to stop is not accurately defined, one only knows that it has to stop at some point.}. It is expected that the kinetic freezeout occurs later than chemical freezeout so that the temperature for chemical freezeout is always larger ($T_{ch} > T_{kin}$). After kinetic freezeout, hadrons are free to stream towards the detector.  
    \item When the system reaches the state of kinetic freezeout, the space-time hypersurface of the system is called the {\it freezeout hypersurface}. The momentum space distribution of the final identified hadrons is then obtained by converting the fluid dynamical information from each hypersurface cell into the local phase-space distributions of hadrons, through {\it particlization} method. 
\end{itemize}
Fig~\ref{fig: t-z digram diff stages HI} shows the space-time evolution and different stages of heavy-ion collision. It is clearly visible that in the hadron gas phase there exists checmical and kinetic freezeout separately.  

It should be noted that after the hydrodynamic evolution, when the system reaches the freezeout state, the hadrons particlized from the freezeout hypersurface, can again undergo secondary elastic or inelastic interactions and cascade decays before finally hitting the detectors. Such interactions and cascade decays can redistribute the momentum distribution of the identified particles. Therefore, modern hydrodynamic simulations take these into account by carrying out hydrodynamic evolution followed by a hadron-cascade stage, where the latter is implemented through a {\it after-burner}. We will revisit this fact at the end of this chapter.       

\subsection{Initial conditions}
\label{initial condition}
At the time of the collision, two nuclei deposit energy (or entropy) at the overlap region which serves as the initial state for the collision event. The initial state condition is of practical importance in heavy-ion collision, in the sense that many final state observables depend largely on the initial state properties~\cite{Bozek:2020drh,Bozek:2021zim,Niemi:2012aj,Gardim:2011xv,Giacalone:2020dln}. A full 3D initial condition would involve a 2D density profile on the transverse plane along with the space-time rapidity ($\eta_s$) distribution of those transverse profiles~\cite{Hirano:2002ds,Schenke:2010rr,Bozek:2011ua,Bozek:2022svy}.  However, as long as we restrict ourselves to the central rapidity region, the space-time evolution of the system can be assumed to be a boost-invariant, homogeneous (or uniform) longitudinal expansion, known as the {\it Bjorken flow} which is a very simplistic yet effective assumption of hydrodynamic expansion first proposed by J.D. Bjorken~\cite{Bjorken:1982qr}, which will be again discussed later. In the Bjorken picture, the boost-invariance ensures that most of the knowledges about the transverse collective properties of the system can be obtained using a 2+1 D hydrodynamic expansion of the system, which will be our primary topic of discussion in this document. For the initial conditions of such boost-invariant hydrodynamic evolution, the two dimensional density profiles are sufficient. This initial density could be obtained from some state of the art initial state models e.g. 2D Glauber model (discussed in Sec.~\ref{glauber model}), saturation based {\it color glass condensate} model (Kharzeev-Levin-Nardi(KLN) model~\cite{Kharzeev:2000ph,Kharzeev:2001yq}, IP-Glasma model~\cite{Schenke:2012wb,Schenke:2012hg}), parametric Glauber-like model (TRENTo model~\cite{Moreland:2014oya}) etc.  In this document we will mostly use the Glauber model which we have already discussed in Sec.~\ref{glauber model} and the parametric TRENTo model to generate 2D initial conditions for hydro-evolution, which we briefly discuss below.

\subsubsection{Two-component Glauber model}
\label{two-component Glauber}
Although in Sec.~\ref{glauber model}, we discussed the basic features of the optical and MC Glauber model, we did not discuss the particle production i.e. how the final state multiplicity scales according to the number of participants or binary collisions. In the simplest picture, the multiplicity of hadrons per unit rapidity in an event scales according to the soft processes i.e. the total number of participants or the number of wounded nucleons :  $\frac{dN_{ch}}{d\eta} = N_{pp} \frac{N_{part}}{2} $, known as the {\it wounded-nucleon model}~\cite{Bialas:1976ed}, where $N_{pp}$ is the average multiplicity in p+p collision. However, later the experimental data have suggested that the total multiplicity in a nucleus-nucleus collision gets contribution from both hard and soft processes. In particular, the multiplicity per unit pseudo-rapidity has two components: the `soft' part is proportional to the number of participants $N_{part}$ and 
the `hard' part is proportional to the number of binary collisions $N_{coll}$~\cite{Kharzeev:2000ph},
\begin{equation}
 \begin{aligned}
    \frac{dN{ch}}{d\eta} = (1-\alpha) \ N_{pp} \ \frac{N_{part}}{2} +  \alpha \ N_{pp} \ N_{coll} \ ,
 \end{aligned}
\label{eq: Two-component Glauber}
\end{equation}
where $0 < \alpha < 1$. The above two models are implemented within $GLISSANDO$~\cite{Broniowski:2007nz}.

\subsubsection{TRENTo model}
\label{trento}
TRENTo which reads as {\it Reduced Thickness Event-by-event Nuclear Topology}, is a parametric non-dynamical effective model for generating initial conditions directly at the thermalization time ($\tau_0$) in high-energy nucleus-nucleus, proton-nucleus and  proton-proton collisions~\cite{Moreland:2014oya}. 

Let us first consider, two protons A and B, separated by an impact parameter b along x-axis, collide with each other having nuclear densities,
\begin{equation}
 \begin{aligned}
    \rho_{A,B} = \rho_{proton}(x\pm b/2,y,z) \ .
 \end{aligned}
\label{eq: proton nuclear densities trento}
\end{equation}
Then the participant thickness associated with each projectile is given by,
\begin{equation}
 \begin{aligned}
    T_{A,B}(x,y)=\int dz \  \rho_{A,B}(x,y,z) \ .
 \end{aligned}
\label{eq: participant thickenss trento}
\end{equation}
The incoming two protons collide with a probability~\cite{dEnterria:2010xip},
\begin{equation}
 \begin{aligned}
    P_{coll} = 1- \exp \bigg[-\sigma_{NN} \int dx \ dy \int dz \rho_A \int dz \rho_B \bigg] \ ,
 \end{aligned}
\label{eq: proton-proton ollision probability trento}
\end{equation}
where the above integral represents the overlap integral of proton thickness function and $\sigma_{NN}$ is the proton-proton inelastic cross-section. To introduce additional event-by-event fluctuations, each proton is assigned a {\it fluctuating} thickness,
\begin{equation}
 \begin{aligned}
     T_{A,B}(x,y)= \omega_{A,B} \ \int dz \ \rho_{A,B}(x,y,z)] \ ,
 \end{aligned}
\label{eq: fluctuated participant thickenss trento}
\end{equation}
where $\omega_{A,B}$ represent independent random weights which are sampled from a gamma distribution with unit mean,
\begin{equation}
 \begin{aligned}
    P_k(\omega) = \frac{k^k}{\Gamma(k)} \omega^{k-1} e^{-k\omega} \ .
 \end{aligned}
\label{eq: gamma fluctuation trento}
\end{equation}
The additional multiplicity fluctuations are introduced by the above gamma weights. 

Next come two primary postulates for the entropy production in this model:
\begin{itemize}
    \item In the collision, the entropy production occurs through the eikonal overlap of $T_A$ and $T_B$.
    \item There exist a scalar field $f(T_A,T_B)$ which converts the projectile thicknesses into entropy deposition i.e $ f \propto dS/dy |_{\tau =\tau_0}$, where $dS/dy |_{\tau =\tau_0}$ is the entropy deposited per unit rapidity at the hydrodynamic thermalization time $\tau_0$.
\end{itemize}
In the TRENTo model, $f$ represents the {\it reduced thickness} having the functional form,
\begin{equation}
 \begin{aligned}
    f=T_R(p;T_A,T_B) \equiv \bigg ( \frac{T_A^p + T_B^p}{2} \bigg )^\frac{1}{p} \ , 
 \end{aligned}
\label{eq: reduced thickness trento}
\end{equation}
named accordingly because the above function takes two thicknesses $T_A$, $T_B$ and `reduces' them to a third thickness, similar to the reduced mass. 
\begin{figure}[ht!]
    \includegraphics[height= 8 cm]{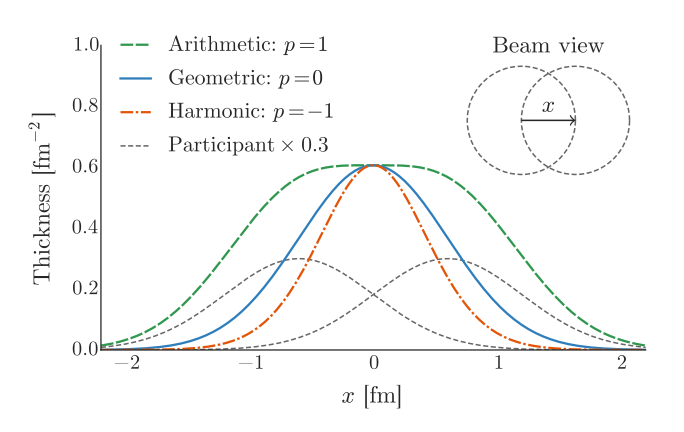}
    \centering
    \caption{ Reduced thickness in the collision of two nucleons at some non-zero impact parameter along x-axis for different values of $p$. The plot shows the cross-sectional view of the overlap of thicknesses. Figure taken from~\cite{Moreland:2014oya}.}
\label{fig: trento reduced thickness}
\end{figure}
The above functional form represents the generalized mean which depending on the value of the parameter $p$, reduces to
\begin{equation}
 \begin{aligned}
    T_R = 
    \begin{cases}
    \max(T_A,T_B), \eqsp{} &p \rightarrow +\infty \\
    (T_A+T_B)/2, \eqsp{} &p = + 1 \ \text{(arithmetic mean)} \\
    \sqrt{T_AT_B}, \eqsp{} &p = 0 \ \text{(geometric mean)} \\
    2T_AT_B/(T_A+T_B), \eqsp{} &p = - 1 \ \text{(harmonic mean)} \\
    \min(T_A,T_B) \ . \eqsp{} &p \rightarrow -\infty 
    \end{cases}
 \end{aligned}
\label{eq: reduced thickness trento for diff p}
\end{equation}
Please note that with $p=1$, the reduced thickness become equivalent to the wounded nucleon model. Fig.~\ref{fig: trento reduced thickness} shows the reduced thickness as a function of the impact parameter for two colliding nucleons for different values of $p$.

Similarly, proton-nucleus and nucleus-nucleus collisions can be treated as a superposition of proton-proton collisions. Let us consider now two colliding nuclei A and B. The position of the nucleons in each projectile is obtained by sampling a nuclear distribution e.g. Woods-Saxon distribution (Sec.~\ref{nuclear distribution}) and then collision probability is sampled for each pair of nucleons from the two projectiles. The nucleons which collide at least once with another nucleon from the other projectile, are called the `participants' and the rest of the nucleons are spectators hence not relevant. Then the fluctuating thickness function of two nuclei reads,
\begin{equation}
 \begin{aligned}
    T_{A,B}(x,y)=\sum_{i=1}^{N_{part}^{A,B}} \omega_i \ \int dz \  \rho^{proton}_{A,B}(x-x_i,y-y_i,z-z_i).
 \end{aligned}
\label{eq: fluctuated participant thickenss trento for nuclei}
\end{equation}
where $\omega_i$ is the weight corresponding to $i^{th}$ participant. Once we have the participant thickness $T_A$ and $T_B$, we can calculate $T_R(p;T_A,T_B)$ using Eq.~(\ref{eq: reduced thickness trento}) and the initial transverse entropy deposition per unit rapidity is given by up to an overall normalization factor,
\begin{equation}
 \begin{aligned}
    \frac{dS}{dy} |_{\tau=\tau_0} \propto T_R(p;T_A,T_B) \ .
 \end{aligned}
\label{eq: entropy per unit rapidity from reduced thickness trento}
\end{equation}

The average charged particle multiplicity ($N_{ch}$) produced in the final state after the hydrodynamic evolution, is to a good approximation, proportional to the average total initial entropy~\cite{Song:2008si} and so to the integrated reduced thickness\footnote{denoting here rapidity as $y'$.}, 
\begin{equation}
 \begin{aligned}
   \frac{dN_{ch}}{d y'} \propto \int dx \ dy \ \frac{dS}{dy'} |_{\tau=\tau_0} \equiv \int dx \ dy \ T_R \ .
 \end{aligned}
\label{eq: Nch from integrated reduced thickness }
\end{equation}
The default value of the parameter $p$ is $0$, the parameter $k$ in Eq.~(\ref{eq: gamma fluctuation trento}) is called the shape parameter having a default value $k=1$. Small values of $k$ ($0<k<1$) correspond to larger multiplicity fluctuations and if $k\gg 1$ then it suppresses the fluctuations.  The proton thickness function in Eq.~(\ref{eq: fluctuated participant thickenss trento for nuclei}) is given by a Gaussian density, 
\begin{equation}
 \begin{aligned}
   \int dz \ \rho_{proton} = \frac{1}{2\pi w^2} \exp \bigg ( - \frac{x^2+y^2}{2 w^2} \bigg ) \ ,
 \end{aligned}
\label{eq: Gaussian proton density }
\end{equation}
where $w^2$ is the effective area, having a default value $w = 0.6$ fm. The latest version of TRENTo also includes the option to work with the constituents (partons) and hence the Gaussian constituent width $v$~\cite{trentoDuke}.

\subsection{Pre-equilibrium}
\label{pre-equilibrium}
Once we have the initial condition in a heavy-ion collision event, the next step is the hydrodynamic evolution of the system once it achieves thermal equilibrium. However, the thermalization is not achieved immediately after the collision, rather there exist a finite time interval between the time of collision ($\tau=0$) and the thermalization time ($\tau=\tau_0$), called the pre-equilibrium phase. In this time interval, the constituents (partons) of the created system just after collision, involve in rapid interaction with each other in order to achieve the equilibrium state. In principle, there could be two limiting cases for the coupling strength inside the QGP medium~\cite{Moreland:2018gsh}: infinitely week coupling where the secondary partons created in the collision free-stream without any interaction and infinitely strong coupling where the inter-particle mean free path becomes extremely small, eventually resulting in a fluid-like system. 

In reality, the initial parton interactions in the preequilibrium phase is governed by a coupling strength lying between the above two extremes, while the system continues to evolve. The simplest choice for the dynamical evolution of the system in pre-equilibrium phase is the free-streaming of partons, governed by the collision-free Boltzmann equation~\cite{Moreland:2018gsh,JETSCAPE:2020mzn} : $p^\mu \partial_\mu f(x,p) = 0$. Usually the free-streaming time is taken to be $0<\tau_{fs} \lesssim 1$ fm/c.   
IP-Glasma~\cite{Schenke:2012wb,Schenke:2012hg} is the one of the most popular state-of-the-art models which takes into account the pre-equilibrium dynamics until hydrodynamics begin. The IP-Glasma model is a CGC based~\cite{McLerran:1993ka,McLerran:1993ni,Krasnitz:2000gz,Iancu:2003xm,Lappi:2003bi,Gale:2012rq}, impact parameter dependent saturation model (IP-Sat) which provides the initial condition for heavy-ion collision by taking into account not only the fluctuations in nucleons' positions but also the quantum fluctuations of color charges. Besides, there are other transport models e.g. URQMD~\cite{Bass:1998ca,Bleicher:1999xi,Steinheimer:2007iy}, AMPT~\cite{Zhang:1999bd,Lin:2004en,Pang:2012he,Bhalerao:2015iya}, KoMPoST~\cite{Kurkela:2018wud} and relativistic ADS/CFT models~\cite{vanderSchee:2013pia} which also incorporate the pre-equilibrium dynamics.    

\subsection{Hydrodynamic evolution}
\label{hydro-evolution}
Once the QGP system reaches thermalized state, the hydrodynamic evolution can be started from the thermalization time ($\tau_0$) assuming the system as a fluid-like medium. The hydrodynamic evolution of the system is governed by the hydrodynamic equations of motion described in Sec.~\ref{hydrodynamics}, as well as the equation of state of the system. In reality, this evolution is a fully three dimensional evolution in space-time, but a boost-invariant two dimensional transverse evolution also works as a good approximation, which stems from the Bjorken picture of the hydrodynamic evolution. In this thesis, all the results in the following chapters are based on a boost-invariant two dimensional evolution, where the underlying assumption is the Bjorken flow that we discuss below. 

\subsubsection{Bjorken flow}  
\label{bjorken flow}
Bjorken flow describes the longitudinal expansion (along  the beam-axis ) of the fluid medium created in heavy-ion collision, first proposed by J.D. Bjorken~\cite{Bjorken:1982qr}. Bjorken proposed that in a central collision of two large nuclei, the fluid medium near the collision axis expand homogeneously (or uniformly) in the longitudinal direction (along z-axis). It means that at a given longitudinal distance $z$, all points on the fluid move with a longitudinal velocity $\beta_z = v_z = z/t$ at a given time $t$ in the lab frame, while the pancake-like nuclei recede in opposite directions and the fluid in the midway stays at rest. The velocity is uniform and the expansion is boost-invariant under Lorentz transformation in a sense that if someone boosts the system (say with  velocity $v$ ) along the $z$-direction, all three quantities $v_z,z,t$ change in the new frame but $v_z = z/t$ still hold in the new frame or in other words, in the new frame the fluid expands with a uniform velocity $v_z'=z'/t'$ in the longitudinal direction. Bjorken's prescription for hydrodynamic expansion is supported by the experimentally observed plateau in the distribution of produced particles in rapidity\footnote{In experiment one often measures the charged particle multiplicity distribution in pseudo-rapidity rather than rapidity}~\cite{CMS:2011aqh,ATLAS:2011ag,ALICE:2016fbt}. In the Bjorken picture, one works with the proper time coordinates (Eq.~\ref{eq: proper coordinates}) or the so called {\it Milne coordinates} given by, 
\begin{equation}
 \begin{aligned}
   \tau = \sqrt{t^2 - z^2} \eqsp{and} \eta_s = \frac{1}{2} \ln \frac{t+z}{t-z} \ ,
 \end{aligned}
\label{eq: Milne coordinates }
\end{equation}
where $\tau$ is the proper time and $\eta_s$ is the space-time rapidity. Conversely, $t$ and $z$ are given in terms of Milne coordinates as,
\begin{equation}
 \begin{aligned}
   t= \tau \ \cosh \eta_s \eqsp{and} z= \tau \ \sinh \eta_s \ .
 \end{aligned}
\label{eq: t,z in terms of Milne coordinates }
\end{equation}
Under the Bjorken scaling $v_z=z/t$, the fluid rapidity $y$ in Eq.~(\ref{eq: rapidity}) becomes equal to space-time rapidity $\eta_s$,
\begin{equation}
 \begin{aligned}
   \eta_s = \frac{1}{2} \ln \frac{t+z}{t-z} = \frac{1}{2} \ln \frac{1+z/t}{1-z/t} = \frac{1}{2} \ln \frac{1+v_z}{1-v_z} = \frac{1}{2} \ln \frac{1+p_z/E}{1-p_z/E} = y.  
 \end{aligned}
\label{eq: fluid rapidity equals space-time rapidity Bjorken flow}
\end{equation}
Under the Lorentz boost along $z$-axis, $\tau$ remains constant, $y$ and $\eta_s$ shift by a constant. With the Bjorken's prescription, initial conditions are usually specified at a given proper time $\tau=\tau_0$, rather than a given time $t=t_0$ and the solution of the hydrodynamic equations becomes much simpler.

\subsubsection{Solution of hydrodynamic equations in the Bjorken picture}
\label{solution of hydrodynamics }
Bjorken picture of hydrodynamic expansion describes only one-dimensional flow in the longitudinal direction and no transverse flow\footnote{It should be noted that the Bjorken flow is an oversimplification of the hydrodynamic expansion. In reality, the fluid expands in all possible directions. However, the transverse expansion sets in at a  later time leading to the transverse anisotropic flow and most distinctive collective behavior of the fluid observed in the final state.}. However, it is important for the assumption of boost-invariant expansion of the fluid. The fluid four-velocity in Bjorken picture becomes,
\begin{equation}
 \begin{aligned}
 u^\mu = \gamma (1, v_z) , \eqsp{with} v_z=\frac{z}{t}, \ \gamma = \frac{1}{\sqrt{1-v_z^2}} = \frac{t}{\tau} \ .
 \end{aligned}
\label{eq: fluid four-velocity in Bjorken picture }
\end{equation}
In terms of Milne coordinates,
\begin{equation}
 \begin{aligned}
(t,z)=\tau \ (\cosh \eta_s , \sinh \eta_s) \Rightarrow u^\mu = (\cosh \eta_s , \sinh \eta_s).
 \end{aligned}
\label{eq: fluid four-velocity in Bjorken picture  in milne coordinates}
\end{equation}
For an ideal relativistic fluid, we recall the hydro-equations in Eq.~(\ref{eq: first hydro equation }) and Eq.~(\ref{eq: fifth hydro-equation }),
\begin{equation}
 \begin{aligned}
 D\epsilon+(\epsilon + P)\theta = 0 \eqsp{and} Dn +n \theta = 0 \ ,
 \end{aligned}
\label{eq: Recalling hydro-equations for ideal fluid}
\end{equation}
where $D=u^\mu \partial_\mu$ and $\theta = \partial_\mu u^\mu$. In terms of Milne coordinates they become, 
\begin{equation}
 \begin{aligned}
 D \equiv \frac{\partial}{\partial \tau}  \eqsp{and} \theta = \frac{1}{\tau} \ .
 \end{aligned}
\label{eq: hydro derivative and expansion scalar in Milne coordinates }
\end{equation}
With this, the hydro equations become,
\begin{equation}
 \begin{aligned}
 \frac{ \partial \epsilon}{\partial \tau} = - \frac{\epsilon + P}{\tau} \eqsp{and}  \frac{ \partial n}{\partial \tau} = - \frac{n}{\tau} \ .
 \end{aligned}
\label{eq: hydro equations in Bjorken picture}
\end{equation}
Assuming that the hydrodynamic evolution starts at time $\tau=\tau_0$, when the initial energy density is $\epsilon=\epsilon_0$ and using the EoS, $P=c_s^2 \epsilon$, the first equation in Eq.~(\ref{eq: hydro equations in Bjorken picture}) gives,
\begin{equation}
 \begin{aligned}
 \frac{ \partial \epsilon}{\partial \tau} + (1+c_s^2)\frac{\epsilon }{\tau} =0 
 \eqsp{} \Longrightarrow \eqsp{} \epsilon_{Bjorken} = \epsilon_0 \bigg(\frac{\tau_0}{\tau}\bigg)^{1+c_s^2} \ .
 \end{aligned}
\label{eq: solution of energy density in Bjorken picture}
\end{equation}
Similar to the second part of Eq.~(\ref{eq: hydro equations in Bjorken picture}), Eq.~(\ref{eq: entropy conservation }) gives the equation for entropy density,
\begin{equation}
 \begin{aligned}
 \frac{ \partial s}{\partial \tau} = - \frac{s}{\tau}  \eqsp{} \Longrightarrow \eqsp{} s = \frac{s_0 \tau_0}{\tau}
 \end{aligned}
\label{eq: equation for entropy density in Bjorken picture}
\end{equation}
and so the pressure of the system would evolve as,  $P \propto \frac{1}{\tau^{1+c_s^2}}$.
For the evolution of the temperature, we use Eq.~(\ref{eq: energy density}) with zero baryon chemical potential, 
\begin{equation}
 \begin{aligned}
 \epsilon = -P +Ts \eqsp{} \Longrightarrow \eqsp{} T \propto \frac{1}{\tau^{c_s^2}} \ .
 \end{aligned}
\label{eq: Temperature evolution in Bjorken picture}
\end{equation}
If we consider the medium as a gas of massless quarks and gluons, then $c_s^2 = 1/3$ leads to the relation : 
\begin{equation}
 \begin{aligned}
 P \propto T^{1+\frac{1}{c_s^2}} = T^4, \eqsp{} \epsilon \propto T^4 \eqsp{and} s \propto T^3 \ ,
 \end{aligned}
\label{eq: Temperature dependence of thermodynamic quantities in Bjorken picture in ideal gas limit}
\end{equation}
which we find for an ideal gas. 

\subsection{Hadronization and freezeout : QCD phase transition }  
\label{phase transition}
\begin{figure}[ht!]
    \includegraphics[height=  9 cm]{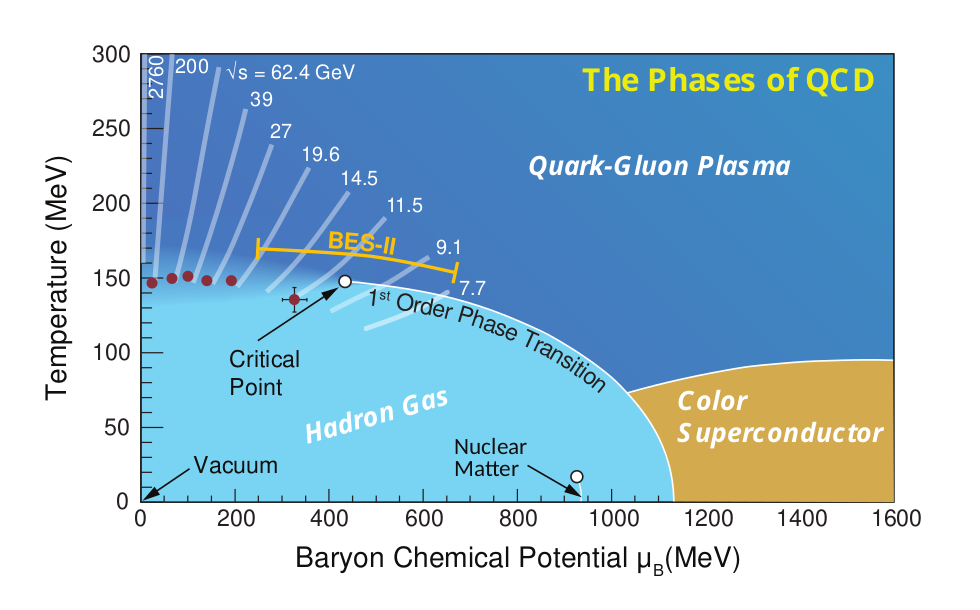}
    \centering
    \caption{Schematic representation of the QCD phase diagram. It is shown that the phase transition and its type depend on the collision energy, probing different  ($T,\mu_B$) regions. The solid curve separating the QGP phase and HRG phase, denotes the {\it first order transition}, and the point where it ends is called {\it critical point or critical end point}. After the critical point there exist a smooth transition region, called the {\it crossover region}. At the LHC or highest RHIC energies, the phase transition from QGP to hadron gas occurs towards the end of the crossover region where $T\sim 150$ MeV and $\mu_B \sim 0$. Figure taken from~\cite{Bzdak:2019pkr}.}
\label{fig: QCD phase diagram}
\end{figure}

Towards the end of hydrodynamic evolution, when the QGP medium has cooled down sufficiently ($T\sim 150 $ MeV), the interaction strength between the constituent partons become large enough that they couple with each other to form bound states or hadrons. At this stage, the system undergoes a smooth phase-transition from the decoupled QGP phase to a strongly coupled hadronic phase, producing a gas of stable hadrons and their unstable resonance particle, called {\it hadron resonance gas} (HRG). This is known as the so-called {QCD phase transition}~\cite{Stephanov:1998dy,Stephanov:1999zu,Fodor:2004nz,Aoki:2006we,Stephanov:2008qz,Bazavov:2011nk,HotQCD:2018pds,Borsanyi:2020fev}  and the corresponding phase diagram is shown in Fig.~\ref{fig: QCD phase diagram}. Please note that the type of phase transition depends on the collision energy which translates into the temperature and baryon chemical potential. It is expected that a smooth transition occurs at a vanishing baryon chemical potential ($\mu_B \sim 0 $) in the {\it crossover region}, which is the  case for collisions at the LHC and highest RHIC energies.
\subsubsection{Freezeout }  
\label{Freezeout}
In the HRG phase, many stable and unstable hadrons are produced, which are further involved in elastic or inelastic collisions and resonance decays. These produce further particles until the system reach a certain state where the production of new particles stops. This is called {\it chemical freezeout}. After chemical freezeout, the system of stable hadrons still interact with each other via elastic collisions until the system reach another state where the elastic collisions also cease and that is known as {\it kinetic freezeout}. When the system reach kinetic freezeout, the space-time hypersurface at that point is called {\it freezeout hypersurface} and the hydrodynamic evolution stops at this stage. The most popular choices for the freeze-out criterion are the constant temperature or constant energy density hypersurface.

\subsubsection{Particlization: Cooper-Frye prescription }  
\label{cooper-frye}
Once the system reaches the kinetic freezeout condition (temperature $T = T_{fo}$ or energy density $\epsilon_{fo}$), stable hadrons stream towards the detector. Therefore, at freezeout one needs to change the description of the system from the `fluid picture' to `particle picture', because at the end those particles are detected by the detectors. We need a method to obtain the momentum distribution of the particles from the freezeout hypersurface at constant temperature or energy density. This is called {\it particlization} and it is commonly given by Cooper-Frye formula~\cite{Cooper:1974mv}, where the basic assumption is that towards the end of the hydrodynamic expansion, the momentum
distribution of the outgoing particles is essentially the momentum distribution of the particles within the fluid, which are treated as independent particles~\cite{Ollitrault:2007du}. In the Cooper-Frye prescription, the momentum distribution of hadron species $i$ with degeneracy $g_i$ is given by \footnote{It should be noted that the Cooper-Frye formula presented here, which involves an integral over the hypersurface, is used only on a smooth surface producing directly the spectra of a particle according to Eq.~(\ref{eq: cooper-frye formula}). We particularly use this formalism in our simulation, as implemented in MUSIC\cite{Schenke:2010nt} hydrodynamics code. This is not exactly {\it particlization} in a literal sense but equivalent to that. Instead of a continuous integral (Eq.~((\ref{eq: cooper-frye formula})) particles can be sampled from from the hypersurface in a discretized way according to particle's momentum through a sampler such as iSS~\cite{issChun}, where the true sense of particlization is realized.},
\begin{equation}
 \begin{aligned}
 E\frac{dN}{d^3p} = \frac{dN}{dy p_T dp_T d\phi_p} = \frac{g_i}{(2\pi)^3} \int_\Sigma f(x,p) p^\mu d^3\Sigma_\mu \ ,
 \end{aligned}
\label{eq: cooper-frye formula}
\end{equation}
where $f(x,p)=f_0(x,p)+\delta f(x,p)$ is the distribution function consisting of the equilibrium part and the dissipative correction. $\Sigma$ is the four dimensional freeze-out hypersurface at $T_{fo}$ (or $\epsilon_{fo}$) given by, 
\begin{equation}
 \begin{aligned}
 \Sigma_\mu = (\tau_f \cosh \eta_s, x , y, \tau_f sinh \eta_s) \ ,
 \end{aligned}
\label{eq: freezeout-hypersurface}
\end{equation}
where $\tau_f$  is the freeze-out time (also called {\it switching time}) determined by the fall of temperature (or energy density) below $T_{fo}$ (or $\epsilon_{fo}$). $d^3\Sigma_\mu$ is the differential freezeout hypersurface element. The equilibrium distribution function is given by Eq.~(\ref{eq: equilibrium distribution function }),
\begin{equation}
 \begin{aligned}
  f(x,p) = \frac{1}{\exp((p^\mu u_\mu-\mu_i)/T_{fo}) \pm 1} \ ,
 \end{aligned}
\label{eq: equilibrium dist. fucntion at freezeout}
\end{equation}
where the $\pm$ sign depends on whether the distribution is Bose-Einstein or Fermi-Dirac distribution, depending on the spin of the hadronic species. The fluid velocity at the hypersurface, resulting from longitudinal and transverse flow is taken into account through the invariant expression $E\equiv E(x) = p^\mu u_\mu$, denoting the local energy of the hypersurface. The dissipative correction could be obtained from different prescriptions~\cite{Monnai:2009ad,Bozek:2009dw,Bozek:2011ua,JETSCAPE:2020mzn}. Among them the most extensively used methods are Grad's 14-moment approximation and Chapman-Enskog expansion discussed in Sec.\ref{dissipative hydro from rel. kin. theory}. If the dissipative effect is considered only due to shear viscosity, then Grad's approximation gives,
\begin{equation}
 \begin{aligned}
  \delta f(x,p)_{\text{shear}} = \frac{f_0 \tilde{f_0}}{2(\epsilon+P)T^2}p^\mu p^\nu \pi_{\mu \nu}
 \end{aligned}
\label{eq: delta f Grad at freezeout only shear}
\end{equation}
and the Chapman-Enskog expansion gives,
\begin{equation}
 \begin{aligned}
  \delta f(x,p)_{\text{shear}} =  \frac{5 f_0 \tilde{f_0}}{2(\epsilon+P)T}\frac{1}{u\cdot p}p^\mu p^\nu \pi_{\mu \nu} \ .
 \end{aligned}
\label{eq: delta f Chapman-Enskog at freezeout only shear}
\end{equation}
In our analysis and the results presented in the subsequent chapters, we limit ourselves to the study of the effect of shear viscosity on the observables under consideration. Therefore, we only need to include the dissipative corrections due to the shear viscosity ($\delta f_{\text{shear}}$) at freezeout. We use relativistic viscous hydrodynamics code MUSIC for our simulation, where the correction due to shear viscosity, at the freezeout, is implemented using the Grad's 14-moment approximation method in Eq.~(\ref{eq: delta f Grad at freezeout only shear})~\cite{Schenke:2010nt,musicManual}.

After the freezeout process, the resulting hadrons could be considered to stream freely to the detector. However, in the modern practice of simulating heavy-ion collision through hydrodynamic framework, instead of particlization from the freezeout hypersurface, people particlize the hypersurface at some temperature, called the {\it particlization temperature}~\cite{Summerfield:2021oex} or the {\it switching temperature} ($T_{sw}$) \footnote{For many such calculations, people take the chemical freezeout tempearture as the  particlization temperature $T_{sw}$}. Then to take into account further inelastic collisions, resonance decays and elastic interactions, known as {\it hadron re-scattering} or {\it hadron cascade}, the produced particles are fed into some hadronic transport models called {\it after burner}. The mean free path between the produced particles is quite large at this stage and the system is evolved through the Boltzman transport equation with collision terms. There exist several state-of-the-art models to take into account hadron rescattering effect or hadronic transport separately, e.g UrQMD~\cite{Bass:1998ca,Bleicher:1999xi}, AMPT~\cite{Zhang:1999bd,Lin:2004en}, SMASH~\cite{SMASH:2016zqf} etc. for the particlized hadrons, to incorporate the hadron cascade until the kinetic freezeout before experimental detection of the particles.      

In this chapter, we have shortly presented a complete picture of ultrarelativistic heavy-ion collision experiments covering the physics behind it, theoretical tools to study it and a description of the hydrodynamic framework which will be the main underlying theory for the results that will be presented in the following chapters. We conclude this chapter by briefly describing the simulation set-up for the results presented in this manuscript.

\subsubsection{Simulation set-up }  
\label{music hydro}
An ultrarelativistic heavy-ion collision comprises many stages and each of these stages can be well described by a certain model. As a result, nowadays it is a common practice in the heavy-ion community to use a hybrid-model-approach to simulate the hydrodynamic framework in heavy-ion collisions. In those approaches, different models corresponding to the different stages are clubbed together~\cite{Shen:2014vra,Ryu:2015vwa,Giacalone:2020byk,JETSCAPE:2020mzn} to properly simulate the collision e.g.  initial condition + pre-equilibrium + hydrodynamic evolution + Cooper-Frye freezeout + hadronic transport $\equiv$ IP-Glasma~\cite{Schenke:2012wb} / Glauber~\cite{Wang:1991hta,Broniowski:2007nz} / TRENTo~\cite{trentoDuke}  + free streaming~\cite{JETSCAPE:2020mzn} / KOMPOST~\cite{Kurkela:2018wud} + MUSIC~\cite{Schenke:2010nt} / VISH2+1~\cite{Song:2007ux} / v-USPhydro~\cite{Noronha-Hostler:2013gga} + iSS~\cite{issChun} + UrQMD~\cite{Bass:1998ca} / SMASH~\cite{SMASH:2016zqf}.  However, the inclusion of all these intermediate stages is not always essential for describing the collective behavior of the final state particles i.e. collective flow, fluctuations and correlation between them.

In our hydrodynamic simulation of ultrarelativistic heavy-ion collision, we do not take into account the pre-equilibrium phase and the rescattering or hadronic transport through a hadronic cascade after-burner. While the effect of a pre-equillibirum phase could be negligible, there could be some effect in the final state from hadronic transport. However, in all our analysis presented in the subsequent chapters we will be mostly dealing with the charged hadrons, rather than any particular species of identified particles. Moreover, we will be always dealing with the event averaged quantity or sometimes the ratio of the observables. In such cases, the effect of hadronic transport is small. One can just follow hydrodynamic freezeout and Cooper-Frye prescription at the final freezeout. We take the initial condition at some thermalization time $\tau_0 \sim 0.6 fm/c$ as the input for hydrodynamic evolution and free streaming after the computation of thermal spectra or momentum distribution, following Cooper-Frye formula at the freezeout temperature ($T_{fo}$). However, we take into account the resonance decays. Our simulations are centered at the LHC and RHIC energies and all the results presented in the subsequent chapters are based on the boost-invariant 2+1D hydrodynamic simulation. The initial conditions for our simulation are generated from the Glauber model~\cite{Broniowski:2007nz} or TRENTo~\cite{trentoDuke} model and for the hydrodynamic simulation we use MUSIC~\cite{Schenke:2010nt,musicManual} hydro code. MUSIC is a simulation package which can simulate both 3+1D and 2+1D  hydrodynamic evolution and it can also operate in different modes. In our calculation, MUSIC does everything from the start of the hydrodynamic evolution to the Cooper-Frye freezeout. We use the default values of the parameter set in the input file of MUSIC~\cite{musicCode} unless otherwise stated for some specific studies. So, our simulation set-up reads Glauber / TRENTo + MUSIC.

%*******************************************************************************
%****************************** Third Chapter **********************************
%*******************************************************************************
\chapter{Collective flow and its fluctuations in heavy-ion collision}

% **************************** Define Graphics Path **************************
\ifpdf
    \graphicspath{{Chapter3/Figs/Raster/}{Chapter3/Figs/PDF/}{Chapter3/Figs/}}
\else
    \graphicspath{{Chapter3/Figs/Vector/}{Chapter3/Figs/}}
\fi

One of the most peculiar and spectacular phenomena observed in high energy heavy-ion collision, is the collective behavior of the final state particles produced in the collisions. This collective nature has been studied extensively and with immense dedication over the past 30 years in theory, as well as in the experiments. The most distinctive feature of this collective phenomena is the {\it collective flow}, specifically the {\it anisotropic flow}~\cite{Ollitrault:1992bk}, which has been the central focus for these theoretical studies~\cite{Ollitrault:1992bk,Voloshin:1994mz,Sorge:1996pc,Poskanzer:1998yz,Bleicher:2000sx,Teaney:2000cw,Huovinen:2001cy,Hirano:2002ds,Bozek:2004dt,Luzum:2008cw,Alver:2010gr,Schenke:2010rr,Bozek:2010bi,Luzum:2010sp,Bozek:2011wa,Schenke:2011bn,Qiu:2011hf,Gardim:2012yp,Heinz:2013th,Ollitrault:2023wjk} and experimental measurements~\cite{STAR:2000ekf,PHENIX:2002hqx,PHENIX:2003qra,STAR:2004jwm,ALICE:2010suc,ATLAS:2012at,CMS:2012zex,ALICE:2016ccg,ATLAS:2018ezv}. The measurement of the anisotropic flow was the first evidence of the thermalized QGP medium produced in heavy-ion collisions. The anisotropic flow originates from the spatial anisotropy in the initial state of the collision leading to a pressure gradient of the fireball, which then translates into the momentum anisotropy of the final state particles. Collective flow can be analyzed or measured through different approaches~\cite{Voloshin:1994mz,Poskanzer:1998yz,Borghini:2000sa,Borghini:2001vi,Borghini:2002vp,Bhalerao:2003yq, Bilandzic:2008nx, Bilandzic:2010jr, Bhalerao:2013ina, Bilandzic:2013kga, STAR:2002hbo, ALICE:2014dwt}. Another very interesting and exotic characteristic of heavy ion collisions is the {\it event-by-event fluctuations} of the collective flow of particles in the final state~\cite{Aguiar:2001ac,Ollitrault:2009ie,Qiu:2011iv,Gardim:2012im,Bzdak:2012tp,Heinz:2013bua,Renk:2014jja,Jia:2014vja,Bhalerao:2014mua,Pang:2015zrq,Xiao:2015dma,Bozek:2017qir,Bozek:2018nne,Bozek:2021mov,Nielsen:2022jms,Samanta:2023qem,Bozek:2023dwp,Zhu:2024tns,PHOBOS:2010ekr,ALICE:2011svq,CMS:2015xmx, ALICE:2016kpq, ALICE:2017lyf, ATLAS:2017rij,ATLAS:2019peb,ALICE:2022dtx,ALICE:2024fcv}, stemming from event-by-event fluctuations in the initial state~\cite{Broniowski:2007ft, Alver:2010dn,Qin:2010pf,Teaney:2010vd,Holopainen:2010gz,Werner:2010aa,Muller:2011bb,Gardim:2011xv,Bhalerao:2011yg,Schenke:2012wb}. In this chapter, we will discuss the basic phenomenology of collective flow: its origin, relation to the final state, its centrality dependence, transverse momentum dependence and most importantly its event-by-event fluctuations.   

\section{Anisotropic flow : from spatial anisotropy to momentum anisotropy}
\label{collective flow}
In a heavy-ion collision, two colliding nuclei deposit energy in the overlap area, creating a transverse density profile at the initial state. The spatial distribution of this initial energy or entropy density in the transverse plane is not isotropic, which creates an anisotropy in the pressure of the fluid. As a result, there exist a pressure gradient $\nabla P$ on the QGP fireball created in the collision. This pressure is propagated to the final state through the viscous hydrodynamic evolution of the fireball, resulting in larger transverse momentum of the particles originating from part of the fluid having larger $\nabla P$ and vice versa. This leads to an anisotropy in the transverse momentum distribution of the final state particles and such flow of particles is known as {\it anisotropic flow}~\cite{Ollitrault:1992bk,Voloshin:1994mz,Poskanzer:1998yz}. Thus anisotropic flow is a characteristic signature of the momentum space azimuthal anisotropy of final state particles.    
\begin{figure}[ht!]
\includegraphics[height=5 cm]{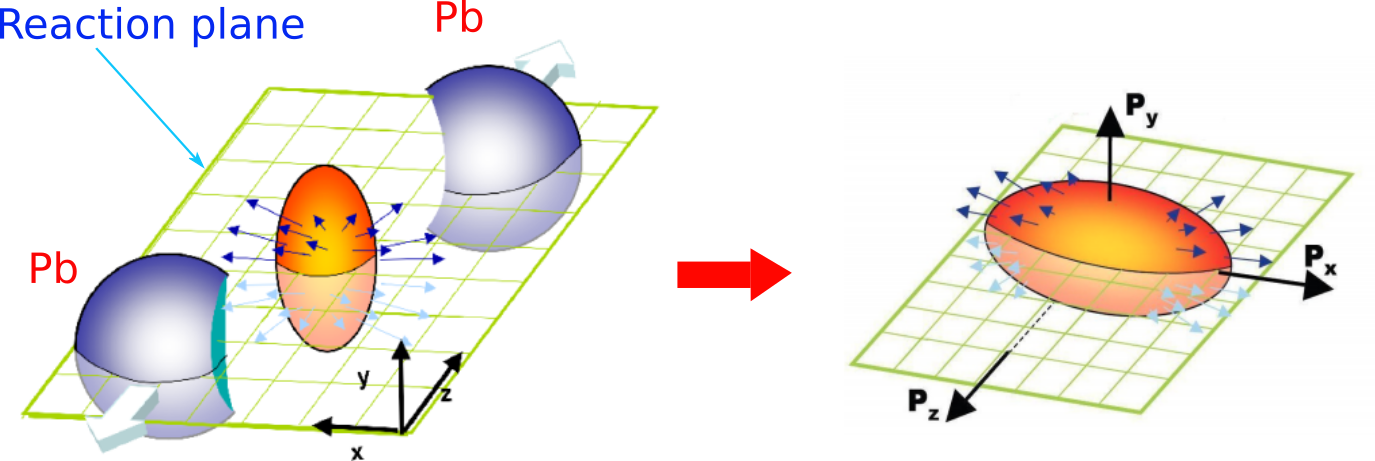}
\centering
\caption{Schematic representation of the almond shaped fireball formation in a non-central Pb+Pb collision and development of momentum anisotropy at the final state. Figure taken from~\cite{Heinz:2008tv} and BNL.}
\label{fig: anisotrpic flow}
\end{figure}
Fig.~\ref{fig: anisotrpic flow} shows the formation of the fireball in a non-central heavy-ion collision and the translation of spatial anisotropy or non-uniform pressure gradient in the initial state, to the azimuthal anisotropy of momentum distribution of final state particles. Before we move into the quantification of anisotropic flow, we need to have knowledge about the characteristic properties of the initial state.  

\subsection{Initial state properties}
\label{Initial geometry}
The nature of the anisotropic flow is largely determined by the initial energy density or entropy density, shape and size of the overlap region at the initial state. In a non-central collision of two identical heavy-nuclei, the overlap area assumes an almond-like shape (Fig.~\ref{fig: origin of elliptic flow}) and the spatial anisotropy of entropy density distribution on the transverse plane (x,y) is characterized with respect to the reaction plane, spanned by the impact parameter and z-axis. Let us denote the initial energy density by $\epsilon(x,y)$ and the entropy density by $s(x,y)$\footnote{In general the initial energy density or entropy density are taken at the thermalization time or at the onset of hydrodynamics, denoted by $s_0(x,y)\equiv s(x,y,\tau_0)$ or $\epsilon_0(x,y)\equiv \epsilon(x,y,\tau_0)$.}. In this manuscript, we will always consider entropy density for our calculations and hydro inputs. The spatial anisotropy of the entropy density distribution can be identified in terms of standard eccentricity~\cite{Sorge:1998mk, Luzum:2013yya},
\begin{equation}
 \begin{aligned}
  \Epsilon_{std} = \frac{\{y^2-x^2\}}{\{y^2+x^2\}}, \eqsp{with} \{\dots\} = \int \dots \ s(x,y)  \ dx \ dy \ .
 \end{aligned}
\label{eq: Standard eccentricity}
\end{equation}

\begin{figure}[ht!]
\includegraphics[height=7 cm]{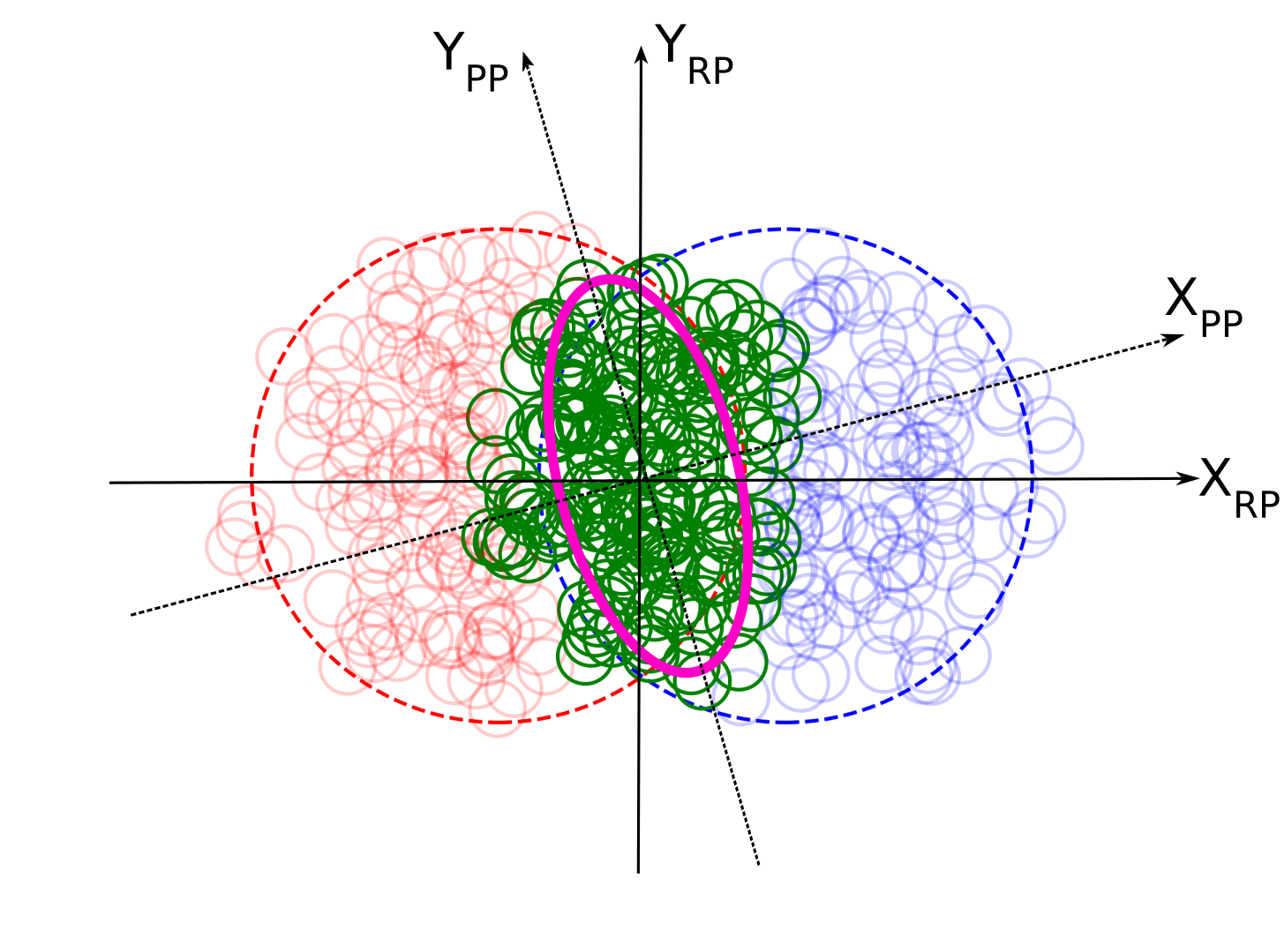}
\centering
\caption{Pictorial depiction of the reaction plane and participant plane in a collision. The area with green circles denote the {\it participant zone} whose principle axis is different than the principle axis of the almond shaped overlap area.}
\label{fig: participant plane and reaction plane}
\end{figure}

In the initial years of flow study, it was believed that the standard eccentricity $\Epsilon_{std}$ of the initial state leads to the formation of anisotropic flow, specifically {\it elliptic flow}, as we will explain later, originating from the ellipsoidal shape in non-central collisions. However, later it was seen that the elliptic flow did not disappear even in central collision especially in small collision systems (e.g. Cu+Cu), where $\Epsilon_{std}$ is small. The reason behind this puzzling behavior was rooted in the fact that event-by-event fluctuations of the initial state geometry are important in central collision to explain the anisotropic flow~\cite{Alver:2008aq}. This phenomena is even more significant in case of {\it triangular flow}~\cite{Alver:2010gr}. In this regard, instead of the almond shaped overlap area of the two nuclei, the region formed by the participant nucleons in each event is more relevant. The plane formed by the principal axis of the {\it participant zone} and the z-axis is called the {\it participant plane} (Fig.~\ref{fig: participant plane and reaction plane}) which is different than reaction plane and its orientation fluctuates event-by-event with respect to reaction plane. The spatial eccentricity which drives the elliptic flow is not the standard eccentricity, rather the eccentricity defined with respect to the participant plane, called {\it participant eccentricity}~\cite{Alver:2010gr},
\begin{equation}
 \begin{aligned}
  \Epsilon_{2}^{part} = \frac{\sqrt{(\{y^2\}-\{x^2\})^2+4\{xy\}^2}}{\{y^2\}+\{x^2\}} \ .
 \end{aligned}
\label{eq: participant eccentricity}
\end{equation}
It is defined with respect to a coordinate system where $\{x\}=0$ and $\{y\}=0$. In terms of polar coordinates ($r,\phi$), the above equation takes the form,
\begin{equation}
 \begin{aligned}
  \Epsilon_{2}^{part} = \frac{\sqrt{\{r^2 \cos(2\phi)\}^2+\{r^2 \sin(2\phi)\}^2}}{\{r^2\}} \ ,
 \end{aligned}
\label{eq: participant eccentricity polar coordinates}
\end{equation}
where $\phi$ denotes the azimuthal angle of the participant nucleon. The elliptic asymmetry of the participant zone, quantified by $\Epsilon_2$, drives the elliptic flow at the final state. Similarly, the triangular asymmetry, which is largely induced by fluctuations, leads to the triangular flow at the final state, for which analogously we can define {\it participant triangularity}~\cite{Alver:2010gr,Alver:2010dn}, 
\begin{equation}
 \begin{aligned}
   \Epsilon_{3}^{part} = \frac{\sqrt{\{r^2 \cos(3\phi)\}^2+\{r^2 \sin(3\phi)\}^2}}{\{r^2\}} \ .
 \end{aligned}
\label{eq: participant triangularity}
\end{equation}

In a generalized way, the eccentricity harmonic coefficients associated with the spatial anisotropy of the participant region can be formulated through the cumulant expansion method~\cite{Teaney:2010vd,Gardim:2011xv},
\begin{equation}
 \begin{aligned}
   \Epsilon_{m,n}e^{\Phi_{m,n}} = -\frac{\{ r^m e^{i n \phi}\}}{\{r^m \} },
 \end{aligned}
\label{eq: eccentricity harmonics general form}
\end{equation}
which accounts for different order moments of r associated with $n^{th}$ harmonic. In this definition, $\Epsilon_2 = \Epsilon_{2,2}$ and $\Epsilon_3 = \Epsilon_{2,3}$. However, it has been found that using the moment of $r^3$ works as a better estimator than $r^2$ for the triangular flow~\cite{Gardim:2011xv}, so that $\Epsilon_3 = \Epsilon_{3,3}$. In general, the conventional eccentricity coefficients are given by~\cite{Luzum:2013yya,Heinz:2013th}, 
\begin{equation}
 \begin{aligned}
  \Epsilon_1 e^{i \Phi_1} =  -\frac{\{ r^3 e^{i \phi_1}\}}{\{r^3 \}} \eqsp{and}  \Epsilon_{n}e^{i n \Phi_{n}} = -\frac{\{ r^n e^{i n \phi}\}}{\{r^n \} } \eqsp{for} n > 1,
 \end{aligned}
\label{eq: conventional eccentricity harmonics general form}
\end{equation}
where $\Epsilon_1$ is known as {\it dipole asymmetry}~\cite{Teaney:2000cw}, which turns out to be responsible for the {\it directed flow}, $\Epsilon_2$ is the {\it quadrupole asymmetry} or ellipticity, $\Epsilon_3$ is the {\it octupole asymmetry} or triangularity of the initial transverse density profile and so on. $\Phi_n$ is the $n^{th}$ order participant plane angle or orientation angle. 
\begin{figure}[ht!]
\hspace{-3cm}\begin{subfigure}{0.33\textwidth}
\centering
\includegraphics[height=6 cm]{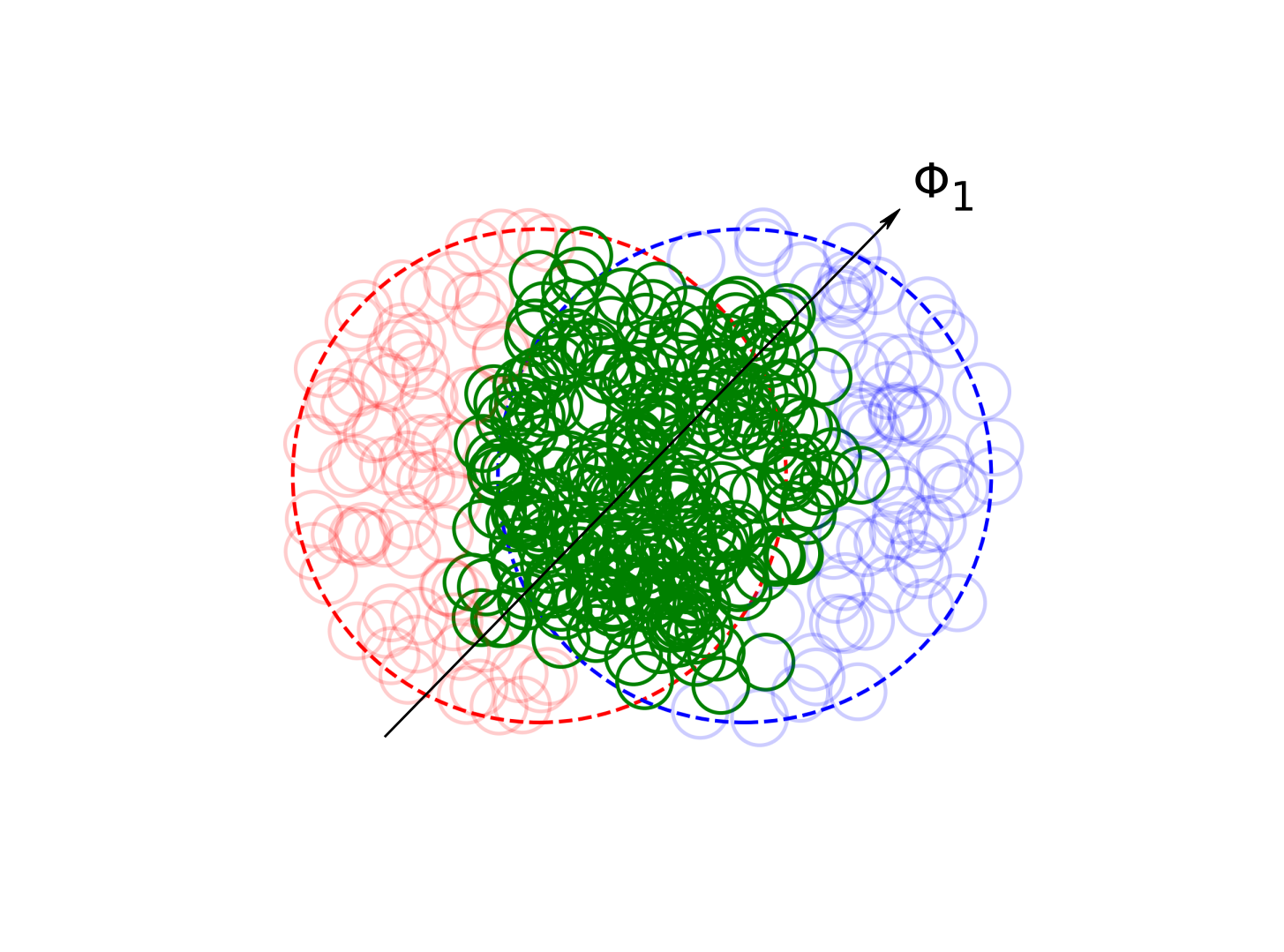}
\end{subfigure}~~
\begin{subfigure}{0.33\textwidth}
\centering
\includegraphics[height=6 cm]{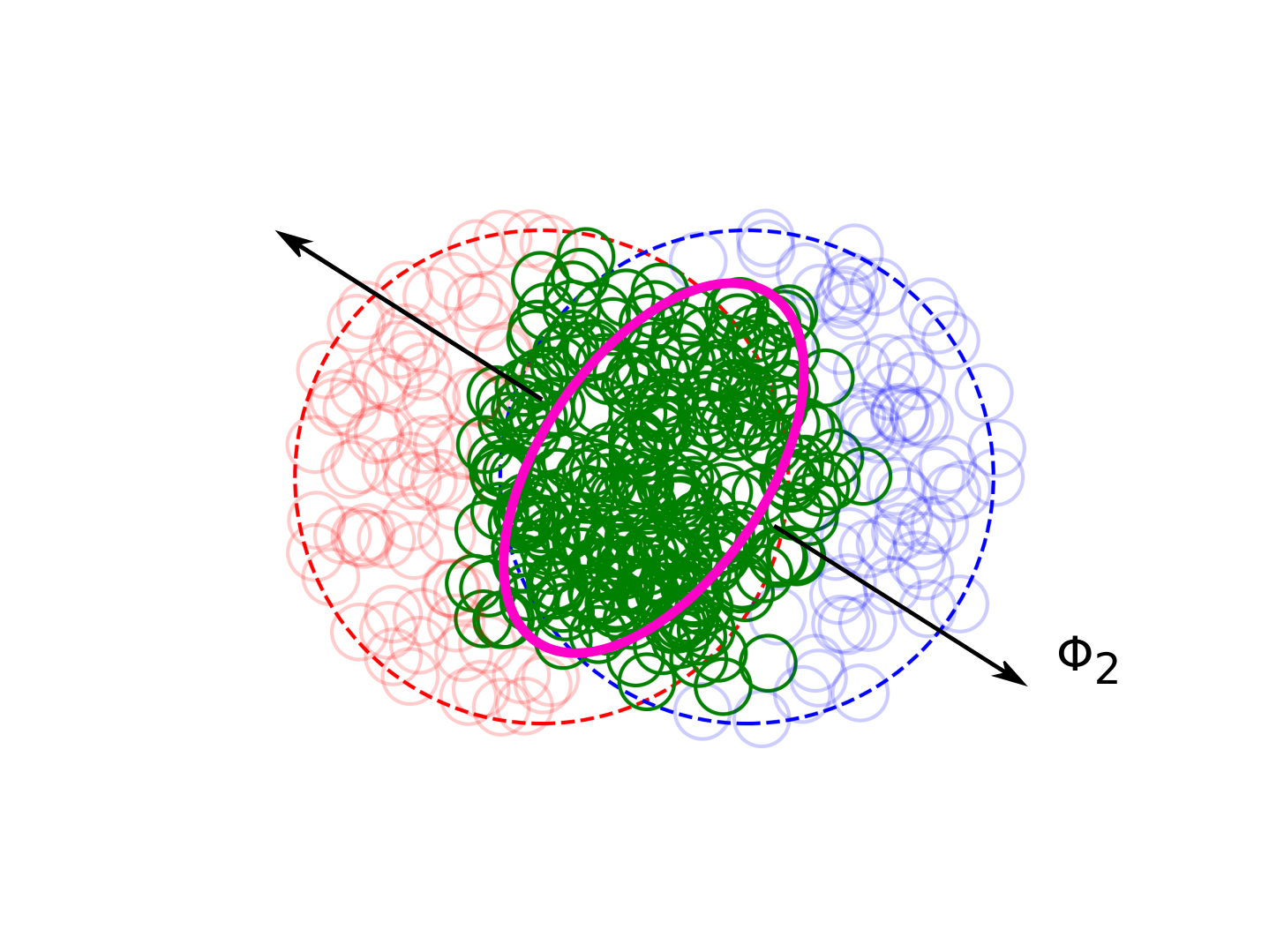}  
\end{subfigure}~~~~~~
\begin{subfigure}{0.33\textwidth}
\centering
\includegraphics[height=6 cm]{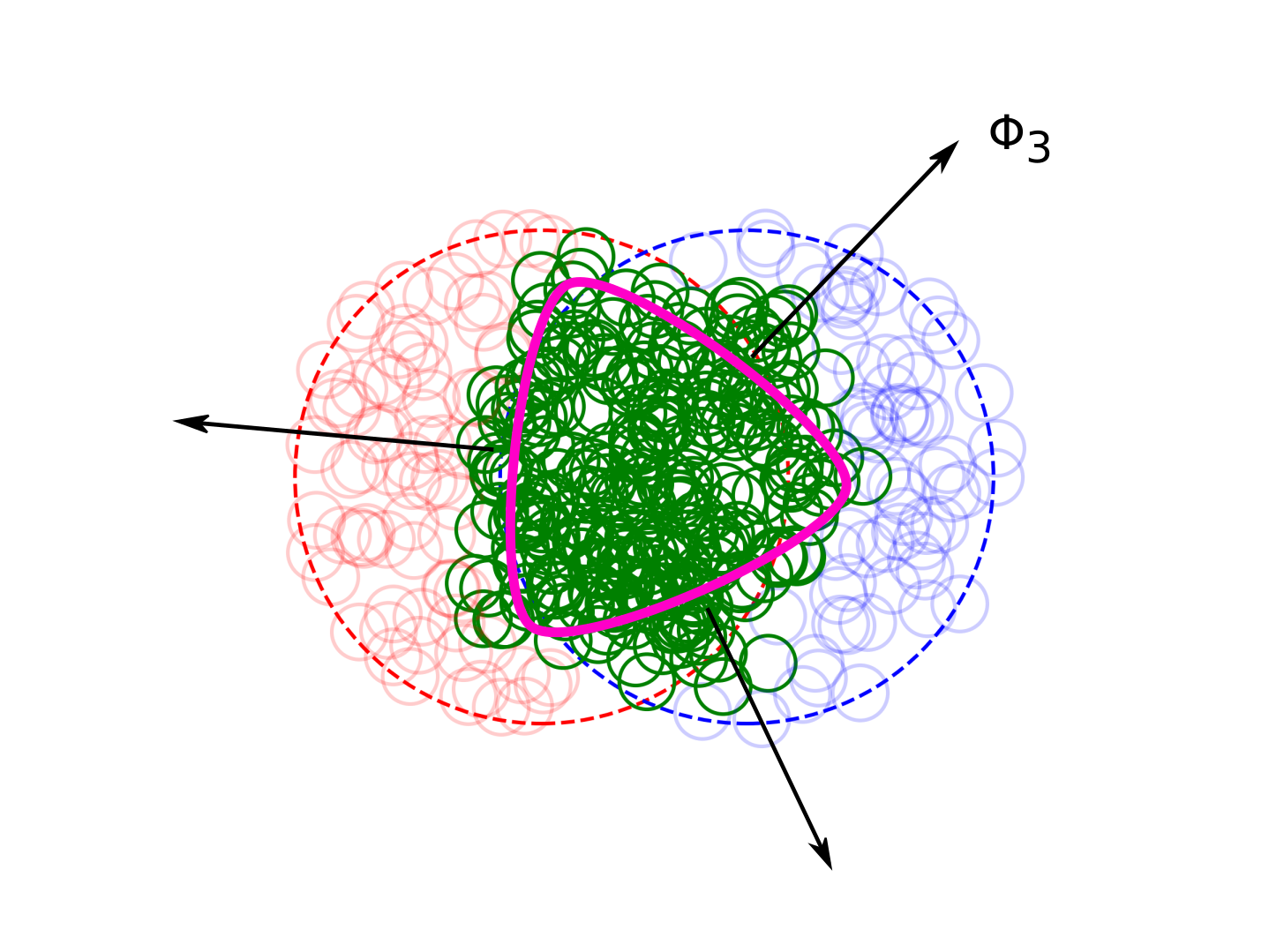}  
\end{subfigure}
\centering
\caption{Pictorial representation of the participant eccentricity harmonics. The principal axes or participant plane angles corresponding to the dipole (left), elliptic (middle) and triangular asymmetry (right) are shown.}
\label{fig: partcipant eccentricities}
\end{figure}

The total entropy and energy per unit rapidity at the initial state is given by, 
\begin{equation}
 \begin{aligned}
  S \propto \int s(r,\phi) \ r \ dr \ d\phi  \eqsp{and} E_i \propto \int \epsilon(r,\phi) \ r \ dr \ d\phi \ ,
 \end{aligned}
\label{eq: total energy and entropy density at initial state}
\end{equation}
and the transverse size of the fireball is quantified by the RMS radius,
\begin{equation}
 \begin{aligned}
  R^2 = \frac{\int r^2 \ s(r,\phi) \ r \ dr \ d\phi }{\int s(r,\phi) \ r\ dr\ d\phi } \ . 
 \end{aligned}
\label{eq: rms radius of final satate}
\end{equation}

\subsection{Particle spectra}
\label{Spectra}
\begin{figure}[ht!]
\begin{subfigure}{0.5\textwidth}
\centering
\includegraphics[height=9 cm]{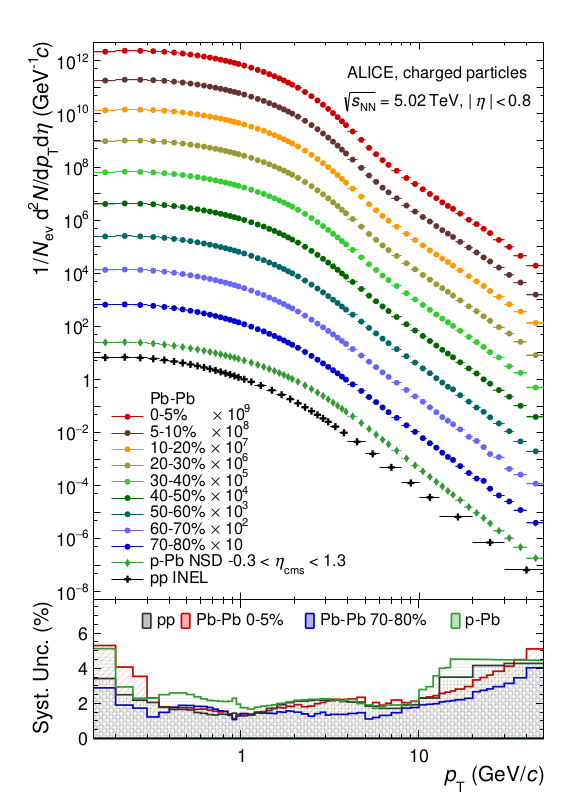}
\end{subfigure}~
\begin{subfigure}{0.5\textwidth}
\centering
\includegraphics[height=9 cm]{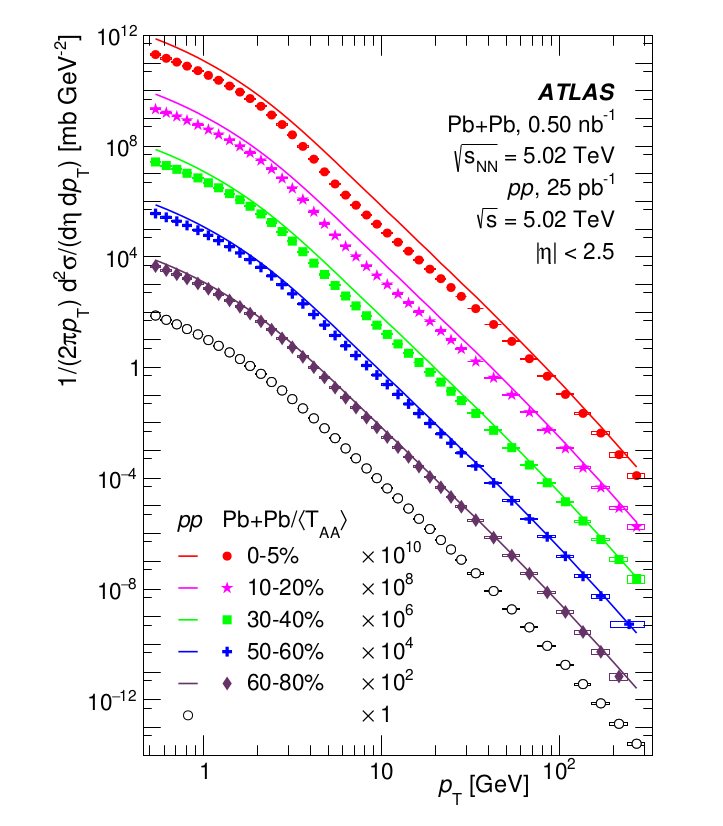}  
\end{subfigure}
\centering
\caption{Event-averaged charged-particle multiplicity spectra measured by the ALICE (left) and ATLAS (right) collaboration for Pb+Pb collision at 5.02 TeV is shown. Figure taken from~\cite{ALICE:2018vuu} and \cite{ATLAS:2022kqu} respectively.}
\label{fig: Data for charged particle spectra}
\end{figure}

To make a quantitative description of the anisotropic flow, first we need to discuss the particle spectra, as it includes a collective effect of all the particles in the final state. At the end of the hydrodynamic evolution, from the freeze-out hypersurface, we obtain the transverse momentum ($p_T$) distribution of the particles through Cooper-Frye freezeout method. In principle, the distribution could be obtained for individual particle species (identified particles), but in this document we will only consider the charged particles ($\pi^+,\pi^-,K^+,K^-,p,\bar{p}$). The distribution is given by,    
\begin{equation}
 \begin{aligned}
  f(p_T,\phi) = \frac{dN}{p_T dp_T d\phi},
 \end{aligned}
\label{eq: charged particle spectra}
\end{equation}
which serves as the probability distribution for the particles carrying transverse momentum $\vec{p_T}$, and $\phi$ ($\equiv \phi_p$) is the azimuthal angle corresponding to the transverse momentum $p_T$. The charged particle spectra from ALICE~\cite{ALICE:2018vuu} and ATLAS~\cite{ATLAS:2022kqu} collaboration for Pb+Pb collision at $\sqrt{s_{NN}}$=5.02 TeV, are shown in Fig.~\ref{fig: Data for charged particle spectra}. In experiments, the charged particle spectra are calculated in pseudorapidity bins\footnote{Although in Eq.~(\ref{eq: charged particle spectra}) we present the two dimensional spectra or just the transverse momentum spectra, which is relevant for boost invariant calculations and description of anisotropic flow, in practice we need to consider the full phase-space ( $y$ or $\eta$, $p_T$ $\phi$) or the full spectra obtained from the freeze-out hypersurface i.e. the spectra is calculated in either rapidity or pseudo-rapidity bins} i.e. $\frac{dN}{p_T dp_T d\phi d\eta}$, whereas in our hydro-calculation we use rapidity bins for the spectra i.e. $\frac{dN}{p_T dp_T d\phi dy}$. For the charged particle spectra, these two have similar meaning and the difference is negligible. However, for the identified particle spectra these two differ because $y$ depends on the mass of the particle whereas $\eta$ does not. For this reason, in experiments the identified particle spectra are usually measured in rapidity bins~\cite{ALICE:2013mez,ALICE:2019hno}. 

The charged particle spectra can be used to calculate the total charged particle multiplicity in an event,
\begin{equation}
 \begin{aligned}
  N_{ch} \equiv N =\int p_T dp_T d\phi f(p_T,\phi) = \int_{p_{min}}^{p_{max}} dp_T \ \frac{dN}{dp_T}.
 \end{aligned}
\label{eq: charged partcile multiplicity}
\end{equation}
Another very important collective observable is the mean transverse momentum per particle in an event, which we denote as $[p_T]$ and it is defined as,

\begin{equation}
 \begin{aligned}
  [p_T] = \frac{1}{N} \int_{p_{min}}^{p_{max}} dp_T \ p_T \ \frac{dN}{dp_T} \ .
 \end{aligned}
\label{eq: mean transverse momentum per particle}
\end{equation}
Event-by-event fluctuations of $[p_T]$ are also of great importance and contain interesting information of the QGP, which we will study in detail in the next chapter.

\subsubsection{Radial flow}
In the present context, it will be useful to mention another type of flow, called the {\it radial flow}, which does not originate from the anisotropy of the initial density profile, but is due to rotationally symmetric collective transverse fluid motion. In the case of rotational symmetric and baryonless fluid, the transverse momentum distribution at zero rapidity ($p_z=0$), for any particle species is given by the Cooper-Frye formula~\cite{Ollitrault:2007du},
\begin{equation}
 \begin{aligned}
  \frac{dN}{2\pi p_T dp_T} \propto \exp\bigg(-\frac{E^*}{T_{fo}}\bigg) = \exp\bigg(-\frac{p^\mu u_\mu}{T_{fo}}\bigg) = \exp\bigg(-\frac{m_T u_0 + p_T v }{T_{fo}}\bigg),
 \end{aligned}
\label{eq: spectra for radial flow}
\end{equation}
where we consider the Maxwell-Boltzman statistics for simplicity. $E^*$ is the energy of the particles in the lab frame, $v$ is the maximum transverse velocity of the fluid at zero-rapidity according to Bjorken scaling and $m_T$ is the transverse mass of the particles eventually depicting the energy of the particles in fluid rest frame with $p_z=0$. If the fluid is at rest in lab frame, $v =0$ and $u_0=1$, which makes the spectra exponential in $m_T$ with slope $1/T_{fo}$ i.e. $\propto \exp(\frac{-m_T}{T_{fo}})$. This means that the particle-spectra are scaled according to the transverse mass of each species but having the same slope $1/T_{fo}$. This describes only the thermal motion of the particles. However, if the fluid moves with a transverse velocity $v$, then on top of the thermal motion, the particles have a collective velocity (the fluid velocity) $v$, which increases the overall kinetic energy of the particles. Due to the collective motion, the heavier particles attain larger kinetic energy and this breaks the $m_T$-scaling of the spectra, because for a given $m_T$, heavier particles have smaller $p_T$. As a result, the slope of the $m_T$-spectra becomes flatter for heavier particles~\cite{Lee:1990sk,Hung:1997du,Huovinen:2001cy}. Such transverse collective flow is known as the radial flow, named because of the radial or axial symmetry.      

\subsection{Flow harmonics}
\label{flow harmonics}
In section~\ref{Spectra}, we have identified the magnitude of the transverse momentum in terms of mean transverse momentum per particle in an event $[p_T]$. Now we want to characterize the directions or the azimuths associated with $p_T$, as the anisotropic flow originates from the anisotropy of the azimuthal distribution of the particles. The azimuthal distribution of the particles is quantified event-by-event in terms of Fourier expansion with respect to the azimuthal angle~\cite{Voloshin:1994mz,Poskanzer:1998yz,Voloshin:2008dg,Bozek:2021mov,Bozek:2021zim},
\begin{equation}
 \begin{aligned} 
  \frac{dN}{p_Tdp_Td\phi} = \frac{dN}{2\pi p_Tdp_T} \bigg( 1+2\sum_{n=1}^{\infty} V_n(p_T) e^{-in\phi} \bigg) \ ,
 \end{aligned}
\label{eq: fourier expansion of flow harmonics w.r.t event plane}
\end{equation}
where the Fourier coefficients $V_n$ are known as $n^{th}$ order {\it harmonic flow coefficients}, which depend on $p_T$ i.e. $V_n = V_n(p_T)$. It could be decomposed as $V_n(p_T)=v_n(p_T)e^{in\Psi_n(p_T)}$ such that $V_n(p_T)$ is interpreted as {\it flow vector} with {\it flow magnitude} $v_n(p_T)$ and {\it flow angle} $\Psi_n(p_T)$\footnote{Please note that $v_n$ and $\Psi_n$ depend on both $p_T$ and $\eta$ i.e.  $v_n = v_n(p_T,\eta)$ and $\Psi_n=\Psi_n(p_T,\eta)$, if instead of only on the transverse plane, we consider full 3D distribution, $\frac{dN}{p_Tdp_Td\phi d\eta}$}. The angle $\phi$ is the azimuthal angle of the particles and $\Psi_n$ is known as the {\it event plane angle}~\cite{Luzum:2013yya, Heinz:2013th, Busza:2018rrf} which serves as a proxy or an estimate of the reaction plane (orientation of which is not known experimentally) in each event and could be determined independently for each harmonic of anisotropic flow. 

\subsubsection{Differential and integrated flow}
The harmonic flow coefficient in Eq.~(\ref{eq: fourier expansion of flow harmonics w.r.t event plane}), $V_n(p_T)$ is a function of the transverse momentum $p_T$ in an event and it also depends on the pseudo-rapidity($\eta$) i.e. $V_n(p_T,\eta)$, when full three dimensional spectra are considered. In that case, it is called $p_T$-differential or $\eta$-differential flow or in general {\it differential flow}. To calculate the flow harmonics over the whole phase space in the transverse plane, $V_n(p_T)$ should be integrated with respect to the distribution,
\begin{equation}
 \begin{aligned}
  V_n = \frac{1}{N} \int_{p_{min}}^{p_{max}} dp_T \ V_n(p_T) \ \frac{dN}{dp_T},
 \end{aligned}
\label{eq: integrated flow in an event}
\end{equation}
where $V_n$ is known as {\it integrated flow} or specifically {$p_T$-integrated flow} in an event. In Fig.~\ref{fig: Integrated and differential flow from cumulant method}, differential and integrated elliptic and triangular flow for various centralities in Pb+Pb collision have been shown. 

\subsection{Different types of flow and their relation to initial anisotropy}
\label{types of flow}
 
For different order $n$, we get different type of flow e.g. $v_2$ is called {\it elliptic flow} , $v_3$ is called {\it triangular flow}, $v_4$ is known as {\it quadrangular flow}, $v_5$ is called {\it pentagonal flow} etc, which are driven by participant eccentricities $\Epsilon_n$ at the initial state. For $n=1$, the flow is known as {\it directed flow} $v_1$ which at the mid-rapidity is driven by dipole asymmetry~\cite{Teaney:2010vd} $\Epsilon_1$ of the initial state.

\subsubsection{Elliptic flow}
\begin{figure}[ht!]
\includegraphics[height=6.5 cm]{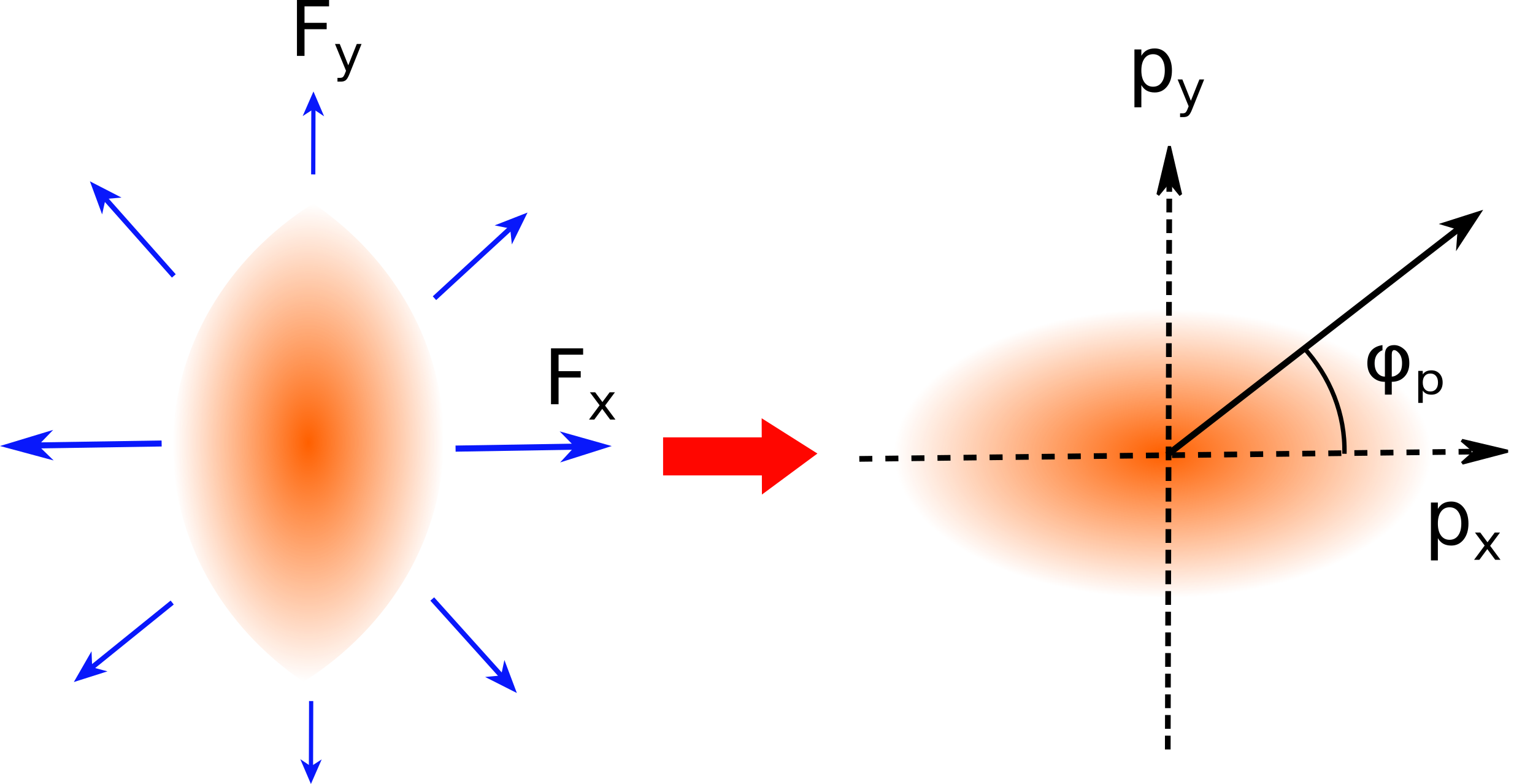}
\centering
\caption{Schematic representation of the origin of elliptic flow in a non central heavy-ion collision. The left hand side shows the formation of almond shaped or elliptic geometry at the initial state, creating a pressure gradient and hence an anisotropic outward force. The right hand side shows the development of the momentum anisotropy at the final state due to the pressure gradient, leading to the elliptic flow of particles.}
\label{fig: origin of elliptic flow}
\end{figure}

Elliptic flow, originally proposed by J-Y. Ollitrault~\cite{Ollitrault:1992bk}, is identified as one of the most peculiar and significant signature of the collective flow~\cite{Sorge:1996pc,Poskanzer:1998yz,Huovinen:2001cy,Teaney:2000cw}. It originates mainly due to the ellipsoidal geometry at the initial state of the collision or quadrupole asymmetry of the initial density profile. If the fireball is elliptic shaped, then it develops a pressure gradients $\nabla P$ which results in larger transverse momentum of the particles that are emitted in the direction of reaction plane because of the larger fluid velocity in that direction ( larger pressure gradient results in larger force : $\vec{F} = - \nabla P$), and smaller momentum for the particles which are emitted in the direction perpendicular to the reaction plane (Fig.~\ref{fig: origin of elliptic flow}). 
This azimuthal anisotropy of the transverse momentum is reflected through the elliptic flow of particles and identified by the harmonic coefficient $v_2$. In non-central collision, the elliptic flow $v_2$ is mainly due to the elliptic geometry of the initial state, which is characterized by the eccentricity or ellipticity $\Epsilon_2$ (Eq.~(\ref{eq: conventional eccentricity harmonics general form})). There exist direct phenomenological relation between the two~\cite{Alver:2010gr,Gardim:2011xv},
\begin{equation}
 \begin{aligned} 
  v_2 e^{2i\Psi_2}= k_2 \Epsilon_2e^{2i\Phi_2} \eqsp{} \Rightarrow \eqsp{} v_2 \simeq k_2\Epsilon_2 \ ,
 \end{aligned}
\label{eq: relation between elliptic flow and eccentricity}
\end{equation}
where it is assumed that the event plane $\Psi_2$ approximately coincides with the participant plane $\Phi_2$. The coefficient $k_2$ is the {\it hydrodynamic response coefficient}, which depends on the properties of the QGP medium. 

\begin{figure}[ht!]
\hspace{-0.3 cm}\begin{subfigure}{0.5\textwidth}
\centering
\includegraphics[height=6 cm]{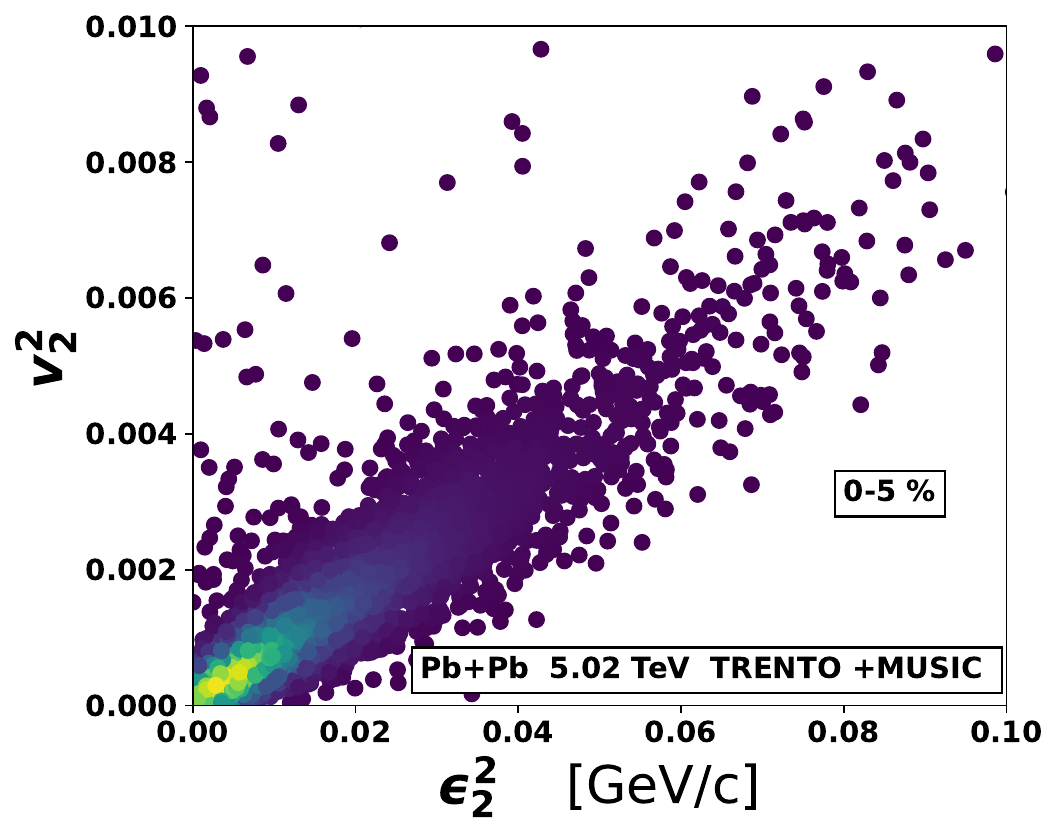}
\end{subfigure}~~
\begin{subfigure}{0.5\textwidth}
\centering
\includegraphics[height=6 cm]{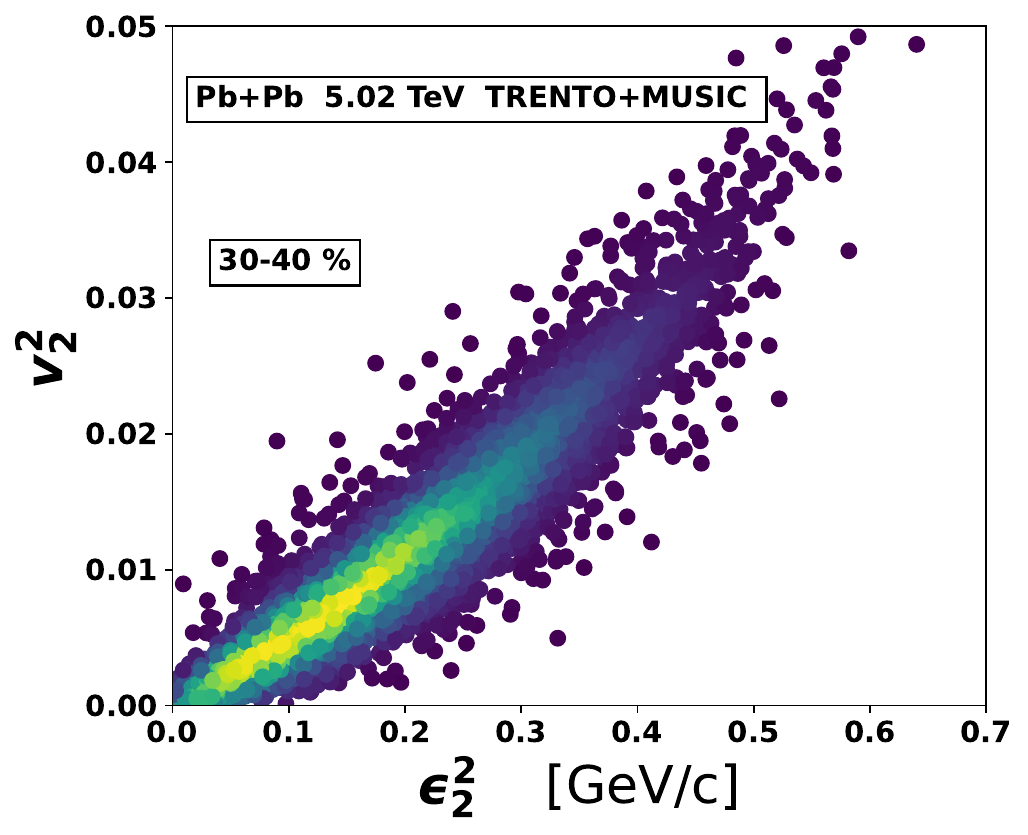}  
\end{subfigure}
\centering
\caption{Scatter plot between $v_2^2$ and $\Epsilon_2^2$ for $0-5 \%$ (left) and $30-40\%$ (right) centrality in Pb+Pb collision at 5.02 TeV with TRENTO initial condition.}
\label{fig: v2 vs eps2 plot}
\end{figure}

However, in central collisions (say, $0-5$ \%), in addition to the elliptic geometry, event-by-event fluctuations of the initial state plays a major role in contributing to the quadrupole asymmetry $\Epsilon_2$, generating elliptic flow in central collision, and this effect become dominant in ultracentral collisions ($0-1$ \%). This might not be obvious at first instance when one thinks of $v_2$ in a single event, however in practice only the event-averaged quantities are important, where fluctuations and its contributions play a crucial role. Generally in central collision, event-by-event fluctuations of $\Epsilon_2$, fluctuations of $\Psi_2$ around $\Phi_2$ and even fluctuations of $v_2$ for a given $\epsilon_2$, all contribute to the event averaged $v_2$.  We will explain this fact while discussing experimental method for the measurement of $v_2$. The scatter plot between $v_2^2$ and $\Epsilon_2^2$ for central and semi-central Pb+Pb collision at 5.02 TeV are shown in Fig.~\ref{fig: v2 vs eps2 plot}. As discussed, the correlation is stronger in case of semi-central collision ($30-40 \%$) where elliptic geometry of the overlap are dominates, whereas the central collision shows significant effect of event-by-event fluctuations.  

\subsubsection{Triangular flow}
Similar to the elliptic flow, the triangular flow $v_3$ originates from the third order participant eccentricity or triangualrity $\Epsilon_3$~\cite{Alver:2010gr,Gardim:2019xjs}, defined in (Eq.~(\ref{eq: conventional eccentricity harmonics general form})). It also follows the approximated phenomenological relation,
\begin{equation}
 \begin{aligned} 
  v_3 \simeq k_3 \Epsilon_3,
 \end{aligned}
\label{eq: relation between triangualar flow and eccentricity}
\end{equation}
where $k_3$ is the hydro-response coefficient for the triangular flow. 
\begin{figure}[ht!]
\hspace{-0.3 cm}\begin{subfigure}{0.5\textwidth}
\centering
\includegraphics[height=6 cm]{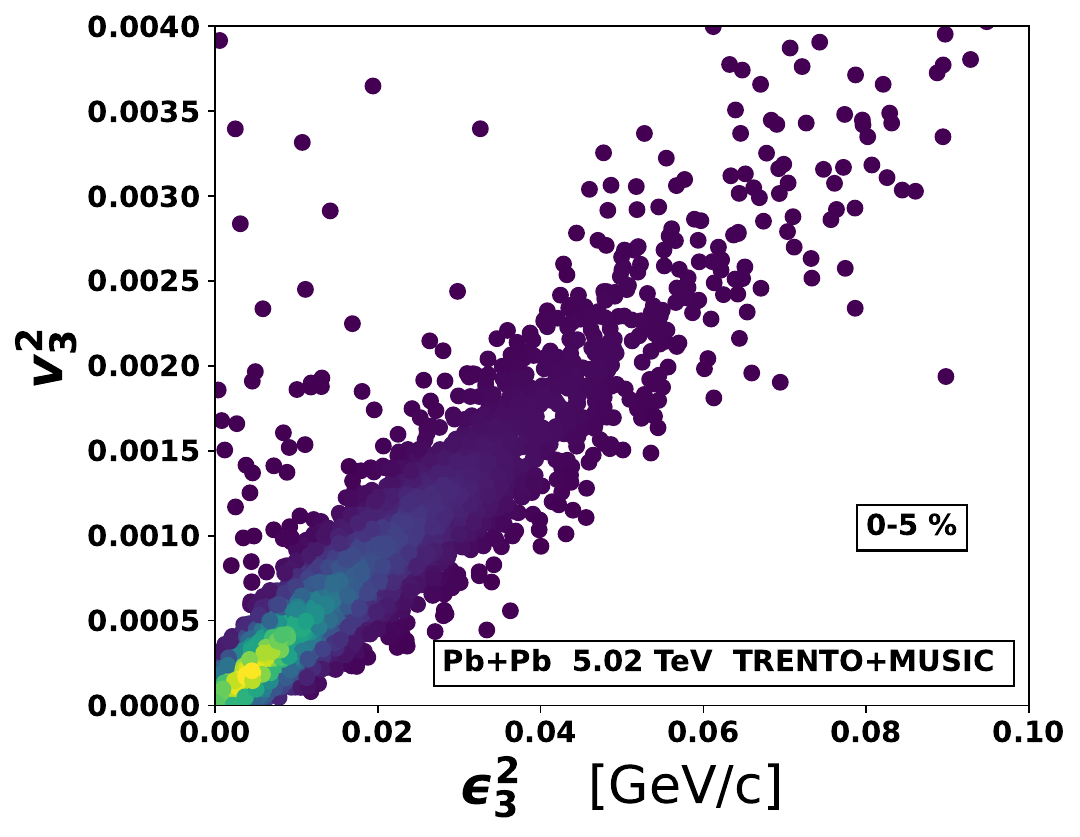}
\end{subfigure}~~
\begin{subfigure}{0.5\textwidth}
\centering
\includegraphics[height=6 cm]{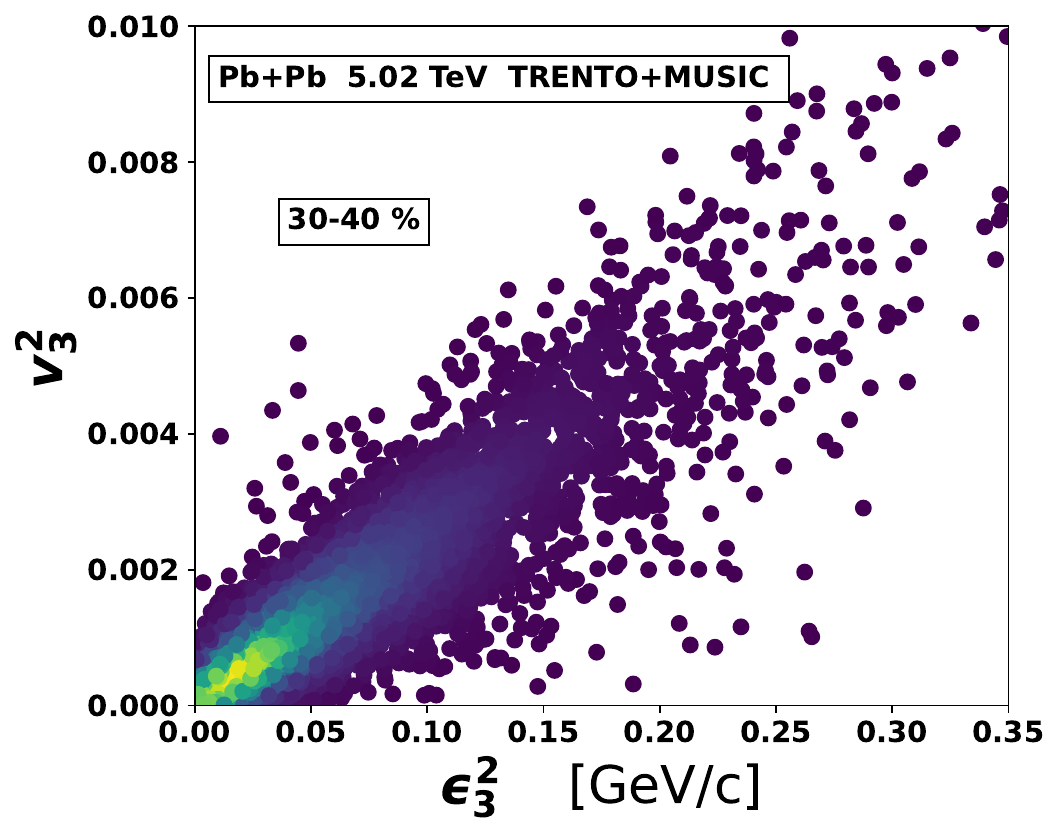}  
\end{subfigure}
\centering
\caption{Scatter plot between $v_3^2$ and $\Epsilon_3^2$ for $0-5 \%$ (left) and $30-40\%$ (right) centrality in Pb+Pb collision at 5.02 TeV with TRENTO initial condition.}
\label{fig: v3 vs eps3 plot}
\end{figure}
The value of the response coefficients $k_n$ can be estimated using the linear relationship between flow and eccentricity~\cite{Gardim:2011xv},
\begin{equation}
 \begin{aligned} 
  v_n e^{in\Psi_n} = k_n \Epsilon_n e^{in\Phi_n} \eqsp{} \Rightarrow \eqsp{} k_n = \frac{\langle v_n \Epsilon_n \cos (n(\Psi_n - \Phi_n)) \rangle}{ \langle \Epsilon_n^2 \rangle} \ ,
 \end{aligned}
\label{eq: estimating response coefficient}
\end{equation}
where the angular bracket $\langle \dots \rangle$ denotes the average over all events\footnote{It should be noted that Eq.~(\ref{eq: relation between elliptic flow and eccentricity}) and Eq.~(\ref{eq: relation between triangualar flow and eccentricity}) only imply a strong correlation (linear) between $v_n$ and $\Epsilon_n$ for $n=2,3$. The ratio $v_n/\Epsilon_n$ (even if they are event averaged values) should never be taken as an estimation of $k_n$ which should always be estimated by Eq.~(\ref{eq: estimating response coefficient}).}. Unlike elliptic flow, the triangularity or octupole asymmetry $\Epsilon_3$, driving the triangular flow $v_3$, is largely dominated by the contribution from event-by-event fluctuations of the initial density profile, even in non-central collision and especially in central collision, as depicted in Fig.~\ref{fig: v3 vs eps3 plot}. Fluctuations at the initial state are more important for the triangular flow.

\subsubsection{Quadrangular and pentagonal flow}
The fourth order ($n=4$) harmonic flow is called quadrangular flow, $v_4$. Similar to the elliptic and triangular flow one could expect that $v_4$ also originates from fourth order harmonic eccentricity of the initial state, $\Epsilon_4$. However, it has been observed that the estimator is only valid for central collision but fails to predict $v_4$ for peripheral events. This happens because in the peripheral collisions, due to the elliptic shape of the participant area, the second order moment $\Epsilon_2$ becomes very dominant or much larger in comparison to higher order moments such as $\Epsilon_4$~\cite{Gardim:2011xv}. As a result, there exist a possibility of non-linear contributions from lower order eccentricities to the flow harmonic $v_4$. The first choice of such contribution would be $\Epsilon_2^2$ by symmetry\footnote{Here we mean rotational symmetry i.e. the quantity $\langle V_4(\Epsilon_n e^{-in\Phi_n})^m \rangle$ must be invariant under rotation.}, so that the actual predictor for $v_4$ is given by, 
\begin{equation}
 \begin{aligned} 
  v_4 e^{4i\Psi_4} = k_4 \Epsilon_4 e^{4i\Phi_4} + k_4' \Epsilon_2^2 e^{4i\Phi_2},
 \end{aligned}
\label{eq: relation between quadrangular flow and eccentricities}
\end{equation}
which serves as a very good estimator for all centralities. For central collisions, both $\Epsilon_4$ and $\Epsilon_2$ are driven by fluctuations and so they are small. As a result, the contribution of $\Epsilon_4$ dominates in terms of eccentricities.  

For $n=5$, the flow is is called pentagonal flow, $v_5$ which also assumes non-linear contribution from the initial state anisotropies, even for central collision.  In addition to $\Epsilon_5$, by symmetry the non-linear contribution comes from $\Epsilon_2\Epsilon_3$ and the estimator for $v_5$ reads~\cite{Gardim:2011xv},
\begin{equation}
 \begin{aligned} 
  v_5 e^{5i\Psi_5} = k_5 \Epsilon_5 e^{5i\Phi_5} + k_5' \Epsilon_2e^{2i\Phi_2}\Epsilon_3 e^{3i\Phi_3},
 \end{aligned}
\label{eq: relation between pentagonal flow and eccentricities}
\end{equation}
which serves as a very good predictor for $v_5$ for all centralities.

\subsubsection{Directed flow}
For $n=1$ the flow is known as directed flow $v_1$ which could be separated into two parts depending on rapidity of the particles $y$ : {\it rapidity-odd} directed flow and {\it rapidity-even} directed flow\cite{Gardim:2011qn}. The rapidity odd $v_1$ is the usual directed flow~\cite{STAR:2003xyj,PHOBOS:2005ylx,STAR:2008jgm} which largely depends on the space-time rapidity profile ($\eta_s$) of the fireball, driven by the tilt of the fireball at the initial phase of the evolution~\cite{Bozek:2010bi,Bozek:2010vz,Chatterjee:2017ahy,Parida:2022ppj}. On the other hand, the rapidity-even $v_1$, which shows a little dependence on $y$, is interesting at the mid-rapidity and it originates from the dipole asymmetry of the initial transverse density profile, characterized by $\Epsilon_1$~\cite{Teaney:2010vd}. 

Event-by-event fluctuations of the initial geometry break the symmetry of the transverse density profile and make it steepest in a particular direction, which is quantified as the dipole asymmetry~\cite{Teaney:2010vd,Luzum:2010fb},
\begin{equation}
 \begin{aligned} 
  \Epsilon_1 e^{i\Phi_1} = -\frac{\{ r^3 e^{i \phi_1}\}}{\{r^3 \}},
 \end{aligned}
\label{eq: dipole asymmetry}
\end{equation}
contributing to $v_1$ in a similar manner as fluctuation-induced quadrupole asymmetry $\Epsilon_2$ contributes to $v_2$ and octupole asymmetry $\Epsilon_3$ contributes to $v_3$. The dipole asymmetry creates a gradient in the transverse density profile, resulting in largest fluid velocity along the direction of the steepest gradient having azimuth $\Phi_1$. As a result, the particles with larger $p_T$ are emitted along the direction of $\Phi_1$ and the small-$p_T$ particles are emitted in the opposite direction~\cite{Luzum:2010fb}. Due to the conservation of total transverse momentum, $v_1$ is positive for high-$p_T$ particles and negative for small-$p_T$ particles giving rise to a specific pattern for the $p_T$ dependence of $v_1$ ~\cite{Teaney:2010vd,Luzum:2010fb}.

\subsection{Methods of flow analysis }
\label{flow analysis}
Anisotropic flow is originally defined as the azimuthal correlation of the outgoing particles with the reaction plane. But experimentally the reaction plane orientation cannot be measured. That is why the flow is represented in terms of event plane angle which we also call the flow angle. If we drop the $p_T$ dependence for a moment or just consider the integrated flow, then we can write only the azimuthal distribution of the particles from Eq.~(\ref{eq: fourier expansion of flow harmonics w.r.t event plane}) as, 
\begin{equation}
 \begin{aligned}
  f(\phi)=\frac{dN}{d\phi} \propto \bigg[1+2\sum_{n=1}^{\infty} v_n e^{in(\phi - \Psi_n)}\bigg], 
 \end{aligned}
\label{eq: fourier expansion of flow harmonics from azimuthal distribution event plane}
\end{equation}
where $\phi$ is the azimuthal angle of the particle and $\Psi_n$ is the event-plane angle. Then the flow harmonics $v_n$ can be estimated as~\cite{Poskanzer:1998yz,Borghini:2000sa},
\begin{equation}
 \begin{aligned}
  v_n = \langle \cos [n(\phi-\Psi_n)] \rangle,
 \end{aligned}
\label{eq: flow harmonics from azimuthal distribution EP }
\end{equation}
where the angular bracket $\langle \dots \rangle$ denotes the average over all particles and over all events. The sine term does not appear because of the reflection symmetry and it cancels out when averaged over events. This method of estimating flow harmonics is called the {\it event-plane method} which is obsolete and no longer in use. Modern experimental methods of flow analysis in heavy-ion collision use the {\it cumulant method} involving multi-particle azimuthal correlations. Below we discuss the theoretical background and experimental implementation of the method~\cite{Borghini:2000sa,Borghini:2001vi,Bilandzic:2010jr}.

\subsubsection{Cumulant method: multi-particle correlations}
As the actual orientation of the reaction plane is unknown in experiments, instead of considering a single particle, if we consider the relative azimuthal angles between the outgoing particles, then it could capture the correlation with the reaction plane and eventually provide an estimate of the flow~\cite{Borghini:2000sa,Borghini:2001vi}. The simplest case would be a two-particle azimuthal correlation which could be expressed as,
\begin{equation}
 \begin{aligned}
\langle \cos[n(\phi_1-\phi_2)] \rangle \equiv \langle e^{in(\phi_1 - \phi_2)} \rangle,
 \end{aligned}
\label{eq: two particle azimuthal correlation}
\end{equation}
which comes from the two-particle azimuthal distribution in an event $f(\phi_1,\phi_2)\equiv\frac{dN}{d\phi_1 d\phi_2} $ and the angular bracket has the similar meaning as before i.e. first an average over all the pairs of particle in an event and then the average over all the events. The two-particle distribution in an event can be decomposed as, 
\begin{equation}
 \begin{aligned}
 f(\phi_1,\phi_2) = f(\phi_1) f(\phi_2) + f_c(\phi_1,\phi_2),
 \end{aligned}
\label{eq: decomposition two particle azimuthal distribution}
\end{equation}
where the first term on the r.h.s. represents the product of the uncorrelated distributions and the second term denotes the correlated distribution. Accordingly, Eq.~(\ref{eq: two particle azimuthal correlation}) can be decomposed as, 
\begin{equation}
 \begin{aligned}
 \langle e^{in(\phi_1 - \phi_2)} \rangle = \langle e^{in\phi_1} \rangle \langle e^{in\phi_2} \rangle + \llangle e^{in(\phi_1 - \phi_2)} \rrangle \ .
 \end{aligned}
\label{eq: decomposition two particle azimuthal correlation}
\end{equation}

The terms $\langle e^{in\phi_1} \rangle$ and $\langle e^{in\phi_2} \rangle$ vanishes because the angles $\phi_1$ and $\phi_2$ are measured in the laboratory frame, assuming that the detector is a `perfect detector' covering the entire acceptance region\footnote{ As the azimuths of the particles are randomly oriented, average over all the events give zero }. The second term $\llangle e^{in(\phi_1 - \phi_2)} \rrangle$ represents the {\it genuine} two-particle correlation which can include contribution from anisotropic flow as well as from other sources e.g. due to global momentum conservation, resonance decay (in which the decay products are correlated), final state coulomb, strong or quantum interactions etc.~\cite{Dinh:1999mn,Borghini:2000cm}, called the {\it non-flow correlations} or `direct' correlations. If the source was isotropic, $\llangle e^{in(\phi_1 - \phi_2)} \rrangle$ would represent only the non-flow correlations. 

The quantity $\llangle e^{in(\phi_1 - \phi_2)} \rrangle$ is known as {\it cumulant} or specifically {\it two-particle cumulant}, which can be obtained using a generalized cumulant expansion method from the cumulant generating function~\cite{Borghini:2000sa,Borghini:2001vi} and written as,  
\begin{equation}
 \begin{aligned}
 c_n\{2\} \equiv \llangle e^{in(\phi_1 - \phi_2)} \rrangle  \ .
 \end{aligned}
\label{eq: two particle cumulant}
\end{equation}
The contribution of the flow to the two-particle cumulant is given by~\cite{Borghini:2001vi},
\begin{equation}
 \begin{aligned}
 v_n\{2\}^2=c_n\{2\} \eqsp{} \Rightarrow \eqsp{}  v_n\{2\} = \sqrt{c_n\{2\}},
 \end{aligned}
\label{eq: flow from two particle cumulant}
\end{equation}
where $v_n\{2\}$ represents the $n^{th}$ order harmonic flow estimated from two-particle cumulant. The advantage of this method lies in its ability to reduce the non-flow correlation, which is achieved by constructing higher order cumulants or multi-particle cumulants. For example, similar to Eq.~(\ref{eq: two particle azimuthal correlation}), if we consider four particle azimuthal correlation, then corresponding four-particle cumulant could be written as~\cite{Borghini:2000sa,Borghini:2001vi}, 
\begin{equation}
 \begin{aligned}
 c_n\{4\} &\equiv \llangle e^{in(\phi_1 +  \phi_2-\phi_3-\phi_4)} \rrangle \\
 &= \langle e^{in(\phi_1 +  \phi_2-\phi_3-\phi_4)} \rangle - \langle e^{in(\phi_1 -\phi_3)} \rangle \langle e^{in(\phi_2 -\phi_4)} \rangle - \langle e^{in(\phi_1 -\phi_4)} \rangle \langle e^{in(\phi_2 -\phi_3)} \rangle \\
 &= \langle e^{in(\phi_1 +  \phi_2-\phi_3-\phi_4)} \rangle - 2 \langle e^{in(\phi_1 -\phi_3)} \rangle^2 \ ,
 \end{aligned}
\label{eq: four particle cumulant}
\end{equation}
\begin{figure}[ht!]
\hspace{-1 cm}\begin{subfigure}{0.5\textwidth}
\centering
\includegraphics[height=9 cm]{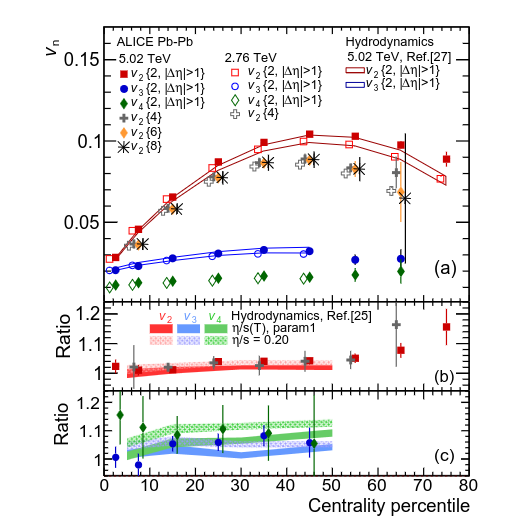}
\end{subfigure}~~~~~~~~~~
\begin{subfigure}{0.5\textwidth}
\centering
\includegraphics[height=9.5 cm]{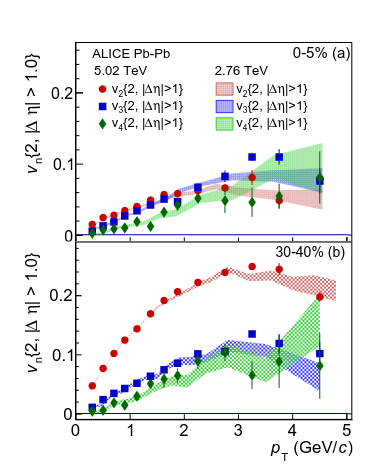}  
\end{subfigure}
\centering
\caption{Measurement of differential and integrated flow using multi-particle cumulant method. The left hand side shows the centrality dependence of the integrated flow cumulants while the right hand side shows differential flow cumulants corresponding to different order harmonics. Figure taken from~\cite{ALICE:2016ccg}.}
\label{fig: Integrated and differential flow from cumulant method}
\end{figure}
where the simplification at the last step is due to the symmetry between $\phi_1$ and $\phi_2$. Please note, in the final expression of $c_n\{4\}$ on the r.h.s, both the first and second term include two-particle non-flow (direct) correlation which is eliminated by the subtraction. Thus by constructing higher order cumulants, we can eliminate the non-flow correlation of sub-leading orders. Similarly one can construct six-particle cumulant $c_n\{6\}$ and corresponding contributions of the flow are given by\cite{Borghini:2001vi},
\begin{equation}
 \begin{aligned}
 -v_n\{4\}^4=c_n\{4\} \eqsp{} \Rightarrow \eqsp{}  v_n\{4\} = (-c_n\{4\})^{1/4} , \\
 \eqsp{and}  4v_n\{6\}^6=c_n\{6\} \eqsp{} \Rightarrow \eqsp{}  v_n\{6\} = (\frac{c_n\{6\}}{4})^{1/6},
 \end{aligned}
\label{eq: flow from four-particle and six-particle cumulant}
\end{equation}
where $v_n\{4\}$ and $v_n\{6\}$ represent the flow estimated from four-particle and six-particle cumulants respectively\footnote{Please note, as 2-particle estimation of flow is non-flow contaminated, it is always larger than 4- or 6-particle estimation. In general the trend follows: $v_n\{2\} > v_n\{4\} \simeq v_n\{6\}$ .}. 

\subsubsection{Experimental method : Q-vector}
Experimentally, the above mentioned cumulant method is used for the estimation of flow harmonics, but in a slightly different way. In experiments, the cumulants are expressed in terms of the moments of the flow vector $Q_n$, called $Q-vector$, in an event, which is defined as~\cite{Bilandzic:2010jr},
\begin{equation}
 \begin{aligned}
 Q_n \equiv \sum_{i=1}^{M} e^{in\phi_i} \ , 
 \end{aligned}
\label{eq: definition of Qn vector}
\end{equation}
where $M$ is the number of particles in the event, used for the analysis. This way, the method is not biased by the interference of various harmonics, does not involve the approximation as used in formalism involving the generating function~\cite{Borghini:2000sa} and can also disentangle the detector effect more efficiently; ideal for experimental realization. As in this approach, the cumulants are calculated without any approximation, directly from the data using the $Q-vector$, it is sometimes referred as the {\it direct cumulant} or {\it Q-cumulant} method~\cite{Bilandzic:2010jr} which is the experimental implementation of the cumulant method discussed just before. 

In the experimental method, first the {\it average} two-particle or four-particle azimuthal correlation are found in a {\it single event}, defined as,
\begin{equation}
 \begin{aligned}
 \overline{2} = \overline{e^{in(\phi_1-\phi_2)}}=& \frac{1}{M(M-1)}\sum_{i\neq j = 1}^M e^{in(\phi_i-\phi_j)},\\
 \overline{4} = \overline{e^{in(\phi_1+\phi_2-\phi_3-\phi_4)}}=& \frac{1}{M(M-1)(M-2)(M-3)}\sum_{i\neq j\neq k\neq l=1}^M e^{in(\phi_i+\phi_j-\phi_k-\phi_l)},
 \end{aligned}
\label{eq: average 2 and 4-particle correlation in an event}
\end{equation}
where the unequal signs between the sum indices denote the removal of self-correlation between the particles and the factor sitting before the sum could be expressed for general k-particle correlator as $\frac{(M-k)!}{M!}$. Then the above correlators are expressed in terms of the moments of the magnitude of $Q_n$-vectors. For example, in case of 2-particle correlator,
\begin{equation}
 \begin{aligned}
 |Q_n|^2 = Q_nQ_n^* = \sum_{i,j=1}^{M}  e^{in(\phi_i-\phi_j)} = M + \sum_{i\neq j}^M e^{in(\phi_i-\phi_j)}
 \end{aligned}
\label{eq: magnitude of Q_n square in terms of two particle correlator}
\end{equation}
and then using Eq.~(\ref{eq: average 2 and 4-particle correlation in an event}), we can write,
\begin{equation}
 \begin{aligned}
 \overline{2}=\frac{|Q_n|^2-M}{M(M-1)} \ . 
 \end{aligned}
\label{eq: two-particle correlator in terms of Q_n square}
\end{equation}
 Then the average over all events~\cite{Bilandzic:2010jr} is found, which gives the two-particle cumulant,
\begin{equation}
 \begin{aligned}
 c_n\{2\}\equiv \langle \cos n(\phi_1-\phi_2) \rangle = \langle \frac{1}{M(M-1)}\sum_{i\neq j = 1}^M e^{in(\phi_i-\phi_j)} \rangle = \langle \overline{2} \rangle \ ,
 \end{aligned}
\label{eq: two-particle cumulant from Q-cumulant}
\end{equation}
from which flow harmonics $v_n\{2\}$ can be estimated using Eq.~(\ref{eq: flow from two particle cumulant}) and which would be of course non-flow contaminated. Please note, here the angular bracket denotes only average over events. Similarly, the fourth moment of the magnitude of $Q_n$ can be written as,
\begin{equation}
 \begin{aligned}
 |Q_n|^4 = Q_nQ_nQ_n^*Q_n^* = \sum_{i,j,k,l=1}^{M}  e^{in(\phi_i+\phi_j-\phi_k-\phi_l)}  \ ,
 \end{aligned}
\label{eq: magnitude of Q_n power 4 in terms of two particle correlator}
\end{equation}
which could involve four different cases for the indices $i,j,k$ and $l$ : i) all are different (4-particle correlator in Eq.~(\ref{eq: average 2 and 4-particle correlation in an event})), ii) three are different iii) two are different and iv) all are same. Expanding all the cases one can obtain~\cite{Bilandzic:2010jr},
\begin{equation}
 \begin{aligned}
 \overline{4}= \frac{|Q_n|^4+|Q_{2n}|^2-2Re[Q_{2n}Q_n^*Q_n^*]-4(M-2)|Q_n|^2+2M(M-3)}{M(M-1)(M-2)(M-3)} \ .
 \end{aligned}
\label{eq: four-particle correlator in terms of Q_n moments}
\end{equation}
Then after taking the average over all events i.e. $\langle \overline{4} \rangle $, the four-particle cumulant can be found following Eq.~(\ref{eq: four particle cumulant}), 
\begin{equation}
 \begin{aligned}
 c_n\{4\}=\langle \overline{4} \rangle - 2\langle \overline{2} \rangle^2 \ .
 \end{aligned}
\label{eq: four-particle cumulant from Q-cumulant}
\end{equation}
and the six-particle cumulant can be obtained in a similar way as~\cite{Borghini:2000sa,Bilandzic:2010jr}, 
\begin{equation}
 \begin{aligned}
 c_n\{6\}=\langle \overline{6} \rangle - 9\langle \overline{2} \rangle \langle \overline{4} \rangle + 12 \langle \overline{2} \rangle^3.
 \end{aligned}
\label{eq: six-particle cumulant from Q-cumulant}
\end{equation}
Finally, the flow harmonics $v_n\{4\}$ and $v_n\{6\}$ can be obtained using Eq.~(\ref{eq: flow from four-particle and six-particle cumulant}), which are non-flow subtracted.

In the present context, let us mention that the $2k$-particle cumulants $c_n\{2k\}$ discussed above, can be expressed as the event average of different powers of the magnitude of harmonic flow $v_n$. Let us consider the 2-particle cumulant from Eq.(\ref{eq: two-particle cumulant from Q-cumulant}) and Eq.~(\ref{eq: average 2 and 4-particle correlation in an event}),
\begin{equation}
 \begin{aligned}
 c_n\{2\} = \langle \overline{2} \rangle = \langle \frac{1}{M(M-1)}\sum_{i\neq j=1}^M e^{in(\phi_i-\phi_j)} \rangle.
 \end{aligned}
\label{eq: two particle cumulant from Q-cumulant revisited}
\end{equation}
Using the definition of flow vector $V_n=v_n e^{in\Psi_n}$ in an event as described in Eq.~(\ref{eq: fourier expansion of flow harmonics from azimuthal distribution event plane}), we can write the quantity within bracket on the r.h.s. of the above equation, as a {\it scalar product} of two flow-vectors in an event,
\begin{equation}
 \begin{aligned}
\frac{1}{M(M-1)}\sum_{i\neq j=1}^M e^{in(\phi_i-\phi_j)} = V_nV_n^* = v_n^2,
 \end{aligned}
\label{eq: two-particle correlation in terms of flow magnitude square in an event}
\end{equation}
where the self-correlation between the particles have been omitted while calculating the scalar product. Then from Eq.~(\ref{eq: two particle cumulant from Q-cumulant revisited}) we can write,
\begin{equation}
 \begin{aligned}
c_n\{2\} = \langle v_n^2 \rangle \ .
 \end{aligned}
\label{eq: two-particle cumulant as event average of flow magnitude square }
\end{equation}
Similarly, using Eqs.~(\ref{eq: four-particle cumulant from Q-cumulant}) and (\ref{eq: six-particle cumulant from Q-cumulant}) the 4-particle and 6-particle cumulants can be expressed as~\cite{Moravcova:2020wnf},
\begin{equation}
 \begin{aligned}
c_n\{4\} &= \langle v_n^4 \rangle - 2 \langle v_n^2 \rangle^2 , \\
\text{and} \eqsp{} c_n\{ 6 \} &= \langle v_n^6 \rangle - 9\langle v_n^2 \rangle \langle v_n^4 \rangle + 12 \langle v_n^2 \rangle^3 \ .
 \end{aligned}
\label{eq: 4-particle and 6-particle cumulants as event average of powers of flow magnitude }
\end{equation}
Similar methods could be applied to obtain the differential flow harmonics~\cite{Borghini:2001vi,Bilandzic:2010jr}. Fig.~\ref{fig: Integrated and differential flow from cumulant method} shows the measured differential and integrated flow using cumulant method. 

\subsubsection{Flow analysis in our simulation:}
Let us also mention the approach we use for flow analysis in simulation, based on which we present results in the current and in the subsequent chapters. In our simulation set-up which uses MUSIC~\cite{Schenke:2010nt} hydro code, we obtain the spectra $dN/2\pi p_T dp_Tdy$ and the differential harmonic flow vectors $V_n(p_T)$ event-by-event as a function of $p_T$. Then in each event we can calculate the integrated flow vector $V_n$, mean transverse momentum per particle $[p_T]$ and number of charged particle $N_{ch}$ using Eqs.(~\ref{eq: integrated flow in an event}), (\ref{eq: mean transverse momentum per particle}) and (\ref{eq: charged partcile multiplicity}) respectively. Then the event averaged flow is calculated using,
\begin{equation}
 \begin{aligned}
v_n\{2\} =\sqrt{\langle V_nV_n^* \rangle} = \sqrt{\langle v_n^2 \rangle} ,
 \end{aligned}
\label{eq: flow in simulation}
\end{equation}
where the angular bracket denotes the average over all events.

\section{Fluctuations of harmonic flow}
\begin{figure}[ht!]
\includegraphics[height=6 cm]{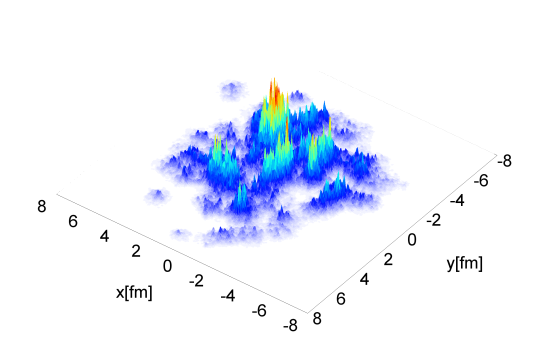}
\centering
\caption{Pictorial representation of fluctuations at the initial state, generated from the MC Glauber model. Figure taken from~\cite{Schenke:2012wb}.}
\label{fig: Initial state fluctuation}
\end{figure}

One of the most exotic and distinctive features of the anisotropic flow is the event-by-event fluctuations which have served as a phenomenon of sheer interest in both theoretical~\cite{Aguiar:2001ac,Ollitrault:2009ie,Qiu:2011iv,Gardim:2012im,Bzdak:2012tp,Heinz:2013bua,Renk:2014jja,Jia:2014vja,Bhalerao:2014mua,Pang:2015zrq,Xiao:2015dma,Bozek:2017qir,Bozek:2018nne,Bozek:2021mov,Nielsen:2022jms,Samanta:2023qem,Bozek:2023dwp,Zhu:2024tns} and experimental \cite{PHOBOS:2010ekr,CMS:2015xmx, ALICE:2016kpq, ALICE:2017lyf, ATLAS:2017rij,ATLAS:2019peb,ALICE:2022dtx,ALICE:2024fcv} studies. In these studies, a wide range of aspects and potential probes of the flow-fluctuation have been explored. The primary reason behind the flow-fluctuations are the event-by-event fluctuations of the initial state~\cite{Broniowski:2007ft,Alver:2010dn,Qin:2010pf,Teaney:2010vd,Holopainen:2010gz,Werner:2010aa,Muller:2011bb,Gardim:2011xv,Bhalerao:2011yg,Schenke:2012wb} (as shown in Fig.~\ref{fig: Initial state fluctuation}), providing the source of the anisotropic flow in each event, along with the other sources e.g. thermal fluctuations. 
Fluctuations in the initial state can originate from different sources e.g. geometry fluctuations or shape fluctuations, quantum fluctuations in the overlap of the wavefunctions of the colliding nucleons etc. Here our primary focus is to discuss event-by-event fluctuations of the harmonic flow coefficients $v_n$ and possible effects of fluctuations in terms of physical observables which could potentially probe the fluctuations. One way to probe such flow fluctuations is to study the ratio of the multi-particle cumulants which in turn could probe the initial state fluctuations in terms of eccentricity cumulants~\cite{Giacalone:2017uqx,ATLAS:2012at}. Another very interesting observable which is our primary focus in this chapter, to study flow-fluctuation, is the {\it flow-factorization breaking coefficient} or simply {\it factorization-breaking coefficient}~\cite{Gardim:2012im,Bozek:2017qir,Bozek:2018nne,Bozek:2021mov,Nielsen:2022jms,Samanta:2023qem}, accounting for the breaking of the factorization between the flow harmonics in two different kinematic bins ($p_T$ or $\eta$ bins), which we discuss below.    

\subsection{Factorization-breaking coefficients}
\label{fact-break coefficient}
As discussed in the previous sections, the harmonic flow coefficients $v_n$ can be measured through two-particle correlations at the lowest order, which eventually measure the flow harmonic square $v_n^2$ as given in Eq.~(\ref{eq: flow from two particle cumulant}). If the two particles are measured in two different $p_T$ bins $p_1$ and $p_2$ (for simplicity we use $p_T\equiv p$), then the factorization between the flow harmonics in those two bins would imply~\cite{Gardim:2012im},
\begin{equation}
 \begin{aligned}
 C_n(p_1, p_2) \equiv \langle \cos n(\phi_1 - \phi_2) \rangle \stackrel{\text{(factorization)}}{=\joinrel=} v_n\{2\}(p_1) \times v_n\{2\}(p_2),
 \end{aligned}
\label{eq: factorization of flow from two particle correlation}
\end{equation}
where $ C_n(p_1, p_2)$ is the event averaged correlation matrix, $\phi_1$ and $\phi_2$ are the azimuthal angles of the two particles in two transverse momentum bins $p_1$ and $p_2$, and the bracket has the usual meaning as in Eq.~(\ref{eq: two particle azimuthal correlation}). In the earlier years of flow-analysis, the factorization relation in Eq.~(\ref{eq: factorization of flow from two particle correlation}) was assumed to hold and was investigated in experiments~\cite{ALICE:2011svq,PHOBOS:2010ekr,ATLAS:2012at}. 
%The factorization can be understood from the two particle transverse momentum distribution in a single event which, assuming for a moment that there is no non-flow correlation, could be written as~\cite{Dinh:1999mn,Gardim:2012im},
%\begin{equation}
% \begin{aligned}
%\frac{dN}{d^2p_1 d^2p_2} \stackrel{\text{(flow)}}{=\joinrel=} \frac{dN}{d^2p_1} \times \frac{dN}{d^2p_2} = \frac{dN}{p_1dp_1d\phi_1} \times \frac{dN}{p_2dp_2d\phi_2},
% \end{aligned}
%\label{eq: factorization of two-particle transverse momentum distribution}
%\end{equation}
%where of course the last two terms in the above expression produce Eq.~(\ref{eq: factorization of flow from two particle correlation}), assuming that the angles $\phi_1$ and $\phi_2$ are taken with respect to the reaction plane\footnote{Please note that the difference between Eq.~(\ref{eq: decomposition two particle azimuthal distribution}) and Eq.~(\ref{eq: factorization of flow from two particle correlation}) is that, in the former case the azimuthal angles of the particles are measured in laboratory frame which is the case in practice, but in the latter case the angles for the particles in each $p_T$bin are considered with respect to the reaction plane for the sake of explanation, so that the correlated part of the distribution (which we didn't show in Eq.~(\ref{eq: factorization of flow from two particle correlation})) would only contain the non-flow correlation, whereas in Eq.~(\ref{eq: decomposition two particle azimuthal distribution}) it contained both flow and non-flow correlation~\cite{Borghini:2000sa,Borghini:2001vi}}. 
However, in practice even if we do not consider any non-flow correlation (which naturally breaks the factorization), there is still considerable factorization-breaking between the flow harmonics in two different kinematic bins~\cite{Gardim:2012im,Bozek:2017qir,Bozek:2018nne,CMS:2015xmx, ALICE:2017lyf, ATLAS:2017rij}, due to event-by-event fluctuations of the flow harmonics originating from the fluctuations of the initial state. Here we will restrict our discussions to transverse momentum dependent flow-fluctuations.

In the case of integrated ($p_T$-averaged) flow, the flow harmonics are obtained from the flow vector $V_n=v_ne^{in\Psi_n}$ using the standard `two-particle cumulant' formula in (Eqs.~(\ref{eq: flow from two particle cumulant}) and Eq.~(\ref{eq: two-particle cumulant as event average of flow magnitude square })),
\begin{equation}
 \begin{aligned}
v_n\{2\} = \sqrt{c_n\{2\}}=\sqrt{\langle V_nV_n^* \rangle} = \sqrt{\langle v_n^2\rangle} \ ,
 \end{aligned}
\label{eq: integrated flow from scalar product of two flow-vectors }
\end{equation}
where the angular bracket denotes the average over all events. Similarly one can define the transverse momentum dependent flow vector (i.e. flow vector measured in a transverse momentum bin), $V_n(p)=v_n(p)e^{in\Psi_n(p)}$, where both {\it flow magnitude} $v_n(p)$ and {\it flow angle} (event-plane angle) $ \Psi_n(p)$ depend on the transverse momentum which we write as $p$ in the present context. Then the transverse momentum dependent or the differential measure of the harmonic flow is given by~\cite{Borghini:2001vi,Bilandzic:2010jr,Samanta:2023qem},
\begin{equation}
 \begin{aligned}
v_n\{2\}(p) =\frac{\langle V_nV_n^*(p) \rangle}{v_n\{2\}}=\frac{\langle V_nV_n^*(p) \rangle}{\sqrt{\langle V_n V_n^* \rangle}},
 \end{aligned}
\label{eq: diff flow using global and pT-dependent flow}
\end{equation}
which is the definition used in the experimental analysis for differential flow, where one particle is taken from a particular transverse momentum bin $p$ and the other particle is taken from the entire $p_T$-acceptance region  or {\it all} the particles detected in the event. Here the flow $V_n$ serves as the {\it reference flow}~\cite{Bilandzic:2010jr} for studying $p_T$-dependent differential flow. However, an alternative definition for the differential flow could be used, which is given by~\cite{Heinz:2013bua,Samanta:2023qem},  
\begin{equation}
 \begin{aligned}
v_n[2](p) =\sqrt{\langle V_n(p)V_n^*(p) \rangle} = \sqrt{\langle v_n(p)^2 \rangle},
 \end{aligned}
\label{eq: diff flow using flow-vectors in same pT bin}
\end{equation}
where both particles are taken within the same transverse momentum bin $p$. Although in principle this definition could be used for the theoretical study, but it is not suitable for the experimental analysis as it might be difficult to find two particles within the same $p_T$-bin at larger $p_T$ , due to limited statistics\footnote{From the particle spectra it is obvious that there are less and less number of particles as we go higher in $p_T$}. This concept will be extremely useful for the upcoming discussions in this section. 

Let us consider two particles in two different transverse momentum bins $p_1$ and $p_2$. Then the event averaged correlation matrix $C_n(p_1,p_2)$ given in Eq.~(\ref{eq: factorization of flow from two particle correlation}) should satisfy two conditions: first, its diagonal elements should be positive and second, the off-diagonal elements must satisfy the Cauchy-Schwarz inequality,
\begin{equation}
 \begin{aligned}
   C_n(p_1,p_1) \geq 0 \eqsp{and} C_n(p_1,p_2)^2 \leq C_n(p_1,p_1) C_n(p_2,p_2) \ .
 \end{aligned}
\label{eq: cauchy-Swarz inequality for the correlatoon matrix}
\end{equation}
If there is factorization between the flow harmonics in different $p_T$-bins, then the inequality saturates to equality condition. In other words, based on this inequality condition, we can construct a correlation coefficient which can quantify the amount of factorization breaking~\cite{Gardim:2012im},
\begin{equation}
 \begin{aligned}
r_n = \frac{C_n(p_1,p_2)}{\sqrt{C_n(p_1,p_1) C_n(p_2,p_2)}},  
 \end{aligned}
\label{eq: correlation coefficient constructed from Cauchy-Swarz inequality}
\end{equation}
which could be written explicitly in terms of the event flow in two bins $V_n(p_1)$ and $V_n(p_2)$ as~\cite{Gardim:2012im,Bozek:2018nne},
\begin{equation}
 \begin{aligned}
r_n(p_1,p_2)=\frac{\langle V_n(p_1) V_n^*(p_2) \rangle}{\sqrt{\langle v_n^2(p_1) \rangle \langle v_n^2(p_2) \rangle}} \ .
 \end{aligned}
\label{eq: first order fact-break coeff between flow vectors in two bins }
\end{equation}
If there is factorization, $r_n = 1$, which is the limiting case and any deviation of $r_n$ from $1$ will correspond to the breaking of the factorization. The correlation coefficient $r_n(p_1,p_2)$ is known as {\it factorization-breaking coefficient}. 

The factorization-breaking coefficient in Eq.~(\ref{eq: first order fact-break coeff between flow vectors in two bins }) is defined as the linear correlation coefficient between two flow vectors which are complex numbers. The effect of event-by-event flow fluctuations is the {\it decorrelation} between these harmonic flow vectors in two different $p_T$-bins. In each event, $V_n(p)$ is a smooth function of $p$. In general, it is expected that the correlation is stronger when $p_1\simeq p_2$ and it decreases as the difference between $p_1$ and $p_2$ increases i.e there is decorrelation which is understood as the deviation of the factorization-breaking coefficient $r_n(p_1,p_2)$ from 1. The decorrelation occurs due to decoherence between the flow vectors, induced by the quantum fluctuations in the wavefunctions of the incoming nucleons, in the initial state. 
\begin{figure}[ht!]
\includegraphics[height=6 cm]{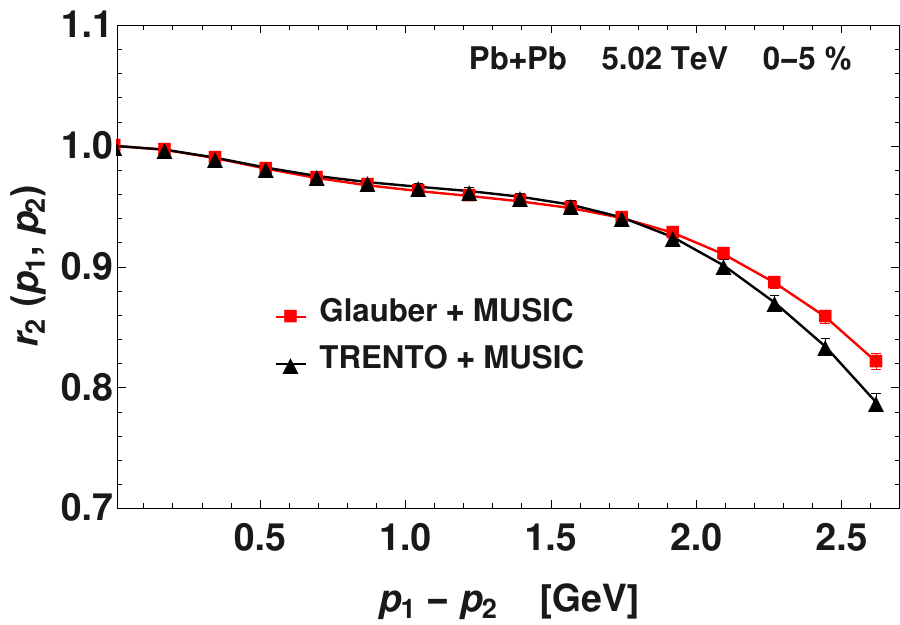}
\centering
\caption{The factorization-breaking coefficient between elliptic flow vectors in two different transverse momentum bins $p_1$ and $p_2$, plotted as a function of $p_1-p_2$ for Pb+Pb collision at 5.02 TeV in $0-5 \%$ centrality. The red squares and the black triangles denote the results obtained in hydrodynamic simulations with Glauber and TRENTO initial conditions respectively.}
\label{fig: fact-break first order p1-p2}
\end{figure}
The factorization-breaking coefficient $r_n(p_1,p_2)$ has been studied in models~\cite{Gardim:2012im,Kozlov:2014fqa,Gardim:2017ruc,Zhao:2017yhj,Bozek:2018nne} and measured in the experiments~\cite{CMS:2013bza,Zhou:2014bba,CMS:2015xmx}. In Fig.~\ref{fig: fact-break first order p1-p2}, we show the results for $r_2(p_1,p_2)$ as a function of the difference $p_1-p_2$, calculated in our model for Pb+Pb collision at $5.02$ TeV, where the flow vector in one $p_T$-bin is kept fixed and correlated with the flow vectors in other bins. As expected, the correlation $r_n \rightarrow 1$ as $p_1$ and $p_2$ are closer to each other and it gradually deviates from $1$ i.e. the decorrelation increases as the difference $p_1-p_2$ increases. 

It should be noted that all the results reported in this and following sections are obtained using a boost invariant viscous hydrodynamic model MUSIC~\cite{Schenke:2010nt} for Pb+Pb collisions at $\sqrt{s_{NN}}=5.02$ TeV. The density distributions in the initial state, used as an input for the hydrodynamic evolution, are obtained from two initial-state models: a two-component Glauber Monte Carlo model~\cite{Bozek:2019wyr} and the TRENTO model~\cite{Moreland:2014oya}. Unless otherwise stated, we use a constant shear viscosity to entropy density ratio $\eta/s=0.08$. 

%It is important to mention that, in a formal sense, it is possible to measure any comprehensive set of moments or cumulants to constrain the underlying multivariate probability distribution function~\cite{Bilandzic:2013kga}. However, in experiments, we are limited by the maximum rank of the flow moments that can be accurately derived from the experimental measurements. When the set of moments reaches its maximum rank, the set of cumulants of higher rank can be utilized~\cite{Mordasini:2019hut}. When utilizing an alternative set of linearly independent moments, it is important to note that lower order correlations may exist within a moment of a specific order. In the following, we employ a specific selection of higher order moments based on their connection to lower order factorization breaking coefficients.

The flow vector $V_n(p)=v_n(p)e^{in\Psi_n(p)}$ has two parts: flow magnitude $v_n(p)$ and flow angle $\Psi_n(p)$, both of which fluctuate event-by-event depending on transverse momentum. Naturally, one would expect flow magnitude factorization breaking (decorrelation) and flow angle decorrelation, where the latter represents the event-by-event difference in the flow angles $\Psi_n(p_1)$ and $\Psi_n(p_2)$ (event-plane angle) at two transverse momenta. The decorrelation between the flow magnitudes at two transverse momentum bins is defined as\cite{Bozek:2018nne},
\begin{equation}
\begin{aligned}
r_n^{v_n}(p_1,p_2)=\frac{\langle |V_n(p_1)| |V_n(p_2)| \rangle}{\sqrt{\langle v_n^2(p_1) \rangle \langle v_n^2(p_2) \rangle}}=\frac{\langle v_n(p_1) v_n(p_2) \rangle}{\sqrt{\langle v_n^2(p_1) \rangle \langle v_n^2(p_2) \rangle}} \ ,
 \end{aligned}
\label{eq: first order fact-break coeff between flow magnitudes in two bins }
\end{equation}
and the corresponding flow angle decorrelation is given by,
\begin{equation}
\begin{aligned}
\langle \cos [n (\Psi_n(p_1)-\Psi_n(p_2))]\rangle \ .
 \end{aligned}
\label{eq: first order flow angle decorrelation in two bins }
\end{equation}
In principle, one could study all these three quantities (flow vector, magnitude and angle decorrelation) in models~\cite{Bozek:2018nne}. The decorrelation  between harmonic flow vectors at two different momenta involves both the decorrelation of the flow magnitudes and of the flow angles
\cite{Heinz:2013bua,Jia:2014ysa,Bozek:2018nne}. Therefore, it is natural to expect that the flow vector decorrelation factorizes into flow magnitude and angle decorrelation i.e.
\begin{equation}
\begin{aligned}
r_n(p_1,p_2) &= \frac{\langle  v_n(p_1) v_n(p_2) \cos [n (\Psi_n(p_1)-\Psi_n(p_2))] \rangle}{\langle v_n^2(p_1) \rangle \langle v_n^2(p_2) \rangle} \\
&\simeq \frac{\langle v_n(p_1) v_n(p_2) \rangle}{\sqrt{\langle v_n^2(p_1) \rangle \langle v_n^2(p_2) \rangle}} \times \langle \cos [n (\Psi_n(p_1)-\Psi_n(p_2))]\rangle,
 \end{aligned}
\label{eq: factorization of flow magnitude and angle decorrelation }
\end{equation}
which however does not hold true in practice~\cite{Bozek:2018nne}. 

\subsubsection{Flow angle and flow magnitude factorization breaking : Need for 2nd order }

It should be noted that experimentally, flow magnitude and angle decorrelation cannot be measured through the formulae presented in Eqs.~(\ref{eq: first order fact-break coeff between flow magnitudes in two bins }) and (\ref{eq: first order flow angle decorrelation in two bins }). The reason is that in experiment, we can only measure the scalar product between two flow-vectors, which involves two-particle correlation. As a result, in the {\it first order} of flow, only flow vector decorrelation (Eq.~(\ref{eq: first order fact-break coeff between flow vectors in two bins })) can be measured, flow magnitude and flow angle decorrelation cannot measured using two-particle correlators. In order to measure the flow magnitude and flow angle decorrelation experimentally, we need to consider four-particle correlators or in other words, we need to construct the factorization-breaking coefficients between the {\it squares} of the flow. A similar method was applied to estimate separately, the flow magnitude and flow angle decorrelation in pseudorapidity bins, using four-particle correlators \cite{Jia:2017kdq}. A similar procedure can be used for the correlation between the harmonic flow in two different transverse momentum bins~\cite{Bozek:2018nne}.

The factorization-breaking coefficients between two flow vectors {\it squared} can be defined as,
\begin{equation}
 \begin{aligned}
r_{n;2}(p_1,p_2)=\frac{\langle V_n(p_1)^2 V_n^*(p_2)^2 \rangle}{\sqrt{\langle v_n^4(p_1) \rangle \langle v_n^4(p_2) \rangle}} \ ,
 \end{aligned}
\label{eq: fact-break coeff between flow vectors squared in two bins}
\end{equation}
which is a four-particle correlator, measuring the breaking of the factorization between two flow vectors in two transverse momentum bins, in {\it second order}.  Please note that in the above definition, the four particle correlator is constructed by taking two particles from the same transverse momentum bin $p_1$ and other two particles from the bin $p_2$. In a similar manner, the factorization-breaking coefficient between flow magnitude {\it squared} can be defined as,
\begin{equation}
 \begin{aligned}
r_{n;2}(p_1,p_2) &=\frac{\langle V_n(p_1)V_n(p_1)^*V_n(p_2)V_n(p_2)^*  \rangle}{\sqrt{\langle v_n^4(p_1) \rangle \langle v_n^4(p_2) \rangle}}\equiv\frac{\langle |V_n(p_1)|^2 |V_n(p_2)|^2  \rangle}{\sqrt{\langle v_n^4(p_1) \rangle \langle v_n^4(p_2) \rangle}}\\
&= \frac{\langle v_n(p_1)^2 v_n^2(p_2)  \rangle}{\sqrt{\langle v_n^4(p_1) \rangle \langle v_n^4(p_2) \rangle}} ,
 \end{aligned}
\label{eq: fact-break coeff between flow magnitudes squared in two bins}
\end{equation}
which in principle can be used in experiments to measure flow magnitude decorrelation. However, it is not possible to define an experimental observable which could directly measure the flow angle correlation between two transverse momentum bins. Only an estimate of the flow angle correlation (or decorrelation) can be obtained by taking the ratio of four-particle flow vector correlator  and flow magnitude correlator~\cite{Jia:2017kdq} given by, 
\begin{equation}
 \begin{aligned}
F_{n}(p_1,p_2)=\frac{\langle V_n(p_1)^2 V_n^*(p_2)^2 \rangle}{\langle v_n(p_1)^2 v_n(p_2)^2  \rangle},
 \end{aligned}
\label{eq: second order flow angle decorrelation in two bins}
\end{equation}
which is the ratio of the flow vector squared (Eq.~(\ref{eq: fact-break coeff between flow vectors squared in two bins})) and flow magnitude squared (Eq.~(\ref{eq: fact-break coeff between flow magnitudes squared in two bins})) factorization-breaking coefficients. 
\begin{figure}[ht!]
\hspace{-0.3cm}\begin{subfigure}{0.5\textwidth}
\centering
\includegraphics[height=5.2 cm]{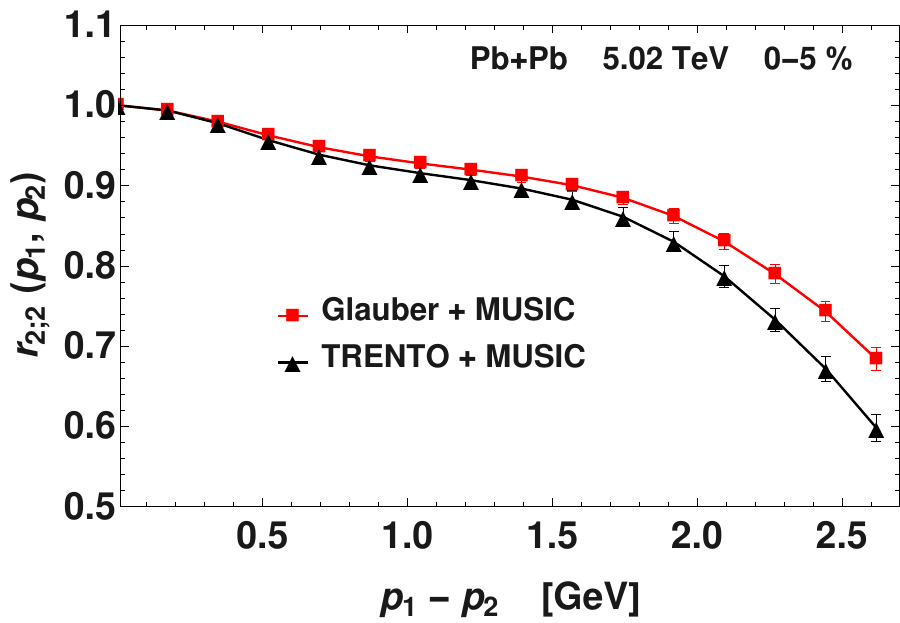}
\end{subfigure}~~
\begin{subfigure}{0.5\textwidth}
\centering
\includegraphics[height=5.2 cm]{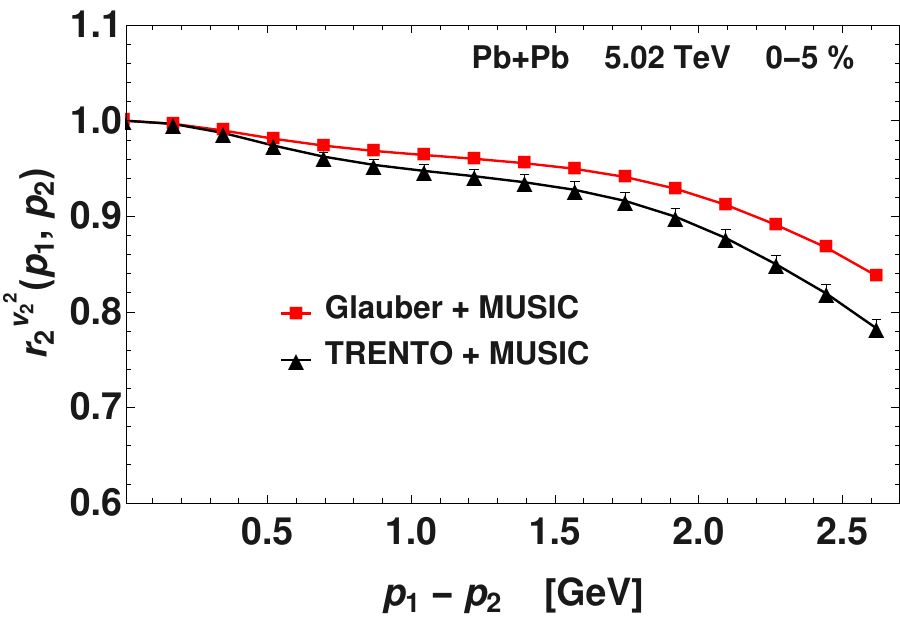}  
\end{subfigure}
\centering
\caption{ Factorization-breaking coefficients between flow vectors squared (left) and flow magnitude squared (right) between two different bins $p_1$ and $p_2$, plotted as a function of $p_1-p_2$ for the elliptic flow in Pb+Pb collision for $0-5 \%$ centrality. The symbols carry similar meaning as Fig.~\ref{fig: fact-break first order p1-p2}.}
\label{fig: flow vector and mag fact-break coeff second order p1-p2}
\end{figure}

Fig.~(\ref{fig: flow vector and mag fact-break coeff second order p1-p2}) shows the flow vector squared and flow magnitude squared factorization-breaking coefficients for the elliptic flow in our model calculation. It could be seen that the flow magnitude decorrelation is smaller than flow vector decorrelation, because the other part of it is given by the flow angle decorrelation. Although formally possible, the experimental implementation of the formulae in Eqs.~(\ref{eq: fact-break coeff between flow vectors squared in two bins}) and (\ref{eq: fact-break coeff between flow magnitudes squared in two bins}) are not feasible because one needs two particle from the same transverse momentum bin, which become very difficult at larger momenta due to low statistics.   

\subsection{Removing experimental difficulty: Taking one flow $p_T$-averaged}
\label{fact-break coefficient with one flow momentum averaged}
To ease the experimental measurement, one could take only one transverse momentum bin and can correlate the particles from that particular bin with the particles from the entire acceptance range. This was originally introduced by the measurement from the ALICE collaboration~\cite{ALICE:2017lyf,ALICE:2022dtx, ALICE:2024fcv}. This way we could partly overcome the limitation from low multiplicity in bins at high transverse momentum. Below we present our model calculations following this method for the factorization-breaking coefficients. The following sections are, for the most part, presentations from the original publications~\cite{Bozek:2021mov, Bozek:2022slu}, coauthored by the author. 
\begin{figure}[ht!]
\includegraphics[height=6 cm]{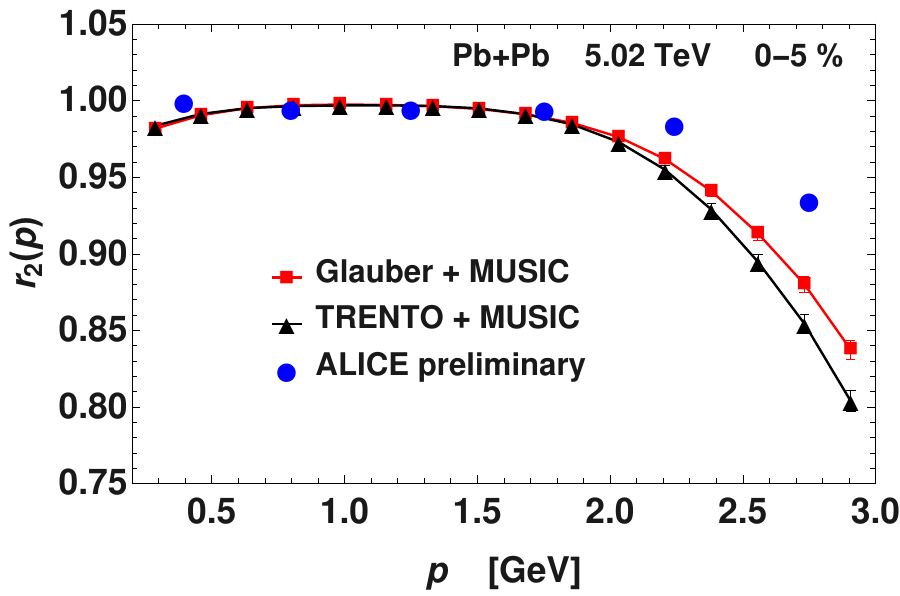}
\centering
\caption{The factorization-breaking coefficient between transverse momentum averaged ($V_2$) and transverse momentum dependent elliptic flow vector( $V_2(p)$) as a function of the transverse momentum $p$ in Pb+Pb collision at 5.02 TeV for $0-5 \%$ centrality. The results obtained in hydrodynamic simulations with Glauber and TRENTO initial conditions are represented by the red squares and the black triangles respectively. The blue dots represent the experimental data from the ALICE collaboration~\cite{ALICE:2022dtx}.}
\label{fig: fact-break first order Vn-Vnp}
\end{figure}
In the first order, the factorization-breaking coefficient can be defined as,
\begin{equation}
 \begin{aligned}
r_n(p)=\frac{\langle V_n V_n^*(p) \rangle}{\sqrt{\langle v_n^2 \rangle \langle v_n^2(p) \rangle}} \ ,
 \end{aligned}
\label{eq: first order fact-break flow vector one bin}
\end{equation}
which measures the correlation between the harmonic flow vector averaged over transverse momentum ($p_T$-averaged flow or global flow) and the flow vector in a transverse momentum bin $p$\footnote{we use $p\equiv p_T$, where $p$ represents a particular $p_T$-bin}. The coefficient $r_n(p)$ represents a two-particle correlator, where one particle comes from the transverse-momentum bin $p$ and the second particle comes from anywhere of the full $p_T$-acceptance range of the detector, which serves as the reference particle (reference flow) much like Eq.~(\ref{eq: diff flow using global and pT-dependent flow}). In fact, the factorization-breaking coefficient $r_n(p)$ could be written using Eq.~(\ref{eq: diff flow using global and pT-dependent flow}) and (\ref{eq: diff flow using flow-vectors in same pT bin}) as,
\begin{equation}
 \begin{aligned}
r_n(p)=\frac{v_n\{2\}(p)}{v_n[2](p)},
 \end{aligned}
\label{eq: first order fact-break in terms of differential flows }
\end{equation}
which is a measure of the difference between two definitions of the differential harmonic flow coefficient~\cite{Heinz:2013bua}. Fig.~(\ref{fig: fact-break first order Vn-Vnp}) shows the factorization-breaking coefficient $r_2(p)$ for the elliptic flow calculated in our model, along with the ALICE data. Our model results can reproduce the qualitative nature of the data well, where both results show significant decorrelation at higher transverse momentum. Quantitatively speaking, at large transverse momentum, the decorrelation is stronger in our model than the data, which could be due to the presence of non-flow correlations in the data. We will return to this point shortly after.

\begin{figure}[ht!]
\hspace{-0.3cm}\begin{subfigure}{0.5\textwidth}
\centering
\includegraphics[height=5.1 cm]{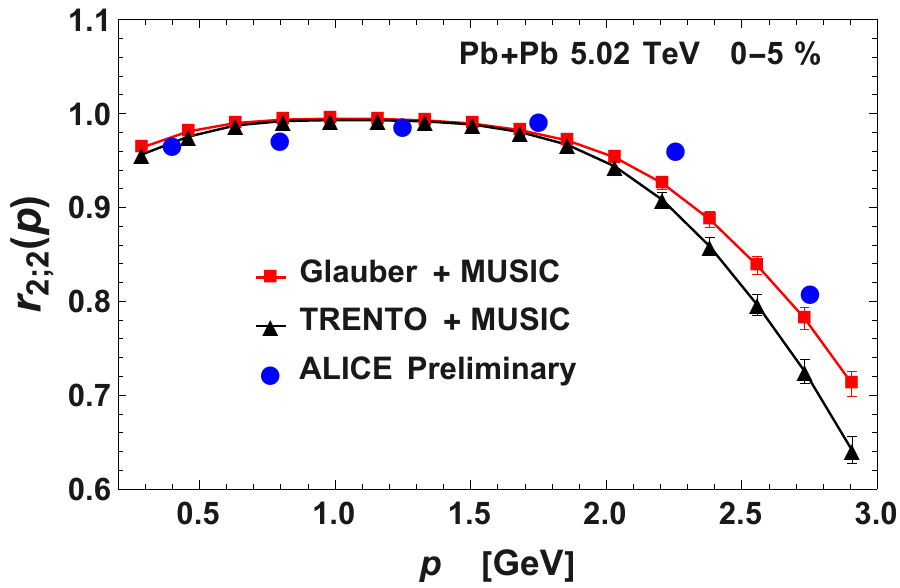}
\end{subfigure}~~~
\begin{subfigure}{0.5\textwidth}
\centering
\includegraphics[height=5.1 cm]{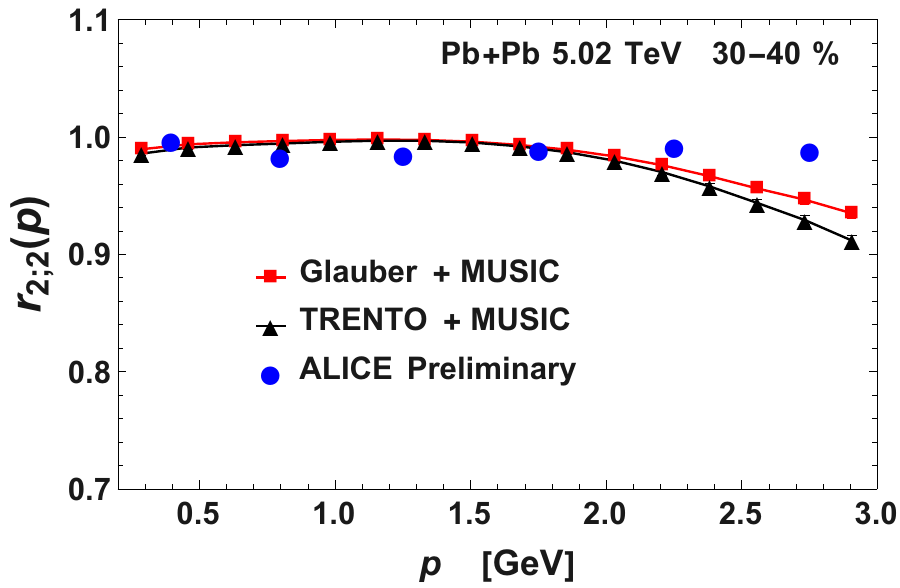}  
\end{subfigure}
\centering
\caption{Factorization-breaking coefficients between momentum averaged ($V_2^2$) and momentum dependent elliptic flow ($V_2(p)^2$) vector squared as a function of the transverse momentum in Pb+Pb collision at 5.02 TeV for $0-5\%$ (left) and $30-40 \%$ (right) centrality. The symbols carry similar meaning as Fig.~\ref{eq: fact-break flow vector squared one bin}. The figure is from the original publication~\cite{Bozek:2021mov}, coauthored by the author.}
\label{fig: flow vector squared fact-break coeff elliptic flow}
\end{figure}

\begin{figure}[ht!]
\hspace{-0.3cm}\begin{subfigure}{0.5\textwidth}
\centering
\includegraphics[height=5.1 cm]{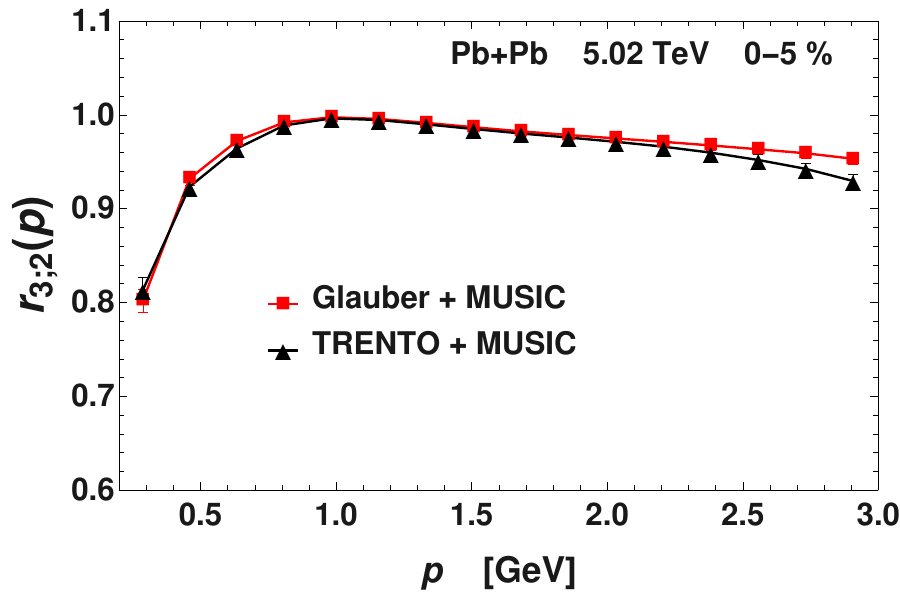}
\end{subfigure}~~~
\begin{subfigure}{0.5\textwidth}
\centering
\includegraphics[height=5.1 cm]{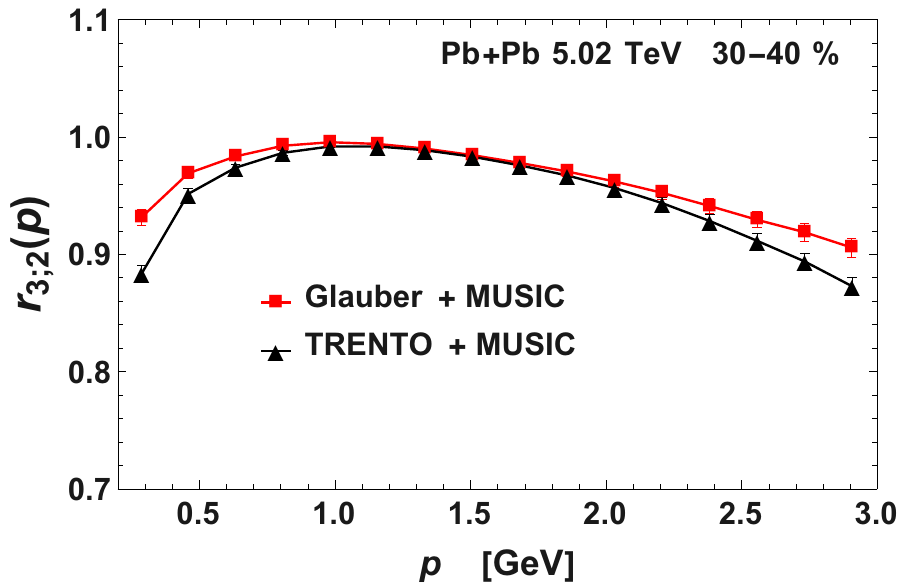}  
\end{subfigure}
\centering
\caption{Flow vector squared factorization-breaking coefficients between momentum averaged and momentum dependent triangular flow in Pb+Pb collision at 5.02 TeV for $0-5\%$ (left) and $30-40 \%$ (right) centrality. Symbols are same as Fig.~\ref{fig: flow vector squared fact-break coeff elliptic flow}. The right panel is from the original publication~\cite{Bozek:2021mov}, coauthored by the author.}
\label{fig: flow vector squared fact-break coeff triangular flow}
\end{figure}

We can use the similar idea to define the factorization-breaking coefficient in second order of flow, which is our ultimate goal. In the second order, we can define the factorization-breaking coefficient between the flow vector squared as, 
\begin{equation}
 \begin{aligned}
r_{n;2}(p)=\frac{\langle V_n^2 V_n^*(p)^2 \rangle}{\sqrt{\langle v_n^4 \rangle \langle v_n^4(p) \rangle}} \ ,
 \end{aligned}
\label{eq: fact-break flow vector squared one bin}
\end{equation}
which is a four-particle correlator, where two of them are from the transverse momentum bin $p$ and the other two are taken globally and eventually represent the correlation coefficient between $V_n^2$ and $V_n^2(p)$. Figs.~\ref{fig: flow vector squared fact-break coeff elliptic flow} and ~\ref{fig: flow vector squared fact-break coeff triangular flow} show the results obtained in the hydrodynamic model, for the factorization-breaking coefficients $r_{n;2}(p)$ in the case of elliptic and triangular flow respectively, for $0-5 \%$ and $30-40 \%$ centrality in each case.
A first observation of the figures reveal that the decorrelation is significantly larger for the centralities where fluctuations dominate i.e. for the elliptic flow in central collisions and for the triangular flow in any centrality. In Fig.~\ref{fig: flow vector squared fact-break coeff elliptic flow}, for the elliptic flow, the model results are similar to the data for $0-5 \%$ centrality and there is large decorrelation between the flow vectors at high transverse momentum. On the contrary, for $30-40 \%$ centrality (semi-central collisions) the decorrelation is much smaller as there is less fluctuations, but the model results show stronger decorrelation than in the data at high transverse momentum, making the difference prominent for $p > 2.0$ GeV. 
A possible reason behind this phenomena could be that our model calculation do not include any contributions from non-flow correlations, while in the experimental data non-flow effects are present, which are usually reduced by using larger rapidity gaps between the measured flow vectors. A precise quantification of the contributions of the remaining non-flow correlation and the genuine difference in flow fluctuations between the model and the experiment, to the observed difference between the model results and the data in Fig.~\ref{fig: flow vector squared fact-break coeff elliptic flow} (right), lies beyond the scope of our present study. 

The decorrelation between the triangular flow vector squared in Fig.~\ref{fig: flow vector squared fact-break coeff triangular flow}, is quite strong for both $0-5 \%$ and $30-40 \%$ centralities due to dominating flow fluctuations. A careful inspection of the two plots discloses another interesting fact: there is a considerable decorrelation even at low transverse momentum, which is not so noticeable in case of elliptic flow. The origin of this effect lies within the definition of the factorization-breaking coefficient $r_{n;2}(p)$ in Eq.~(\ref{eq: fact-break flow vector squared one bin}). Here we correlate the flow vector in a transverse momentum bin $V_n(p)^2$ with the $p_T$-averaged flow vector $V_n$ which correspond to the average transverse momentum over all events, $\langle p_T \rangle \equiv \langle p \rangle \sim 0.8$ GeV. As the difference between $p$ and $ \langle p \rangle$ increases, which could happen for both low and high transverse momentum bins, there could be decorrelations. This effect is small for $V_2$, but dominant in the case of $V_3$.   

\subsubsection{Flow magnitude decorrelation}
Similar to Eq.~(\ref{eq: fact-break flow vector squared one bin}), through a four-particle correlator we can define the factorization-breaking coefficient between the flow magnitude squared with one flow momentum-averaged and another flow in a momentum bin,
\begin{equation}
 \begin{aligned}
r_n^{v_n^2}(p)=\frac{\langle v_n^2 v_n^2(p) \rangle}{\sqrt{\langle v_n^4 \rangle \langle v_n^4(p) \rangle}} \ ,
 \end{aligned}
\label{eq: fact-break flow magnitude squared one bin}
\end{equation}
which is easier to measure in experiment as compared to Eq.~(\ref{eq: fact-break coeff between flow magnitudes squared in two bins}), because now it requires to find only two particles in a transverse momentum bin $p$.
\begin{figure}[ht!]
\hspace{-0.3cm}\begin{subfigure}{0.5\textwidth}
\centering
\includegraphics[height=5 cm]{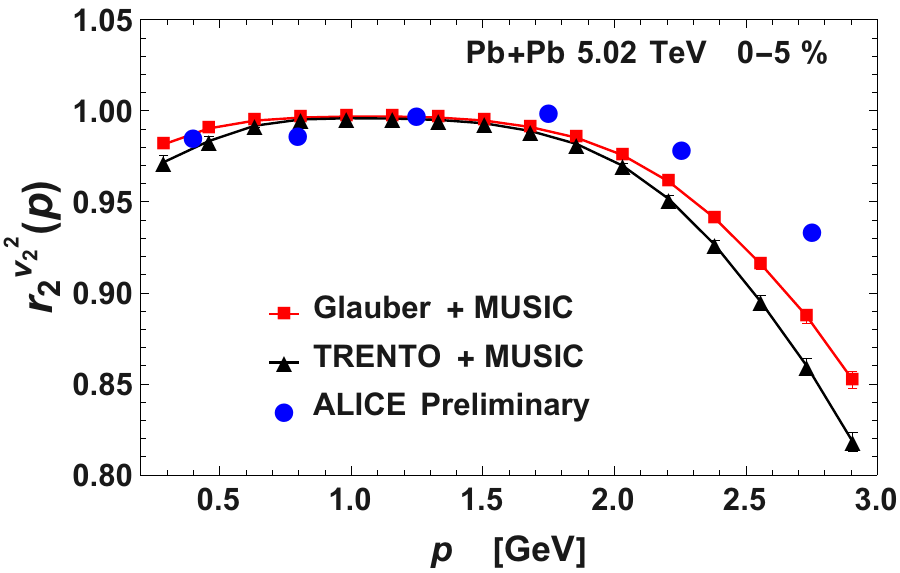}
\end{subfigure}~~~
\begin{subfigure}{0.5\textwidth}
\centering
\includegraphics[height=5 cm]{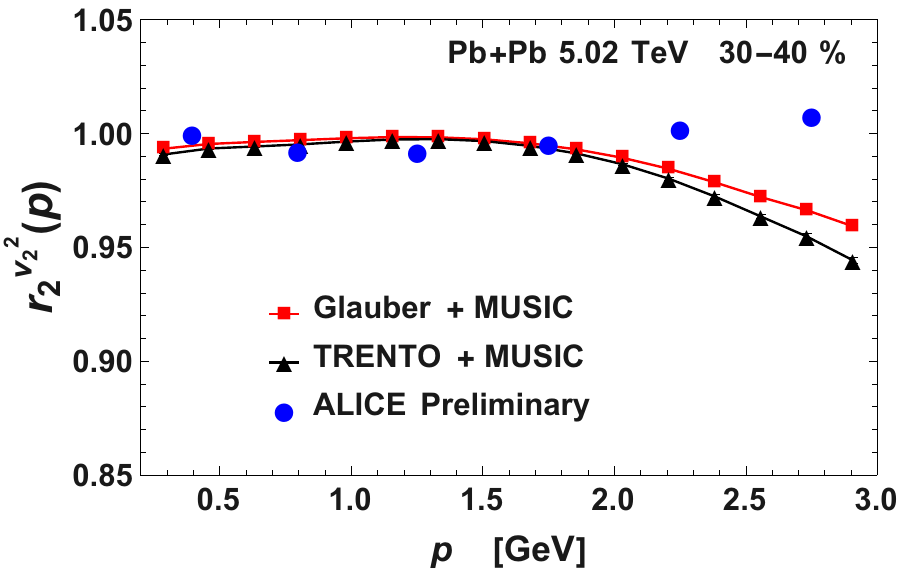}  
\end{subfigure}
\centering
\caption{Factorization-breaking coefficients for the flow magnitude squared between transverse momentum averaged and transverse momentum dependent elliptic flow as a function of the transverse momentum in Pb+Pb collisions for $0-5\%$ (left) and $30-40 \%$ (right) centrality. The symbols carry similar meaning as Fig.~\ref{fig: flow vector squared fact-break coeff elliptic flow}. The figure is from the original publication~\cite{Bozek:2021mov}, coauthored by the author.}
\label{fig: flow magnitude squared fact-break coeff elliptic flow}
\end{figure}
Fig.~\ref{fig: flow magnitude squared fact-break coeff elliptic flow} shows the results obtained in the hydrodynamic model and comparison with the data, for the factorization-breaking coefficients between magnitude squared of elliptic flow, $r_2^{v_2^2}(p)$ for $0-5 \%$ and $30-40 \%$ centrality. Similar to the flow vector factorization breaking coefficients, the  magnitude factorization breaking coefficients show that the experimental data lie above the simulation results at high transverse momentum. This discrepancy can be partly attributed to the contribution of non-flow correlations. The disparity is particularly noticeable for the $30-40$\% centrality, where certain data points even surpass 1, indicating significant dominance of the non-flow correlations. 

A second observation reveals that the flow magnitude decorrelation accounts for roughly one-half of the flow vector decorrelation, i.e.
\begin{equation}
 \begin{aligned}
[1-r_{n;2}(p)] \simeq 2[1-r_n^{v_n^2(p)}] \ .
 \end{aligned}
\label{eq: relation between flow vector and magnitude decorrelation}
\end{equation}
This suggests that the other half of the flow vector decorrelation can be approximately attributed to the flow-angle decorrelation, as we will see shortly. This relation regarding the flow vector, magnitude and angle decorrelation, as an outcome of the event-by-event flow fluctuations is expected in a random toy model~\cite{Bozek:2023dwp}(Appendix.~\ref{a: flow decorr toy model}). Although it is generally possible to have any proportions of angle and magnitude decorrelation contributing to total vector decorrelation, previous experimental measurements~\cite{ATLAS:2017rij} and model calculations~\cite{Bozek:2017qir} have shown that flow decorrelation in rapidity exhibits a roughly equal strength of angle and magnitude decorrelation. The results for the triangular flow are qualitatively similar, which we do not show here. 
\begin{figure}[ht!]
\includegraphics[height=6 cm]{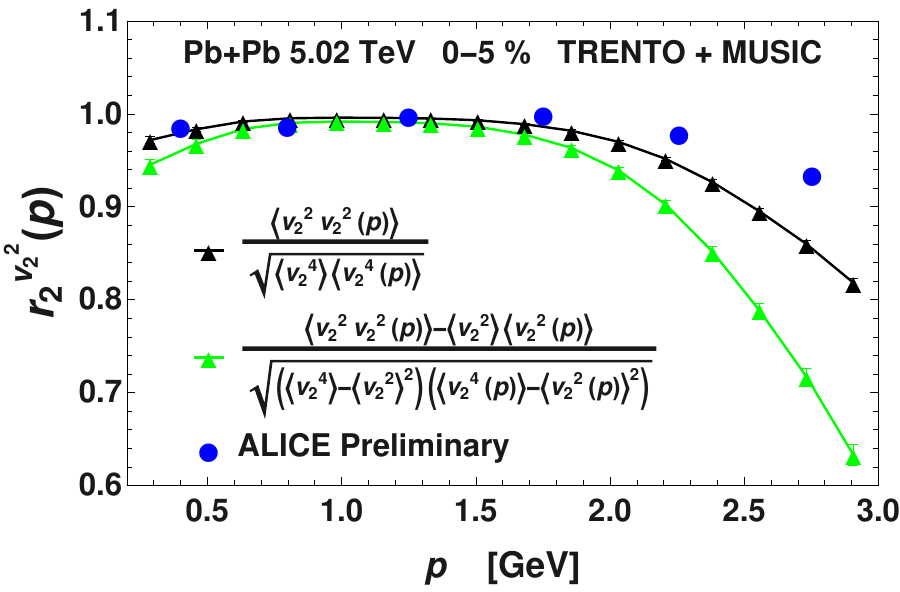}
\centering
\caption{Comparison between different definitions of flow magnitude factorization-breaking coefficient for the elliptic flow, in Pb+Pb collision at 5.02 TeV in $0-5 \%$  centrality with TRENTO initial condition. The black triangles represent our usual definition of flow magnitude squared factorization-breaking coefficients given in Eq.~\ref{eq: fact-break flow magnitude squared one bin}, while the green triangles denote the Pearson correlation coefficient between $v_2^2$ and $v_2^2(p)$. The blue dots represent the ALICE data for the flow magnitude decorrelation.}
\label{fig: flow-magnitude fact-break comparison}
\end{figure}

In the present context, let us point out that the correlation coefficient in Eq.~(\ref{eq: fact-break flow magnitude squared one bin}) which is a measure of the flow magnitude squared factorization-breaking, is not equal to or not to be confused with the Pearson correlation coefficient between $v_n^2$ and $v_n^2(p)$,
\begin{equation}
 \begin{aligned}
\frac{\langle v_n^2 v_n^2(p) \rangle-\langle v_n^2\rangle \langle v_n^2(p)\rangle }{\sqrt{(\langle v_n^4 \rangle-\langle v_n^2 \rangle^2) (\langle v_n^4(p) \rangle-\langle v_n^2(p) \rangle^2)}} \ .
 \end{aligned}
\label{eq: Pearson's correlation between flow magnitude squares}
\end{equation}
Fig.~\ref{fig: flow-magnitude fact-break comparison} shows the comparison between these two definitions of the correlation coefficient for elliptic flow, along with the data for flow magnitude decorrelation. It could be clearly seen that the correlation coefficient in Eq.~(\ref{eq: fact-break flow magnitude squared one bin}) follows the trend of the data and much closer to it, hence provide true measure of flow magnitude decorrelation, whereas the Pearson correlation coefficient in Eq.~(\ref{eq: Pearson's correlation between flow magnitude squares}) shows completely different behavior and huge departure from the data.  

Please note that the extraction of the flow vectors squared and flow magnitudes squared factorization breaking coefficients from the experimental data of the ALICE collaboration, involves several considerations. The experiment does not directly measure these flow vector and flow magnitude factorization coefficients (Eq.~\ref{eq: fact-break flow vector squared one bin} and \ref{eq: fact-break flow magnitude squared one bin})~\cite{NielsenIS2021,ALICE:2022dtx,ALICE:2024fcv}. The primary challenge lies in extracting the four-particle correlator $ \langle  v_n^4(p)\rangle$, sitting at the denominator of the correlation formulae, because all four particles  are from a narrow transverse momentum bin. On the other hand, the fourth moment $\langle v_n^4 \rangle$ which involves four particles from anywhere in the full acceptance, can be measured. The results presented by the ALICE Collaboration for the correlators, are with different scaling~\cite{NielsenIS2021}\footnote{Please note that when we performed our analysis, the data from the ALICE collaboration was in the preliminary form, as presented in the IS2021 conference in Rehovot~\cite{NielsenIS2021}. In the original publication~\cite{ALICE:2022dtx}, which came after our publication~\cite{Bozek:2021mov}, the data for magnitude squared correlation was presented with the scaling presented in Eq.~(\ref{eq: estimate of flow magnitude squared fact-break from ALICE data}), so is followed in a recent preprint with more systematic study~\cite{ALICE:2024fcv}}, e.g. for flow magnitude squared correlations,
\begin{equation}
 \begin{aligned}
\frac{\langle v_n^2 v_n^2(p) \rangle}{\langle v_n^2 \rangle \langle v_n^2(p) \rangle} \ .
 \end{aligned}
\label{eq: flow magnitude squared correlation ALICE}
\end{equation}
Upon dividing this scaled correlator by $\langle v_n^4\rangle/\langle v_n^2 \rangle^2$~\cite{NielsenIS2021}, an estimate of the flow magnitude squared factorization-breaking coefficient,
\begin{equation}
 \begin{aligned}
r_{n}^{v_n^2}(p) \simeq \frac{\langle v_n^2 v_n^{ 2}(p)\rangle \langle v_n^2 \rangle }{{\langle v_n^4 \rangle \langle v_n^2(p) \rangle}},
 \end{aligned}
\label{eq: estimate of flow magnitude squared fact-break from ALICE data}
\end{equation}
or the flow vector squared factorization-breaking coefficient,
\begin{equation}
 \begin{aligned}
r_{n;2}(p) \simeq \frac{\langle V_n^2 V_n^{\star  }(p)^{2}\rangle \langle v_n^2 \rangle }{{\langle v_n^4 \rangle \langle v_n^2(p) \rangle}} 
 \end{aligned}
\label{eq: estimate of flow vector squared fact-break from ALICE data}
\end{equation}
can be obtained. The difference between Eqs.~(\ref{eq: estimate of flow vector squared fact-break from ALICE data}) and (\ref{eq: estimate of flow magnitude squared fact-break from ALICE data}) and the factorization-breaking coefficients defined in Eqs.~(\ref{eq: fact-break flow vector squared one bin}) and (\ref{eq: fact-break flow magnitude squared one bin}) is a factor,
\begin{equation}
 \begin{aligned}
\sqrt{\frac{\langle v_n^4(p) \rangle \langle v_n^2\rangle^2  }{\langle v_n^4 \rangle \langle v_n^2(p)\rangle^2 }} \ .
 \end{aligned}
\label{eq: factor differentiating def. of fact-break coeff}
\end{equation}
In our hydrodynamic simulations, we have verified that the deviation of this factor from unity is less than $6 \times 10^{-3}$ for transverse moemntum ranging from $0.5 \ \text{GeV}$ to $ 3.0 \ \text{GeV}$. The experimental values for the flow vector and magnitude squared correlations depicted in the figures are determined using the formulae given in Eqs.~(\ref{eq: estimate of flow vector squared fact-break from ALICE data}) and (\ref{eq: estimate of flow magnitude squared fact-break from ALICE data}).

\subsubsection{Flow angle decorrelation}
As discussed earlier, an experimental observable which could directly measure the flow angle correlation (or decorrelation) cannot be defined. Therefore, the flow angle decorrelation between the momentum averaged flow and momentum dependent flow could be defined following the similar prescription as in Eq.~(\ref{eq: second order flow angle decorrelation in two bins}), i.e. by taking the ratio of flow vector squared (Eq.~\ref{eq: fact-break flow vector squared one bin}) and flow magnitude squared (Eq.~\ref{eq: fact-break flow magnitude squared one bin}) factorization-breaking coefficients,
\begin{equation}
 \begin{aligned}
F_n(p) = \frac{r_{n;2}(p)}{r_n^{v_n^2}(p)} = \frac{\langle V_n^2 V_n^*(p)^2 \rangle}{\langle v_n^2 v_n(p)^2  \rangle} \ ,
 \end{aligned}
\label{eq: flow angle decorrelation one bin}
\end{equation}
which could be easily measured in the experiments. 

\begin{figure}[ht!]
\hspace{-0.3cm}\begin{subfigure}{0.5\textwidth}
\centering
\includegraphics[height=5.1 cm]{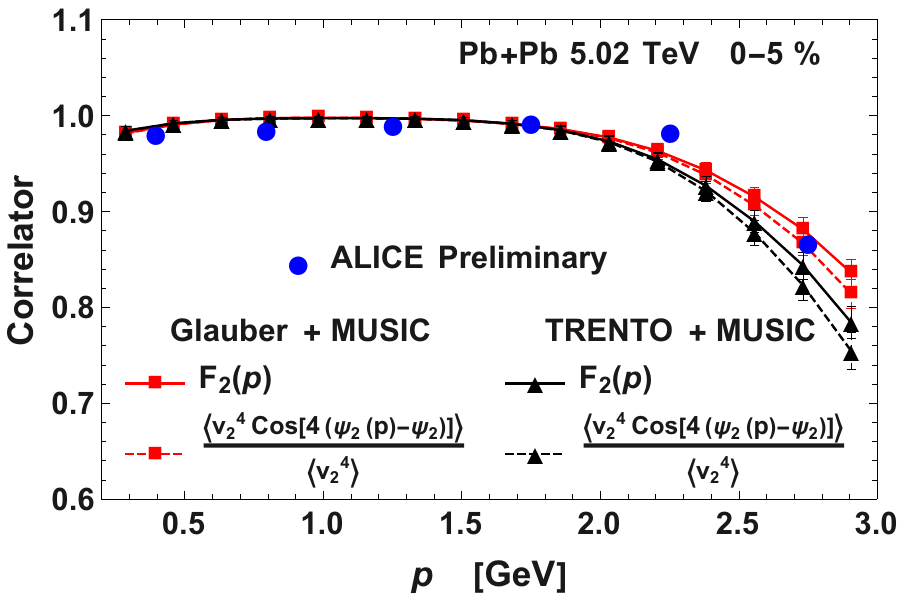}
\end{subfigure}~~
\begin{subfigure}{0.5\textwidth}
\centering
\includegraphics[height=5.1 cm]{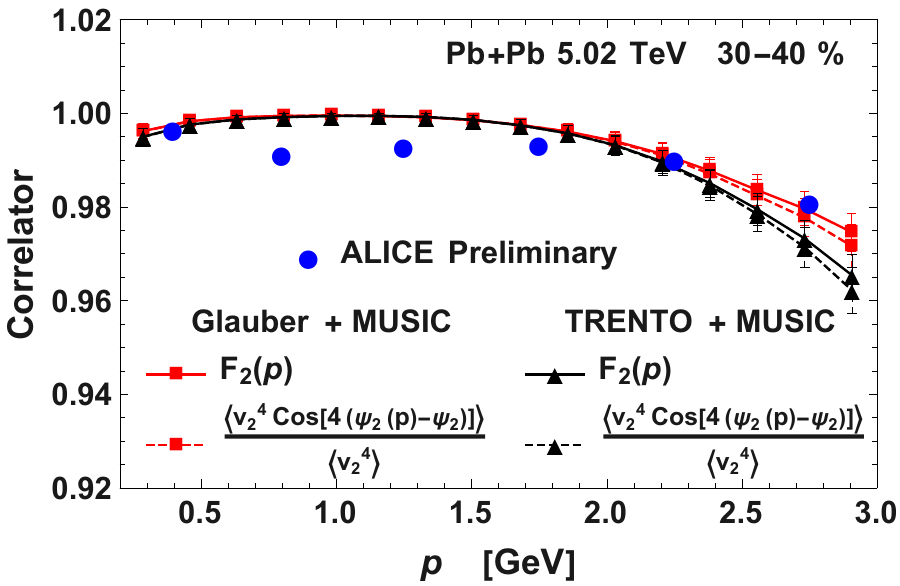}  
\end{subfigure}
\centering
\caption{Flow angle decorrelation as a function of the transverse momentum for the elliptic flow in Pb+Pb collision at 5.02 TeV for $0-5\%$ (left) and $30-40 \%$ (right) centrality. The red squares and black triangles denote the results obtained with the initial conditions from the Glauber and TRENTO model respectively. The solid lines denote the estimate of the flow angle correlation (or decorrelation) that can be measured in experiments while the dashed lines denote the actual flow-angle correlation. The blue dots represent the corresponding ALICE data for flow angle decorrelation. The figure is from the original publication~\cite{Bozek:2021mov}, coauthored by the author.}
\label{fig: flow angle decorrelation elliptic flow}
\end{figure}

\begin{figure}[ht!]
\includegraphics[height=6 cm]{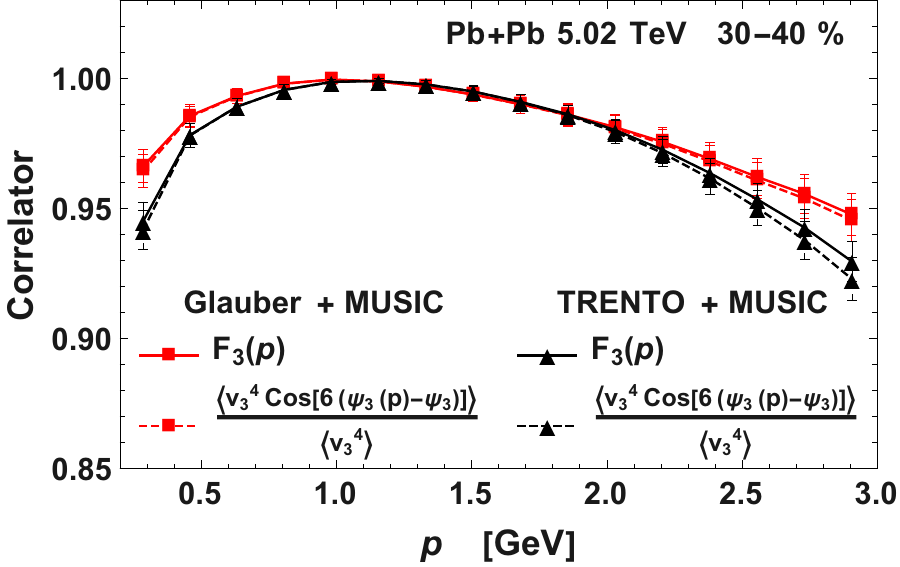}
\centering
\caption{ Flow angle decorrelation for the triangular flow with $30-40 \%$ centrality in Pb+Pb collision at 5.02 TeV. The symbols have similar meaning as Fig.~\ref{fig: flow angle decorrelation elliptic flow}. The figure is from the original publication~\cite{Bozek:2021mov}, coauthored by the author.}
\label{fig: flow angle decorrelation triangular flow}
\end{figure}

Figs.~\ref{fig: flow angle decorrelation elliptic flow} and \ref{fig: flow angle decorrelation triangular flow} show the results for the flow angle decorrelation obtained in hydrodynamic simulations compared to the ALICE data, for elliptic and triangular flow respectively. The model simulations predict a noticeable flow angle decorrelation between the global flow $V_n$ and the differential, momentum dependent flow $V_n(p)$, for both the elliptic and triangular flow. For the elliptic flow in central collision ($0-5\%$), the simulation results are consistent with the preliminary data from the ALICE collaboration. However, for the centrality $30-40\%$, both the model results and experimental data show a smaller angle decorrelation in comparison to the central collision. Once again, this can be linked to the unavoidable presence of the non-flow correlations which are expected to be relatively more important in off-central collisions. The difference in the decorrelation strength between the central and semi-central collisions arises from the global correlation of the elliptic flow with the initial geometry in non-central collisions.

Moreover, as claimed earlier, from Fig.~\ref{fig: flow angle decorrelation elliptic flow} it could be seen that the flow angle decorrelation is roughly one half of the flow vector decorrelation (Fig.~\ref{fig: flow vector squared fact-break coeff elliptic flow}). Therefore, it is quite safe to infer the following theorem (at least for the central collisions) based on our analysis, \\

{\centering \it Transverse momentum dependent flow vector decorrelation approximately amounts to the summation of flow magnitude and flow angle decorrelation (Appendix.~\ref{a: flow decorr toy model})} : 
\begin{equation}
 \begin{aligned}
[1-r_{n;2}(p)] \simeq [1-r_n^{v_n^2(p)}] + [1- F_n(p)] \ .
 \end{aligned}
\label{eq: relation between flow vector, magnitude and angle decorrelation}
\end{equation}

In Fig.~\ref{fig: flow angle decorrelation triangular flow}, similar to the past situations, there exist flow decorrelation on either side of the average transverse momentum ($\langle p \rangle$), indicating the peculiar characteristics of the triangular flow, coming from the dominance of fluctuations.   

The formula in Eq.~(\ref{eq: flow angle decorrelation one bin}) provides a way to estimate the flow angle decorrelation in experiments. It measures the quantity,
\begin{equation}
 \begin{aligned}
\frac{\langle v_n^2 v_n(p)^2 \cos[2n(\Psi_n(p)-\Psi_n)]\rangle}{\langle v_n^2 v_n(p)^2  \rangle},
 \end{aligned}
\label{eq: estimate of flow angle decorrelation one bin}
\end{equation}
providing an estimate of the true measure of the flow angle decorrelation,
\begin{equation}
 \begin{aligned}
\frac{\langle v_n^4 \cos[2n(\Psi_n(p)-\Psi_n)]\rangle}{\langle v_n^4  \rangle},
 \end{aligned}
\label{eq: true flow angle decorrelation one bin}
\end{equation}
which involves only the angle decorrelation. The above expression for the actual flow angle decorrelation is based on the implicit assumption that the decorrelations between the flow magnitudes in the numerator and the denominator cancel out. In other words, the transverse momentum  dependence of the flow magnitudes in Eq.~(\ref{eq: estimate of flow angle decorrelation one bin}) vanishes, so that we could get,
\begin{equation}
 \begin{aligned}
F_n(p) \approx \frac{\langle v_n^4 \cos[2n(\Psi_n(p)-\Psi_n)]\rangle}{\langle v_n^4  \rangle} \ .
 \end{aligned}
\label{eq: approx equality between the estimate and actual flow angle decorrelation one bin}
\end{equation}

The above formula (Eq.~(\ref{eq: true flow angle decorrelation one bin})) cannot be directly applied in experimental analysis. However, in our model, we can assess the similarity between the results obtained with the two formulae in Eqs.~(\ref{eq: estimate of flow angle decorrelation one bin}) and (\ref{eq: true flow angle decorrelation one bin}), specifically addressing the validity of Eq.~(\ref{eq: approx equality between the estimate and actual flow angle decorrelation one bin}). This is equivalent of examining whether the momentum dependence of the magnitude and flow angle in the numerator of Eq.~(\ref{eq: estimate of flow angle decorrelation one bin}) factorizes in the hydrodynamic model simulations. The results from hydrodynamic model calculated using both formulae are presented in Figs.~\ref{fig: flow angle decorrelation elliptic flow} and \ref{fig: flow angle decorrelation triangular flow}. It can be clearly seen that the two formulae produce similar results, which implies that the experimental measure (Eq.~(\ref{eq: flow angle decorrelation one bin})) can be used to estimate the {\it weighted} flow angle decorrelation. It should be noted that the angle correlation is weighted by the fourth power of flow magnitude i.e. only the $v_n^4$ weighted or {\it magnitude weighted} flow angle decorrelation can be measured in the experiment.
\begin{figure}[ht!]
\hspace{-0.3cm}\begin{subfigure}{0.5\textwidth}
\centering
\includegraphics[height=5.1 cm]{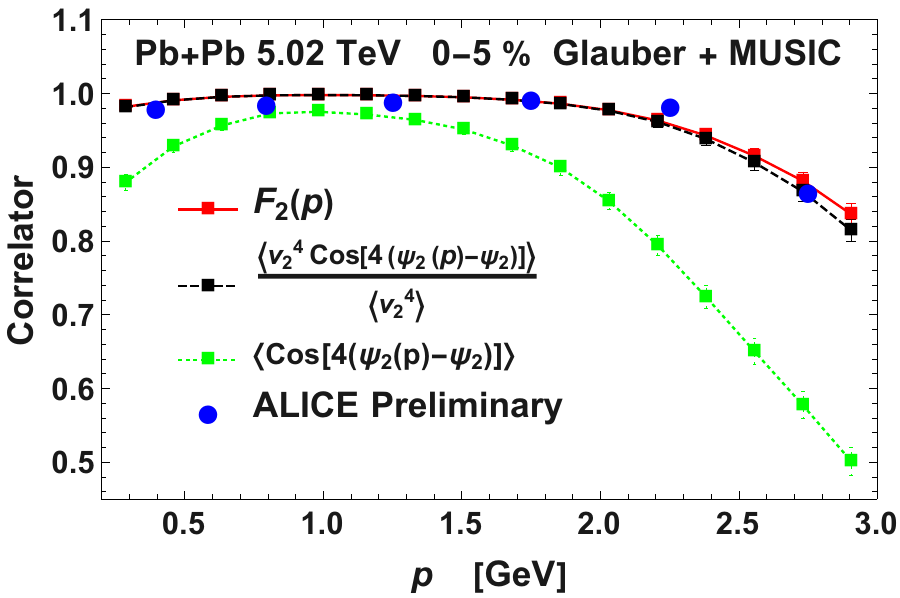}
\end{subfigure}~~~
\begin{subfigure}{0.5\textwidth}
\centering
\includegraphics[height=5.1 cm]{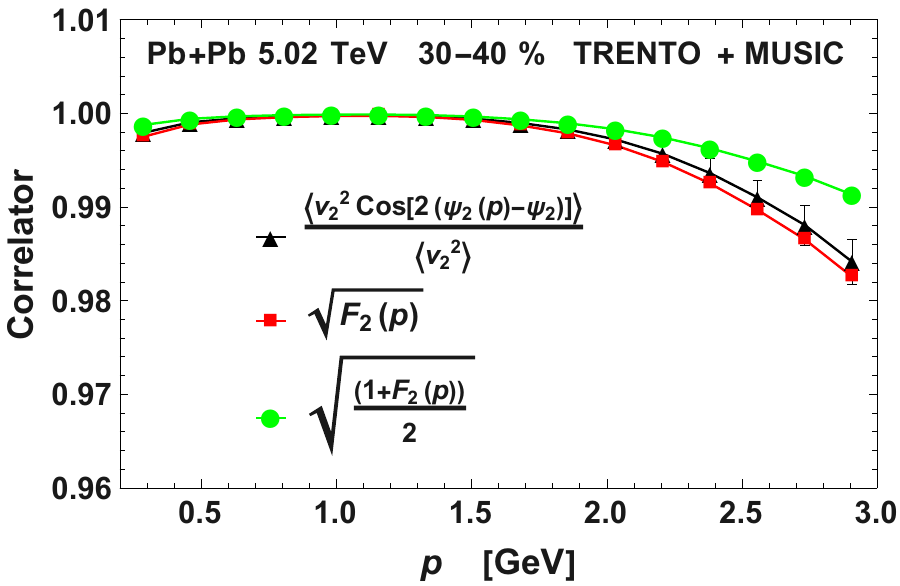}  
\end{subfigure}
\centering
\caption{Left: Comparison between different definitions of flow angle correlation along with the data for Pb+Pb collision in centrality $0-5 \%$. The red and black squares denote the experimental measure and actual flow angle correlations respectively. The green squares represent the simple angle correlation without any weights of flow magnitudes. The blue dots denote the ALICE data. Right:  The flow angle correlation between the first moment of flow vectors $V_2(p)$ and $V_2$ as a
function of the transverse momentum for centrality $30-40$\% (black triangles), compared to its approximation by $\sqrt{F_2(p)}$ (red squares) and its upper limit $\sqrt{(1+F_2(p))/2}$ (green dots). The figure is from the original publications~\cite{Bozek:2021mov, Bozek:2022slu}, coauthored by the author.}
\label{fig: flow angle correlation comparison and approx. definition}
\end{figure}

Another interesting fact in the present context is that the simple average of the cosine between the angles,
\begin{equation}
 \begin{aligned}
\langle \cos [2n(\Psi_n(p)-\Psi_n)]\rangle
 \end{aligned}
\label{eq: average between the cosines one bin}
\end{equation}
is not a measure of the angle decorrelation. Fig.~(\ref{fig: flow angle correlation comparison and approx. definition}) (left) shows the comparison between different definitions of the angle decorrelation compared with the ALICE data. It can be seen that Eq.~(\ref{eq: average between the cosines one bin}) provides very different results. However, such results are quite expected. In events where flow harmonics are large (large $|V_n|$ and large $|V_n(p)|$), the random decorrelation between two vectors is relatively small, making the angle decorrelation minimal. For a detailed discussion of the effect we refer interested readers to Ref.~\cite{Bozek:2017qir}.

It is to be noted that in the experiment, only the angle decorrelation between
the flow vectors {\it squared } can be measured. However, in the model one can check how it is related to the angle correlation between first order of flow i.e. between first moments of flow vectors : 
\begin{equation}
 \begin{aligned}
\frac{\langle v_n^2 \cos\left[ n \left( \Psi_n(p)-\Psi_n\right) \right]\rangle}{\langle v_n^2\rangle} \ .
 \end{aligned}
\label{eq: angle correlation between first order of flow}
\end{equation}
In Fig.~\ref{fig: flow angle correlation comparison and approx. definition} (right), we show the hydrodynamic model results for the flow angle correlation between the first moments and we find that it can be approximated as the square root of the angle decorrelation between the flow vectors squared $F_n(p)$, i.e.
\begin{equation}
 \begin{aligned}
\frac{\langle v_n^2 \cos\left[ n \left( \Psi_n(p)-\Psi_n\right) \right]\rangle}{\langle v_n^2\rangle} \simeq \sqrt{F_n(p)} \ .
 \end{aligned}
\label{eq: relation between first and second order flow angle decorrelation}
\end{equation}
A similar relation was found for the correlators of higher moments of flow vectors in pseudorapidity bins~\cite{ATLAS:2017rij}. In Fig.~\ref{fig: flow angle correlation comparison and approx. definition} (right), we also show the upper limit for the flow angle correlation, given by $\sqrt{(1+F_n(p))/2}$, as proposed by the ALICE Collaboration \cite{NielsenIS2021,ALICE:2022dtx}.

\subsection{Mixed-flow factorization-breaking: measure of non-linearity}
\label{mixed-flow fact break}
It is a general characteristic of flow harmonics $V_n$ that as we go higher order in $n$ ($n>2$), the flow vectors get contributions from the subsequent lower orders of flow, identified as the non-linear flow correlation~\cite{Bhalerao:2011yg,Teaney:2012ke}. The major contributing factors behind such non-linear behavior in the final state are the similar non-linear relations of the higher order eccentricities to the lower orders (Eqs.~(\ref{eq: relation between quadrangular flow and eccentricities}) and (\ref{eq: relation between pentagonal flow and eccentricities})) at the initial state, as discussed in Sec.~\ref{types of flow}.  As a natural consequence, the correlations between mixed flow harmonics or specifically, the correlations between the event planes corresponding to different orders of harmonic flow could provide very good measure of nonlinearities in the hydrodynamic expansion, as well as of the correlations in the initial state~\cite{Bhalerao:2011yg,Teaney:2012ke,Jia:2012ju,Qiu:2012uy,Luzum:2013yya,Teaney:2013dta,Bhalerao:2013ina,Qian:2016fpi,Giacalone:2016afq,Giacalone:2018wpp,Jia:2012sa,ALICE:2017fcd,CMS:2019nct,ALICE:2020sup}. 

Typically these studies involve correlators between the flow vectors of different orders or moments of flow, which are averaged over transverse momentum. However, the momentum dependent correlation between flow harmonics of different orders could reveal interesting measure of differential non-linear response of the medium and put additional constraints on the initial state models~\cite{Qian:2017ier,Bozek:2017qir}. Such studies involve higher order moments of mixed-flow harmonics in bins of transverse momentum i.e. many-particle correlator in small bins, which again cannot be measured easily in experiments due to limited statistics. 
\begin{figure}[ht!]
\hspace{-0.2cm}\begin{subfigure}{0.5\textwidth}
\centering
\includegraphics[height=4.4 cm]{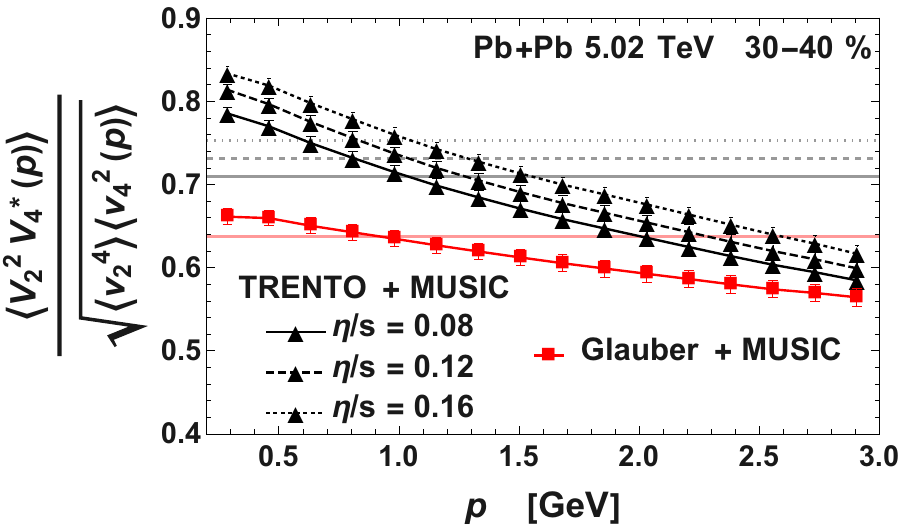}
\end{subfigure}~~
\begin{subfigure}{0.5\textwidth}
\centering
\includegraphics[height=4.4 cm]{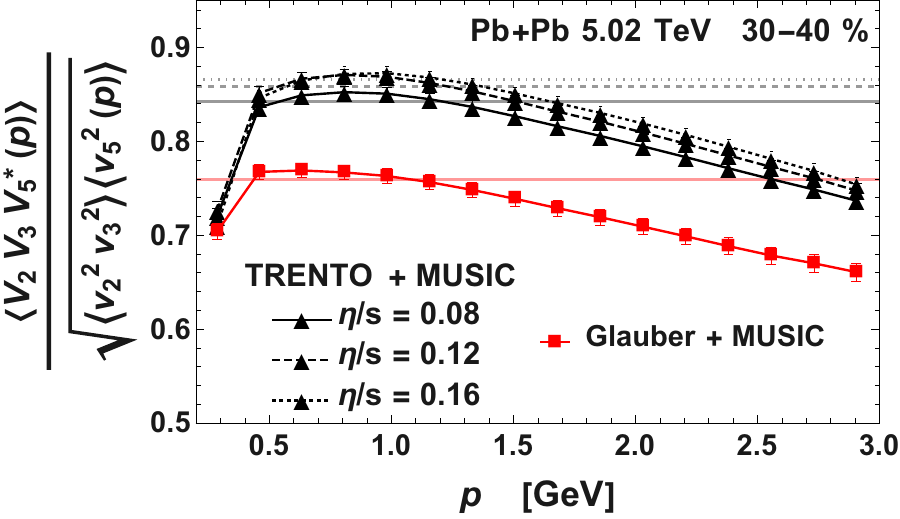}  
\end{subfigure}
\centering
\caption{Mixed-flow correlation between $V_2^2$ and $V_4(p)$ (left), and between $V_2V_3$ and $V_5(p)$ (right) as a function of transverse momentum in Pb+Pb collision at 5.02 TeV with $30-40 \%$ centrality. The results from the hydrodynamic model using Glauber and TRENTO model initial conditions are represented  by the red squares and black triangles respectively. The solid, dashed  and dotted black lines represent results with $\eta/s=0.08,\ 0.12, \ 0.16$ respectively for the TRENTO initial conditions. Corresponding correlation coefficients between the momentum averaged flow vector $V_2^2$ and $V_4$, and between $V_2V_3$ and $V_5$ are denoted by the horizontal lines. The figure is from the original publication~\cite{Bozek:2021mov}, coauthored by the author.}
\label{fig: V2sq-V4 and V2V3-V5 correlation}
\end{figure}

Following the similar method as before, we can construct a correlation coefficient (in this context the name correlation coefficient suits more, as used in the literature as well, instead of factorization-breaking coefficient) in the first order, between the mixed flow harmonics, with only one of the flow harmonics restricted to a transverse momentum bin through a generalized formula,
\begin{equation}
 \begin{aligned}
\frac{\langle  V_k V_l V_m^*(p)\rangle}{\sqrt{\langle v_k^2 v_l^2 \rangle \langle v_m^2(p) \rangle}},
 \end{aligned}
\label{eq: mixed-flow correlation general form}
\end{equation}
with $m=k+l$. For $m\neq k+l$, the cumulant in the numerator becomes zero by symmetry and no physical information can be extracted in that case. 

Owing to the non-linear relationships described in Eqs.~(\ref{eq: relation between quadrangular flow and eccentricities}) and (\ref{eq: relation between pentagonal flow and eccentricities}), we can write the non-linear relationships for the quadrangular and pentagonal flow in terms of elliptic and triangular flow as,
\begin{equation}
 \begin{aligned}
V_4 = V_4^L + \chi_{4,22} V_2^2 \eqsp{and} V_5 = V_5^L + \chi_{5,23} V_2V_3 \ ,
 \end{aligned}
\label{eq: non-linear relations between flow harmonics}
\end{equation}
where $V_4^L$ and $V_5^L$ correspond to the linear contributions with $\chi_{4,22}$ and $\chi_{5,23}$ as the {\it non-linear response coefficients } respectively. Therefore, we can construct the correlation coefficients between $V_2^2-V_4$ and $V_2V_3-V_5$, to measure the non-linearity. In particular, the momentum-dependent correlations would read,
\begin{equation}
 \begin{aligned}
\frac{\langle V_2^2 V_4^*(p) \rangle}{\sqrt{\langle v_2^4 \rangle \langle v_4^2(p) \rangle}} \eqsp{and} \frac{\langle V_2V_3 V_5^*(p) \rangle}{\sqrt{\langle v_2^2v_3^2 \rangle \langle v_4^2(p) \rangle}} \ .
 \end{aligned}
\label{eq: V2sq-V4p and V2V3-V5p correlation}
\end{equation}

In Fig.~\ref{fig: V2sq-V4 and V2V3-V5 correlation}, we show the results for the correlation coefficients measuring the nonlinear coupling between $V_4(p)$ and $V_2^2$ (left) and between $V_5(p)$ and $V_3 V_2$ (right) as a function of the transverse momentum.  We present the correlation coefficients for semi-central ($30-40\%$) collisions only, where the nonlinear components in $V_4$ and $V_5$ have dominant contributions. It could be seen that the correlations are the largest for small transverse momentum and the decorrelation gradually increases with increasing transverse momentum in case of $V_2^2-V_4(p)$ correlation. For $V_2V_3-V_5(p)$ correlation, there are decorrelations on the either side of average transverse momentum, similar to the triangular flow. Additionally, for mixed-flow correlations we show results corresponding to different $\eta/s$ with TRENTO initial conditions. The results show a weak dependence on the shear viscosity of the medium. The transverse momentum dependence of the mixed-flow correlations gives an additional constraint on the initial state and on the hydrodynamic evolution.
\begin{figure}[ht!]
\hspace{-0.2cm}\begin{subfigure}{0.5\textwidth}
\centering
\includegraphics[height=4.4 cm]{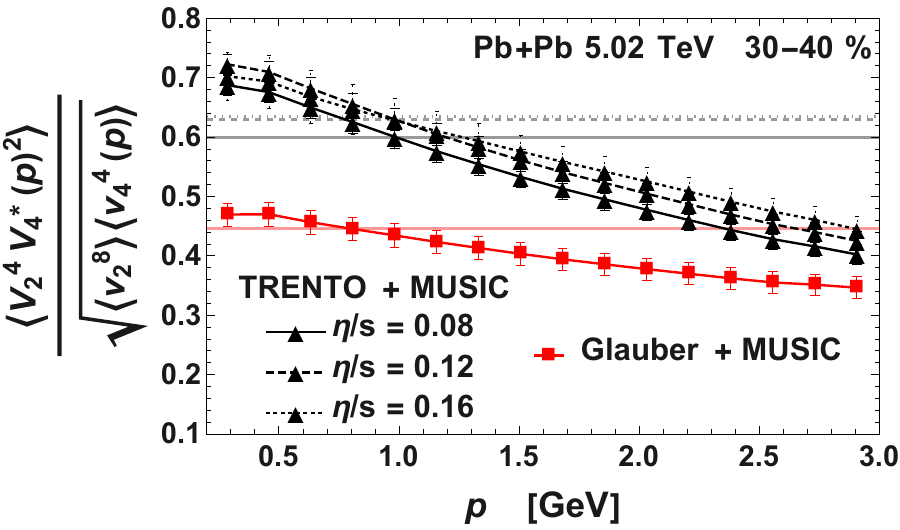}
\end{subfigure}~~
\begin{subfigure}{0.5\textwidth}
\centering
\includegraphics[height=4.4 cm]{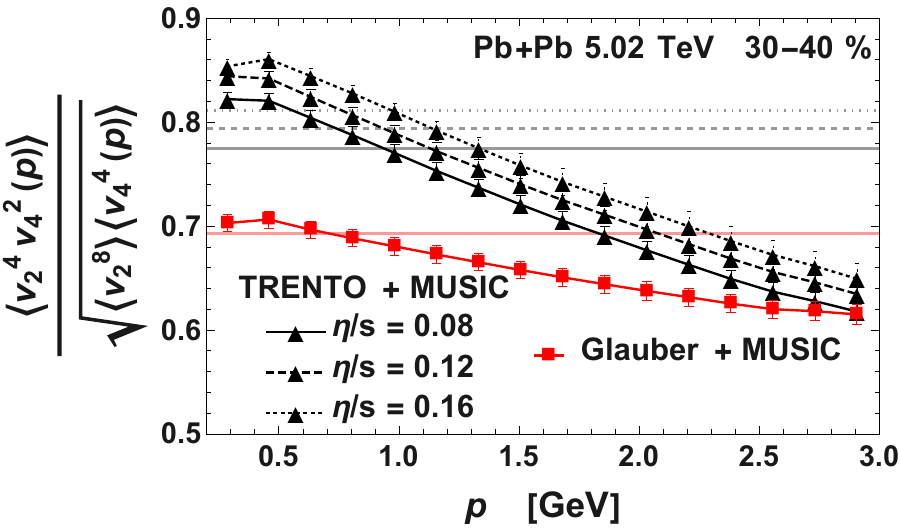}  
\end{subfigure}
\centering
\caption{Correlation coefficients between flow vector squared $V_2^4$ and $V_4^2(p)$ (left) and flow magnitudes squared $v_2^4$ and $v_4^2(p)$ (right) as a function of the transverse momentum in Pb+Pb collision at 5.02 TeV with $30-40\%$ centrality. The symbols have similar meaning as Fig.~\ref{fig: V2sq-V4 and V2V3-V5 correlation}. The figure is from the original publication~\cite{Bozek:2021mov}, coauthored by the author.}
\label{fig: V24-V4sq flow vector and magnitude correlation}
\end{figure}

In order to extract the momentum-dependent flow angle correlation (or event-plane correlation) between mixed flow harmonics, separately from the magnitude decorrelation, similar to the previous methods, correlators involving higher powers of flow harmonics must be taken into account. For example, if we consider the non-linear coupling between $V_4$ and $V_2$, the correlation coefficient between the flow vector squared is given by,
\begin{equation}
 \begin{aligned}
\frac{\langle V_2^4 V_4^*(p)^2 \rangle}{\sqrt{\langle v_2^8 \rangle \langle v_4^4(p) \rangle}}, 
 \end{aligned}
\label{eq: V24-V4sqp flow vector correlation}
\end{equation}
while the correlation coefficient  or factorization-breaking coefficient between the non-linear flow magnitude squared can be constructed as,
\begin{equation}
 \begin{aligned}
\frac{\langle v_2^4 v_4(p)^2 \rangle}{\sqrt{\langle v_2^8 \rangle \langle v_4^4(p) \rangle}}, 
 \end{aligned}
\label{eq: v24-v4sqp flow magnitude correlation}
\end{equation}
and then the flow angle correlation can be estimated from the ratio of the above two as,
\begin{equation}
 \begin{aligned}
\frac{\langle V_2^4 V_4^*(p)^2 \rangle}{\langle v_2^4 v_4(p)^2 \rangle} \ .
 \end{aligned}
\label{eq: v24-v4sqp flow angle correlation}
\end{equation}
\begin{figure}[ht!]
\includegraphics[height=6 cm]{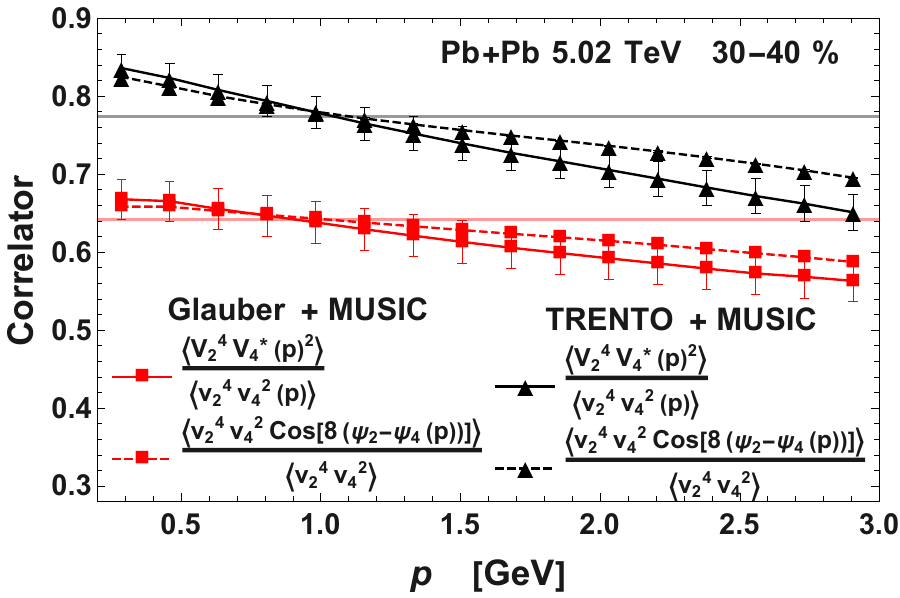}
\centering
\caption{Flow angle decorrelation between $V_2^4$ and $V_4^2(p)$ as a function of transverse momentum in Pb+Pb collision at 5.02 TeV with $30-40 \%$ centrality. The red squares and the black triangles denote the hydrodynamic results obtained with Glauber and TRENTO model initial conditions respectively. The solid lines denote the estimate of the flow angle decorrelation while the dashed lines represent the actual measure of flow angle correlation between the mixed harmonics. Corresponding angle correlations between the momentum-averaged flow $V_2^4$ and $V_4^2$, are indicated by the horizontal lines. The figure is from the original publication~\cite{Bozek:2021mov}, coauthored by the author.}
\label{fig: flow angle decorrelation V24-V4sqp}
\end{figure}

The correlation coefficients between flow vector and magnitude squared are shown in Fig.~\ref{fig: V24-V4sq flow vector and magnitude correlation}. It should be noted that the correlation between higher powers of the flow vectors (Fig.~\ref{fig: V24-V4sq flow vector and magnitude correlation} (left)) is smaller than between the lower powers of the respective flow vectors (Fig.~\ref{fig: V2sq-V4 and V2V3-V5 correlation} (left)). Moreover, it could be seen that the flow magnitude decorrelation accounts for about one half of the flow vector decorrelation shown in Fig.~\ref{fig: V24-V4sq flow vector and magnitude correlation}. Therefore, the relation described by Eq.~(\ref{eq: relation between flow vector and magnitude decorrelation}) is also valid for non-linear mixed-flow correlators. 

\begin{figure}[ht!]
\includegraphics[height=6 cm]{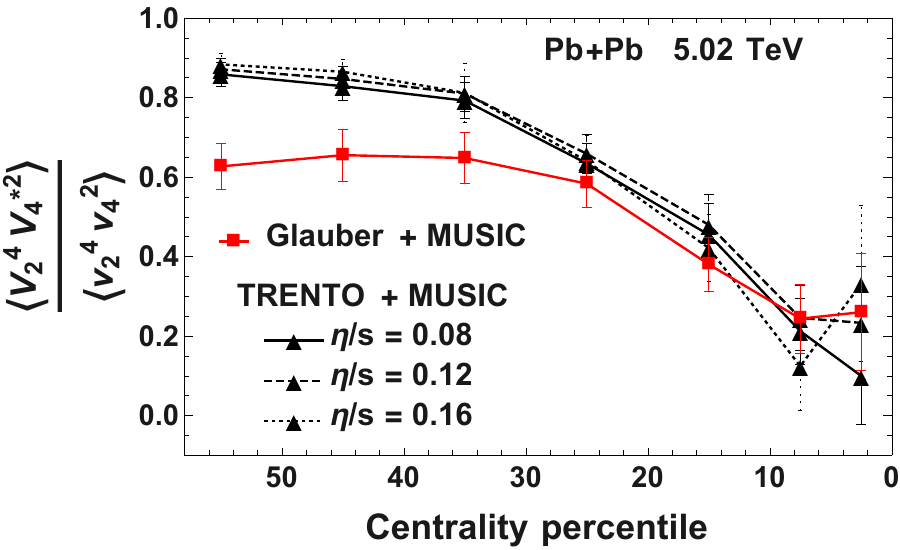}
\centering
\caption{Centrality dependence of the flow angle correlation between $V_2^4$ and $V_4^2$ in Pb+Pb collision at 5.02 TeV. The hydrodynamic results with Glauber and TRENTO model are denoted with red squares and black triangles respectively. For the TRENTO model initial conditions, the solid, dashed  and dotted lines represent results with the viscosities, $\eta/s=0.08,\ 0.12,$ and $ 0.16$ respectively. The figure is from the original publication~\cite{Bozek:2021mov}, coauthored by the author.}
\label{fig: centrality dependence of V24-V4sq flow angle decorrelation}
\end{figure}

In Fig.~(\ref{fig: flow angle decorrelation V24-V4sqp}), we show results for the flow-angle correlation for mixed-harmonics, which involve six-particle correlators and could be extracted from the experimental data. It could be observed that the flow angle decorrelation is again approximately one half of the flow vector decorrelation (Fig.~\ref{fig: V24-V4sq flow vector and magnitude correlation}). We notice that the flow angle correlation (Eq.~\ref{eq: v24-v4sqp flow angle correlation}) defined as the ratio of the correlation coefficient between flow vectors and of the factorization breaking  coefficient between the flow magnitudes, serves as a good approximation for the flow angle correlation weighted with the powers of flow magnitudes,
\begin{equation}
 \begin{aligned}
\frac{\langle v_4^2 v_2^4 \cos [ 8( \Psi_4(p)- \Psi_2 ) ] \rangle}{\langle  v_4^2 v_2^4 \rangle},
 \end{aligned}
\label{eq: v24-v4sqp actual flow angle correlation}
\end{equation}
which is the true measure of the flow angle correlation as discussed previously. For completeness, in Fig.~\ref{fig: centrality dependence of V24-V4sq flow angle decorrelation}, we show the centrality dependence of flow angle decorrelation between momentum averaged flow vectors $V_4^2$ and $V_2^4$. Alongside the momentum dependent flow angle correlation depicted in Fig.~\ref{fig: flow angle decorrelation V24-V4sqp}, this can serve as an extra experimental observable, showing sensitivity to various models of initial conditions.

\subsection{Experimental measurements and removing non-flow correlation}
\label{Removing non-flow correlation}
As seen in the discussions above, the simulation results for centrality $30-40$\% do not reproduce the experimental data of the ALICE Collaboration. This is because in semi-central collisions, in general we find stronger decorrelation in our model, while in the experiment the harmonic flow correlation is somewhat larger and sometimes surpassing $1$. This might be attributed to a large contribution of non-flow correlations in the measured data. 

 In order to minimize the non-flow correlations, an estimate of correlation measures can be constructed involving four harmonic flow vectors in separated pseudorapidity bins. For example, using four bins in pseudorapidity located at $-\eta_{F}, -\eta, \eta, \eta_{F}$, the factorization coefficient for flow vectors squared can be estimated as,
\begin{equation}
 \begin{aligned}
 r_{n;2}(p)\simeq 
 \frac{\langle V_n(-\eta_{F}) V_n^\star(-\eta,p)V_n^\star(\eta,p) V_n(\eta_{F})  \rangle \langle V_n^\star(-\eta) V_n(\eta) \rangle }{\langle V_n(-\eta_{F}) V_n^\star(-\eta)V_n^\star(\eta) V_n(\eta_{F})  \rangle \langle V_n^\star(-\eta,p) V_n(\eta,p) \rangle } \ , 
 \end{aligned}
\label{eq: experimental estimation of flow vector squared fact-break coeff}
\end{equation}
the factorization breaking coefficient between flow vector magnitudes can be defined as
\begin{equation}
 \begin{aligned}
 r_{n;2}^{v_n^2}(p)\simeq 
 \frac{\langle V_n(-\eta_{F}) V_n(-\eta,p)V_n^\star(\eta,p) V_n^\star(\eta_{F})  \rangle \langle V_n^\star(-\eta) V_n(\eta) \rangle }{\langle V_n(-\eta_{F}) V_n(-\eta)V_n^\star(\eta) V_n^\star(\eta_{F})  \rangle \langle V_n^\star(-\eta,p) V_n(\eta,p) \rangle } \ , 
 \end{aligned}
\label{eq: experimental estimation of flow magnitude squared fact-break coeff}
\end{equation}
and the flow angle correlation is estimated from the ratio the two quantities above
\begin{equation}
 \begin{aligned}
F_n(p) \simeq \frac{\langle V_n(-\eta_{F}) V_n^\star(-\eta,p)V_n^\star(\eta,p) V_n(\eta_{F})  \rangle\langle V_n(-\eta_{F}) V_n(-\eta)V_n^\star(\eta) V_n^\star(\eta_{F})  \rangle }{\langle V_n(-\eta_{F}) V_n(-\eta,p)V_n^\star(\eta,p) V_n^\star(\eta_{F})  \rangle\langle V_n(-\eta_{F}) V_n^\star(-\eta)V_n^\star(\eta) V_n(\eta_{F})  \rangle },
 \end{aligned}
\label{eq: experimental estimation of flow angle decorrelation}
\end{equation}
similar to~\cite{ATLAS:2017rij}. The four-particle correlators in these formulae involve flow vectors at different pseudorapidity and transverse momentum. Consequently, in the result we would observe a combination of flow decorrelation in transverse momentum as well as in pseudorapidity. However, assuming the longitudinal decorrelation factorizes from the transverse momentum decorrelation, it eventually cancels between the numerator and the denominator, i.e. we assume,
\begin{equation}
 \begin{aligned}
 \langle V_n(\eta, p) V_n^*(\eta, p)\rangle \simeq \sqrt{ \langle V_n(\eta) V_n^*(\eta)\rangle  \langle V_n( p) V_n^*( p)\rangle } \ .
 \end{aligned}
\label{eq: factorization of flow between pT and eta}
\end{equation}

In the above formulae for experimental measure, we also use the approximation
\begin{equation}
 \begin{aligned}
\frac{\sqrt{\langle v_n^4(p)\rangle}}{\langle v_n^2(p) \rangle } \simeq
\frac{\sqrt{\langle v_n^4\rangle}}{\langle v_n^2 \rangle } \ ,
 \end{aligned}
\label{eq: approximation between pT-dependent and pT-averaged flow}
\end{equation}
as discussed earlier in Eq.~(\ref{eq: factor differentiating def. of fact-break coeff}). By using this approximation, we remove the difficulty of measuring a four-particle correlator in the experiment, where all flow vectors are defined in a narrow transverse momentum bins, as in the denominators of Eqs.~(\ref{eq: fact-break flow vector squared one bin}) and (\ref{eq: fact-break flow magnitude squared one bin}). Note that the flow vectors at forward and backward rapidities $\pm \eta_{F}$ do not require the measurement of the transverse momenta of the particles and can be measured using the forward/backward calorimeters. Only the flow vectors $V_n(\pm \eta, p)$ necessitate the measurement of the individual particles' transverse momenta. For this purpose, two bins  well separated in pseudorapidity within the central rapidity region of the detector's acceptance can be utilized. Thus employing a simple approximation (Eq.~(\ref{eq: approximation between pT-dependent and pT-averaged flow})) and using well-separated bins in pseudorapidity, the experimental difficulty in measurement of multi-particle correlators due to limited statistics can be surmounted and the influence of non-flow correlations can be substantially reduced.

%*******************************************************************************
%****************************** Fourth Chapter **********************************
%*******************************************************************************
\chapter{Transverse momentum fluctuations in ultracentral collisions}

% **************************** Define Graphics Path **************************
\ifpdf
    \graphicspath{{Chapter4/Figs/Raster/}{Chapter4/Figs/PDF/}{Chapter4/Figs/}}
\else
    \graphicspath{{Chapter4/Figs/Vector/}{Chapter4/Figs/}}
\fi

In a heavy-ion collision event, a large number of particles (around $35000$ hadrons in a head-on collision of two $^{208}$Pb nuclei at $5.02$ TeV at the LHC) are emitted at the final state, which share the total energy of the initial fireball in terms of individual momentum, specifically transverse momentum ($p_T$) which is particularly interesting in the present context. In each such event, one can define a { \it mean transverse momentum per charged particle}, which we denote as $[p_T]$ and it is defined as $[p_T] \equiv (\sum p_T)/N_{ch}$. Similar to the fluctuations of anisotropic flow, discussed in the previous chapter, event-by-event fluctuations of transverse momentum per particle $[p_T]$, are of great interest and can serve as an excellent and even more direct probe of the QGP matter produced in the collision. For collisions with same $N_{ch}$, the transverse momentum per particle, $[p_T]$ fluctuates from event to event. There exist of course trivial {\it statistical} fluctuations of $[p_T]$, due to averaging over a finite number of charged particles, but the measured fluctuations are larger than that. The excess fluctuations are known as true {\it dynamical} fluctuations of $[p_T]$, which we are interested in. They originate due to event-by-event fluctuations of the distribution of the source in the initial state of collisions, known as {\it shape fluctuation} which could have a geometrical origin as well as a quantum or intrinsic origin reflecting the randomness in positions of the nucleons in colliding nuclei. The dynamical fluctuations of $[p_T]$ have been studied over the years in theoretical models~\cite{Liu:1998xf,Voloshin:1999yf,Korus:2001au,Ferreiro:2003dw,Gavin:2003cb,Broniowski:2005ae,Broniowski:2009fm,Bozek:2012fw,Bozek:2017elk,Gardim:2020sma,Schenke:2020uqq,Giacalone:2020lbm,Bhatta:2021qfk} and have been measured in experiments~\cite{PHENIX:2002aqz,PHENIX:2003ccl,STAR:2003cbv,ALICE:2014gvd,Tripathy:2022vwb,ALICE:2023tej}. In this chapter, we will focus on the characteristics of $[p_T]$ in ultracentral~\cite{CMS:2013bza,Plumari:2015cfa,Shen:2015qta,Carzon:2020xwp,Liu:2022kvz,Giannini:2022bkn,Kuroki:2023ebq} Pb+Pb collisions at the LHC. The intriguing behavior of average transverse momentum ( which we denote as $\langle p_T\rangle \equiv \langle [p_T]\rangle  $, averaged over events) in ultracentral collisions provides a sensitive probe of hydrodynamics and could help in extracting important physical quantities e.g. speed of sound in the QGP medium with utmost precision~\cite{Gardim:2019brr}. After subtracting the trivial statistical fluctuations, the dynamical fluctuations of $[p_T]$ are very small (below $1 \%$)~\cite{ALICE:2014gvd}. However, recent measurements by the ATLAS collaboration~\cite{ATLAS:2022dov, ATLAS:2023xpw}, studying the variation of $[p_T]$ as a function of $N_{ch}$, shows a very striking and peculiar pattern of $[p_T]$-fluctuation in the ultracentral (high $N_{ch}$) regime. This motivated us to perform a meticulous and careful study of $[p_T]$-fluctuation in ultracentral Pb+Pb collisions, which ultimately proves to offer great physical significance and direct probe to the formation of the QGP fluid. The following sections are, for the most part, presentations from the original publications~\cite{Samanta:2023amp,Samanta:2023kfk}, coauthored by the author.

\section{Variance of $[p_T]$-fluctuation}
\label{Variance}
Fluctuations of the mean transverse momentum per particle $[p_T]$ observed in heavy-ion collisions can have many features. In general, fluctuations can be characterized by different orders of cumulant: mean, variance, skewness, kurtosis etc. Let us first consider the variance, which is the primary quantity when one talks about fluctuations.

\subsection{Strange behavior of the ATLAS data}
\label{ATLAS data}
\begin{figure}[ht!]
\begin{subfigure}{0.5\textwidth}
\centering
\includegraphics[height=6 cm]{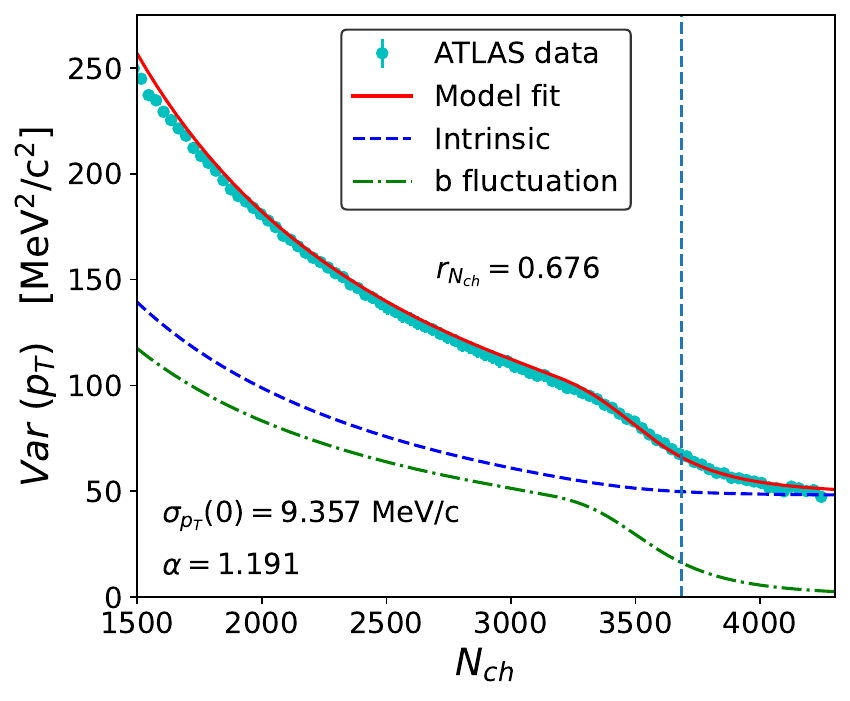}
\end{subfigure}~~
\begin{subfigure}{0.5\textwidth}
\centering
\includegraphics[height=6 cm]{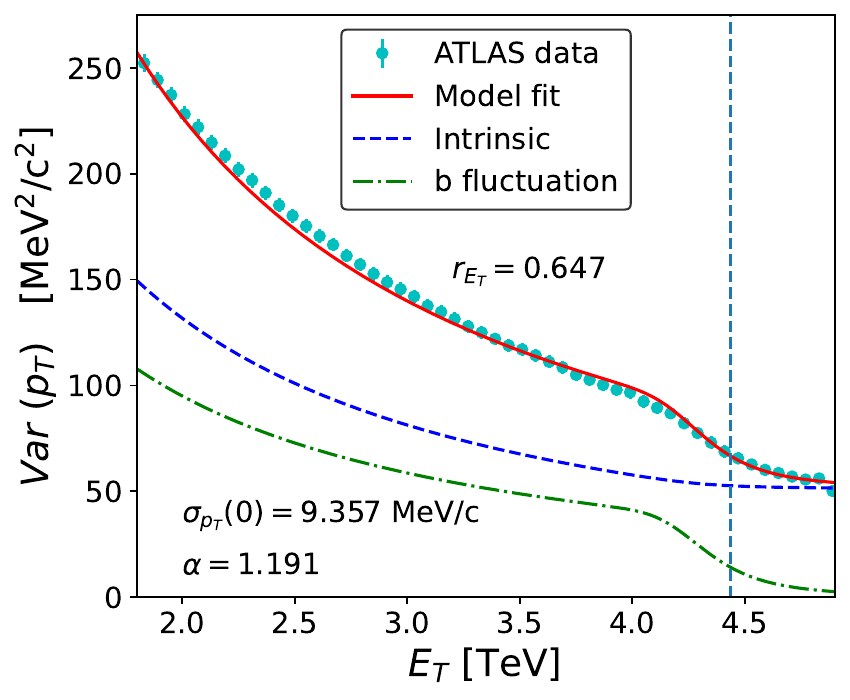}  
\end{subfigure}
\centering
\caption{Variance of the transverse momentum per particle $[p_T]$ as a function of the centrality estimators $N_{ch}$ (left) and $E_T$ (right). Symbols represent the ATLAS data~\cite{ATLAS:2022dov}. The model fit to the data are represented by the solid red lines. We show separately the two contributions to the variance as per Eq.~(\ref{eq: mean and variance of pt}) from our model calculation : the contribution from intrinsic and impact parameter fluctuations represented by dashed and dashed-dotted lines respectively. The sum of the two contributions is the solid red line. The figure is from the original publication~\cite{Samanta:2023amp}, coauthored by the author.}
\label{fig: varpt}
\end{figure}

We start with the ATLAS data for $[p_T]$-fluctuation~\cite{ATLAS:2022dov, ATLAS:2023xpw} presenting the variance as a function of the centrality estimator which can be either the charged particle multiplicity $N_{ch}$ or the transverse energy deposited on the forward calorimeter $E_T$, shown in Fig.~\ref{fig: varpt}. The ATLAS collaboration at the LHC detects the charged particles through an inner detector that covers an angular range of approximately $10^\circ<\theta< 170^\circ$ (where $\theta$ represents the angle between the collision axis and the particle's direction). The detector measures the transverse momenta of these particles, given by $p_T\equiv p\sin\theta$, where $p$ is the magnitude of total momentum. The analysis includes all charged particles detected in a specific interval of $p_T$. In particular, results in two $p_T$ intervals: $0.5<p_T<5$~GeV$/c$ and $0.5<p_T<2$~GeV$/c$ are presented. Let us first consider the results for the interval with larger upper $p_T$-cut i.e.$0.5<p_T<5$~GeV$/c$, the default interval in our analysis. Later we will discuss the effect of $p_T$-cut dependence of the fluctuations. 

A first observation on the plots reveals that after subtracting trivial statistical fluctuations, the remaining dynamical fluctuations~\cite{STAR:2003cbv} are very small, below 1\% in central Pb+Pb collisions~\cite{ALICE:2014gvd}. We focus on these small dynamical fluctuations in this chapter. The left panel of Fig.~\ref{fig: varpt} displays their variance as a function of $N_{ch}$. It is seen that the variance decreases as $N_{ch}$ increases i.e as we approach towards more central events. The most striking phenomenon is a sudden steep decrease, by a factor of $\sim 2$, over a narrow interval of $N_{ch}$ around $3700$. Other models of the collision in which the Pb+Pb collision is treated as a superposition of independent nucleon-nucleon collisions fail to reproduce this behavior. For example, in the HIJING model~\cite{Wang:1991hta,Gyulassy:1994ew}, the decrease of the variance is found to be proportional to $1/N_{ch}$~\cite{ALICE:2014gvd,Bhatta:2021qfk} for all $N_{ch}$. In the following, we argue that the impact parameter, $b$, plays a crucial role behind this phenomenon. It will be shown that the relation between multiplicity $N_{ch}$ and $b$ is not one-to-one, while $[p_T]$ depends on both quantities.

\subsection{Hydro vs HIJING results at fixed b}
\label{hydro vs hijing}
\begin{figure}[ht!]
\begin{subfigure}{0.5\textwidth}
\centering
\includegraphics[height=3.5 cm, valign=c]{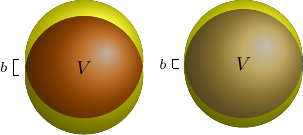}
\end{subfigure}~~
\begin{subfigure}{0.5\textwidth}
\centering
\includegraphics[height=7 cm, valign=c]{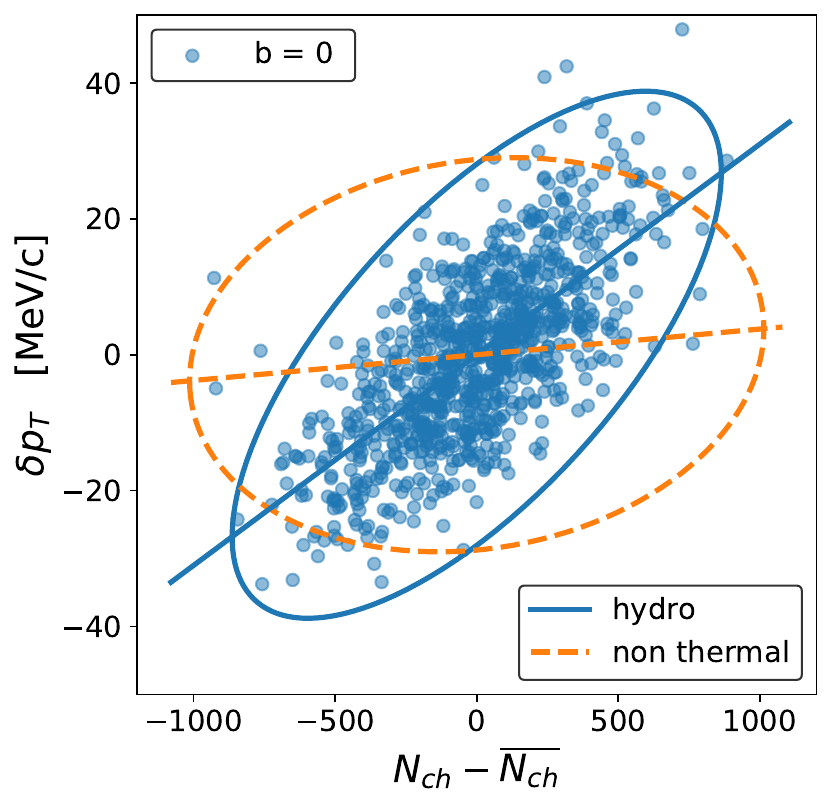}  
\end{subfigure}
\centering
\caption{Left: Pictorial depiction of Pb+Pb collisions at fixed multiplicity but different impact parameters: $b=1.8$~fm (left) and $b=1.0$~fm (right). The impact parameters correspond to centrality fractions $c_b\simeq 1.5$~\% and $c_b\simeq 0.5$~\% respectively. The difference between these two values is typically the spread of $c_b$ at fixed multiplicity. A larger value of $b$ corresponds to a smaller collision volume, resulting in a larger density, which we symbolically denote with a darker color.  Right: The scatter plot between the charged particle multiplicity $N_{ch}$ and the transverse momentum per particle $[p_T]$ in Pb+Pb collisions at 5.02~TeV and $b=0$. The symbols and the solid lines represent the results obtained from hydrodynamic simulations for 1000 events. The  dashed lines correspond to the results of $1.4\times 10^6$ collisions simulated with HIJING~\cite{Gyulassy:1994ew} (individual points are not shown). The results are obtained using the same kinematic cuts used in the ATLAS analysis. Instead of plotting $N_{ch}$ and $[p_T]$ themselves, we plot the differences $N_{ch}-\overline{N_{ch}}$ and $\delta p_T\equiv [p_T]-\overline{p_{T0}}$,  where $\overline{N_{ch}}=6662$ and $\overline{p_T}=1074$~MeV$/c$ are the event averaged values. The straight lines indicate the average value $\overline{\delta p_T}(N_{ch},b=0)$ (Eq.~\ref{eq: mean and variance of pt}), and the ellipses are 99\% confidence ellipses evaluated by assuming that the distribution is a correlated Gaussian using Eq.~(\ref{eq: correlated Gaussian}). The left panel is from the original publication~\cite{Samanta:2023kfk} and the right panel is a modification of the figure in the original publication~\cite{Samanta:2023amp}, coauthored by the author.}
\label{fig: b-fluct and pt-Nch scatter plot}
\end{figure}

In order to illustrate the dependence of $[p_T]$ on $N_{ch}$, we simulate 1000 Pb+Pb collisions at 5.02 TeV at fixed impact parameter $b=0$, using relativistic viscous hydrodynamics, and evaluate $N_{ch}$ and $[p_T]$ for every collision. Our hydro-simulation set up remains same as before and discussed in Appendix.~\ref{a: hydro vs hijing}. The right panel of Fig.~\ref{fig: b-fluct and pt-Nch scatter plot} displays the distribution of $[p_T]$ and $N_{ch}$ at $b=0$. One can see that both the quantities exhibit significant dynamical fluctuations and span finite ranges. In particular, for $N_{ch}$, the fluctuations around the mean extend up to $\sim 14\%$, whereas it is around $\sim 3\%$ for $[p_T]$. These fluctuations originate from different sources of quantum fluctuations: from the fluctuations in the positions of nucleons at the time of impact~\cite{Miller:2007ri}, from the partonic content of the nucleons~\cite{Gelis:2010nm} as well as from the process of particle production.\footnote{Please note that we only consider spherical nuclei. For deformed nuclei, the fluctuations in their orientations must be considered, which in turn affect both the multiplicity~\cite{STAR:2015mki} and the momentum per particle~\cite{Giacalone:2019pca}.} These fluctuations are taken into account~\cite{Aguiar:2001ac} in modern hydrodynamic simulations by implementing a different initial density profile (the initial condition of hydrodynamic equations) for each collision event. The second observation in Fig.~\ref{fig: b-fluct and pt-Nch scatter plot} is that there is a significant positive correlation between $[p_T]$ and $N_{ch}$ in hydrodynamics, which carries a crucial importance in our analysis. 

\textbf{\textit{Thermodynamic interpretation of the correlation:}} The correlation observed in Fig.~\ref{fig: b-fluct and pt-Nch scatter plot} could be interpreted as a natural consequence of local thermalization. Thermalization is an underlying assumption of the hydrodynamic description of the QGP evolution. If we fix the impact parameter $b$, we essentially fix the collision volume $V$ (left panel of Fig.~\ref{fig: b-fluct and pt-Nch scatter plot}). Then at fixed volume, larger $N_{ch}$ implies a larger density $N_{ch}/V$. In hydrodynamics, as the system is locally thermalized, a larger density corresponds to a higher initial temperature. It should be noted that, relativity plays an essential role in this interpretation. In non-relativistic thermodynamics, density and temperature are independent variables and if we heat a system at constant volume, the density remains unchanged, because the number of particles is conserved. On the other hand, in a relativistic system, particles can be both created (by converting kinetic energy into mass) and destroyed. In that case, a larger temperature implies a higher density. 
Therefore, a higher density of the system implies a higher temperature, resulting in a higher energy per particle in the final state of hydrodynamic evolution, which eventually means a larger momentum per particle $[p_T]$~\cite{Gardim:2019xjs}. In the previous sentence, we mean that the larger energy per entropy in the initial state translates into larger energy per particle in the final state, while maintaining consistency with the thermodynamic picture.  

The above phenomenon could be understood from the perspective of collective flow as well. The larger density at fixed collision volume increases the overall magnitude of the outward pressure in the fireball. This is another way of realizing the thermalization of the system manifested through the equation of state (pressure and energy density are related). The larger pressure, through collective flow, results in larger transverse momentum per particle in the final state after the hydrodynamic evolution.    

In order to illustrate that the positive correlation between $[p_T]$ and $N_{ch}$ is not trivial but caries a greater physical significance, we also display results of simulations using the HIJING model~\cite{Gyulassy:1994ew} in the right panel of Fig.~\ref{fig: b-fluct and pt-Nch scatter plot}. In HIJING which is a non-thermal model, particles do not interact after they are produced. We see that the corresponding correlation is much smaller (individual points are not shown) by a factor of $\sim 10$. However, on should note that while thermalization always implies a positive correlation, the converse statement does not hold. As an example, in the color-glass condensate picture of high-energy collisions, such a correlation could be already present at the level of particle production, since both the momentum per particle and the particle density increase with the saturation scale~\cite{Gelis:2010nm}. More details about the analysis of hydrodynamic and HIJING simulations are given in Appendix~\ref{a: hydro vs hijing}

\subsubsection{$[p_T]$-fluctuation at fixed $N_{ch}$ : Effect of impact parameter fluctuations}

Let us now discuss how thermalization plays an important role on the observed $[p_T]$ fluctuations in the data. First, note that the experimental analysis is performed at fixed $N_{ch}$ which is traditionally used as an estimator of centrality, whereas our hydrodynamic simulation is done at fixed $b$ ($=0$ to be specific in this case). 
Both choices are driven by practical considerations. Experimentally, the impact parameter $b$ cannot be measured. On the other hand, in simulations, $b$ can be fixed before starting, and the final multiplicity $N_{ch}$ is obtained only at the end. Therefore, in order to interpret experimental results, we need to provide explanations at fixed $N_{ch}$, with $b$ varying. Let us understand the phenomenon of thermalization and its effect on $[p_T]$, as explained in the paragraph preceding the last, at fixed $N_{ch}$ and fluctuating $b$. If $N_{ch}$ is fixed and $b$ fluctuates, then a larger $b$ results in a smaller collision volume $V$ and so a larger density $N_{ch}/V$, hence larger temperature and eventually larger $[p_T]$ on average in the final state. The opposite scenario occurs if $b$ is smaller, as shown in the left panel of Fig.~\ref{fig: b-fluct and pt-Nch scatter plot}. Thus thermalization and collective flow of the QGP medium have direct consequences on $[p_T]$ of the particles and at fixed multiplicity, impact parameter fluctuations contribute to transverse momentum fluctuations. This is the main underlying physics behind our analysis, as we will see that the contribution of impact parameter fluctuations gradually disappears in the ultracentral regime causing the sharp decline observed in the data.     

\subsection{Modelling the correlation : Two dimensional Gaussian }
\label{correlated Gaussian}
To understand the data, we first need to model the correlation between $[p_T]$ and $N_{ch}$ as seen in Fig.~\ref{fig: b-fluct and pt-Nch scatter plot}. The figure shows that even both $b$ and $N_{ch}$ are fixed, $[p_T]$ can still fluctuate. Instead of $b$, let us use the centrality fraction $c_b\simeq \pi b^2/\sigma_{\rm Pb}$~\cite{Das:2017ned} (where $\sigma_{\rm Pb}$ is the inelastic cross section of the Pb+Pb collision) as an equivalent variable throughout this chapter, where $c_b$ lies between $0$ and $1$. We assume that the joint probability distribution of $N_{ch}$ and $[p_T]$ at fixed $c_b$, given by $P([p_T],N_{ch}|c_b)$, is a two dimensional correlated Gaussian. The choice of this Gaussian ansatz can be justified through the following arguments. Within a hydrodynamic model, fluctuations in $N_{ch}$ and $[p_T]$ originate from fluctuations in the initial density profile. When the impact parameter is fixed, these density fluctuations stem from quantum fluctuations which may occur either in the wave functions of colliding nuclei~\cite{PHOBOS:2006dbo,Miller:2007ri,Gelis:2010nm} or in the dynamics of the collision. At ultrarelativistic energies, the fluctuations at different locations on the transverse plane are independent due to causality. Therefore, it can be thought that a large number of such independent contributions result in the fluctuations of $N_{ch}$ and $[p_T]$. According to the central limit theorem, these fluctuations can be treated as approximately Gaussian.

The two-dimensional Gaussian distribution (as will be shown shortly) is characterized by five parameters: The mean and variance (or equivalently standard deviation) of $N_{ch}$ and of $[p_T]$, which we denote by $\overline{p_T}(c_b)$, $\overline{N_{ch}}(c_b)$, ${\rm Var}(p_T|c_b)$,  ${\rm Var}(N_{ch}|c_b)$ respectively and the Pearson correlation coefficient (or the covariance) $r_{N_{ch}}(c_b)$ between $[p_T]$ and $N_{ch}$. We expect this correlation to be positive as shown in Fig.~\ref{fig: b-fluct and pt-Nch scatter plot}. We now explain how they are obtained.
\begin{figure}[ht!]
\hspace{-0.3cm}\begin{subfigure}{0.5\textwidth}
\centering
\includegraphics[height=5.2 cm]{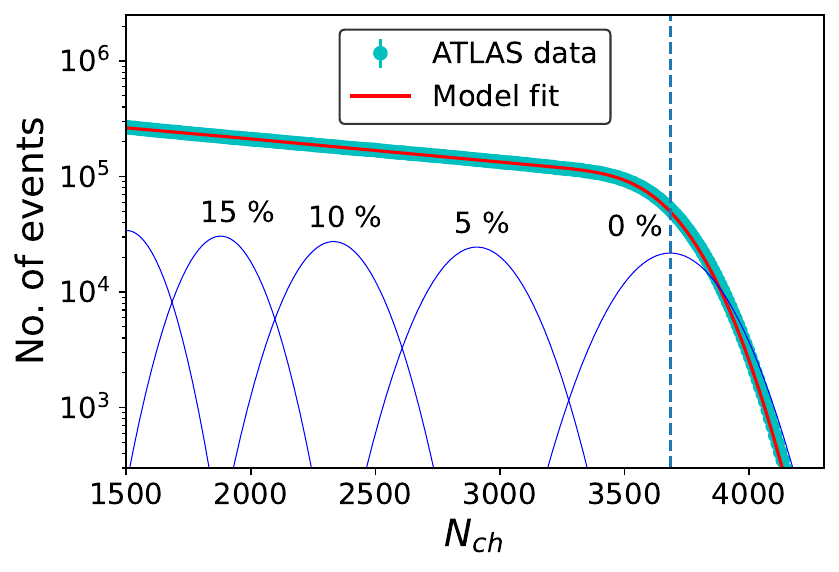}
\end{subfigure}~~~
\begin{subfigure}{0.5\textwidth}
\centering
\includegraphics[height=5.2 cm]{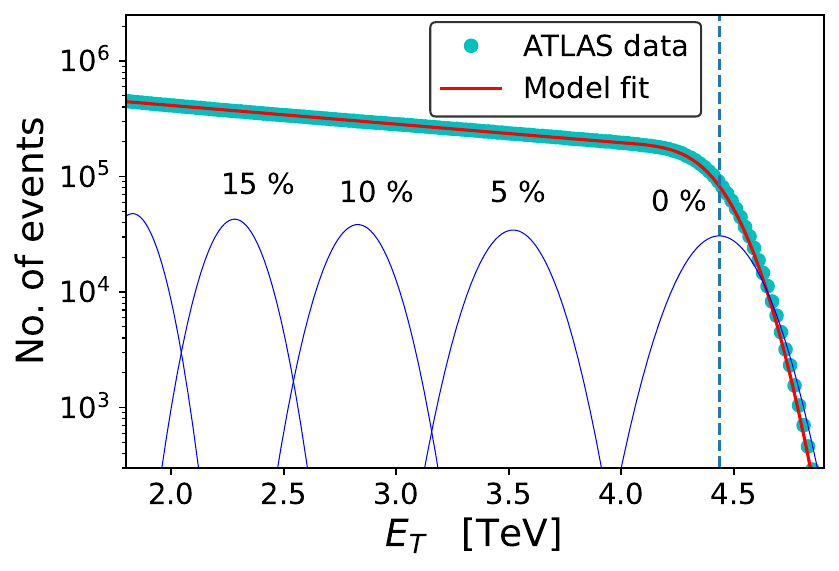}  
\end{subfigure}
\centering
\caption{Histogram of the charge particle multiplicity $N_{ch}$ (left), measured by ATLAS, and of transverse energy $E_T$ (right), deposited at the forward and backward calorimeters. The fits using superposition of Gaussians are denoted by the solid lines (Eqs.~(\ref{eq: Nch dist at fixed b}) and (\ref{eq: marginal distribution of Nch})). Contributions of collisions at fixed impact parameter $b$, given by Eq.~(\ref{eq: Nch dist at fixed b}) are shown corresponding to centrality fractions $0$, $5\%$, $10\%$, $15\%$ by thin blue lines. The ``knee'' is denoted the vertical dashed line, defined as the average value of  $N_{ch}$ or $E_T$ at $b=0$. The figure is from the original publication~\cite{Samanta:2023amp}, coauthored by the author.}
\label{fig: Nch and Et dist}
\end{figure}

\subsubsection{Constructing ``knee'' of the $N_{ch}$-distribution}

Let us discuss how we obtain $\overline{N_{ch}}(c_b)$ and $\sigma_{N_{ch}}(c_b)$. First, without any microscopic modeling, precise information can be obtained about the probability distribution of impact parameter at fixed $N_{ch}$, given by $P(c_b|N_{ch})$~\cite{Das:2017ned}. To achieve this, we first solve the inverse problem: Finding the probability distribution of $N_{ch}$ at fixed $c_b$, given by $P(N_{ch}|c_b)$. Then we apply Bayes' theorem:
\begin{equation}
\begin{aligned}
P(c_b|N_{ch})P(N_{ch})=P(N_{ch}|c_b)P(c_b) \ \ \Rightarrow \ \  P(c_b|N_{ch})=\frac{P(N_{ch}|c_b)P(c_b)}{P(N_{ch})} \ ,
 \end{aligned}
\label{eq: Bayes' theorem for Nchdist}
\end{equation}
where $P(c_b)\simeq 2\pi b/\sigma_{Pb}$ depicting the probability distribution of $b$. As explained earlier, collisions at the same impact parameter vary due to quantum fluctuations, which result in fluctuations of $N_{ch}$. In nucleus-nucleus collisions, these fluctuations are sufficiently small to be approximated as Gaussian. The Gaussian distribution is characterized by the mean, $\overline{N_{ch}}(c_b)$, and the variance, ${\rm Var}(N_{ch}|c_b)$ (or equivalently $\sigma_{N_{ch}}(c_b)$). The distribution is given by, 
\begin{equation}
\begin{aligned}
P(N_{ch}|c_b)=\frac{1}{\sqrt{2\pi{\rm Var}(N_{ch}|b)}}\exp\left(-\frac{\left(N_{ch}-\overline{N_{ch}}(c_b)\right)^2}{2{\rm Var}(N_{ch}|c_b)}\right) \ .
\end{aligned}
\label{eq: Nch dist at fixed b}
\end{equation}

In experiment, one measures the marginal distribution $P(N_{ch})$ which is obtained after integrating Eq.~(\ref{eq: Nch dist at fixed b}) over $c_b$ within $0<c_b<1$ :
\begin{equation}
\begin{aligned}
P(N_{ch})=\int_0^1 P(N_{ch}|c_b)dc_b \ ,
\end{aligned}
\label{eq: marginal distribution of Nch}
\end{equation}
shown in Fig.~\ref{fig: Nch and Et dist} (left). In the figures, we display values of $N_{ch}$ larger than some threshold that corresponds to 20\% centrality, which can be considered as fairly central collisions on which our analyses focus. The distribution $P(N_{ch})$ shows a mild variation up to $N_{ch}\sim 3500$, after which it declines sharply. We assume that $\overline{N_{ch}}(c_b)$ is a smooth function of $c_b$, and parametrize it as the exponential of a polynomial. A third-degree polynomial provides an excellent fit to $P(N_{ch})$ within this range: 
\begin{equation}
\begin{aligned}
\overline{N_{ch}}(c_b)=\overline{N_{ch}}(0)\exp\left(-\sum_{k=1}^3 a_k c_b^k\right) \ .
\end{aligned}
\label{eq: paremetrization of cb dependence of Nch}
\end{equation}
Similarly, we assume that the variance ${\rm Var}(N_{ch}|c_b)$ encounters a smooth variation with $c_b$. The parameters are obtained by fitting Eq.~(\ref{eq: marginal distribution of Nch}) as a superposition of Gaussians,  to the distribution $P(N_{ch})$ measured by ATLAS in Pb+Pb collisions. The fit is shown in Fig.~\ref{fig: Nch and Et dist}. We normalize the probability distribution $P(N_{ch})$ using  the centrality calibration provided by the ATLAS collaboration, that 40\% of events have $N_{ch}>705$. The fit is in agreement with data within 2\%.

Through fitting we can precisely reconstruct $\overline{N_{ch}}(c_b)$ and ${\rm Var}(N_{ch}|c_b=0)$~\cite{Yousefnia:2021cup}. The {\it knee} of the distribution, representing the mean value of $N_{ch}$ for collisions at $c_b=0$, is accurately reconstructed, and denoted with a vertical line on Fig.~\ref{fig: Nch and Et dist}. Above this knee, the rapid decline of $P(N_{ch})$ provides direct access to ${\rm Var}(N_{ch}|c_b=0)$. It is important to note that the variance can only be reconstructed at $c_b=0$, and we assume ${\rm Var}(N_{ch}|c_b)/\overline{N_{ch}}(c_b)$ to be constant by default. Additionally, we have tested two alternative scenarios: one assuming constant ${\rm Var}(N_{ch}|c_b)$ and another assuming constant ${\rm Var}(N_{ch}|c_b)/\overline{N_{ch}}(c_b)^2$. We have checked the robustness of our results with respect to these assumptions; the quality of the fit remains equally good and the fit parameters are essentially unchanged, as summarized in Table~\ref{tab: fit parameters Nch and Et dist}. Thus performing the fit to the $N_{ch}$ distribution, along with our assumption, we obtain two parameters: $\overline{N_{ch}}(c_b)$ and ${\rm Var}(N_{ch}|c_b)$, out of the five parameters of the correlated Gaussian. 

In our analysis, the events with multiplicities above the knee are termed as ultracentral collisions~\cite{Luzum:2012wu,CMS:2013bza}. These events constitute a small fraction of the total, approximately $0.35\%$. However, ATLAS has observed enough collisions that for a few events $N_{ch}$ exceeds the knee by $20\%$, corresponding to 4 standard deviations. It should be noted that the Poisson fluctuations contribute only by 15\% to the variance~\cite{Yousefnia:2021cup}, indicating that $N_{ch}$ fluctuations are predominantly due to dynamical factors.  
\begin{table}
\centering
\begin{tabular}{|c|| c | c| }
\hline
$N$&$N_{ch}$&$E_T$\cr
\hline
$\overline{N}(b=0)$&$3683\pm 4$&$4.435\pm 0.003$~TeV\cr
$\sqrt{{\rm Var}(N|b=0)}$&$168.1\pm 0.1$&$0.1433\pm 0.0001$~TeV\cr
$a_1$&$4.31\pm 0.02$&$4.18\pm 0.01$\cr
$a_2$&$-4.19\pm 0.03$&$-3.45\pm 0.01$\cr
$a_3$&$10.21\pm 0.09$&$8.54\pm 0.05$\cr
\hline
\end{tabular}
\vskip 2mm
\caption{\label{tab: fit parameters Nch and Et dist} 
Values of the fit parameters for Pb+Pb collisions at  $\sqrt{s_{NN}}=5.02$~TeV. The central value for each parameter is obtained with the assumption that the variance is proportional to the mean. The error bars reflect the changes after considering alternate scenario with the variance being constant, or proportional to the square of the mean.}
\end{table}

The main source of error in determining the impact parameter from data lies in the global normalization, due to the difficulty in experimentally estimating which fraction of the cross-section is detected~\cite{ALICE:2013hur}. Since we are interested in ultracentral collisions, this issue can be ignored. When we mention using the 20\% most central events, we refer to the 20\% most central of the events that are actually observed in the detector. The overlapping circles in Fig.~\ref{fig: 2d correlated dist} are pictorial depictions of the colliding Pb nuclei, having a radius $R=6.62$~fm. The values of $b$ are calculated from the inelastic cross section of Pb+Pb collisions, $\sigma_{PbPb}=767$~fm$^2$. Using Eqs.~(\ref{eq: Nch dist at fixed b}) and (\ref{eq: marginal distribution of Nch}), the probability distribution of the impact parameter at fixed $N_{ch}$ is obtained from Eq.(~\ref{eq: Bayes' theorem for Nchdist}):
\begin{equation}
\begin{aligned}
P(c_b|N_{ch})=\frac{1}{P(N_{ch})}P(N_{ch}|c_b),
\end{aligned}
\label{eq: dist of cb using Bayes' theorem}
\end{equation}
where $P(c_b)=1$ have been used, because the probability distribution of $c_b$ is uniform by construction. The distributions $P(c_b|N_{ch})$ becomes narrower as we move towards ultracentral collisions, illustrated in Ref.~\cite{Das:2017ned}.

\textbf{\textit{Impact parameter dependence of other parameters:}} For the 30\% most central collisions, the average transverse momentum is largely independent of centrality~\cite{ALICE:2018hza}. Therefore, we assume $\overline{p_T}(c_b)$ is independent of $c_b$, and we denote its value by $\overline{p_{T0}}$ ($\equiv \langle p_T \rangle$). We decompose $[p_T]=\overline{p_{T0}}+\delta p_T$, and we only model the distribution of $\delta p_T$, as we show below. Since we only consider fluctuations around $\overline{p_{T0}}$, our results are independent of its value. The variance ${\rm Var}(p_T|c_b)$ may vary with the impact parameter, but this dependence should be smooth. For statistical fluctuations, the variance is proportional to $1/N_{ch}$. For the $c_b$ dependence, we assume a more general form for ${\rm Var}(p_T|c_b)$ which behaves like a power law of the mean multiplicity:
\begin{equation}
\begin{aligned}
{\rm Var}(p_T|c_b)\equiv\sigma_{p_T}^2(c_b) =\sigma_{p_T}^2(0)\bigg(\frac{\overline{N_{ch}}(0)}{\overline{N_{ch}}(c_b)}\bigg)^{\alpha} \ ,
 \end{aligned}
\label{eq: c_b dependence of varpt}
\end{equation}
where $\sigma_{p_T}(0)$ and $\alpha$ are constants. Finally, for the sake of simplicity, we also assume that the correlation coefficient $r_{N_{ch}}$ does not vary with impact parameter. 

\subsection{$\rm{Var}(p_T | N_{ch})$ from the correlated Gaussian :}

Once we fix the impact parameter dependence of the parameters of correlated Gaussian distribution, we can fit the remaining three parameters ( $\overline{N_{ch}}(c_b)$ and ${\rm Var}(N_{ch}|c_b)$ are already fixed ) $\sigma_{p_T}(0)$,$\alpha$ and $r_{N_{ch}}$ to the ATLAS data for variance. To do so, we need an analytic expression for $\rm{Var}(p_T | N_{ch})$ which is obtained from the correlated Gaussian distribution at fixed $c_b$, $ P(\delta p_T, N_{ch} | c_b)$ given by,

\begin{equation}
\begin{aligned}
P(\delta p_T,N_{ch} | c_b)=&\frac{1}{2\pi\sqrt{(1-r^2){\rm Var}(p_T){\rm Var}(N_{ch})}} \cr &\times\exp\bigg[\frac{1}{1-r^2}\left(-\frac{(\delta p_T)^2}{2{\rm Var}(p_T)}
-\frac{\left(N_{ch}-\overline{N_{ch}}\right)^2}{2{\rm Var}(N_{ch})}
+\frac{r\left(N_{ch}-\overline{N_{ch}}\right)\delta p_T}{\sqrt{{\rm Var}(N_{ch}){\rm Var}(p_T)}} 
\right)\bigg ] \ ,
\end{aligned}
\label{eq: correlated Gaussian}
\end{equation}
where the impact parameter ($c_b$) dependence on the right hand side is implicit, we have used $r$ to represent $r_{N_{ch}}$ and as explained above instead of $[p_T]$, we use $\delta p_T\equiv [p_T]-\overline{p_{T0}}$. The linear correlation between $[p_T]$ and $N_{ch}$ can be understood as,
\begin{equation}
\begin{aligned}
\int_{-\infty}^{\infty}\int_{-\infty}^{\infty}{\delta p_T(N_{ch}-\overline{N_{ch}})P(\delta p_T,N_{ch})dN_{ch}d\delta p_T}=r\sqrt{{\rm Var}(N_{ch}){\rm Var}(p_T)} \ .
\end{aligned}
\label{linearcorr from corr Gaussian}
\end{equation}

A characteristic of the two-dimensional Gaussian distribution is that its marginal distributions which are obtained by integrating over one of the variables, are also Gaussians. Integrating Eq.~(\ref{correlated Gaussian}) over $\delta p_T$, one recovers Eq.~(\ref{eq: Nch dist at fixed b}). Similarly, integrating Eq.~(\ref{correlated Gaussian}) over $N_{ch}$, one obtains the distribution of $\delta p_T$ at fixed $c_b$: 
\begin{equation}
\begin{aligned}
P(\delta p_T|c_b)=\frac{1}{\sqrt{2\pi{\rm Var}(p_T|c_b)}}\exp\left(-\frac{(\delta p_T)^2}{2{\rm Var}(p_T|c_b)}\right) \ .
\end{aligned}
\label{eq: pt dist at fixed cb}
\end{equation}
Another interesting property of the two-dimensional Gaussian is that if one of the variables is fixed, let us say $N_{ch}$, the distribution of the other variable, e.g. $\delta p_T$, is also Gaussian. The distribution of  $\delta p_T$ at fixed $N_{ch}$ and $c_b$ is defined by, 
\begin{equation}
\begin{aligned}
 P(\delta p_T|N_{ch},c_b)&=\frac{P(\delta p_T,N_{ch}|c_b)}{P(N_{ch}|c_b)} \cr
  &=\frac{1}{\sqrt{2\pi\kappa_2(c_b)}}\exp\bigg[-\frac{\left(\delta p_T-\kappa_1(c_b)\right)^2}{2\kappa_2(c_b)}\bigg] \ ,
\end{aligned}
\label{eq: pt dist at fixed Nch and fixed cb}
\end{equation} 
where we omit the dependence on $N_{ch}$ in the right-hand side. The coefficients $\kappa_1(c_b)$ and $\kappa_2(c_b)$ represent the mean and the variance of $\delta p_T$ at fixed $N_{ch}$ {\it and\/} $c_b$, given by:
\begin{equation}
\begin{aligned}
\kappa_1(c_b)\equiv \overline{\delta p_T}(N_{ch},c_b)&=r\sqrt{\frac{{\rm Var}(p_T|c_b)}{{\rm Var}(N_{ch}|c_b)}}\left(N_{ch}-\overline{N_{ch}}(c_b)\right)\cr
&\equiv r\frac{\sigma_{p_T}(c_b)}{\sigma_{N_{ch}}(c_b)}\left(N_{ch}-\overline{N_{ch}}(c_b)\right) \ , \cr 
\eqsp{and}\kappa_2(c_b)\equiv {\rm Var}(p_T|N_{ch},c_b)&= \left(1-r^2\right) {\rm Var}(p_T|c_b) \equiv \left(1-r^2\right)\sigma_{p_T}^2(c_b) \ ,
\end{aligned}
\label{eq: mean and variance of pt at fixed Nch and fixed cb}
\end{equation}
where we omit $c_b$-dependence of $r$ as per our assumption. 

The $n^{th}$ order moment is obtained by multiplying Eq.~(\ref{eq: pt dist at fixed Nch and fixed cb}) with $\delta p_T^n$ and integrating over $\delta p_T$. The first and second order moments are given by : 
\begin{equation}
\begin{aligned}
  \langle \delta p_T \rangle = \kappa_1 \eqsp{and} \langle \delta p_T^2 \rangle=\kappa_1^2+\kappa_2 \ ,
  \end{aligned}
\label{eq: first and 2nd moment of pt}
\end{equation}
where the dependencies on $c_b$ and $N_{ch}$ are implicit. These moments are then averaged over $c_b$, to obtain only the $N_{ch}$-dependence which can be compared to the data. The mean and variance of $p_T$ at fixed $N_{ch}$ are finally constructed using the $c_b$-average of the moments in Eq.~(\ref{eq: first and 2nd moment of pt}) as,
\begin{equation}
\begin{aligned}
  \langle \delta p_T | N_{ch} \rangle &= \langle \kappa_1 \rangle_{c_b} = \langle \overline{\delta p_T}(N_{ch},c_b) \rangle_{c_b} \ , \cr 
 {\rm Var}(p_T | N_{ch}) &= \langle \kappa_1^2+\kappa_2 \rangle_{c_b} - \langle \kappa_1 \rangle_{c_b}^2 = \bigg(\langle \kappa_1^2 \rangle_{c_b}- \langle \kappa_1 \rangle_{c_b}^2\bigg)+\langle \kappa_2 \rangle_{c_b}  \cr
  &= \bigg(\langle \overline{\delta p_T}(N_{ch},c_b)^2 \rangle_{c_b}- \langle \overline{\delta p_T}(N_{ch},c_b) \rangle_{c_b}^2\bigg) + \langle {\rm Var}(p_T|N_{ch},c_b) \rangle_{c_b} \ ,
\end{aligned}
\label{eq: mean and variance of pt}
\end{equation}
with,
\begin{equation}
\begin{aligned}
\langle \dots \rangle_{c_b}\equiv \int_0^1 \dots P(c_b|N_{ch}) dc_b = \frac{1}{P(N_{ch})}\int_0^1 \dots P(N_{ch}|c_b) dc_b  \ ,
\end{aligned}
\label{eq: cb average}
\end{equation}
where we have used Eq.~(\ref{eq: Bayes' theorem for Nchdist})\footnote{$c_b$ is the cumulative distribution of $b$, therefore, $P(c_b)=1$ by construction.} in the last line. Putting the explicit expressions of $\overline{\delta p_T}(N_{ch},c_b)$ and ${\rm Var}(p_T|N_{ch},c_b)$ from Eq.~(\ref{eq: mean and variance of pt at fixed Nch and fixed cb}) into Eq.~(\ref{eq: mean and variance of pt}), we obtain the multiplicity dependence of mean (average) of $\delta p_T$ and variance of $p_T$. The first term of the expression for variance in Eq.~(\ref{eq: mean and variance of pt}), denoted within bracket, stems from the variation of $\overline{\delta p_T}(N_{ch},c_b)$ with impact parameter, reflecting the contribution of $b$-fluctuation to $[p_T]$-fluctuation. We refer to the second term as the intrinsic variance (or intrinsic fluctuations), in the sense that it is not a by-product of impact parameter fluctuations. We will see that both terms are of comparable magnitudes below the knee, and most importantly, the first term explains the peculiar pattern (steep fall) observed for large $N_{ch}$. The variance ${\rm Var}(p_T | N_{ch})$ is fitted to the ATLAS data in Fig.~\ref{fig: varpt} (left) with three parameters $\sigma_{p_T}(0)$, $\alpha$ and $r_{N_{ch}}$, details of which are provided in Appendix~\ref{a: fitting var data}. The fitted values of these parameters are shown on the figure. 

\subsection{Fit results: Thermalization and predictions for mean $\langle \delta p_T\rangle$}
\label{fit results}
\begin{figure}[ht!]
\hspace{-0.3cm}\begin{subfigure}{0.5\textwidth}
\centering
\includegraphics[height=5.8 cm]{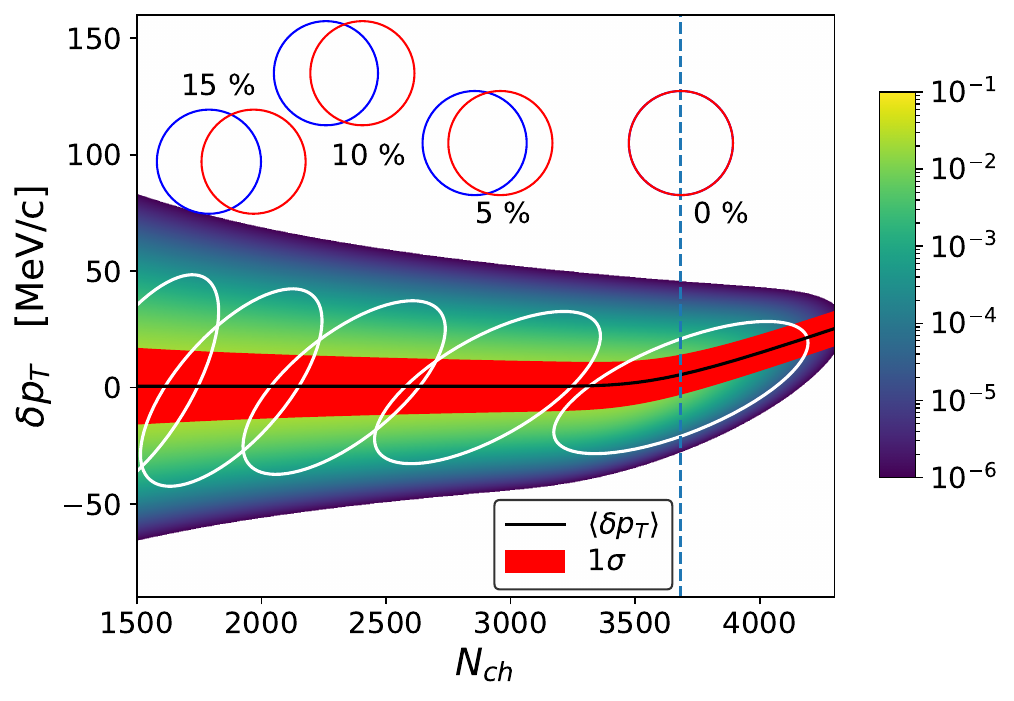}
\end{subfigure}~~~
\begin{subfigure}{0.5\textwidth}
\centering
\includegraphics[height=5.8 cm]{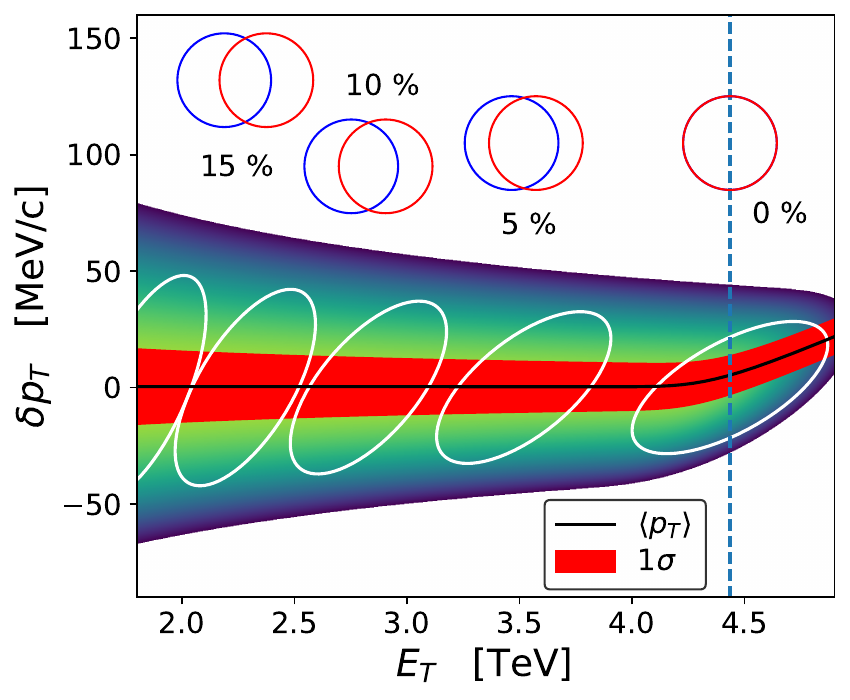}  
\end{subfigure}
\centering
\caption{Joint distribution of $\delta p_T$ and $N_{ch}$ (left) or $E_T$ (right), obtained from the fit of our model and integrating Eq.~(\ref{eq: correlated Gaussian}) over $c_b$. The white curves are 99\% confidence ellipses at fixed impact parameters, obtained from the correlated Gaussian distribution in Eq.~(\ref{correlated Gaussian}) at fixed values of $c_b$. Schematic representation of two nuclei colliding with these impact parameters are also shown. The black line is the mean value of $\delta p_T$ (Eq.~(\ref{eq: mean and variance of pt})), and the red band is the 1-$\sigma$ band, representing square-root of the red line in Fig.~\ref{fig: varpt}. The figure is from the original publication~\cite{Samanta:2023amp}, coauthored by the author.}
\label{fig: 2d correlated dist}
\end{figure}

Let us first examine the distribution of $\delta p_T$ and $N_{ch}$ returned by our fit, shown in the left panel of Fig.~\ref{fig: 2d correlated dist}. The white curves represent 99\% confidence ellipses at fixed impact parameter~\cite{Yousefnia:2021cup}, which are drawn using the probability given by Eq.~(\ref{eq: pt dist at fixed cb}). 
It can be seen that they are tilted with respect to the horizontal axis, similar to the hydrodynamic calculation of Fig.~\ref{fig: b-fluct and pt-Nch scatter plot}. This tilt reflects the positive correlation between $\delta p_T$ and $N_{ch}$, parameterized by $r_{N_{ch}}$. As explained above, this correlation is a natural consequence of thermalization. The width of the $\delta p_T$ distribution for fixed $N_{ch}$ can be attributed to two contributions. A part of it comes from the fact that several ellipses contribute for a given $N_{ch}$ (first term in Eq.~(\ref{eq: mean and variance of pt})), and the rest is due to the vertical width of a single ellipse (second term in Eq.~(\ref{eq: mean and variance of pt})). 

The left panel of Fig.~\ref{fig: varpt} displays the data and the model fit, along with the two terms of Eq.~(\ref{eq: mean and variance of pt}). There exist equal contributions from the two terms below knee. However, above knee the first term ($b$-fluctuation) gradually disappears. Thus our model precisely explains the observed steep decrease of the variance around the knee which comes from the first term, namely, from impact parameter fluctuations at fixed $N_{ch}$, whose effect becomes negligible in ultracentral collisions. The magnitude of this term is essentially determined by the correlation coefficient $r_{N_{ch}}$, which is thus constrained by data. 

As a corollary, we also predict a small increase in the average transverse momentum $\langle \delta p_T \rangle$, represented by a black line in Fig.~\ref{fig: 2d correlated dist}, in ultracentral collisions. This effect had been predicted a while ago~\cite{Gardim:2019brr,Nijs:2021clz} and has recently been observed by CMS collaboration~\cite{Bernardes:2023nnf}. Note that our model calculation quantitatively predicts this increase.

\subsubsection{Using different centrality estimator: Transverse energy $E_T$}

A particularity of the ATLAS analysis is the use of an alternative centrality estimator, in addition to $N_{ch}$:  the transverse energy $E_T$\footnote{The quantity used as the centrality estimator can be generically denoted as $N$, can be either $N_{ch}$ or $E_T$.}. This is defined as energy multiplied by $\sin\theta$ and is measured in two calorimeters located symmetrically on either side of the collision point, covering approximately the angular ranges $1^\circ<\theta< 5^\circ$ and $175^\circ<\theta< 179^\circ$. Our analyses and results are presented for the observables of interest with both centrality estimators $N_{ch}$ and $E_T$, presented on the left and right panel of the figures respectively. The variance analysis can be repeated by selecting events based on $E_T$, rather than $N_{ch}$, as illustrated in the right panel of Fig.~\ref{fig: Nch and Et dist}. Similarly, our model calculation can be repeated by replacing $N_{ch}$ with $E_T$ everywhere. This serves as a valuable and rigorous check for the validity of our approach. 

Although the distributions of $N_{ch}$ and $E_T$ are similar in shape (Fig.~\ref{fig: Nch and Et dist} (right)), it could be seen that the decline above the knee is steeper for $E_T$ than for $N_{ch}$. Please also note, only $0.26\%$ of events fall above the knee for $E_T$, while for $N_{ch}$ it is $0.35\%$. It is interesting to note, despite this difference, the observed decrease in variance around the knee remains consistent for both estimators, as measured by ATLAS (Fig.~\ref{fig: Nch and Et dist} (right)). The parameters $\sigma_{ p_T}(0)$ and $\alpha$, determining the  impact parameter dependence of the variance of $[p_T]$, should not depend on whether the events are classified according to $N_{ch}$ or $E_T$. Therefore, we find the values that show best simultaneous agreement with $N_{ch}$ and $E_T$-based data, discussed in detail in Appendix~\ref{a: fitting var data}. 

However, the Pearson correlation coefficient $r_{E_T}$ between $[p_T]$ and $E_T$ is independently fitted and does not necessarily match $r_{N_{ch}}$. Note that the correlation $r_{N_{ch}}$ pertains to the correlation between $[p_T]$ and $N_{ch}$ for the {\it same\/} particles, whereas $r_{E_T}$ pertains to the correlation between $[p_T]$ and the $E_T$ measured in different angular windows separated by rapidity interval. Therefore we expect $r_{E_T}<r_{N_{ch}}$, which is confirmed by the fit from our analysis. The similarity in the values indicates that particle depositions in different $\theta$ windows are strongly correlated.

\subsection{Effect of $p_T$ interval on variance}
Another peculiar aspect of the ATLAS analysis is its study of how $[p_T]$-fluctuations  change with the  $p_T$ interval. The default analysis involves all particles in the range $0.5<p_T<5$~GeV$/c$ (particles with $p_T<0.5$~GeV$/c$ are not detected, and those with $p_T>5$~GeV$/c$ are likely to be associated with jets, and  not relevant to the collective behavior). Additionally, in this section we present results for the particles in the range $0.5<p_T<2$~GeV$/c$, which excludes about $7\%$ of the particles as considered in the previous case. 
\begin{figure}[ht!]
\begin{subfigure}{0.5\textwidth}
\centering
\includegraphics[height=6 cm]{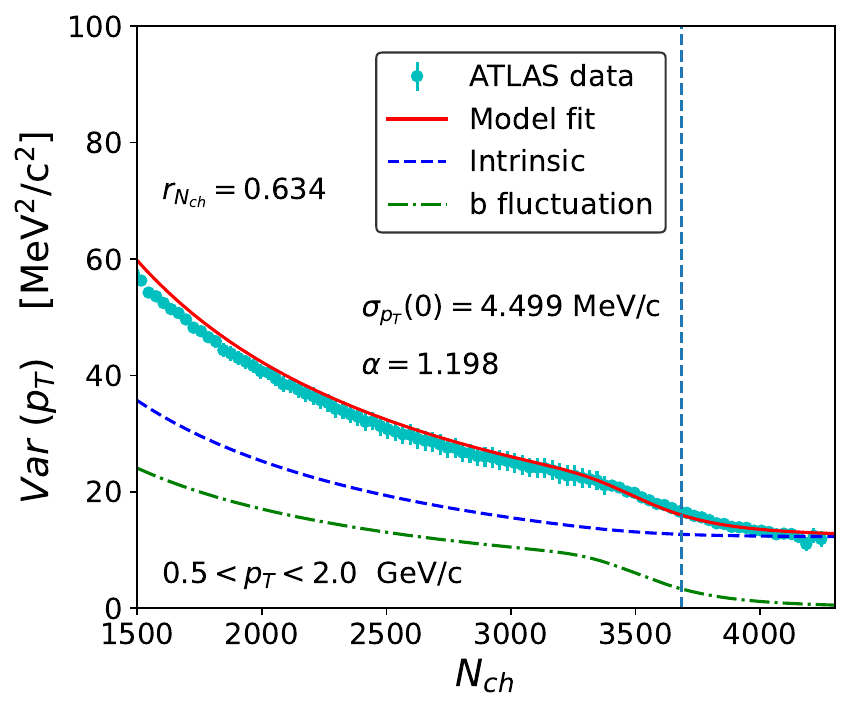}
\end{subfigure}~~~
\begin{subfigure}{0.5\textwidth}
\centering
\includegraphics[height=6 cm]{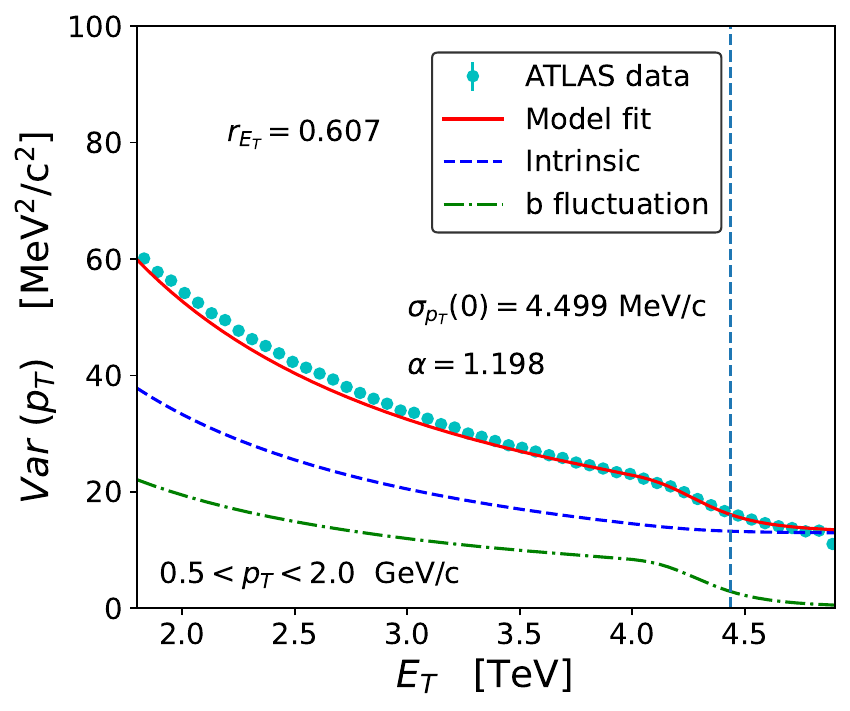}  
\end{subfigure}
\centering
\caption{Same as Fig.~\ref{fig: varpt}, but for the transverse momentum range of the particles in: $0.5<p_T<2$~GeV$/c$. One sees that there is a factor of $\sim 4$ reduction in the magnitude of variance in comparison to Fig.~\ref{fig: varpt} with $0.5<p_T<5$~GeV$/c$, thus showing the effect of $p_T$ interval on $p_T$-fluctuation.}
\label{fig: varpt lowpt}
\end{figure}
The remarkable effect is that the variance decreases by a factor $\sim 4$, when comparing these intervals in Figs.~\ref{fig: varpt} and \ref{fig: varpt lowpt}. 
Strikingly, the phenomenon is  also observed in the hydrodynamic simulations, where the magnitude of  $\delta p_T$ is reduced by a factor of approximately 2 for the smaller $p_T$-interval (see Fig.~\ref{fig: pt-Nch scatter plot for different pt interval}). 
Numerically, the variance representing the average value of $\delta p_T^2$, decreases by a factor $4.3\pm 0.2$, aligning with the ATLAS result.
\begin{figure}[ht!]
\includegraphics[height=7 cm]{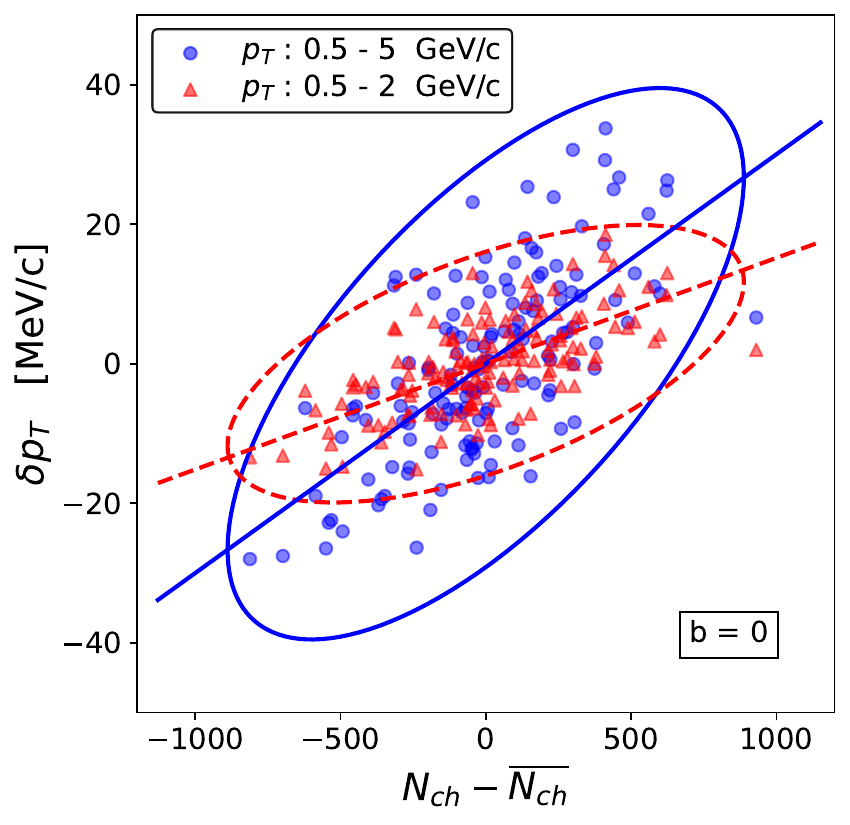}  
\centering
\caption{Distribution of $\delta p_T$ and $N_{ch}$ for two different $p_T$-intervals: $0.5<p_T<2.0$ GeV and $0.5<p_T<5.0$ GeV, denoted by red and blue symbols respectively, similar to Fig.~\ref{fig: b-fluct and pt-Nch scatter plot}. The results are shown for 150 events in Pb+Pb collision at 5.02~TeV and $b=0$ obtained from hydrodynamics. The solid and red curves denote the 99\% confidence ellipses. For the interval $0.5-5$~GeV$/c$, the value of the average (Eq.~(\ref{eq: mean pt from mean distribution})) $\overline{p_{T0}}$ is $1074$~MeV$/c$. As one removes particles with higher values of $p_T$, the average value $\overline{p_{T0}}$ decreases to $970$~MeV$/c$ for $0.5-2$~GeV$/c$ interval. The straight lines indicate the average value $\overline{\delta p_T}(N_{ch},b=0)$. The values of the standard deviations (Eq.~(\ref{eq: pt dependence of fluctuation of pt})) obtained for the two intervals are 6 MeV/c and 13 MeV/c for $0.5-2$~GeV$/c$ and $0.5-5$~GeV$/c$ respectively, showing factor of 2 increase (i.e. factor of 4 increase in variance) and reflecting consistency with ATLAS data.}
\label{fig: pt-Nch scatter plot for different pt interval}
\end{figure}

We provide an estimate how the variance depends on the $p_T$ selection. Fluctuations in fluid velocity cause global fluctuations in the $p_T$ distribution, where the tail of the distribution is largely affected. Thus in hydrodynamics, event-by-event $[p_T]$ fluctuations arise from transverse fluid velocity fluctuations. The momentum distribution of particles follows a boosted Boltzmann distribution, where $p_T$ appears in the exponent. Thus relative change in the $p_T$ distribution $f(p_T)\equiv\frac{dN}{dp_T}$ due to a small change in the fluid velocity is linear in $p_T$~\cite{Gardim:2019iah}: 
\begin{equation}
\begin{aligned}
f(p_T)=\langle f(p_T)\rangle \bigg(1+x(p_T-\langle p_T \rangle)\bigg ), 
\end{aligned}
\label{eq: pt dist as a deviation from mean dist}
\end{equation}
where $\langle f(p_T)\rangle$ denotes the event average of the $p_T$ distributions, and $\langle p_T \rangle \equiv \langle [p_T]\rangle$ given by,
\begin{equation}
\begin{aligned}
\langle p_T \rangle \equiv \frac{\int p_T \langle f(p_T)\rangle dp_T}{\int \langle f(p_T)\rangle dp_T}
\end{aligned}
\label{eq: mean pt from mean distribution}
\end{equation}
is the average $p_T$. Here $x$ is a random variable fluctuating event to event around zero. For a class of events with the same multiplicity, the integral of $f(p_T)-\langle f(p_T)\rangle$ must vanish, making the relative fluctuation proportional to $p_T-\langle p_T \rangle$, instead of just $p_T$. The fluctuations in transverse momentum per particle are obtained by integrating the spectrum (\ref{eq: pt dist as a deviation from mean dist}) over the $p_T$ range used in the analysis:
\begin{equation}
\begin{aligned}
\sigma_{p_T}=x\frac{\int_{p_{\rm min}}^{p_{\rm max}} (p_T-\langle p_T \rangle)^2\langle f(p_T)\rangle dp_T}{\int_{p_{\rm min}}^{p_{\rm max}} \langle f(p_T)\rangle dp_T} \ .
\label{eq: pt dependence of fluctuation of pt}
\end{aligned}
\end{equation}
We assume the fluctuations are small enough to replace $f(p_T)$ with the average distribution $\langle f(p_T)\rangle$ in the denominator. 

The dependence of the right-hand side of Eq.~(\ref{eq: pt dependence of fluctuation of pt}) on the upper bound $p_{\rm max}$ can be evaluated using spectra measured by ALICE in central Pb+Pb collisions at the same energy~\cite{ALICE:2018vuu}, in place of $\langle f(p_T)\rangle$. We find that lowering $p_{\rm max}$ from $5$ down to $2$~GeV$/c$ reduces the right-hand side of Eq.~(\ref{eq: pt dependence of fluctuation of pt}) by a factor $2.05$, leading to a decrease in variance by a factor $4.23$, consistent with ATLAS observations. 

One should note, in this model, the only variable parameter in Eq.~(\ref{eq: pt dependence of fluctuation of pt}) is the overall factor $x$, which sets the fluctuation magnitude. 
This can be checked in hydrodynamics. 
First, it could be checked by eye that in Fig.~\ref{fig: pt-Nch scatter plot for different pt interval}, symbols of different types appear in pairs with the same  $N_{ch}$. 
%(the centrality estimator is always the charged multiplicity in the interval $0.5<p_T<5$~GeV$/c$, even if the analysis of $[p_T]$-fluctuation uses a different interval). 
Each pair represents one collision event, and the proportionality factor $x$ in Eq.~(\ref{eq: pt dependence of fluctuation of pt}) fluctuates event-by-event. 
The change in $\delta p_T$ \footnote{Please note that $\langle \delta p_T\rangle$ is the average shown in Fig.~\ref{fig: 2d correlated dist}.} from one symbol to the another in the same pair is roughly the same for all events, as suggested by Eq.~(\ref{eq: pt dependence of fluctuation of pt}). More quantitatively, the Pearson correlation between the two $\delta p_T$ values for each event (i.e. between $\delta p_T (0.5<p_T<5)$ and $\delta p_T (0.5<p_T<2)$) is about $ \sim 0.97$, close to its maximum value $1$, implied by Eq.~(\ref{eq: pt dependence of fluctuation of pt}). The observed dependence of the variance on the $p_T$ selection is another layer of evidence supporting the hydrodynamic origin of $[p_T]$ fluctuations. 

\subsubsection{Importance of b-fluctuation}
It is interesting to note that the impact parameter is a classical quantity. One can calculate its quantum uncertainty which is negligible:
Heisenberg's principle gives $\delta b\equiv \hbar/P\sim 4\times 10^{-7}$~fm for a Pb+Pb collision at the LHC, insignificant compared to the range spanned by $b$, of order $15$~fm.\footnote{Note that in event-by-event simulations, the impact parameter is correctly defined only if each nucleus is recentered after randomly drawing nucleon positions. The recentering correction is larger by orders of magnitude than the quantum uncertainty. It is however {\it not\/} implemented in the simulations shown in Fig.~\ref{fig: b-fluct and pt-Nch scatter plot}, but this does not alter the conclusions we draw from the figure.} The {\it only\/} classical quantity characterizing a collision is the impact parameter, and collisions with the same impact parameter differ only by quantum fluctuations. As the collision occurs at high energy, a single quantum fluctuation can produce a large number of particles, which promotes such fluctuations to the status of a classical fluctuations. (Elliptic flow in central collisions~\cite{PHOBOS:2006dbo} and triangular flow~\cite{Alver:2010gr} are driven by a similar mechanism.) The effect studied here, involves a subtle interplay between classical fluctuations of impact parameter, and quantum fluctuations of the collision multiplicity. 

\section{Non-Gaussian features of $[p_T]$-fluctuation : Skewness and Kurtosis}
\label{skewness and kurtosis}
In the previous section, we discussed the leading order cumulant of $[p_T]$-fluctuation i.e. variance. However, one could also think of higher order cumulants, namely skewness and kurtosis of $[p_T]$-fluctuation, which can also exhibit interesting features in ultracentral collisions. The observed decrease in the variance above the knee is caused by the decrease of impact parameter fluctuations with the increase in multiplicity at the ultracentral regime. In this section, we will show that the same mechanism is also responsible for strong non-Gaussian characteristics of the fluctuations of $[p_T]$ in ultracentral collisions.\footnote{It has already been observed that hydrodynamic calculations imply a significant skewness of $[p_T]$ fluctuations~\cite{Giacalone:2020lbm}, but the crucial role of impact parameter has not been studied.} Using the same model of $[p_T]$ fluctuations outlined above, we present robust quantitative, parameter-independent predictions for skewness and excess kurtosis, which characterize standard measures of the non-Gaussianity. 

\subsection{Non-Gaussianity from a simplified model}
\label{non-gaussian from simplified model}
We begin by explaining the origin of non-Gaussian fluctuations based on a simplified model, in which $[p_T]$ is a single-valued function of $N_{ch}$ and $c_b$. In other words, $[p_T]$ does not fluctuate if we fix both $N_{ch}$ and $c_b$. Given that the variation of $c_b$ at fixed $N_{ch}$ is small, the dependence of $[p_T]$ on $c_b$ can be linearized:\footnote{Note that observables depend quadratically on $c_b$ for small $b$ due to symmetry reasons~\cite{Pepin:2022jsd}, which precludes a dependence of the type $\sqrt{c_b}$ for small $b$.}
\begin{equation}
\begin{aligned}
[p_T]=p_T^{\rm min}+\lambda c_b,
\end{aligned}
\label{eq: pt linearized}
\end{equation}
where $p_T^{\rm min}$ and $\lambda$ depend on $N_{ch}$. This approximation in turn means that the correlation coefficient between $[p_T]$ and $N_{ch}$, $r$ assumes a value close to $1$ \footnote{This could be understood if we consider the hydro results in Fig.~\ref{fig: b-fluct and pt-Nch scatter plot} (right). If $[p_T]$ (or equivalently $\delta p_T$) does not fluctuate at fixed $c_b$ and $N_{ch}$, then the values of $\delta p_T$ coincides with the average $\overline{\delta p_T}(N_{ch},c_b)$ given in Eq.~(\ref{eq: mean and variance of pt at fixed Nch and fixed cb}) making $r\rightarrow1$.}.  
\begin{figure}[ht!]
\includegraphics[height=12 cm]{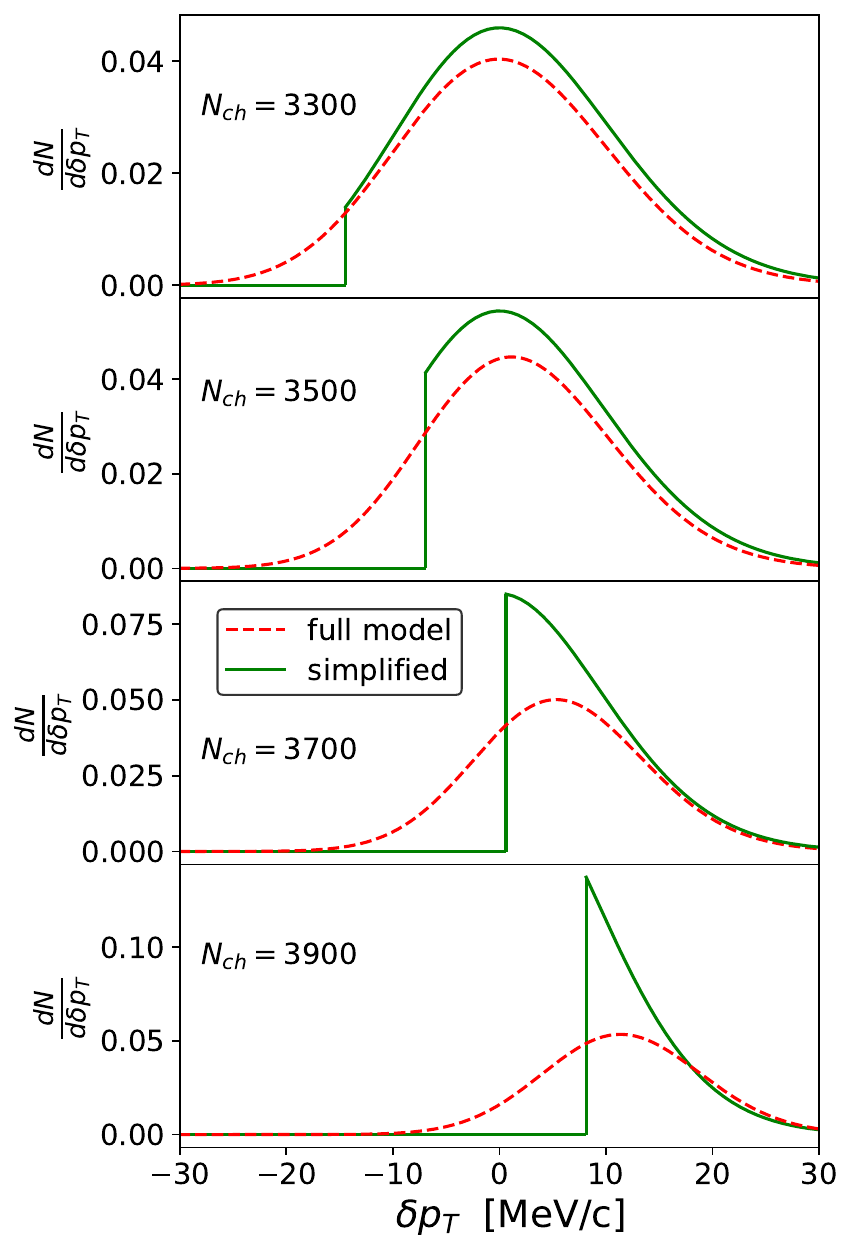}  
\centering
\caption{Probability distribution of $[p_T]$ at fixed multiplicity $N_{ch}$, for various values of $N_{ch}$. If the centrality is defined according to $N_{ch}$, these values of $N_{ch}$ correspond to the centrality fractions, from top to bottom, $2.2\%$, $1.1\%$, $0.3\%$ and $0.04\%$. Similar to previous figures, we plot the distribution of $\delta p_T$ rather than $[p_T]$ . The distributions obtained with the simplified model are denoted by the solid lines, where $[p_T]$ only depends on $N_{ch}$ and impact parameter (Eq.~(\ref{eq: linearized mean})) and it does not fluctuate if both are fixed. The results with a more realistic model are represented by the dashed lines, assuming Gaussian fluctuations of $[p_T]$ at fixed $N_{ch}$ and $c_b$, obtained from Eq.~(\ref{eq: pt dist fixed Nch}), which is referred as ``full model'' on the figure. The figure is from the original publication~\cite{Samanta:2023kfk}, coauthored by the author.}
\label{fig: pt dist simplified model}
\end{figure}
The probability distribution of $[p_T]$ at fixed $N_{ch}$, is then determined by the probability distribution of $c_b$ at fixed $N_{ch}$. The latter one, denoted by $P(c_b|N_{ch})$, is obtained using Eq.~(\ref{eq: Bayes' theorem for Nchdist}) and from the probability distribution of $N_{ch}$ at fixed $c_b$ (Eq.~(\ref{eq: Nch dist at fixed b})), which we have assumed above to be Gaussian. For ultracentral collisions where $c_b\ll 1$, one can neglect the dependence of $\sigma_{N_{ch}}$ on $c_b$, and then the variation of the mean can be linearized as:
\begin{equation}
\begin{aligned}
\overline{N_{ch}}(c_b)=N_{\rm knee}-\beta c_b,
\end{aligned}
\label{eq: linearized mean}
\end{equation}
where $N_{\rm knee}\equiv \overline{N_{ch}}(0)$ and $\beta$ determines how mean multiplicity decreases with centrality. As was done earlier, the values of these parameters can be obtained by fitting the measured distribution of $N_{ch}$, $P(N_{ch})$(Fig.~(\ref{fig: Nch and Et dist})). From the numerical fit, we obtain the values of the parameters as: $N_{\rm knee}=3680$, $\sigma_{N_{ch}}=168$, $\beta=18300$, similar to Table~\ref{tab: fit parameters Nch and Et dist}. 

With the linear approximation, the probability distribution of $c_b$ for fixed $N_{ch}$ is then given by,
\begin{equation}
\begin{aligned}
P(c_b|N_{ch})=\frac{P(N_{ch}|c_b)}{P(N_{ch})}
\propto& \exp\left(-\frac{\left(N_{ch}-N_{\rm knee}+\beta c_b\right)^2}{2\sigma_{N_{ch}}^2}\right),
\end{aligned}
\label{eq: dist of b at fixed Nch linear approx}
\end{equation}
where we have used Eqs.~(\ref{eq: Bayes' theorem for Nchdist}) and (\ref{eq: linearized mean}). Eq.~(\ref{eq: dist of b at fixed Nch linear approx}) indicates that the distribution of $c_b$ is Gaussian, with a width $\sigma_{N_{ch}}/\beta\simeq 0.9\%$. However, this distribution is not a full Gaussian, but rather a {\it truncated\/} Gaussian distribution, which is truncated on the left because of the boundary condition $c_b\ge 0$~\cite{Broniowski:2001ei,Das:2017ned}. Consequently, Eq.~\ref{eq: pt linearized} suggests that the probability distribution of $[p_T]$ is also a truncated Gaussian, subject to the boundary condition $[p_T]\ge p_T^{\rm min}$. The truncated Gaussian distribution for $[p_T]$ is obtained from Eq.~(\ref{eq: dist of b at fixed Nch linear approx}), with the condition that $[p_T]$ is given by Eq.~(\ref{eq: pt linearized}). This results in the solid curves shown in Fig.~\ref{fig: pt dist simplified model}, which are truncated on the left due to the lower limit of $[p_T]$ ($c_b$).

This truncation has several effects which lead to peculiar patterns for the cumulants of $[p_T]$-fluctuation in the ultracentral regime. First, the distribution of $[p_T]$ becomes narrower (Fig.~\ref{fig: pt dist simplified model}), which results in a decrease of the variance, which we have discussed and is reflected in Figs.~\ref{fig: varpt} and \ref{fig: varpt lowpt}. Second, the truncation gives rise to non-Gaussian features such as skewness and kurtosis, which we discuss below.

\subsection{Skewness and kurtosis from the full model}
\label{skewness and kurtosis full mdoel}
In reality, in addition to the effect of impact parameter fluctuations, $[p_T]$ fluctuates even if both $c_b$ and $N_{ch}$ are fixed, as seen in Fig.~\ref{fig: b-fluct and pt-Nch scatter plot} and expressed via Eq.~(\ref{eq: pt dist at fixed Nch and fixed cb}), which we referred as the intrinsic fluctuations previously. The distribution of $\delta p_T$ at fixed $N_{ch}$ is obtained by averaging Eq.~(\ref{eq: pt dist at fixed Nch and fixed cb}) over the impact parameter: 
\begin{equation}
\begin{aligned}
P(\delta p_T|N_{ch})&=&\int_0^1 P(\delta p_T|N_{ch},c_b)P(c_b|N_{ch})dc_b\cr
&=&\frac{1}{P(N_{ch})}\int_0^1 P(\delta p_T,N_{ch}|c_b)dc_b,
\end{aligned}
\label{eq: pt dist fixed Nch}
\end{equation}
where we have used Eqs.~(\ref{eq: Bayes' theorem for Nchdist}) and (\ref{eq: pt dist at fixed Nch and fixed cb}) to arrive at the last line. 
The distributions $P(\delta p_T|N_{ch})$ are shown by the dashed lines in Fig.~\ref{fig: pt dist simplified model} for selected values of $N_{ch}$ close to the knee. The solid lines in this figure, representing the truncated Gaussian, are obtained by setting the correlation coefficient to its maximum value $r=1$, corresponding to the simplified model in Eq.~(\ref{eq: linearized mean}). Note that when $r\to 1$, the variance $\kappa_2(c_b)$ approaches zero and  $P(\delta p_T|N_{ch},c_b)$ collapses into a Dirac delta function $\delta(\delta p_T-\kappa_1(c_b))$, which implies that  $\delta p_T$ is solely determined by $N_{ch}$ and $c_b$.

Similar to Eq.~(\ref{eq: first and 2nd moment of pt}), the third and fourth moment of $\delta p_T$ is obtained by multiplying Eq.~(\ref{eq: pt dist at fixed cb}) with $\delta p_T^3$ and $\delta p_T^4$ respectively and integrating over $\delta p_T$: 
\begin{equation}
\begin{aligned}
\langle \delta p_T^3\rangle&=\kappa_1^3+3\kappa_2\kappa_1 \ , \cr
\langle \delta p_T^4\rangle&=\kappa_1^4+6\kappa_2\kappa_1^2+3\kappa_2^2, 
\end{aligned}
\label{3rd and 4th moment}
\end{equation}
where again the dependence on $c_b$ on the right-hand side is implicit. Averaging the above moments over $c_b$ and using the third and fourth order cumulant formula we find the skewness and kurtosis \footnote{Note we do not consider intrinsic skewness and  consider only excess kurtosis, because our model is based on a Gaussian fluctuation model of $b$. For a Gaussian distribution (Eq.~(\ref{eq: pt dist at fixed Nch and fixed cb})), skewness ($\kappa_3(c_b,N_{ch})$) is zero and kurtosis ($\kappa_4(c_b,N_{ch})$) is 3. Therefore, our results represent an underestimation of these quantities and will always lie below the measured values.}:  
\begin{equation}
\begin{aligned}
{\rm Skew}(p_T|N_{ch})&=\bigg(\langle\kappa_1^3\rangle-3\langle\kappa_1^2\rangle\langle\kappa_1\rangle+2\langle\kappa_1\rangle^3 \bigg )
+3\bigg(\langle\kappa_2\kappa_1\rangle-\langle\kappa_2\rangle \langle\kappa_1\rangle\bigg) \ , \cr
{\rm Kurt}(p_T|N_{ch}) &=\bigg ( \langle\kappa_1^4\rangle-4\langle\kappa_1^3\rangle\langle\kappa_1\rangle+6\langle\kappa_1^2\rangle\langle\kappa_1\rangle^2-3\langle\kappa_1\rangle^4\bigg )\cr
&+6\bigg(\langle\kappa_2\kappa_1^2\rangle-\langle\kappa_2\rangle \langle\kappa_1^2\rangle-2\langle\kappa_2\kappa_1\rangle\langle\kappa_1\rangle+2\langle\kappa_2\rangle\langle\kappa_1\rangle^2\bigg) \cr
&+3\bigg(\langle\kappa_2^2\rangle-\langle\kappa_2\rangle^2\bigg) \ ,
\end{aligned}
\label{eq: skew and kurt of pt}
\end{equation}
where angular brackets denote averages over $c_b$ : $\langle \dots \rangle \equiv \langle \dots \rangle_{c_b}$ and the $N_{ch}$ dependence on the right hand side has been omitted. The expression for skewness has two terms: the first term involves $\kappa_1$ only, and the second term is proportional to the correlation between $\kappa_1$ and $\kappa_2$. The excess kurtosis has three terms: two of which are similar to skewness and the third term is proportional to the variance of $\kappa_2$. The skewness and the kurtosis capture the non-Gaussian characteristics of the event-by-event fluctuations of $[p_T]$. In our model, we assume Gaussian fluctuations for a fixed impact parameter. Therefore, any non-Gaussianities in our model results originate from fluctuations in the impact parameter. In the absence of impact parameter fluctuations, each term in the above expressions of ${\rm Skew}(p_T)$ and ${\rm Kurt}(p_T)$ would be zero.

\subsubsection{Two parametrization : DUKE vs JETSCAPE}

Similar to the analysis of variance, the dependence of the mean multiplicity, $\overline{N_{ch}}(c_b)$ and the standard deviation $\sigma_{N_{ch}(c_b)}$ on $c_b$ can be inferred from the experimental distribution $P(N_{ch})$(Fig.~\ref{fig: Nch and Et dist}). However, because existing data do not constrain the $c_b$ dependence of $\sigma_{N_{ch}}$, we avoid making assumptions, and instead adopt a more realistic approach. We borrow information from state-of-the-art models tuned to experimental data via Bayesian analyses. Specifically, we use the Maximum A Posteriori parameter sets from two sources: one from the Duke group \cite{Moreland:2018gsh} and another from the JETSCAPE collaboration ( which use the Grad viscous correction to the distribution function at freezeout)~\cite{JETSCAPE:2020mzn}. The JETSCAPE analysis, tuned to a larger dataset that includes several collision energies, differ from the Duke analysis which is particularly tuned to 5.02~TeV data ( the same energy used in our analysis) and accounts for the nucleon substructure, which may potentially affect the fluctuations. We evaluate $\sigma_{N_{ch}}(c_b)$ based on both models, the method of which is explained in detail in Appendix~\ref{a: duke vs jetscape}. The prediction by the DUKE parametrization is an increase in $\sigma_{N_{ch}}$ between $b=0$ and $b=3.5$~fm, whereas the JETSCAPE model predicts a slight decrease. The difference between these two models provides an estimate of the errors in our predictions. 

\subsection{Results : Predictions for skewness and kurtosis}
\begin{figure}[ht!]
\begin{subfigure}{0.5\textwidth}
\centering
\includegraphics[height=5.3 cm]{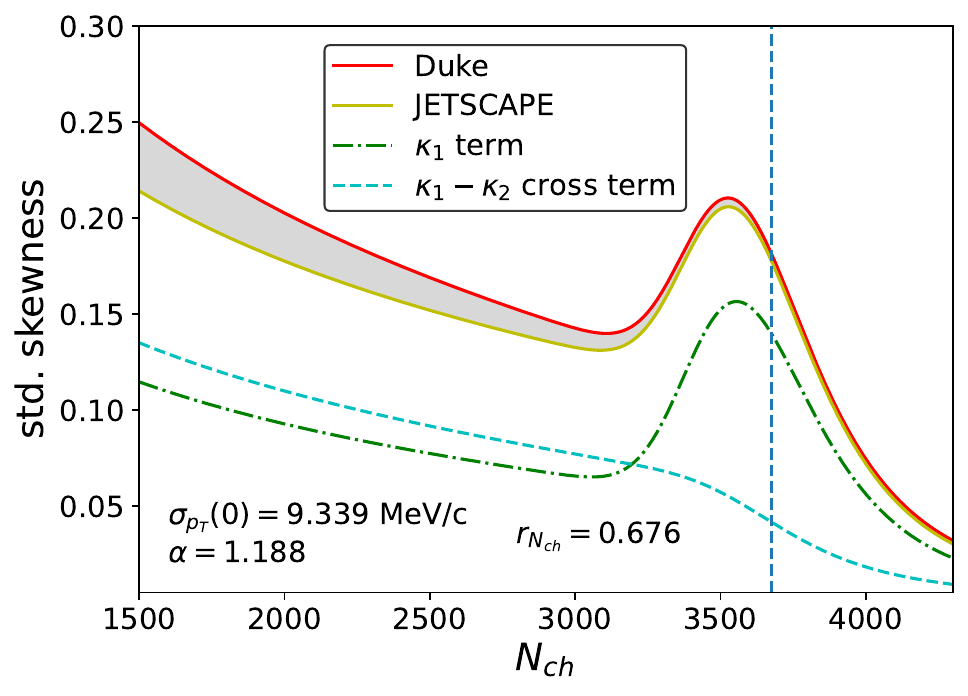}
\end{subfigure}~~
\begin{subfigure}{0.5\textwidth}
\centering
\includegraphics[height=5.3 cm]{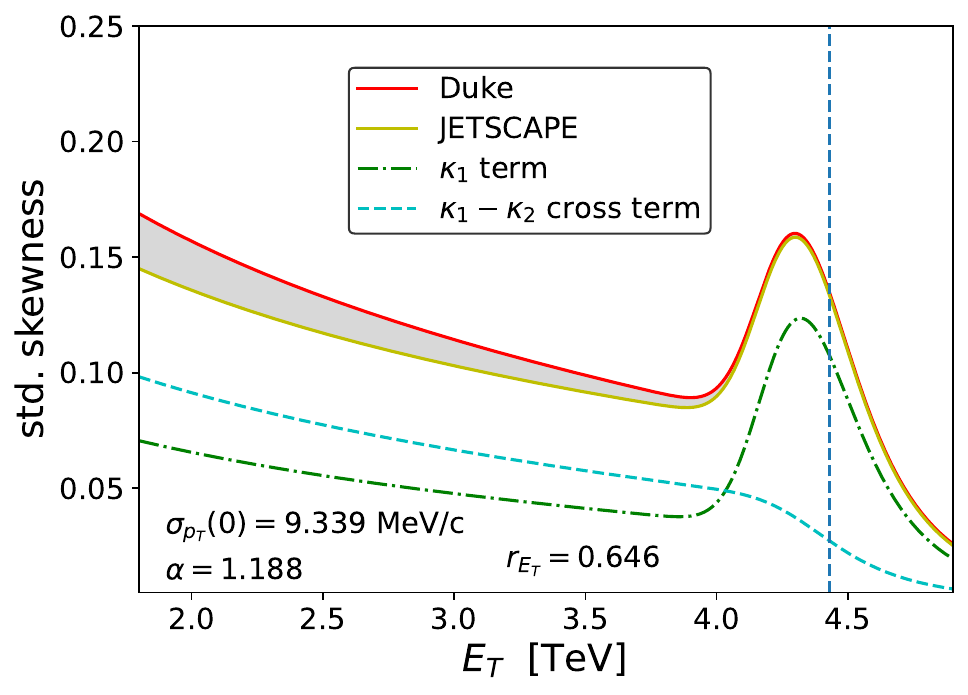}  
\end{subfigure}
\centering
\caption{Predictions for the standardized skewness as a function of $N_{ch}$(left) and $E_T$(right), based on our model fit to the variance data from ATLAS, and using the Duke and JETSCAPE parametrizations of the centrality dependence of $\sigma_{N_{ch}}$ (Appendix~\ref{a: duke vs jetscape}). The difference between the two parametrization is sizable and it is shown as a gray shaded band, thereof serves as error in our prediction. The various terms in Eq.~(\ref{eq: skew and kurt of pt}), contributing to skewness, are shown for the Duke parametrization only. The figure is from the original publication~\cite{Samanta:2023kfk}, coauthored by the author.}
\label{fig: skew pt}
\end{figure}

\begin{figure}[ht!]
\begin{subfigure}{0.5\textwidth}
\centering
\includegraphics[height=5.3 cm]{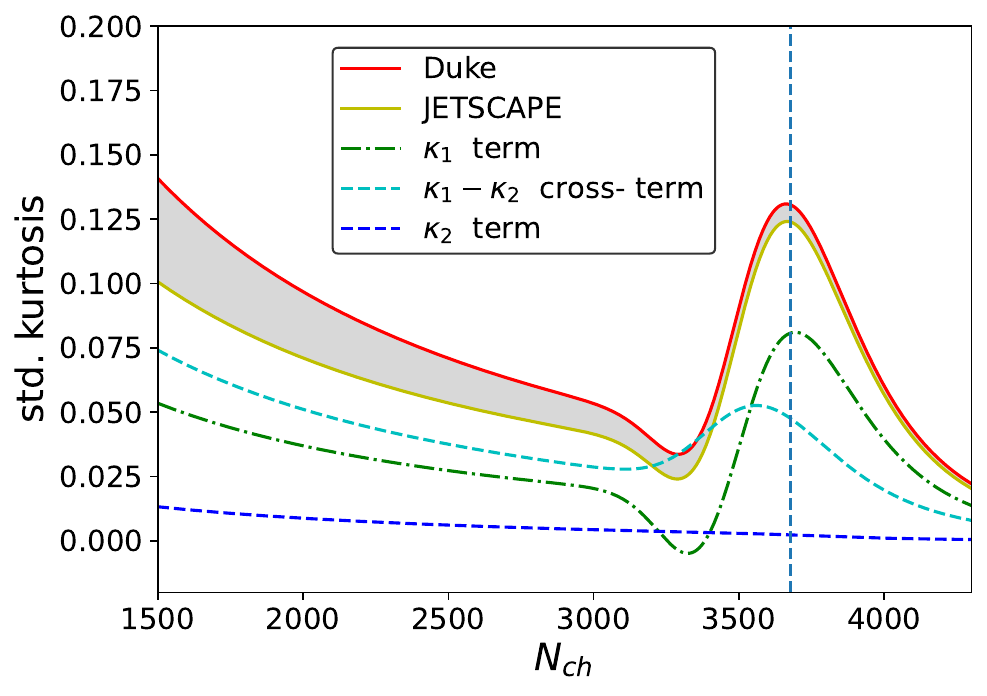}
\end{subfigure}~~
\begin{subfigure}{0.5\textwidth}
\centering
\includegraphics[height=5.3 cm]{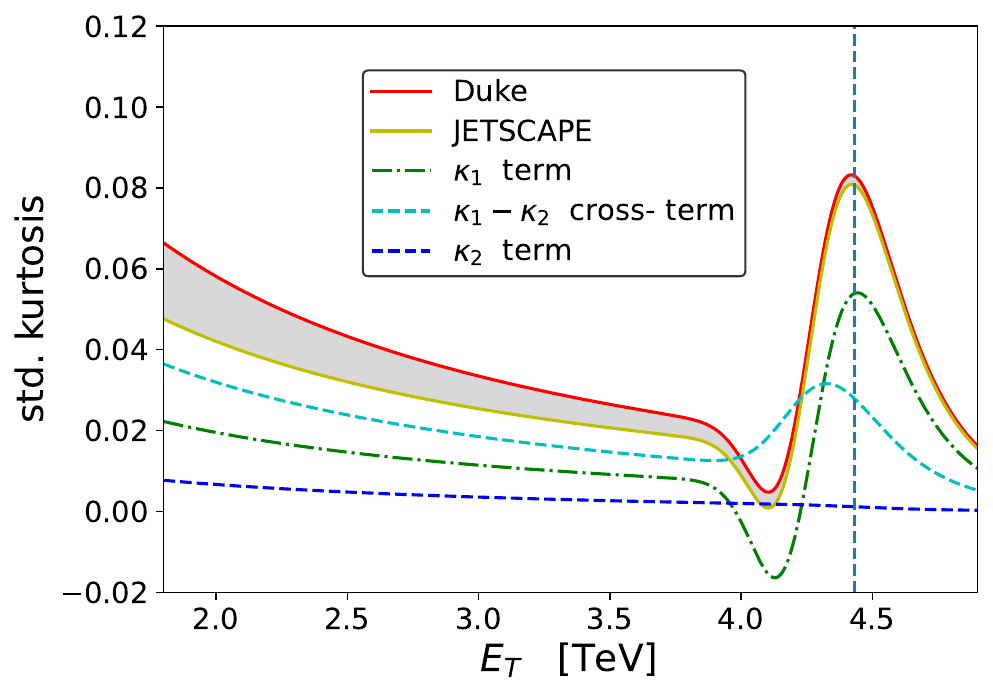}  
\end{subfigure}
\centering
\caption{Same as Fig.~\ref{fig: skew pt} but for the predictions for the standardized excess kurtosis. The figure is from the original publication~\cite{Samanta:2023kfk}, coauthored by the author.}
\label{fig: kurt pt}
\end{figure}

We find quantitative predictions for the skewness and kurtosis of $[p_T]$-fluctuations based on Eq.~(\ref{eq: skew and kurt of pt}) and using the fit parameters obtained by fitting the variance data in Fig.~\ref{fig: varpt}. Our quantitative predictions are displayed in Figs. ~\ref{fig: skew pt} and \ref{fig: kurt pt}. Note that the fit parameters $\sigma_{p_T}(0)$ and $\alpha$  are not the same as Fig.~\ref{fig: varpt}, because here we take different $c_b$ dependence for $\sigma_{N_{ch}}$ based on the two different parametrizations: DUKE and JETSCAPE. However the correlation coefficient $r$ remain unchanged in Figs.~\ref{fig: varpt} and \ref{fig: skew pt}. One can see that, the results on the skewness and the kurtosis exhibit sharp variations around the knee. In particular, our model predicts an increase of the skewness below the knee (such an increase has already been seen by the ALICE collaboration~\cite{Saha:2022bxf}, as will be discussed below), followed by a steep decrease above the knee. 
The kurtosis, on the other hand, has first a minimum below the knee, followed by a maximum roughly at the knee and then it encounters sharp decrease. It is interesting to note that, these peculiar patterns come from the terms involving $\kappa_1$, also caused the sharp decrease in variance around knee, and are actually inherited from the truncated Gaussian, reflecting the effect of $b$-fluctuation at the ultracentral regime. Note that it is possible to calculate the cumulants of the truncated Gaussian (Fig.~\ref{fig: pt dist simplified model}) analytically. Quantitatively speaking, the maximum skewness occurs around $N_{ch}\simeq N_{\rm knee}-\sigma_{N_{ch}}\simeq 3510$, while the kurtosis reaches a minimum at $N_{ch}\simeq N_{\rm knee}-2\sigma_{N_{ch}}\simeq 3340$, followed by a maximum at $N_{ch}\simeq N_{\rm knee}\simeq 3680$. These numerical values correspond to the structures seen in our model results.

Our predictions show a little dependency on whether one adopts the Duke or JETSCAPE parametrization to characterize the centrality dependence of the multiplicity fluctuations. In fact, the primary limitation of our model lies in assuming a Gaussian distribution for $[p_T]$ at fixed $N_{ch}$ and $b$. Since $[p_T]$ is inherently positive, it naturally exhibits a positive skewness $\kappa_3$ and a positive excess kurtosis $\kappa_4$. These factors additionally give positive contributions to ${\rm Skew}(p_T|N_{ch})$ and ${\rm Kurt}(p_T|N_{ch})$ in Eq.~(\ref{eq: skew and kurt of pt}), in the form of $\langle\kappa_3\rangle$ and $\langle\kappa_4\rangle$. One should consider our predictions as lower bounds both for the skewness and for the kurtosis. Precise quantitative predictions of the values of these additional terms would require extensive hydrodynamic simulations with high statistics. However, we can safely comment that these additional contributions should have a smooth dependence on $N_{ch}$, resulting in a positive offset from our predictions. 
The sharp {\it variations\/} of the skewness and kurtosis around the knee in Figs.~\ref{fig: skew pt}  and \ref{fig: kurt pt} are robust, quantitative predictions. 
\begin{figure}[ht!]
\begin{subfigure}{0.5\textwidth}
\centering
\includegraphics[height=5.2 cm]{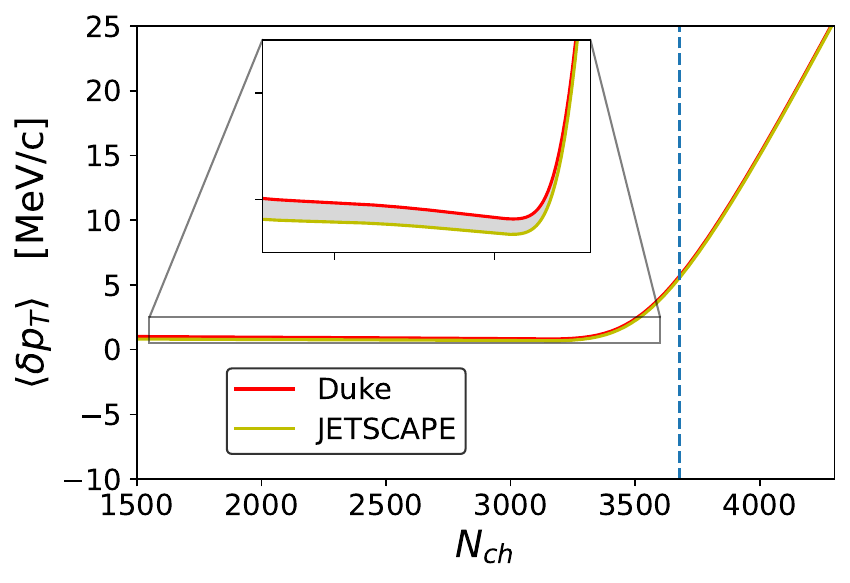}
\end{subfigure}~~
\begin{subfigure}{0.5\textwidth}
\centering
\includegraphics[height=5.2 cm]{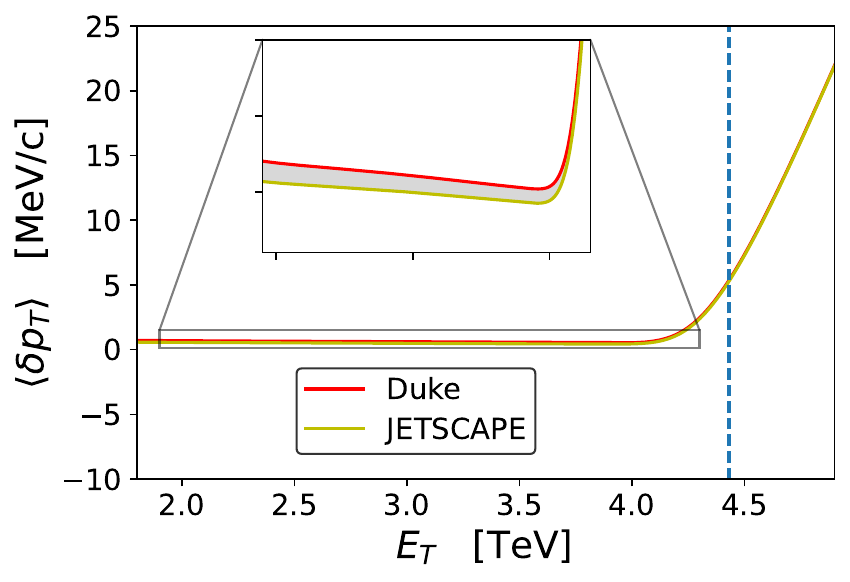}  
\end{subfigure}
\centering
\caption{Predictions for average $\langle \delta p_T \rangle$, based on our model, using DUKE and JETSCAPE parametrization. The difference is negligible for the mean, unlike skewness and kurtosis. The figure is from the original publication~\cite{Samanta:2023kfk}, coauthored by the author.}
\label{fig: mean pt}
\end{figure}

Like the results for the variance, here also we present predictions based on both centrality estimators $N_{ch}$(left) and $E_T$(right). However, $E_T$ turns out to be a better centrality estimator than $N_{ch}$, because of having smaller impact parameter fluctuations~\cite{Pepin:2022jsd,Yousefnia:2021cup}. 
In our model, all the non-Gaussianities arise from the impact parameter fluctuations and hence, both the skewness and the kurtosis are expected to be smaller if the centrality is determined by $E_T$ instead of $N_{ch}$. This is exactly seen in our predictions (Figs.~\ref{fig: skew pt} and \ref{fig: kurt pt}). Experimental verification of these predictions will be essential in assessing the importance of impact parameter fluctuations.
Similar to Fig.~\ref{fig: 2d correlated dist}, here also we present predictions on the increase of the average $\langle \delta p_T\rangle$ based on two different parameterizations displayed in Fig.~\ref{fig: mean pt}.  

\subsubsection{Comparison with ALICE measurements}
Finally, let us compare our results with the recent results on the skewness and kurtosis from the ALICE collaboration~\cite{ALICE:2023tej}. ALICE collaboration uses the amplitude deposited in scintillators located at forward rapidities as the centrality estimator. This is qualitatively similar to the $E_T$-based centrality determination by ATLAS. The skewness measurements are performed in the central pseudorapidity ($\eta$) region, again analogous to the ATLAS analysis but with a narrower $\eta$ interval. The centrality binning by ALICE is significantly coarser compared to ATLAS, with each point covering a $5\%$ interval. Our analysis focuses on approximately the $20\%$ most central collisions. Therefore, our predictions can only be compared with the last four data points from ALICE, corresponding to the ranges (in TeV) $1.6<E_T<2.1$, $2.1<E_T<2.7$, $2.7<E_T<3.5$, and $E_T>3.5$. 
\begin{figure}[ht!]
\begin{subfigure}{0.5\textwidth}
\centering
\includegraphics[height= 6.5 cm]{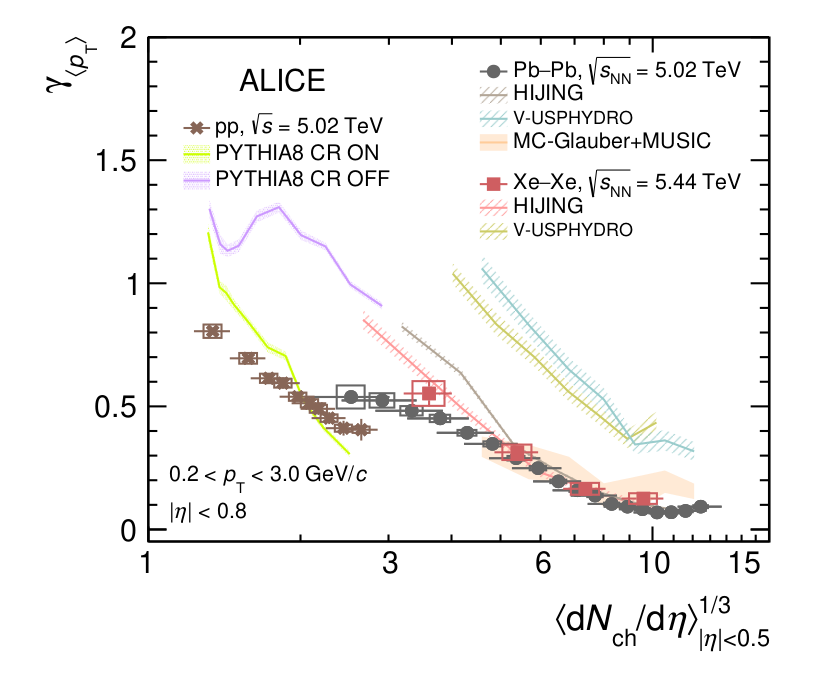}
\end{subfigure}~~
\begin{subfigure}{0.5\textwidth}
\centering
\includegraphics[height=6.5 cm]{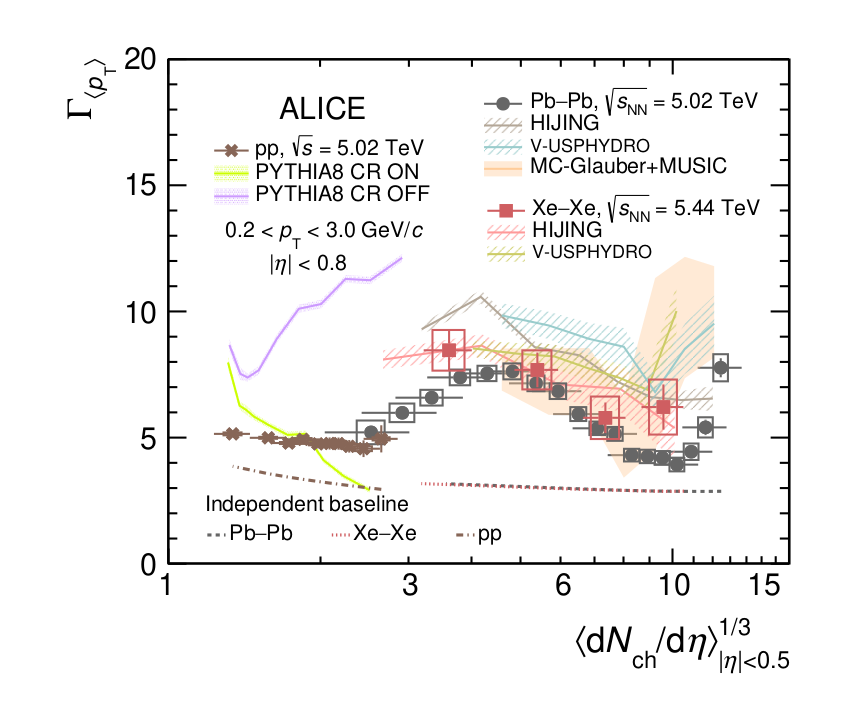}  
\end{subfigure}
\centering
\caption{Measurement of the standardized (left) and intensive (right) skewness of $[p_T]$-fluctuation by ALICE for Pb+Pb. Figure taken from \cite{ALICE:2023tej}.}
\label{fig: skew ALICE}
\end{figure}

The quantities shown by ALICE are both standardized and intensive skewness~\cite{Giacalone:2020lbm} shown in Fig.~\ref{fig: skew ALICE}. The standardized skewness measured by ALICE for the last four points (ultracentral, centrality range of our analysis) is within $0-0.1$, with a slight increase for the last data point, consistent with our prediction. In our prediction, the skewness starts to rise after $E_T\sim4.0$ TeV, for which data from ALICE is not available. The intensive skewness, on the other hand, is obtained by multiplying the standardized skewness with the average value, $\langle p_T\rangle$, and normalizing by the standard deviation ${\rm Var}(p_T)^{1/2}$. We have not evaluated this quantity because we are not provided with the values of $\langle p_T\rangle$ by the ATLAS collaboration. Given the $p_T$ range covered by ATLAS, an approximate estimate for $\langle p_T\rangle$ might be around $\sim 1$~GeV$/c$. We then predict that the intensive skewness remains relatively constant and approaches approximately $10$ within the interval $1.8<E_T<3.5$~TeV. In comparison, ALICE reports the values for the intensive skewness (Fig.~\ref{fig: skew ALICE}, right) that rise from (slightly below) 4 to (slightly above) 5 over a similar range. It is important to note that direct comparison of the absolute values is not straightforward due to the difference in the $p_T$ coverage: ALICE covers $0.2<p_T<3$~GeV$/c$, whereas ATLAS considers $0.5<p_T<5$~GeV$/c$ \footnote{The dependence of $\sigma_{p_T}(c_b)$ on the $p_T$ interval is not trivial.  How fluctuations of the $p_T$ spectrum depends on $p_T$ is at present not known, and assessing it would require to measure the quantity $v_0(p_T)$ introduced in Ref.~\cite{Schenke:2020uqq}. Recently we have found that the observable $v_0(p_T)$ carries much more significance in relation to $[p_T]$-fluctuation. One can actually capture the $p_T$-interval dependence of the fluctuations through $v_0(p_T)$. Work on this particular phenomenon is in progress.}.

Interestingly in the ALICE data, the intensive skewness is close to $8$, in the most central bin, which is significantly higher than in the previous bins. This last point of the ALICE data corresponds to the $E_T$ interval: $E_T>3.5$~TeV, over which our model predicts a rise followed by a fall of the intensive skewness, peaking at a value $\sim 18$, below the knee. ALICE measurements (not shown) on standardized kurtosis (lies between 3.02 - 3.07 for the right most four points) are consistent with our predictions for excess kurtosis. It is tempting to interpret the ALICE results corroborating with our predictions and possibly as a first confirmation. It will be useful if the ALICE analysis is repeated in finer centrality bins, specifying the values of the centrality estimator in each bin and if they can provide measurements for more ultracentral events.

\chapter{Transverse momentum-harmonic flow correlations}

% **************************** Define Graphics Path **************************
\ifpdf
    \graphicspath{{Chapter5/Figs/Raster/}{Chapter5/Figs/PDF/}{Chapter5/Figs/}}
\else
    \graphicspath{{Chapter5/Figs/Vector/}{Chapter5/Figs/}}
\fi
In Chapter~3, we discussed fluctuations of harmonic flow ($v_n$) and in Chapter~4, we discussed fluctuations of the mean transverse momentum per particle ($[p_T]$) in heavy-ion collisions. We have seen that both of these final state quantities are largely determined by the properties of the initial state. The harmonic flow is directly related to the initial spatial anisotropy or eccentricities ($\Epsilon_n$)\cite{Alver:2010gr,Gardim:2011xv}, whereas the mean transverse momentum of the particles is related to the size of the initial fireball ($R$)~\cite{Broniowski:2009fm,Schenke:2020uqq}. Events with a smaller size of the interaction region have larger energy density gradients, resulting in a larger transverse push during the expansion and hence larger $[p_T]$ at the final state. As a result, the event-by-event fluctuations of $v_n$ and $[p_T]$ are governed by the event-by-event fluctuations of the initial state. The fluctuations of the  harmonic flow coefficients could be due to fluctuations of the shape of the initial fireball, as well as due to  dynamical fluctuations in the expansion dynamics. Analogously, fluctuations of $[p_T]$ can be related to the fluctuations of the size of the fireball as well as fluctuations of the initial entropy or energy ($S$ or $E_i$)~\cite{Giacalone:2020dln,Bozek:2021zim}. Therefore, naturally one can expect that these two final state quantities, $[p_T]$ and $v_n$ are correlated. 

The Pearson correlation coefficient between the mean transverse momentum per particle and harmonic flow coefficient, $\rho([p_T],v_n^2)$, first introduced by P. Bo\.zek~\cite{Bozek:2016yoj}, hence sometimes referred as Bo\.zek coefficient~\cite{Giacalone:2020awm}, serves as an excellent tool to study the correlation between collective observables at the final state and a fine probe to the correlation present in the initial state~\cite{Bozek:2020drh, Schenke:2020uqq, Giacalone:2020dln, Giacalone:2020byk,Bozek:2021zim}. Another significant importance of this correlator $\rho$ is that it can be used as a fine tool to study nuclear structure and deformation in high energy heavy-ion collisions~\cite{Bally:2021qys,Giacalone:2020awm,Jia:2021wbq,Giacalone:2021clp}, which we will discuss in detail in the next chapter. Moreover it can be used in Bayesian analysis to put precise constraints on the initial state and medium properties of the QGP~\cite{Bernhard:2016tnd,JETSCAPE:2020mzn,Nijs:2020roc}. The observable $\rho([p_T],v_n^2)$ has been measured in experiments~\cite{ATLAS:2019pvn,ALICE:2021gxt,ATLAS:2022dov,Tuo:2023tye,Yan:2023ugh} and been studied extensively in models over the past few years~\cite{Bozek:2020drh, Schenke:2020uqq, Giacalone:2020dln,Giacalone:2021clp,Giacalone:2020byk,Bozek:2021zim, Zhang:2021phk, Lim:2021auv, Magdy:2021ocp, Magdy:2021cci}. 

In this chapter, at first we discuss the centrality dependence of $\rho([p_T],v_n^2)$ and higher order correlations between transverse momentum and different orders of harmonic flow through normalized symmetric cumulants~\cite{Bilandzic:2013kga,Mordasini:2019hut,Moravcova:2020wnf} which put additional constraints on the initial state properties and correlations. In the second part, we include momentum dependence within the harmonic flow and construct the correlation coefficient between $[p_T]$ and differential flow $v_n(q)$\footnote{We do not use the notation $p$ here and use $q$ instead, to denote transverse momentum bins to avoid confusion.}. This provides the momentum dependent measure of the correlation coefficient and further constrain the initial state parameters. The following sections are, for the most part, presentations from the original publications~\cite{Bozek:2021zim,Samanta:2023rbn}, coauthored by the author.

\section{Correlation between $[p_T]$ and integrated flow $v_n$}
\label{pt-v2 correlation}
Let us first consider the correlation between mean transverse momentum per particle $[p_T]$ and squares of the harmonic flow $v_n^2$, which could be constructed as Pearson correlation coefficient~\cite{Bozek:2016yoj,Bozek:2020drh} at the lowest order or as higher order correlations involving different orders of flow through the {\it symmetric cumulants}~\cite{Mordasini:2019hut,Moravcova:2020wnf}. 

\subsection{Pearson correlation coefficient : $\rho([p_T],v_n^2)$}
\label{rho pt-v2}
The Pearson correlation coefficient, first proposed in~\cite{Bozek:2016yoj}, between the mean transverse momentum per particle and the harmonic flow, can be used to measure the event-by-event correlation and it is defined as,
\begin{equation}
\begin{aligned}
\rho([p_T],v_n^2)=\frac{Cov([p_T],v_n^2)}{\sqrt{Var([p_T]) Var(v_n^2)}} \ ,
\end{aligned}
\label{eq: rho pt-vn def}
\end{equation}
where the covariance is given by,
\begin{equation}
\begin{aligned}
Cov([p_T],v_n^2) =\langle [p_T] V_n V_n^\star \rangle - \langle [p_T]  \rangle \langle  V_n V_n^\star \rangle \ ,
\end{aligned}
\label{eq: cov pt-vn}
\end{equation}
and the variances in the denominator as,
\begin{equation}
\begin{aligned}
Var([p_T]) = \langle [p_T]^2   \rangle -  \langle [p_T]  \rangle^2 \eqsp{and} Var(v_n^2) = \langle( V_n V_n^\star)^2   \rangle -  \langle V_n V_n^\star  \rangle^2.
\end{aligned}
\label{eq: varpt and varvn}
\end{equation}
The Pearson correlation coefficient $\rho([p_T],v_n^2)$ is particularly robust as it does not depend on the hydrodynamic response coefficient $k_n$ and essentially insensitive to the medium properties. Moreover, it is insensitive to statistical fluctuations and picks up only genuine correlation between mean transverse momentum and harmonic flow coefficients. The covariance in the numerator of the correlation coefficient in Eq.~(\ref{eq: rho pt-vn def}) involves three particle correlations, whereas the variance of the flow harmonic in Eq.~(\ref{eq: varpt and varvn}) is a four particle correlator. The experimental measurement for the covariance and the variances in Eq.~(\ref{eq: rho pt-vn def}) involves up to three or four sums over particles in the event, with self-correlations excluded, as discussed in \cite{Bozek:2016yoj}.
\begin{figure}[ht!]
\includegraphics[height=6 cm]{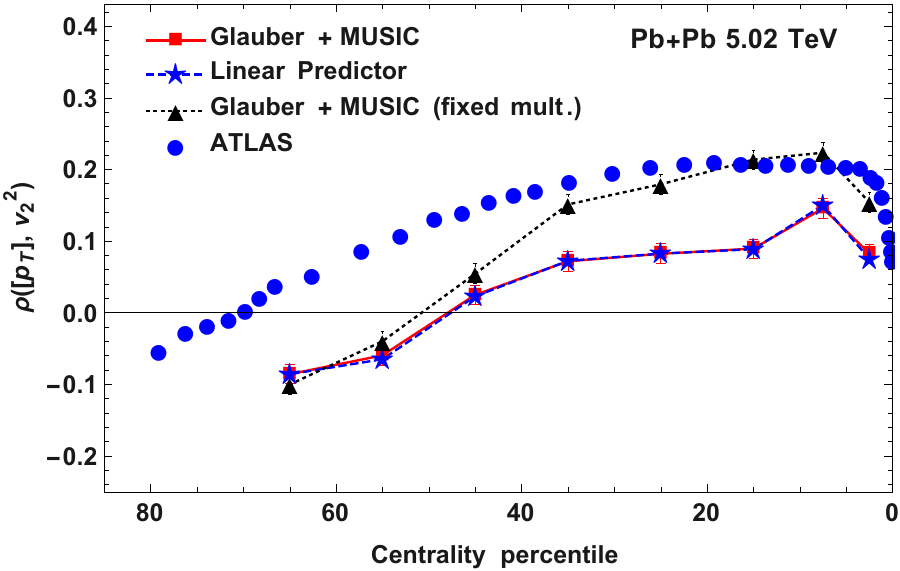}  
\centering
\caption{The Pearson correlation coefficient between the mean transverse momentum per particle $[p_T]$ and the elliptic flow squared $v_2^2$ in Pb+Pb collisions at 5.02 TeV as a function of centrality. The experimental data shown are from the ATLAS collaboration~\cite{ATLAS:2022dov} (blue points). The results of the hydrodynamic simulations with Glauber initial conditions are denoted by the red squares, the black triangles represent the results for the correlation coefficient corrected for multiplicity fluctuations (Eq.~\ref{eq: mult-correction}) and the star symbols with the blue dashed line represent the correlation coefficient obtained from the linear predictor (Eq.~\ref{eq: linear predictor}). The figure is from the original publication~\cite{Bozek:2021zim}, coauthored by the author.}
\label{fig: pearcorr pt-v2}
\end{figure}

To calculate these quantities in our model, here also we simulate Pb+Pb collisions at $\sqrt{s_{NN}}=5.02$~TeV using the boost invariant version of MUSIC \cite{Schenke:2010nt} with the initial energy densities obtained from the two-component Glauber Monte Carlo model~\cite{Bozek:2019wyr} in each event. The details of the model for the initial density can be found in \cite{Bozek:2016yoj}. Unless otherwise specified, we use a constant shear viscosity to entropy ratio $\eta/s=0.08$ for the hydrodynamic evolution.

Fig.~\ref{fig: pearcorr pt-v2} shows the model results for the centrality dependence of the transverse momentum-elliptic flow correlation coefficient $\rho([p_T],v_2^2)$ along with the experimental data from the ATLAS collaboration~\cite{ATLAS:2022dov}. The model calculations for $\rho([p_T],v_2^2)$ follow a qualitatively similar centrality dependence as compared to the ATLAS data~\cite{ATLAS:2022dov}. In particular, the correlation coefficient $\rho([p_T],v_2^2)$ decreases in the most central collisions as well as in peripheral collisions. Both the data and the model calculations show a sign change in peripheral collisions. However, It can be seen  that the change of sign for $\rho([p_T],v_2^2)$ occurs at different centralities, with 
the model calculation changing sign in more central collisions as compared to the data.

\begin{figure}[ht!]
\hspace{-0.35 cm}\begin{subfigure}{0.5\textwidth}
\centering
\includegraphics[height=4.9 cm]{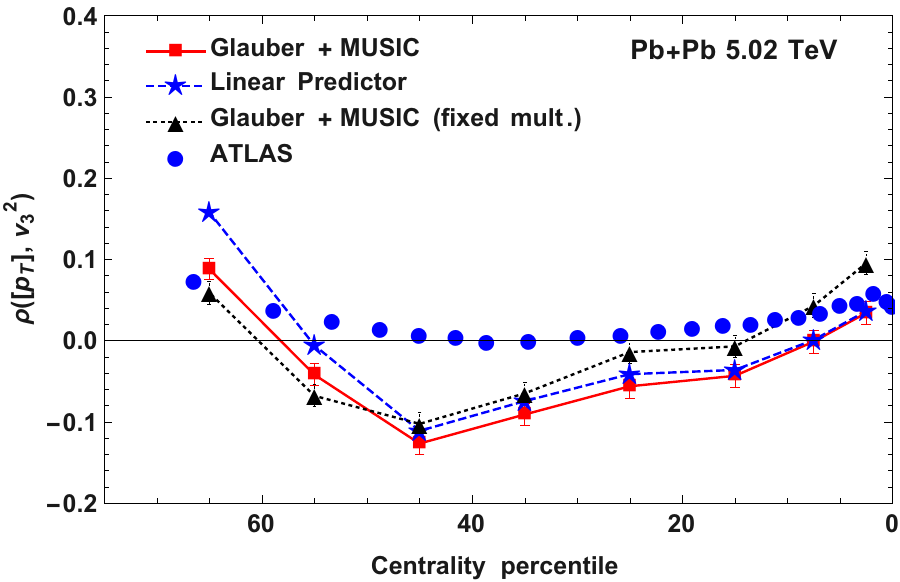}
\end{subfigure}~~~
\begin{subfigure}{0.5\textwidth}
\centering
\includegraphics[height=4.9 cm]{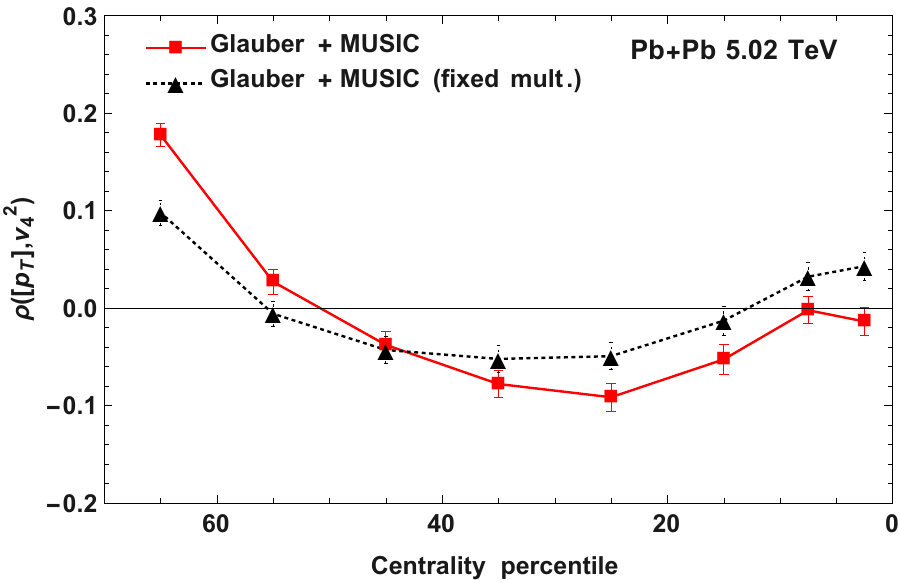}  
\end{subfigure}
\centering
\caption{Left: The Pearson correlation coefficient between the mean transverse momentum per particle and the triangular flow in Pb+Pb collisions at 5.02 TeV as a function of centrality. Right: The same for quadrangular flow. The symbols are same as Fig.~\ref{fig: pearcorr pt-v2}. The left panel of the figure is from the original publication~\cite{Bozek:2021zim}, coauthored by the author.}
\label{fig: pearcorr pt-v3 and pt-v4}
\end{figure}

The correlation coefficient for the triangular flow  $\rho([p_T],v_3^2)$ and the quadrangular flow $\rho([p_T],v_4^2)$ are shown in Fig.~\ref{fig: pearcorr pt-v3 and pt-v4}. The measured correlation $\rho([p_T],v_3^2)$ is small as compared to our model results. 
Moreover, our simulation results cannot describe the full features of the experimental data. Such discrepancies observed between the data and simulations results for $\rho([p_T],v_n^2)$ may stem from some underlying physics of the dynamics in the model or it could indicate that we are missing in our model calculations some essential correlations present in the initial state. For peripheral collisions with very low multiplicities, some correlations may be attributed to the presence of non-flow contributions and/or the initial flow, but why a significant discrepancy for $\rho([p_T],v_n^2)$ occur at the mid-central collisions, is unclear. The correlation coefficient $\rho([p_T],v_4^2)$ is the prediction based on our model.

\subsubsection{Correction for multiplicity fluctuations: Partial correlation}
\label{partial corr}

Along with the mean transverse momentum per particle and harmonic flow, the multiplicity also fluctuates event-by-event. As a result, the correlation between $[p_T]$ and $v_n^2$ may partially originate from the correlations of these quantities with the event multiplicity $N$ (We denote multiplicity in an event by $N\equiv N_{ch}$). Usually the experimental analysis is performed in narrow bins of centrality, where such residual correlation can be significantly reduced. To tackle this effect in model calculations, the dependence of the variance or covariance of the observables of interest on the fluctuations of a third variable, (e.g. here the multiplicity) are taken into account by calculating the partial variance or covariance \cite{Olszewski:2017vyg}. The {\it partial correlation coefficient} between $[p_T]$ and $v_n^2$ is given by, 
\begin{equation}
\rho([p_T],v_n^2 \bullet N)=\frac{\rho([p_T],v_n^2)-\rho([p_T],N)\rho(N,v_n^2)}{\sqrt{1-\rho([p_T],N)^2}\sqrt{1-\rho(v_n^2,N)^2}} \ ,
\label{eq: partial corr}
\end{equation}
which provides an estimate of the correlation coefficient in a centrality bin at fixed multiplicity~\cite{Bozek:2019wyr}. Figs.~\ref{fig: pearcorr pt-v2} and \ref{fig: pearcorr pt-v3 and pt-v4} show the results for the partial correlation coefficients, denoted by black dashed line. One can see that there is sizable correction due to multiplicity fluctuations for the elliptic flow. Moreover, the model results obtained for $\rho([p_T],v_2^2)$ after correcting for multiplicity, is closer to the experimental data in comparison to the uncorrected one (red solid lines). However, for $\rho([p_T],v_3^2)$ and $\rho([p_T],v_4^2)$, the correction are not so significant and both of the calculations lie close to each other.  

In general, if an observable $O$ has an approximately linear dependence on the multiplicity in a given bin, then the correction for multiplicity fluctuations to $O$ can be implemented as \cite{Schenke:2020uqq}, 
\begin{equation}
\tilde{O}= O -\frac{Cov(O,N)}{Var(N)}\left(N-\langle N \rangle \right) \ ,
\label{eq: mult-correction}
\end{equation}
where $\tilde{O}$ denotes the multiplicity-corrected observable. In principle, one can use the above equation for $[p_T]$ and $v_n^2$ and calculate $\rho([p_T],v_n^2)$. This is equivalent of using the formula for the partial correlation coefficient presented in Eq.~(\ref{eq: partial corr}). Later in this section, we use the corrected observables $\tilde{O}$ while estimating the higher order cumulants/correlations to remove the effects of multiplicity fluctuations. 

In order to compare with the experimental data, obtained in narrow bins of multiplicity, the model calculations should always be corrected for the
multiplicity fluctuations and only then it should be compared. If the experimental multiplicity bins are wide or a different centrality estimator (as discussed in the previous chapter) is used for defining centrality bins, a correction for the observables using Eq.~(\ref{eq: mult-correction}) should be done (where $N$ will denote the centrality estimator) in order to establish a consistent description of correlations and cumulants between different experiments and model calculations.

\subsection{Mapping to initial state: Linear predictor}
\label{linear predictor}

\begin{figure}[ht!]
\hspace{-0.15 cm}\begin{subfigure}{0.5\textwidth}
\centering
\includegraphics[height=6 cm]{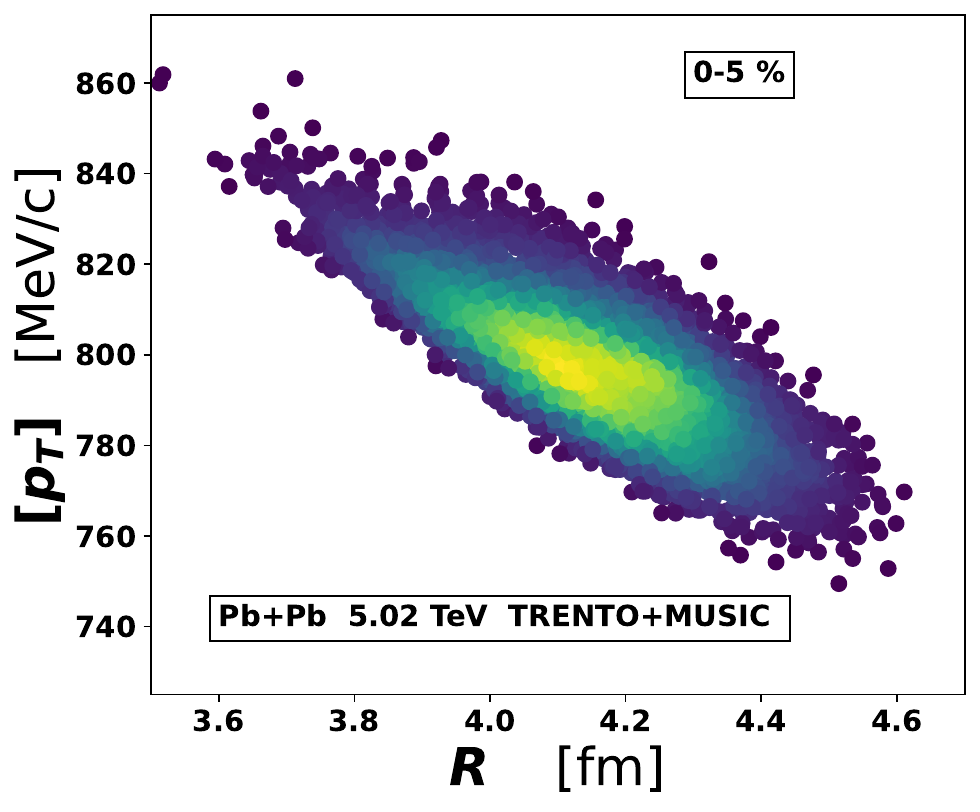}
\end{subfigure}~~~
\begin{subfigure}{0.5\textwidth}
\centering
\includegraphics[height=6 cm]{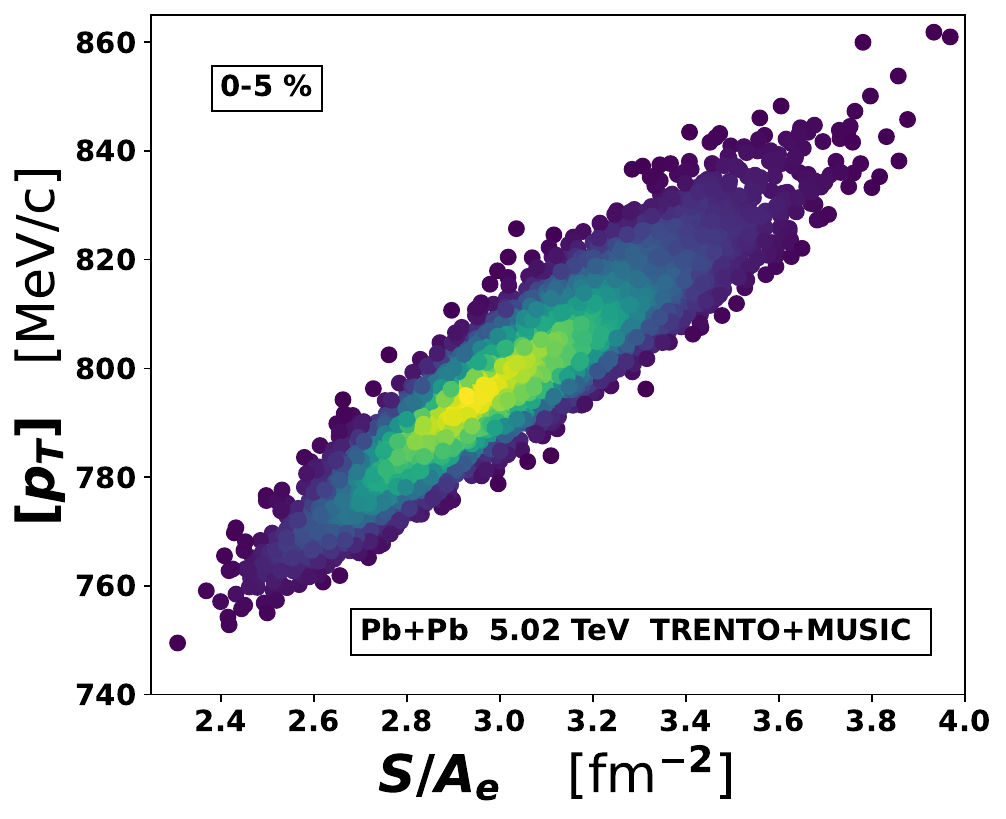}  
\end{subfigure}
\centering
\caption{Left: Scatter plot between event-by-event mean transverse momentum per particle and transverse size of the initial source for $0-5\%$ centrality in Pb+Pb collision at 5.02 TeV obtained with TRENTO initial condition. Right: Similar scatter plot between mean transverse momentum per particle and total entropy per unit elliptic area.}
\label{fig: scatter plot pt-R and pt-S}
\end{figure}

The collective observables in the final state of heavy-ion collisions are largely determined by the initial conditions of the collision. In particular, the mean transverse momentum per particle $[p_T]$ and the harmonic flow coefficients $v_n$ are strongly correlated to these properties of the initial state\cite{Gardim:2011xv,Qiu:2011iv,Niemi:2012aj,Broniowski:2009fm}. As discussed in Chapter 3, the harmonic flow coefficients can be mapped to the spatial anisotropy or eccentricities $\Epsilon_n$ of the initial density distribution (Eqs.~(\ref{eq: relation between elliptic flow and eccentricity}),(\ref{eq: relation between triangualar flow and eccentricity}) and (\ref{eq: relation between quadrangular flow and eccentricities})), therefore resulting in a strong correlation as seen in Figs.~\ref{fig: v2 vs eps2 plot} and \ref{fig: v3 vs eps3 plot}. The mean transverse momentum per particle, on the  other hand, can be related to the the RMS size of the transverse profile $R$ (Eq.~\ref{eq: rms radius of final satate})~\cite{Mazeliauskas:2015efa,Bozek:2017elk}, the total initial entropy $S$ (Eq.~\ref{eq: total energy and entropy density at initial state}) as well as the initial eccentricities. Fig.~\ref{fig: scatter plot pt-R and pt-S} shows the scatter plot between the mean transverse momentum and transverse size $R$ on the left, and between mean transverse momentum and total entropy per unit elliptic area $S/A_e$ on the right, where $A_e = \pi R^2 \sqrt{1-\epsilon_2^2}$ denoting the elliptic area of the initial transverse profile. The figure shows a strong correlation (negative) between $[p_T]$ and $R$, as well as a significant correlation (positive) with the total entropy $S$ and ellipticity of the initial state $\epsilon_2$. 

The anti-correlation between the mean transverse momentum per particle $[p_T]$ and the transverse size $R$ observed in Fig.~\ref{fig: scatter plot pt-R and pt-S} (left) can be understood by the same thermodynamic argument presented in Chapter-4. If we fix the multiplicity (i.e. initial entropy), then a smaller transverse size implies a smaller collision volume, and hence larger density. Then by relativistic thermodynamics, it means larger temperature and eventually larger pressure gradients which results in larger energy per particle or larger transverse momentum per particle $[p_T]$. The reverse scenario occurs when $R$ is large which produce smaller $[p_T]$. Similar argument can be made for explaining the positive correlation between $[p_T]$ and $S$ at fixed collision volume, this is equivalent to dividing the event-by-event total entropy by the transverse area. At fixed collision volume, a larger entropy means a larger density and hence by the same principle of relativistic thermodynamics, it produces larger transverse momentum per particle at the final state resulting in the strong positive correlation observed in Fig.~\ref{fig: scatter plot pt-R and pt-S} (right). As an alternative way, $[p_T]$ can be also predicted from the initial energy per rapidity ($E_i$)~\cite{Giacalone:2020dln} or the energy weighted entropy  \cite{Schenke:2020uqq}. 

The correlation present in the initial state between these quantities can be captured by constructing an appropriate predictor for $[p_T]$ and $v_n$. It was proposed~\cite{Bozek:2020drh} that in order to construct a predictor for the Pearson correlation coefficient between the mean transverse momentum per particle and the elliptic or triangular flow, the initial state eccentricities should also be included in the predictor for the transverse momentum $[p_T]$. According to Ref.~\cite{Bozek:2020drh}, such an {\it improved predictor} can well describe the transverse momentum-harmonic flow correlation $\rho([p_T],v_n^2)$, in connection to the full hydrodynamic simulation. Therefore, for our analysis we use general linear predictors for $[p_T]$ and $v_n$, based on moments of the initial density\footnote{Please note $[p_T]$ can be simultaneously related to $R$, $S$ and also $\epsilon_n$. In Fig.~\ref{fig: scatter plot pt-R and pt-S} (right), we show the correlation between $[p_T]$ and $S/A_e$ which contains all of these quantities in terms of entropy density and hence the correlation is remarkably strong. But in Eq.~(\ref{eq: linear predictor}) we construct a general predictor from individual quantities, in order to capture the genuine correlations between them through the Pearson's correlation coefficient and the symmetric cumulants.}, given by
\begin{equation}
\begin{aligned}
\hat{v}_2^2 & =  k_2 \epsilon_2^2 + \alpha_2 \delta R + \beta_2 \delta S \,  \\
\hat{v}_3^2 & = k_3 \epsilon_3^2 + \alpha_3 \delta R + \beta_3 \delta S \ , \\
\hat{[p_T]} & =  \langle \hat{[p_T]}\rangle + \alpha_p \delta R + \beta_p \delta S + \gamma_p \delta \epsilon_2^2 +\lambda_p \delta \epsilon_3^2   \ \ \ ,
\end{aligned}
\label{eq: linear predictor}
\end{equation}
where the {\it hat} symbol is used to denote the observables predicted from initial state in order to distinguish from the one obtained from hydrodynamic simulations. For any observable $O$, we have $\delta O =O -\langle O \rangle $, where $\langle \dots \rangle$ denotes the event average of the observable and the coefficients sitting before each moments can be thought as hydrodynamic response coefficients. One can also write the linear predictor in Eq.~(\ref{eq: linear predictor}) in a generalized form,
\begin{equation}
\delta \hat{O}_i= L^j_i \delta M_j \ ,
\label{eq: gen predictor}
\end{equation}
where $M_i$ is a set of moments of the initial density and $L^j_i$ are the response coefficients. In our analysis we optimize each of the observables in Eq.~(\ref{eq: linear predictor}) separately. Only after the parameters or the response coefficients are fixed for a linear predictor, the cumulants and correlation are calculated between the predicted observables. Using Eq.~(\ref{eq: gen predictor}), the covariance or the correlations between the final state observables can be expressed as a linear transformation of the correlation at the initial state,
\begin{equation}
\langle \delta O_i \dots \delta O_j \rangle = L^s_i \dots L^k_j \langle
\delta M_s \dots \delta M_k \rangle \ .
\label{eq: linear transformation}
\end{equation}

Figs.~\ref{fig: pearcorr pt-v2} and \ref{fig: pearcorr pt-v3 and pt-v4} show the results for the correlation $\rho([p_T],v_n^2)$ calculated using the linear predictors in Eq.~(\ref{eq: linear predictor}), presented by blue lines. The results are in fair agreement with the corresponding correlations calculated from hydrodynamic simulations. This shows the robustness of the linear predictors and how well it can capture the initial state correlations. This also establishes the fact that the Pearson correlation coefficient (and aslo the higher order cumulants, as we will see shortly) involving the final state observables can be understood as a linear hydrodynamic response of the correlations present in the initial state. Note that the correlation $\rho([p_T],v_4^2)$ cannot be predicted using the set of linear predictor used in our analysis. Therefore in Fig.~\ref{fig: pearcorr pt-v3 and pt-v4} (right), we only present the results obtained from hydrodynamics. 

\subsection{Higher order correlations: Symmetric cumulants}
\label{SC pt-v2}
Pearson correlation coefficient $\rho([p_T],v_n^2)$ represents the leading (lowest) order correlation between mean transverse momentum per particle and the harmonic flow coefficient, and it involves flow harmonic of a specific order. However, in order to understand additional information on such correlations and interplay between transverse momentum and different orders of flow harmonics, one needs to look into higher order correlations. Study of such higher order correlation coefficient is also useful to understand the higher order and mixed correlations present in the initial state. A correlation of order larger than 2, between mean transverse momentum and flow harmonics cannot be constructed using Pearson correlation coefficient and one needs to resort different constructions such as {\it symmetric cumulants (SC)}. Such symmetric cumulants between the magnitudes of the harmonic flow of different orders have been studied \cite{Bilandzic:2013kga}. We adopt similar methodology in order to construct SC between $[p_T]$ and $v_n's$.

\subsubsection{2nd order normalized symmetric cumulants (NSC)}
We start with the second order symmetric cumulants in analogy to the Pearson correlation coefficients. The second order SC is simply the covariance between two observables
\begin{equation}
 SC(A,B)=\langle A B \rangle -  \langle  A \rangle  \langle  B \rangle \equiv Cov(A,B)  \ .
\label{eq: 2nd order SC}
\end{equation}
Then the {\it normalized symmetric cumulant} (NSC) is obtained by scaling the above equation by the individual mean of the observables involved in SC. 
\begin{figure}[ht!]
\hspace{-0.35 cm}\begin{subfigure}{0.5\textwidth}
\centering
\includegraphics[height=4.9 cm]{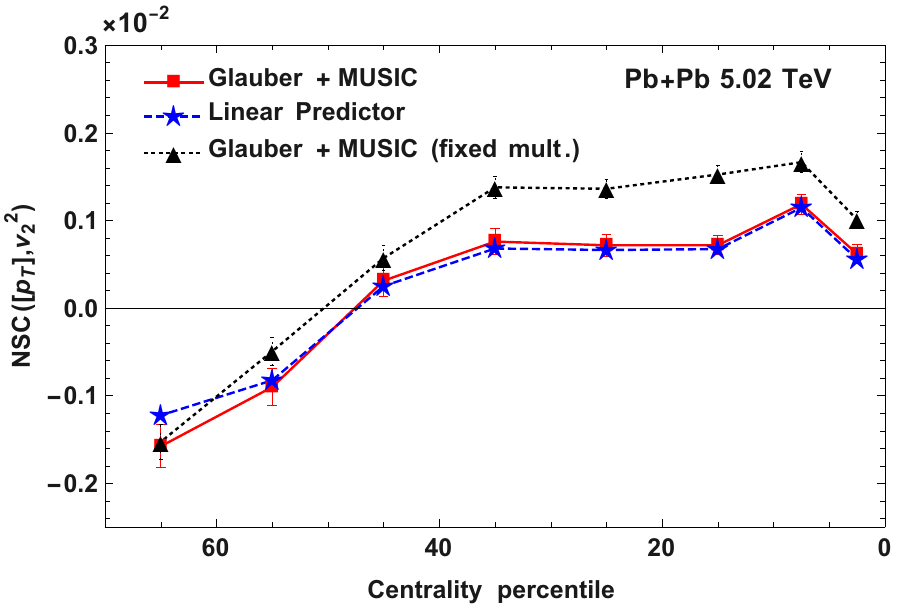}
\end{subfigure}~~~
\begin{subfigure}{0.5\textwidth}
\centering
\includegraphics[height=4.9 cm]{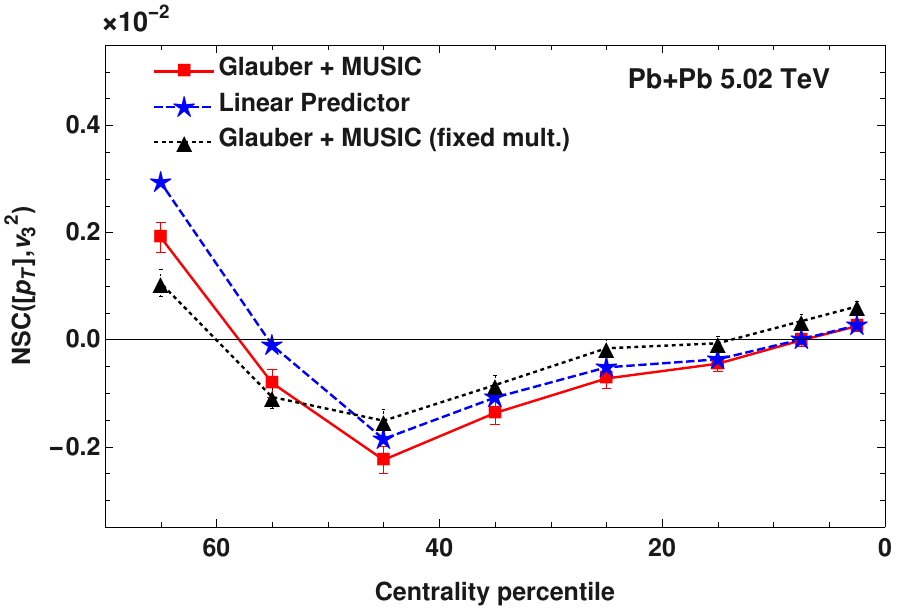}  
\end{subfigure}
\centering
\caption{Left: Normalized symmetric cumulant between mean transverse momentum per particle and elliptic flow coefficient in Pb+Pb collision at 5.02 TeV as a function of centrality. The red squares denote the results obtained from hydrodynamics and the black triangles denote the results corrected for multiplicity fluctuations. The blue stars represent the cumulant obtained from the linear predictor. Right: Same for the triangular flow.}
\label{fig: NSC pt-v2 and pt-v3}
\end{figure}
In case of correlation between transverse momentum and harmonic flow of a particular order, NSC can be defined as,
\begin{equation}
NSC([p_T],v_n^2)=\frac{\langle [p_T] v_n^2 \rangle - \langle [p_T] \rangle \langle v_n^2 \rangle}{\langle [p_T] \rangle \langle v_n^2 \rangle} \ .
\label{eq: NSC pt-vn}
\end{equation}
The normalized symmetric cumulants $NSC([p_T],v_n^2)$ for the elliptic and triangular flow are shown in Fig. \ref{fig: NSC pt-v2 and pt-v3} . $NSC([p_T],v_n^2)$ represents the event by event correlations between the mean transverse momentum and the harmonic flow  and this information is contained in the covariance of the two observables in the numerator. This implies that $NSC([p_T],v_n^2)$ basically carries the same  information as the correlation coefficient $\rho(p_T,v_n^2)$, which is reflected in the results of Fig.~\ref{fig: NSC pt-v2 and pt-v3}. However, the orders of magnitude of the correlation are changed because of the changed normalization. 
Furthermore, it should be noted that the experimental extraction of normalized symmetric cumulant is simpler than the Pearson correlation coefficient $\rho$. This is because the denominator of Eq.~(\ref{eq: NSC pt-vn}) involves at most a two particle correlator, whereas the denominator in Eq.~(\ref{eq: rho pt-vn def}) requires the measurement of three or
four particle correlators. The definition in Eq.~(\ref{eq: NSC pt-vn}) is favourable by experiments and also methods for reducing non-flow effects can be implemented, even in small collision systems \cite{Zhang:2021phk}.

\subsubsection{Third and fourth order NSC }
Next, we move to the constructions of higher order NSC, which may contain additional  information on correlations between mean transverse momentum and harmonic flow. Such higher order cumulants of only harmonic flow of different orders have been studied~\cite{Bilandzic:2013kga,Mordasini:2019hut,Moravcova:2020wnf}. We implement the same methodology for cumulants of $[p_T]$ and $v_n^2$. In general, the $n$-th order normalized symmetric cumulant involves only the genuine correlation between $n$ observables, where all the lower order correlations are subtracted. This way it is similar to the construction of multi-particle flow cumulants discussed in Chapter-3.
The third and fourth order symmetric cumulants for scalar observables are defined as \cite{Mordasini:2019hut},
\begin{equation}
\begin{aligned}
 SC(A,B,C)  &=  \langle A B C  \rangle-\langle A B\rangle \langle C\rangle-\langle A C \rangle \langle B \rangle-\langle B C \rangle \langle A \rangle+2 \langle A\rangle \langle B\rangle\langle C\rangle \ , \\
\eqsp{and} SC(A,B,C,D) & = \langle A B C D \rangle-\langle A B C\rangle \langle D\rangle-\langle A B D \rangle \langle C \rangle -\langle A C D \rangle \langle B \rangle-\langle B C D \rangle \langle A \rangle \\
&-\langle A B\rangle \langle C D\rangle-\langle A C \rangle \langle B D\rangle 
 -\langle B C \rangle \langle A D \rangle \\
 &+2\bigg( \langle A B\rangle \langle C\rangle \langle D\rangle +\langle A C\rangle \langle B\rangle \langle D\rangle  + \langle A D\rangle \langle C\rangle \langle B\rangle \\
 &+\langle B C\rangle \langle A\rangle \langle D\rangle +\langle B D \rangle \langle A\rangle \langle C\rangle +\langle C D \rangle \langle A\rangle \langle B\rangle \bigg) \\
&-6 \langle A\rangle \langle B\rangle\langle C\rangle\langle D\rangle \ , 
\end{aligned}
\label{eq: SC third and fourth order}
\end{equation}
\begin{figure}[ht!]
\includegraphics[height=6 cm]{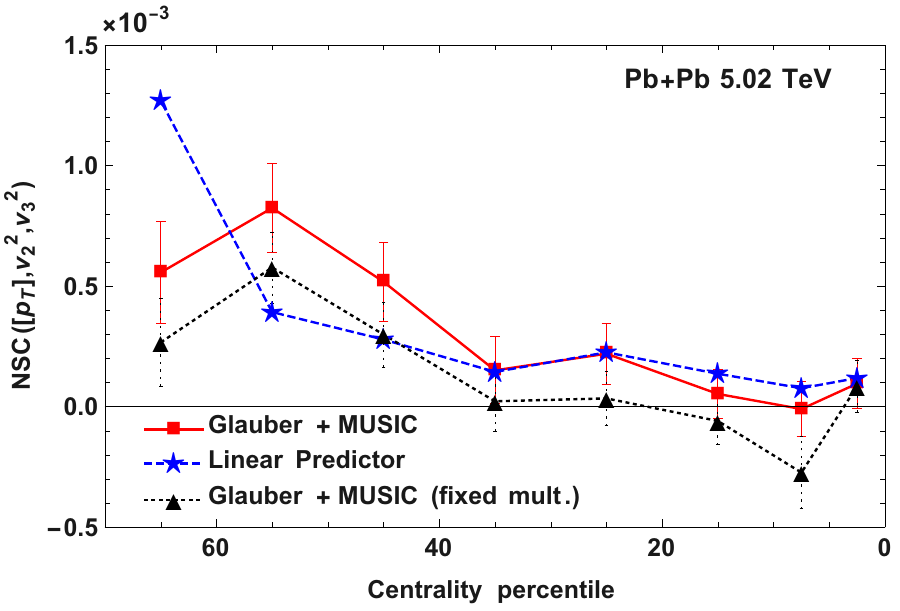}  
\centering
\caption{Third order normalized symmetric cumulant between mean transverse momentum per particle, elliptic flow and triangular flow coefficient in Pb+Pb collision at 5.02 TeV as a function of centrality. The symbols carry similar meaning as Fig.~\ref{eq: NSC pt-vn}. The figure is from the original publication~\cite{Bozek:2021zim}, coauthored by the author.}
\label{fig: NSC pt-v2-v3 }
\end{figure}
and the corresponding normalized symmetric cumulants are given by
\begin{equation}
NSC(A,B,C)=\frac{SC(A,B,C)}{\langle A\rangle\langle B\rangle\langle C\rangle} \eqsp{and} NSC(A,B,C,D)=\frac{SC(A,B,C,D)}{\langle A\rangle\langle B\rangle\langle C\rangle\langle D\rangle} \ .
\label{eq: NSC third and fourth order}
\end{equation}

\begin{figure}[ht!]
\hspace{-0.35 cm}\begin{subfigure}{0.5\textwidth}
\centering
\includegraphics[height=4.9 cm]{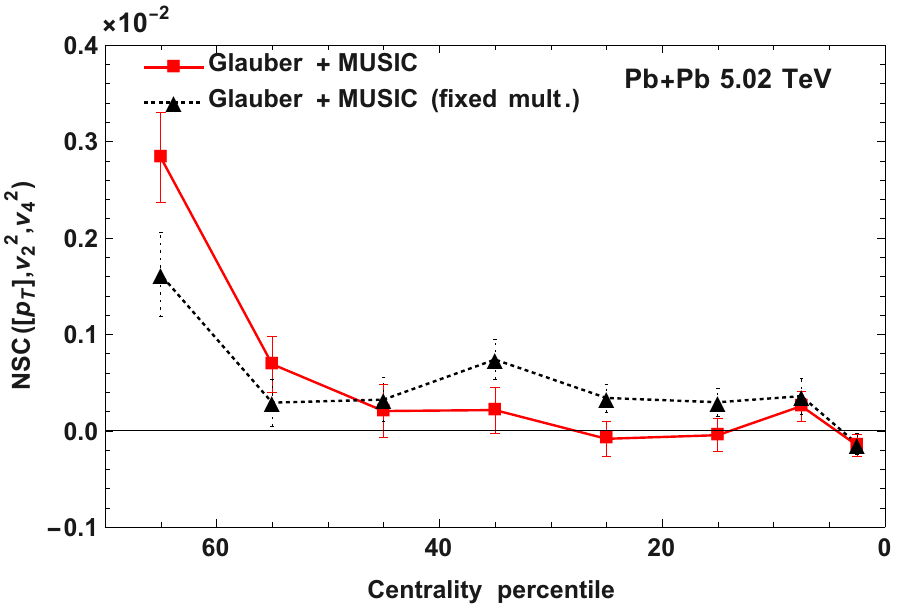}
\end{subfigure}~~~
\begin{subfigure}{0.5\textwidth}
\centering
\includegraphics[height=4.9 cm]{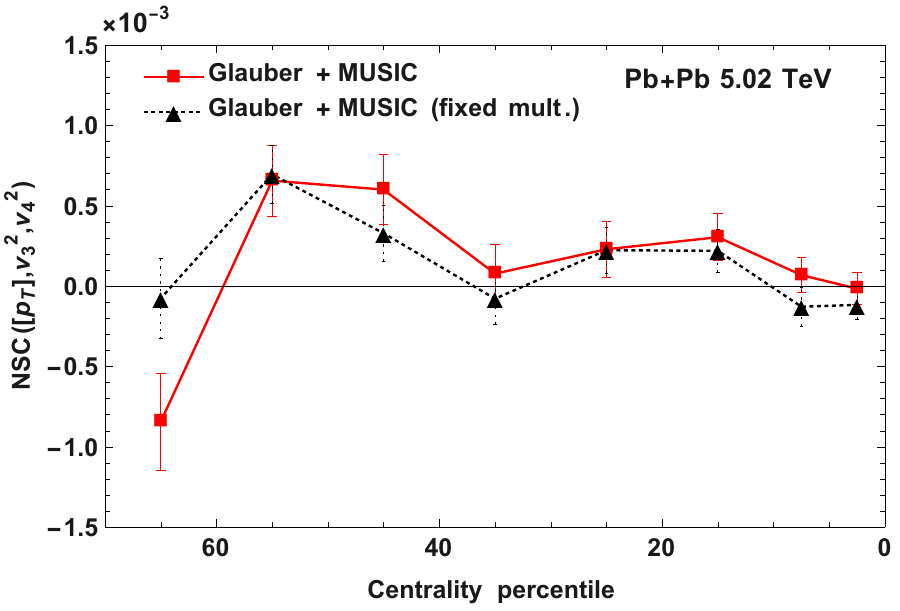}  
\end{subfigure}
\centering
\caption{Third order normalized symmetric cumulant between $[p_T]$, $v_2^2$ and $v_4^2$ in the left and between $[p_T]$, $v_3^2$ and $v_4^2$ in the right for Pb+Pb collision at 5.02 TeV as a function of collision centrality. The red squares and black triangles denote the results obtained from the hydrodynamic simulation without and with corrections for multiplicity fluctuations respectively. The figure is from the original publication~\cite{Bozek:2021zim}, coauthored by the author.}
\label{fig: NSC pt-v2-v4 and pt-v3-v4}
\end{figure}

In Figs. \ref{fig: NSC pt-v2-v3 } and \ref{fig: NSC pt-v2-v4 and pt-v3-v4}, the simulation results for the third order normalized symmetric cumulants $NSC([p_T],v_2^2,v_3^2)$, $NSC([p_T],v_2^2,v_4^2)$ and $NSC([p_T],v_3^2,v_4^2)$ are presented. As noted earlier, for the higher order symmetric cumulants, the effect of multiplicity fluctuations is reduced by correcting the observables for multiplicity using Eq.~(\ref{eq: mult-correction}). 
A first observation is that for all the third order normalized symmetric cumulants, the magnitude of the correlation is much smaller as compared to $\rho$, which confirms the fact that it measures only genuine third order correlations. The cumulants $NSC(p_T,v_2^2,v_3^2)$ and $NSC(p_T,v_2^2,v_4^2)$ increase for the peripheral collisions, whereas the cumulant $NSC(p_T,v_3^2,v_4^2)$ show a decrease in peripheral collisions. For $NSC(p_T,v_2^2,v_3^2)$, the linear predictor (\ref{eq: linear predictor}), based on the initial correlations only, describes the full hydrodynamic calculation within centrality range: 0-50\%. Like in the previous case, the cumulants involving $v_4$ cannot be predicted using our linear predictor, so that we only present the hydro results for them and they could serve as a precise measure of nonlinearities between harmonic flow of different orders with subtle interplay between average transverse momentum and flow.  
\begin{figure}[ht!]
\includegraphics[height=6 cm]{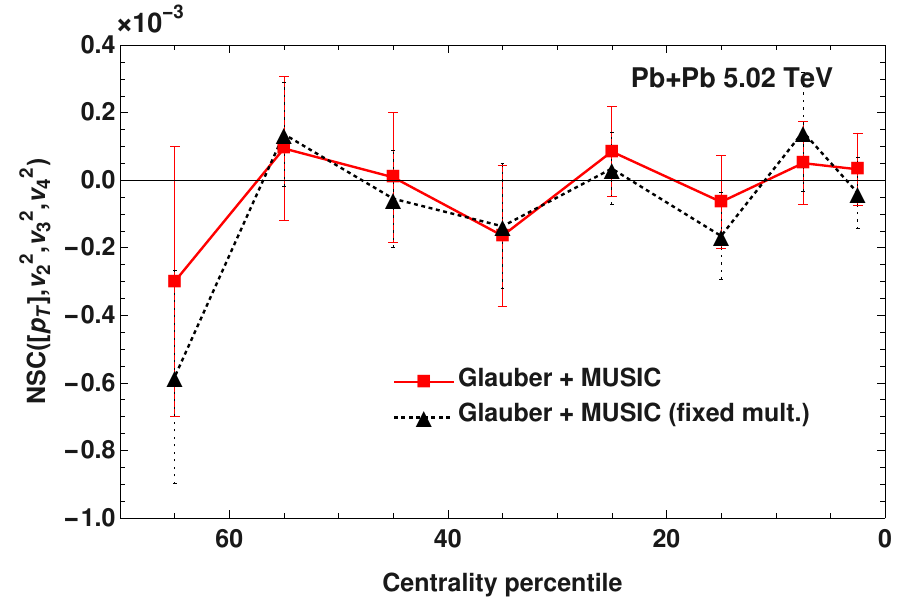}  
\centering
\caption{Fourth order normalized symmetric cumulant between $[p_T]$, $v_2^2$, $v_3^2$ and $v_4^2$ in Pb+Pb collision at 5.02 TeV as a function of centrality. The symbols have similar meaning as Fig.~\ref{fig: NSC pt-v2-v4 and pt-v3-v4}.}
\label{fig: NSC pt-v2-v3-v4 }
\end{figure}
The fourth order normalized symmetric cumulant $NSC(p_T,v_2^2,v_3^2,v_4^2)$ is presented in Fig.~\ref{fig: NSC pt-v2-v3-v4 }. The results are compatible with zero within the statistical accuracy of our calculation.

\begin{figure}[ht!]
\hspace{-0.35 cm}\begin{subfigure}{0.5\textwidth}
\centering
\includegraphics[height=4.9 cm]{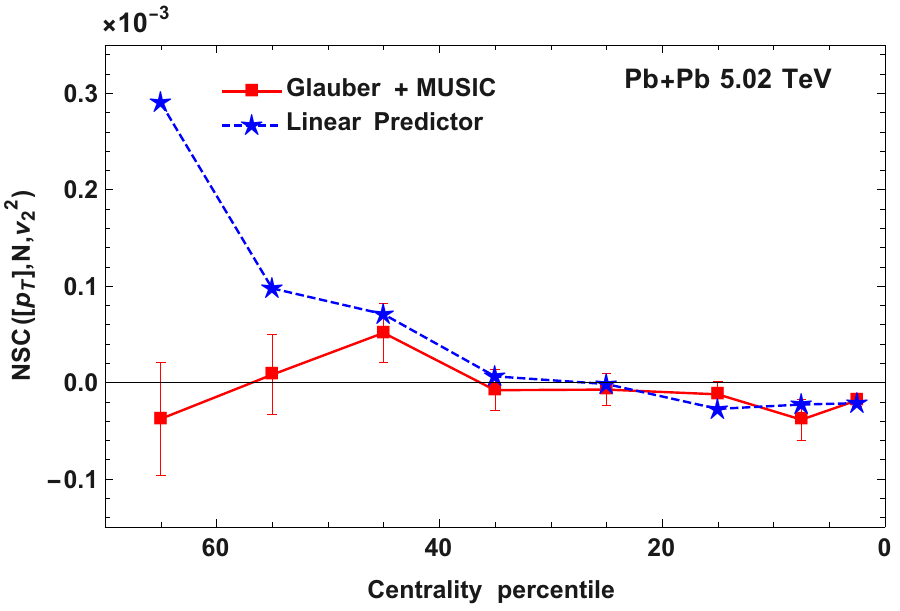}
\end{subfigure}~~~
\begin{subfigure}{0.5\textwidth}
\centering
\includegraphics[height=4.9 cm]{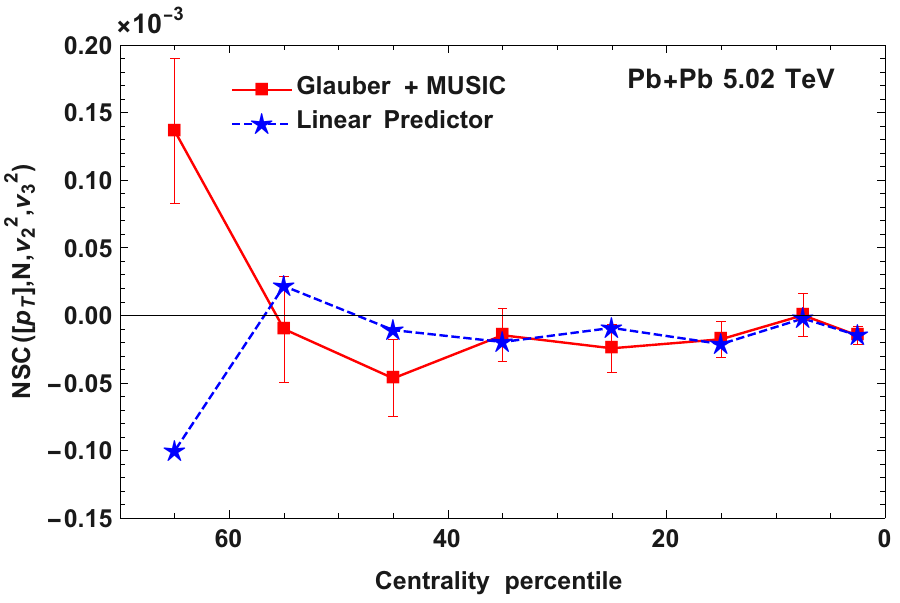}  
\end{subfigure}
\centering
\caption{Left: Third order normalized symmetric cumulant between mean transverse momentum per particle, multiplicity and elliptic flow coefficients in Pb+Pb collision at 5.02 TeV as a function of centrality. The red squares and blue stars represent the results obtained from the hydrodynamic simulation and the linear predictor respectively. Right: Fourth order normalized symmetric cumulant between $[p_T]$, $N$, $v_2^2$ and $v_3^2$. with symbols carrying same meaning.}
\label{fig: NSC pt-n-v2 and pt-n-v2-v3}
\end{figure}

For completeness, we also study the third and fourth order NSC between $[p_T]$, $N$ and $v_n^2$, the results of which are shown in Fig.~\ref{fig: NSC pt-n-v2 and pt-n-v2-v3}. Such correlations involving the multiplicity (or any centrality estimator) as one of the observable could be sensitive to the fluctuations in the entropy deposition at the initial state and the correlation of it with its moments. Our simulation results and the linear predictor show opposite behavior for the peripheral collisions which opens the scope for further investigation of the peripheral behavior of such higher order cumulants. Please note that the results presented in this case, for the observables $[p_T]$ and $v_n^2$, are not corrected for multiplicity fluctuations, because the cumulant involve multiplicity as an observable within itself. Normalized symmetric cumulants involving multiplicity might not carry much significance for spherical nuclei collisions but have a greater importance for deformed nuclei collisions, where fluctuations of multiplicity is more significant and can probe deformed structure of the nucleus. Such effects will be discussed in detail in the next chapter. 

\subsubsection{Change of normalization: Scaled symmetric cumulants (SSC)}
\label{scaled symmetric cumulants}

The denominator of the normalized symmetric cumulants in Eq.~(\ref{eq: NSC third and fourth order}) involve averages of the observables for which the cumulant is calculated.
However, with such definitions the interpretation of the results become less obvious than the Pearson correlation coefficient $\rho$. The average transverse momentum in a collision can depend on many factors \cite{Bozek:2012fw} such as the freeze-out procedure, the bulk viscosity, the preequilibrium flow and even the experimental range for transverse momentum. 
\begin{figure}[ht!]
\includegraphics[height=6 cm]{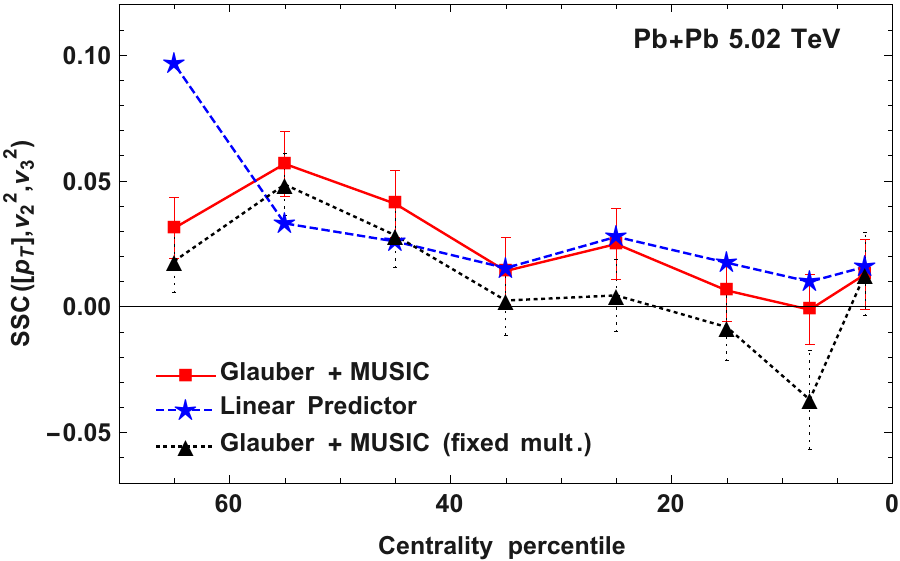}  
\centering
\caption{Third order scaled symmetric cumulant between mean transverse momentum per particle, elliptic flow and triangular flow coefficients in Pb+Pb collision at 5.02 TeV as a function of centrality. Symbols have similar meaning as Fig.~\ref{fig: NSC pt-v2-v3 }. The figure is from the original publication~\cite{Bozek:2021zim}, coauthored by the author.}
\label{fig: SSC pt-v2-v3 }
\end{figure}
Moreover, the linear predictor for the mean transverse momentum per particle in an event Eq.~(\ref{eq: linear predictor}) can only predict the deviations from the average and the absolute value for the average transverse momentum is not provided by the experiments. To overcome all these limitations, a modification for the normalization of the SC could be necessary. 

Such an alternative normalization for the symmetric cumulants can be the standard deviations of the observables involved in the denominator instead of mean. With that, we construct the {\it scaled symmetric cumulants} given by,
\begin{equation}
\begin{aligned}
SSC(A,B,C)&=\frac{SC(A,B,C)}{\sqrt{Var(A)Var(B)Var(C)}} \ ,\\
\eqsp{and} 
SSC(A,B,C,D) &= \frac{SC(A,B,C,D)}{\sqrt{Var(A)Var(B)Var(C)Var(D)}} \ .
\end{aligned}
\label{eq: SSC third and fourth order}
\end{equation}
The scaled symmetric cumulant has two-fold advantages. First the prediction of SSC from the initial state does not require the input on the value of the average transverse momentum, which is sometimes not known from either the simulations or the experiments. Second, the values of the scaled symmetric cumulants can be relatively well predicted using the linear hydrodynamic response (Eq.~(\ref{eq: linear predictor})).

Fig.~\ref{fig: SSC pt-v2-v3 } shows results for the scaled symmetric cumulant between $[p_T]$, $v_2^2$ and $v_3^2$. The results display a very similar behavior as the normalized symmetric cumulants presented in Fig.~\ref{fig: NSC pt-v2-v3 }. The only noticeable difference is in the order of magnitudes. The numerical values for SSC are larger as compared to NSC because of the change in normalization. It could be noted that the sole change of the normalization from the average (mean) transverse momentum to its standard deviation results in a change by a factor in the range\footnote{This effect can be related to the 1\% relative dynamical fluctuations of $[p_T]$ discussed in the last chapter}: $20 - 100$.

\section{Momentum dependent correlation: Between $[p_T]$ and differential flow $v_n(q)$}
\label{pt-v2pt correlation}

In the previous section, we discussed the correlations and cumulants between mean transverse momentum per particle $[p_T]$ and {\it integrated} flow $v_n$. One can think of generalizing this class of observables to momentum dependent correlation coefficients, by introducing transverse momentum dependence within the flow harmonics. In particular, such correlation coefficient would read as a correlation between mean transverse momentum per particle in an event $[p_T]$ and the harmonic flow coefficient in a transverse momentum bin, denoted by $v_n(q)$ (here, $q$ denotes a particular transverse momentum bin). Such momentum dependent correlations can provide useful insights in many aspects. Apart from providing a complementary statistical information of the event-by-event distribution of those particular observables, $[p_T]-v_n(q)$ correlations could shed light on several interesting issues related to heavy-ion collisions, which serve as motivating factors for engaging in such studies, as outlined below. The momentum dependent correlation coefficient between mean transverse momentum and differential harmonic flow 
\begin{itemize}
\item  could help us to understand the observed $p_T$-cut dependence of the correlation coefficients such as $\rho([p_T],v_n^2)$~\cite{ATLAS:2019pvn},
\item could provide useful insights on specific modes in the initial state which can be related to the final state transverse momentum and harmonic flow \cite{Mazeliauskas:2015efa},
\item could show sensitivity to nucleon width or {\it granularity} in the initial state~\cite{Giacalone:2021clp}, which plays a significant role for momentum dependent flow.
\item could provide a measure of the correlation between transverses momentum and harmonic flow independent of the shape of the momentum dependence of the particular flow harmonics,
\item if used for identified particles, it could also test the hadronization mechanism and its possible dependence on the transverse expansion.
\item or, could help in identifying the correlations between mean transverse momentum and harmonic flow from other models such as the color glass condensate dynamics \cite{Altinoluk:2020psk}.

\end{itemize}

In this section, we propose possible definitions for the momentum dependent correlation coefficient between the mean transverse momentum and the harmonic flow. We mostly focus on the momentum dependent construction for Pearson correlation coefficients. Similar momentum dependent higher order cumulants can also be studied but lies beyond the scope of our current study. We also explore the sensitivity of such correlation coefficients to the granularity in the initial state and medium properties (e.g. shear viscosity), and propose new covariance that could be measured in experiments. We propose alternate simplified expressions for the momentum dependent correlation that could be used in experimental analyses with lesser difficulty. Like the previous cases, the simulation results are presented for Pb+Pb collisions at 5.02 TeV energy, which are obtained from the boost invariant relativistic viscous hydrodynamics code MUSIC, with initial conditions generated from the TRENTO or Glauber initial condition model.

\subsection{Pearson correlator: $\rho([p_T],v_n(q)^2) \equiv \rho([p_T],V_n(q)V_n(q)^\star)$ }
\label{rho pt-v2pt}
The momentum dependent construction for the Pearson correlation coefficient between mean transverse momentum per particle and the harmonic flow coefficient in a transverse momentum bin can be defined as,
\begin{equation}
  \rho\left([p_T],V_n(q)V_n(q)^\star\right) = \frac{Cov\left([p_T],V_n(q)V_n(q)^\star\right)}{\sqrt{Var\left([p_T]\right)Var\left(V_n(q)V_n(q)^\star\right)}} \ ,
\label{eq: pearcorr pt-v2pt}
\end{equation}
where the covariance and variance in the above expression are defined similarly as Eqs.~(\ref{eq: cov pt-vn}) and (\ref{eq: varpt and varvn}). The correlation coefficient in Eq.~(\ref{eq: pearcorr pt-v2pt}) is a function of the transverse momentum, denoted by $q$ in order to distinguish from $[p_T]$ which is not a variable. The quantity $V_n(q)V_n(q)^\star$ denotes the differential harmonic flow in the transverse momentum bin $q$ in an event and is equivalent to $v_n(q)^2$ (Eq.~\ref{eq: diff flow using flow-vectors in same pT bin}). We write it explicitly in order to distinguish it from   $V_nV_n(q)^\star$ (Eq.~\ref{eq: diff flow using global and pT-dependent flow})  which will be discussed shortly.

\begin{figure}[ht!]
\hspace{-0.35 cm}\begin{subfigure}{0.5\textwidth}
\centering
\includegraphics[width= 6.5 cm]{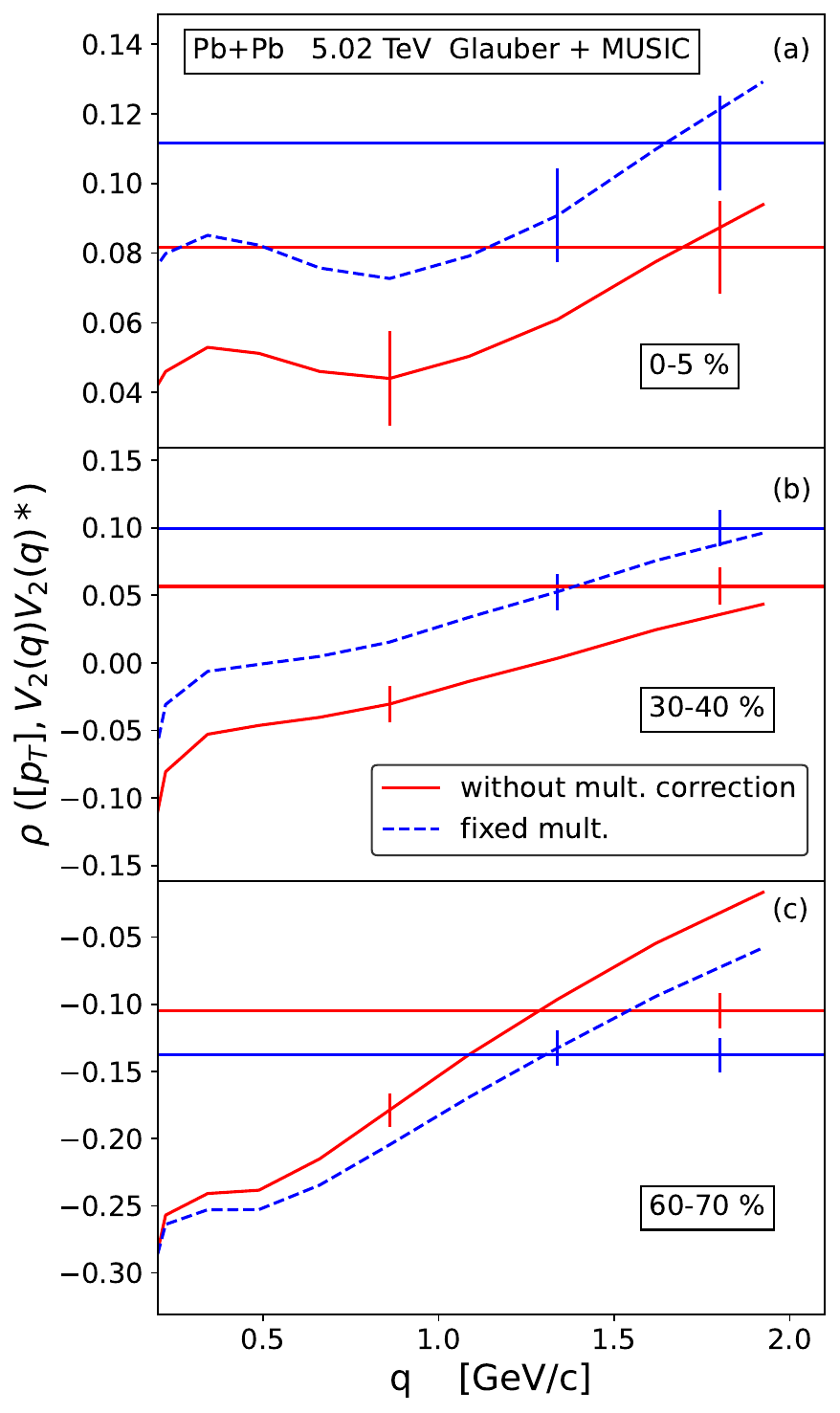}
\end{subfigure}~~~
\begin{subfigure}{0.5\textwidth}
\centering
\includegraphics[width= 6.5 cm]{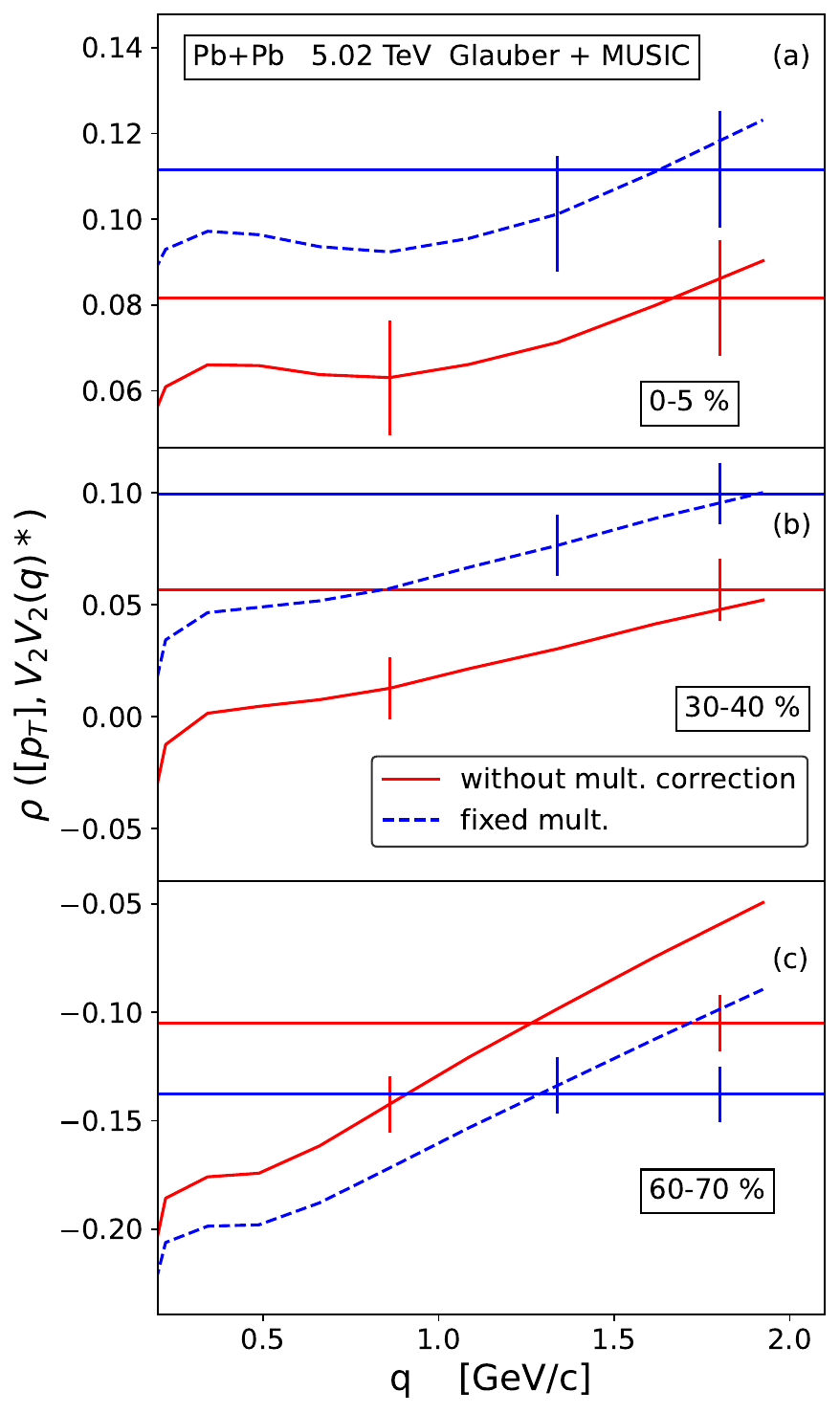}  
\end{subfigure}
\centering
\caption{Left: Momentum dependent Pearson correlation between mean transverse momentum per particle and elliptic flow, $\rho([p_T],V_2(q)V_2(q)^\star)$ in Pb+Pb collision at 5.02 TeV for three different centralities: 0-5 \%, 30-40 \% and 60-70 \%. The red solid lines denote the results obtained from hydrodynamic simulations with Galuber initial condition and the blue dashed lines represent the results corrected for multiplicity fluctuations. The horizontal lines denote the correlation coefficients between the momentum averaged flow, $\rho([p_T], v_2^2)$, serving as the baselines for the momentum dependent curves. Right: Same but for the other definition of the momentum dependent correlation $\rho([p_T],V_2V_2(q)^\star)$, where one of the flow harmonics is momentum averaged. The lines and symbols have same meaning as left plot. The figure is from the original publication~\cite{Samanta:2023rbn}, coauthored by the author.}
\label{fig: rho pt-v2pt}
\end{figure}

\begin{figure}[ht!]
\hspace{-0.35 cm}\begin{subfigure}{0.5\textwidth}
\centering
\includegraphics[width= 6.5 cm]{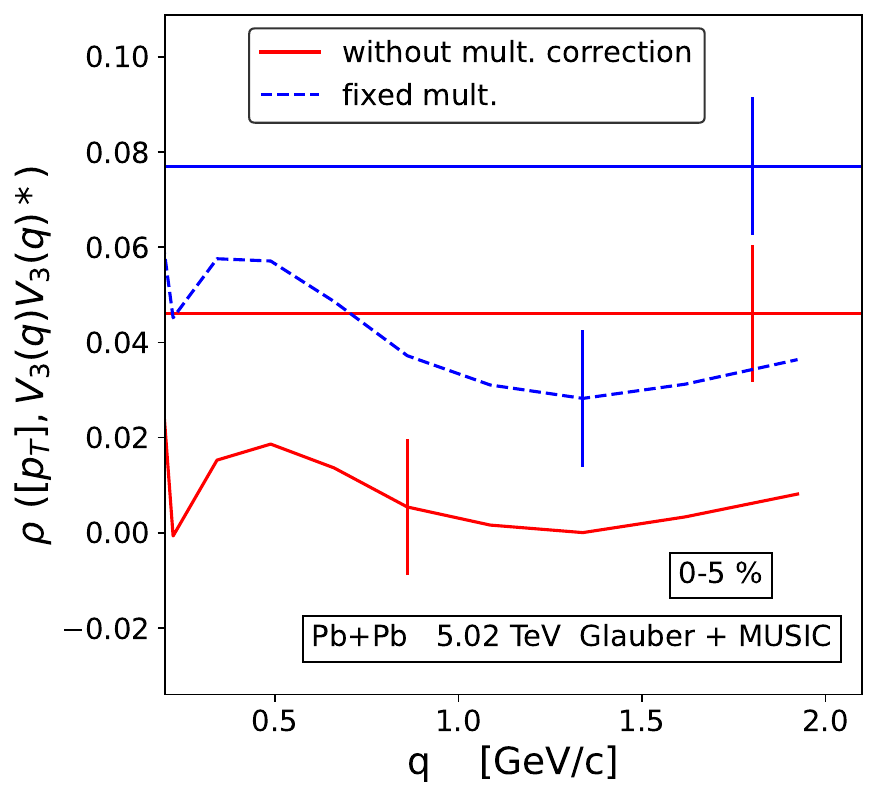}
\end{subfigure}~~~
\begin{subfigure}{0.5\textwidth}
\centering
\includegraphics[width= 6.5 cm]{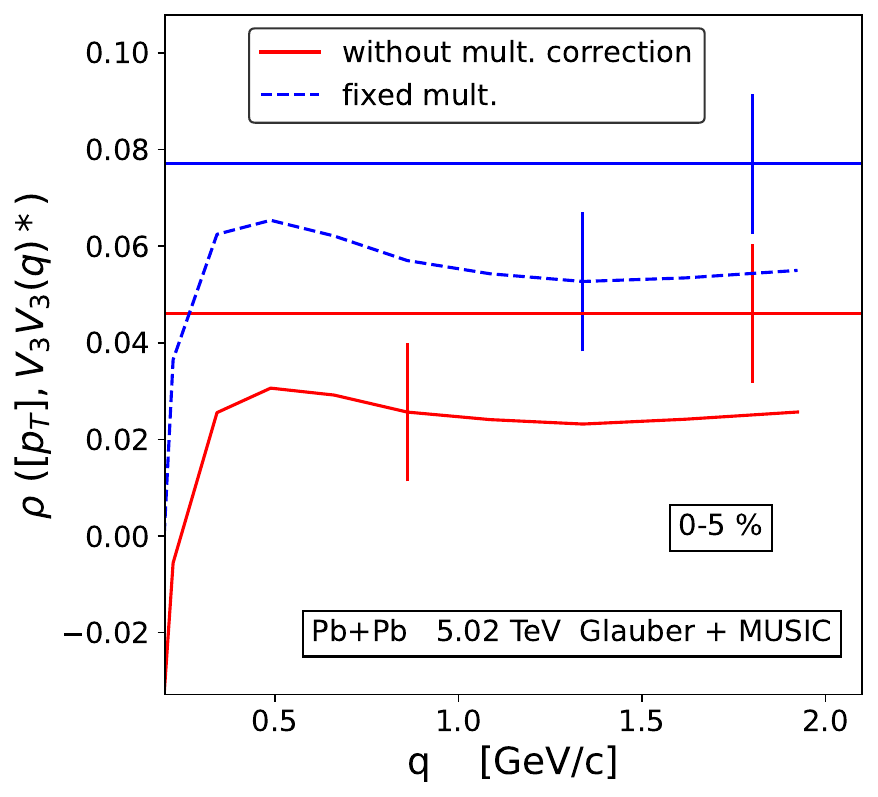}  
\end{subfigure}
\centering
\caption{Same as Fig.~\ref{fig: rho pt-v2pt} but for the triangular flow and only for 0-5 \% centrality. The figure is from the original publication~\cite{Samanta:2023rbn}, coauthored by the author.}
\label{fig: rho pt-v3pt}
\end{figure}

Figs.~\ref{fig: rho pt-v2pt} and \ref{fig: rho pt-v3pt} (left) show the results for the correlation coefficient $\rho([p_T],V_n(q)V_n(q)^\star)$ for the elliptic and the triangular flow. For the elliptic flow, the results are shown for three centralities: $0-5 \%$, $30-40 \%$ and $60-70 \%$, whereas for the triangular flow results are shown only for $0-5 \%$. In the figures, the hydrodynamic results are presented only up to $q=2$ GeV, where hydrodynamics is more applicable. However, it should be noted that the measurements at higher $q$ could be interesting to study non-flow effects, correlations originating from the color-glass condensate etc. Please also note, in order to maintain clarity, the error bars are shown on the figures for specific points to show statistical errors in our simulation results.
Fig.~\ref{fig: rho pt-v2pt} shows that for the elliptic flow,  the correlation coefficients $\rho([p_T],V_2(q)V_2(q)^\star)$ display strong dependence on the transverse momentum $q$ in all centralities. This momentum dependence can be attributed to experimentally observed dependence of the momentum independent correlation coefficient, $\rho([p_T],v_2^2)$, on the transverse momentum cuts
\cite{ATLAS:2019pvn}. Similar dependence on $q$ for triangular flow in Fig.~\ref{fig: rho pt-v3pt} is weak as compared to the elliptic flow.  The momentum dependent  coefficient, $\rho\left([p_T],V_n(q)V_n(q)^\star\right)$, measures the correlation between the mean transverse momentum per particle and the fraction of total (momentum averaged or integrated) harmonic flow at a definite transverse momentum $q$ in an event. Therefore, it does not depend on the specific shape of the $q$-dependence of the event averaged harmonic flow (Eq.~(\ref{eq: diff flow using flow-vectors in same pT bin})) given by $\langle v_n(q)^2\rangle$. The importance of this momentum dependent construction of the Pearson correlation coefficient is that it removes two significant limitations: dependence on transverse momentum cuts and $q$-dependence of harmonic flow (averaged over events). 

For completeness, in those figures we also present results for the momentum independent correlation coefficient $\rho([p_T],v_n^2)$, plotted as the horizontal solid lines, serving as the baselines for the momentum dependent coefficients. It is important to note that the correlation coefficient for the momentum averaged or integrated flow, $\rho([p_T],v_n^2)$, is not simply equal to the momentum average of the correlation coefficient for momentum dependent or differential flow $\rho\left([p_T],V_n(q)V_n(q)^\star\right)$. In particular, it turns out that the observed $q$-dependence of the correlation coefficient as seen in Figs.~\ref{fig: rho pt-v2pt} and \ref{fig: rho pt-v3pt} is due to its construction which involves the ratio of two average quantities, namely the covariance and the variance, which are individually momentum dependent. Later in this section, we discuss this in detail by directly comparing the two covariances, $Cov\left([p_T],V_n(q)V_n(q)^\star\right)$ and $Cov\left([p_T],v_n^2\right)$.

Like the momentum averaged case, the momentum dependent Pearson correlation coefficient between mean transverse momentum and harmonic flow is impacted by the multiplicity fluctuations in a specific centrality bin, as in our calculation we use wide bins of centrality. Therefore, we correct both quantities $[p_T]$ and $V_n(q)V_n(q)^\star$ for the multiplicity fluctuations in each centrality bin using Eq.~(\ref{eq: mult-correction}). In Figs.~\ref{fig: rho pt-v2pt} and \ref{fig: rho pt-v3pt}, we also show the results for the correlation coefficient after multiplicity-correction or at fixed multiplicity, denoted by the dashed lines. It is seen that the corresponding correction is numerically sizable and hence once again emphasizes the necessity of such corrections whenever studying similar correlation coefficients.  In the following parts, unless otherwise specified, we always use quantities corrected for multiplicity fluctuations.

\subsubsection{Experimental definition : $\rho([p_T],V_nV_n(q)^\star)$ }

The correlation coefficient in Eq.~(\ref{eq: pearcorr pt-v2pt}) is difficult to use in experiments. An alternative definition for the momentum dependent correlation coefficient between the mean transverse momentum and the harmonic flow can be,
\begin{equation}
  \rho\left([p_T],V_nV_n(q)^\star\right) = \frac{Cov\left([p_T],V_nV_n(q)^\star\right)}{\sqrt{Var\left([p_T]\right)Var\left(V_nV_n(q)^\star\right)}} \ ,
  \label{eq: pearcorr pt-v2pt expt}
\end{equation}
which is constructed in analogy to the experimental definition of transverse momentum dependent harmonic flow (Eq.~(\ref{eq: diff flow using global and pT-dependent flow})). The difference between the correlation coefficients in Eq.~(\ref{eq: pearcorr pt-v2pt}) and Eq.~(\ref{eq: pearcorr pt-v2pt expt}) is similar to the difference between Eq.~(\ref{eq: diff flow using flow-vectors in same pT bin}) and Eq.~(\ref{eq: diff flow using global and pT-dependent flow}). The above correlation coefficient $\rho([p_T],V_nV_n(q)^\star)$ is relatively easier to measure experimentally. The denominator of Eq.~(\ref{eq: pearcorr pt-v2pt expt}), $Var\left( V_n V_n(q)^\star\right)$ is a four particle correlator where only two of them are restricted to a particular transverse momentum bin, unlike  $Var\left( V_n(q) V_n(q)^\star\right)$, where all the four particles are from same bin hence more difficult to measure in experiment especially for larger value of $q$. However, the correlation coefficient in Eq.~(\ref{eq: pearcorr pt-v2pt expt}) does not have such a simple interpretation as the coefficient in Eq.~(\ref{eq: pearcorr pt-v2pt}). The results for $\rho([p_T],V_nV_n(q)^\star)$, from hydrodynamic simulations, are presented in the right panels of Figs.~\ref{fig: rho pt-v2pt} and \ref{fig: rho pt-v3pt}. The qualitative behavior of the transverse momentum dependence for the correlation coefficient remains similar to Eq.~(\ref{eq: pearcorr pt-v2pt}) for all centralities with, however, weaker dependence for $\rho\left([p_T],V_nV_n(q)^\star\right)$, in comparison to $\rho\left([p_T],V_n(q)V_n(q)^\star\right)$. This shows that even though we change the definition for the momentum dependent correlation coefficient for the sake of experimental measurement, there is no significant effect on the overall results.

\subsection{Constraining granularity in the initial state}
\label{granularity}
Granularity in the initial state is defined by the nucleon width $w$ and it is identified by the size of the region  where the nucleons deposit energy (entropy) at the time of the collision. If $w$ is small then the initial state is more granular and a larger $w$ indicates a less granular initial state. The information on $w$ is important because it defines the nucleon wave function which is considered as Gaussian as discussed in Chapter-2. It has been seen that the correlation coefficient between the mean transverse momentum and harmonic flow is sensitive to the nucleon width or the granularity of the initial state for the hydrodynamic evolution~\cite{Bozek:2016yoj,Giacalone:2021clp}. Therefore, we study the same thing for momentum dependent correlation coefficient between transverse momentum and harmonic flow, where granularity of initial state can have more significant effect. 
\begin{figure}[ht!]
\hspace{-0.35 cm}\begin{subfigure}{0.5\textwidth}
\centering
\includegraphics[width= 6.5 cm]{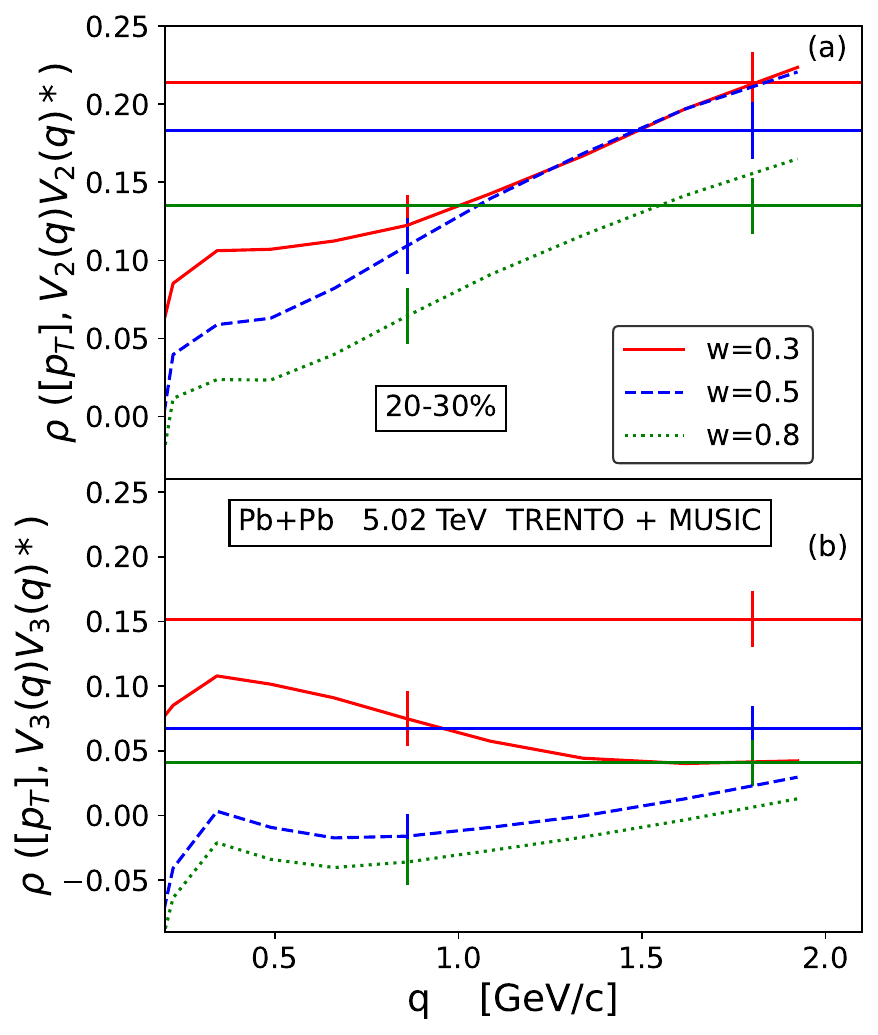}
\end{subfigure}~~~
\begin{subfigure}{0.5\textwidth}
\centering
\includegraphics[width= 6.5 cm]{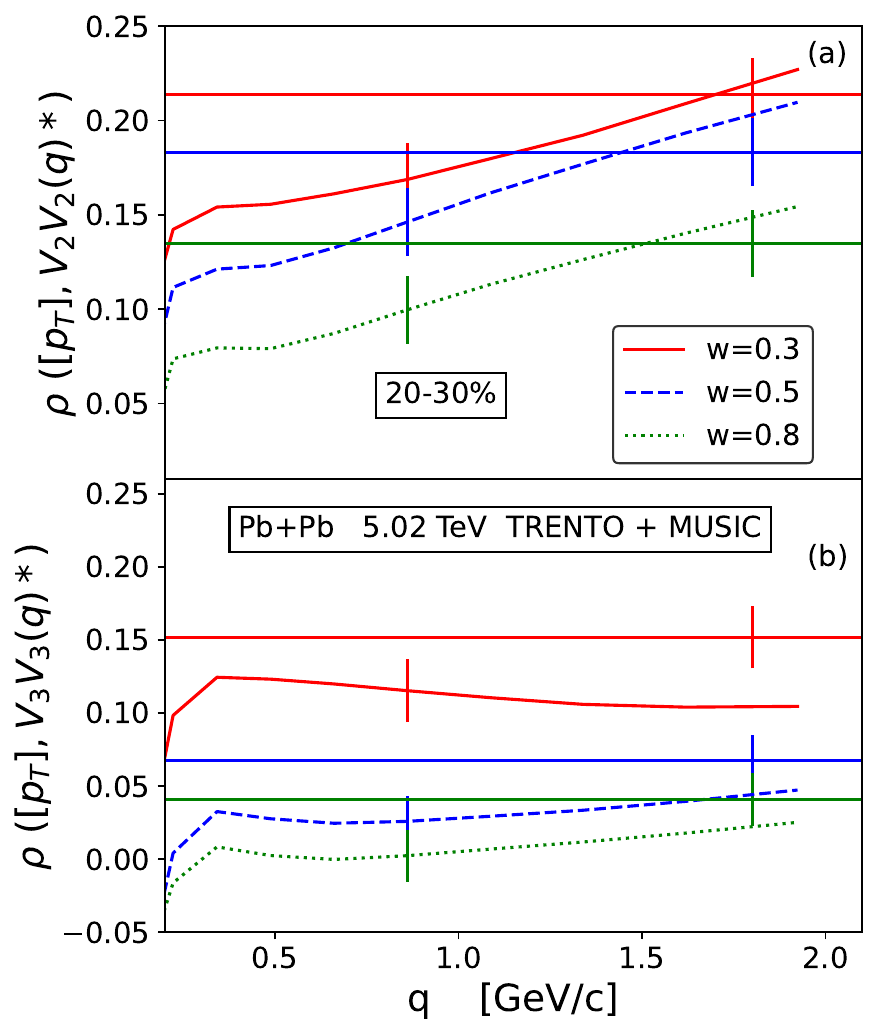}  
\end{subfigure}
\centering
\caption{Left: Pearson correlation coefficient $\rho([p_T], V_n(q)V_n(q)\star)$ between the mean transverse momentum per particle and momentum dependent elliptic flow (a) and triangular flow (b) obtained with TRENTO initial condition in Pb+Pb collision at 5.02 TeV with 20-30 \% centrality for different granularity of the initial state: $w=0.3,0.5$ fm and $0.8 $ fm denoted by red, blue and green colors respectively. The horizontal lines denote the baselines corresponding to the correlation between the momentum averaged flow. Right: Same but for the other definition of momentum dependent correlation : $\rho([p_T], V_nV_n(q)\star)$. The figure is from the original publication~\cite{Samanta:2023rbn}, coauthored by the author.}
\label{fig: rho pt-vnpt nucleon width comparison}
\end{figure}
In models, the granularity of the initial state can be modified by changing the nucleon size, defined by the wave function of the nucleon. In our analysis, we use TRENTO model to obtain the initial state for this particular study, where the granularity is changed by changing the 2D Gaussian width $w$ (described in \ref{trento}) associated to each nucleon in TRENTO. Experimental results suggest that the size of the region where the nucleon deposit energy is small, which corresponds to an initial state with high granularity~\cite{Giacalone:2021clp}. In our analysis, we study this size dependence by changing the nucleon width in TRENTO with three different values:  $w=0.3$, $0.5$, and $0.8$ fm.

Fig.~\ref{fig: rho pt-vnpt nucleon width comparison} (left) shows the results for the momentum dependent correlation coefficient  $\rho([p_T],V_n(q)V_n(q)^\star)$, for $20$-$30$\% centrality for different nucleon widths. For the elliptic flow, the correlation coefficient shows a strong dependence on the transverse momentum $q$ for all granularity. However, the increase of the correlation coefficient with $q$ is less steep for the initial state with a more granularity i.e. with smaller $w$. The effect becomes dominant for the triangular flow (panel (b). For triangular flow, the correlation even decreases with $q$ for most granular state having $w=0.3$ fm. Specifically, the correlation coefficient shows different patterns of transverse momentum dependence within the range: $q=0$-$1.5$ GeV for different $w$. The momentum independent correlation coefficients  (baselines), on the other hand, follow a particular dependence on the granularity; the correlation decreases as $w$ increases or the granularity decreases.   

The correlation coefficients,  $\rho([p_T],V_nV_n(q)^\star)$, is shown on the right panel of Fig.~\ref{fig: rho pt-vnpt nucleon width comparison}. The coefficients show a quite similar behavior for the transverse momentum dependence as $\rho([p_T],V_n(q)V_n(q)^\star)$. Again for the triangular flow, the initial state with higher granularity shows less steep dependence on $q$ and furthermore for this correlation coefficient the difference between the initial states with different $w$ is the strongest in the range $q=0$-$1.5$ GeV. Therefore, it would be interesting to verify our model predictions in experiments, particularly for these correlation coefficients $\rho([p_T],V_n(q)V_n(q)^\star)$ or  $\rho([p_T],V_nV_n(q)^\star)$, which will put precise constraints on the parameters of the initial state in the  hydrodynamic modeling of heavy-ion collisions~\cite{Giacalone:2021clp}. A comparison of the results in Fig.~\ref{fig: rho pt-vnpt nucleon width comparison} indicates that the momentum dependent correlation coefficient $\rho([p_T],V_nV_n(q)^\star)$ serves as a better candidate to probe the granularity whereas its experimental measurement is easier than $\rho([p_T],V_n(q)V_n(q)^\star)$. We will again come to the effect of granularity later while discussing the momentum dependent covariances.

\begin{figure}[ht!]
\hspace{-0.35 cm}\begin{subfigure}{0.5\textwidth}
\centering
\includegraphics[width= 6.5 cm]{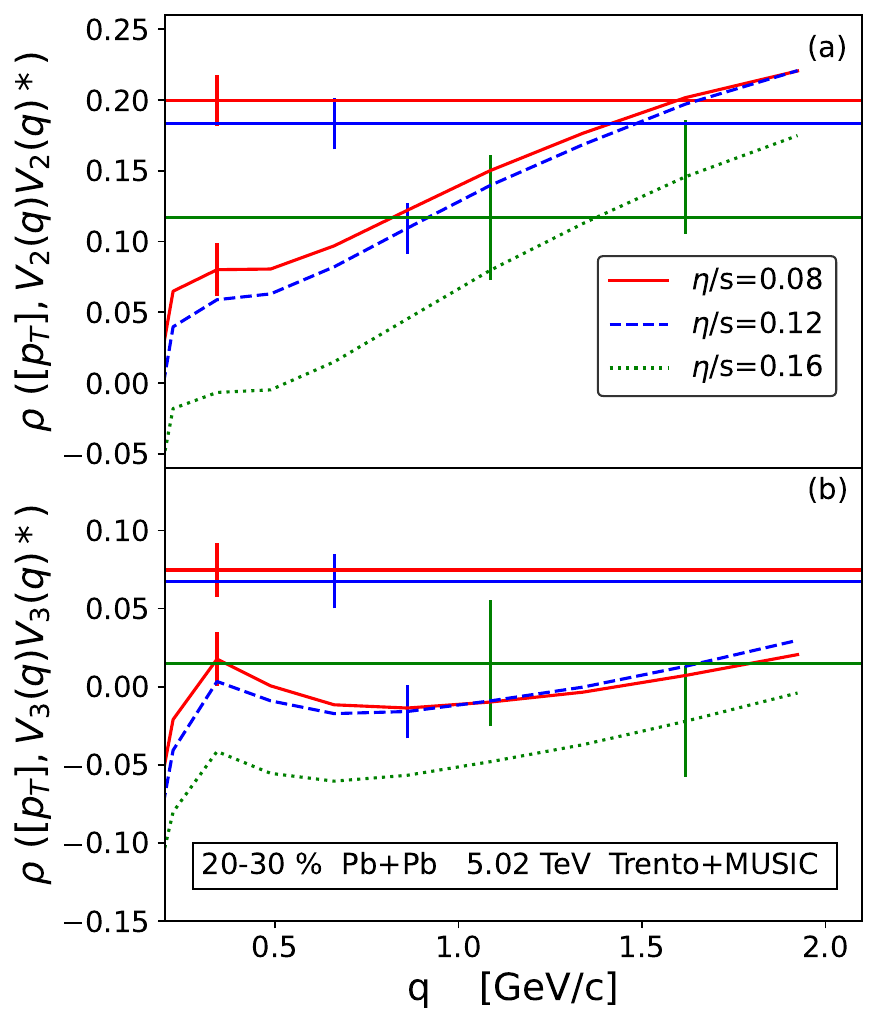}
\end{subfigure}~~~
\begin{subfigure}{0.5\textwidth}
\centering
\includegraphics[width= 6.5 cm]{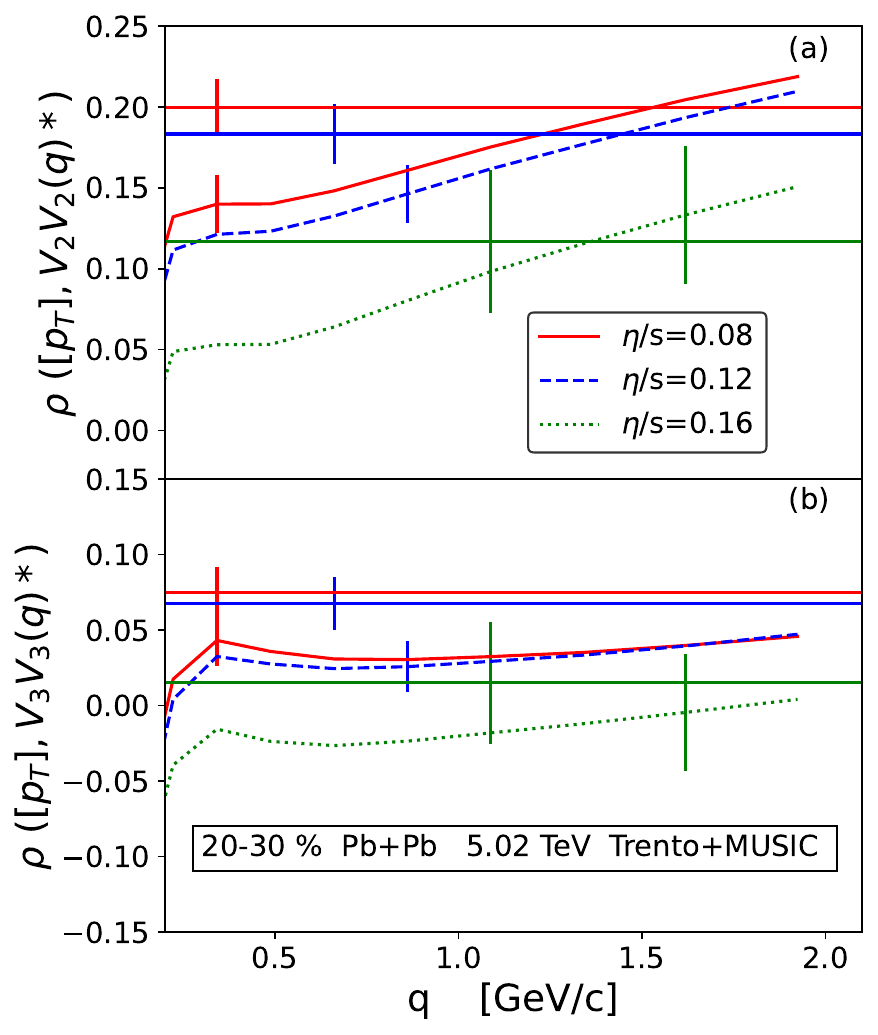}  
\end{subfigure}
\centering
\caption{Same as Fig.~\ref{fig: rho pt-vnpt nucleon width comparison} but for the comparison of three different values of shear viscosity to entropy density ratio $\eta/s$: 0.08, 0.12, 0.16 denoted by red, blue and green colors respectively. The figure is from the original publication~\cite{Samanta:2023rbn}, coauthored by the author.}
\label{fig: rho pt-vnpt shear viscosity comparison}
\end{figure}

\subsubsection{Viscosity dependence}

In order to present a comprehensive picture, discussing the dependence of this momentum dependent correlation coefficient on medium properties such as shear or bulk viscosity remain indispensable. In Fig.~\ref{fig: rho pt-vnpt shear viscosity comparison}, the correlation coefficients for three different values of the shear viscosity to entropy density ratio $\eta/s$ are presented for both definitions and for both elliptic and triangular flow. The qualitative nature of the transverse momentum dependence of the correlation coefficient remain similar for different values of shear viscosity. The dependence on the shear viscosity is rather much weaker in comparison to the granularity. The change in shear viscosity causes a overall shift of the curves, without altering the specific shape for the momentum dependence. The shift is, however, much smaller in magnitude; there is practically not much difference between $\eta/s=0.08$ and $\eta/s=0.12$ for both momentum dependent correlation coefficients and their baselines. Therefore, the momentum dependent correlation coefficient is not specifically sensitive to shear viscosity according to Fig.~(\ref{fig: rho pt-vnpt shear viscosity comparison}); the shape of the curves are similar for different values of shear viscosity. The correlation coefficient between mean transverse momentum and the harmonic flow might not be an ideal candidate to probe the viscosity of the QGP medium. We have checked that the momentum dependence on the bulk viscosity of the considered correlation coefficients is similar as the dependence on shear viscosity.

\subsection{Addressing experimental challenges: Alternative definitions}
\label{alternative pt-dependent correlators}

The experimental measurement of the momentum dependent correlation coefficient, $\rho([p_T],$ $V_n(q)V_n(q)^\star)$ is difficult for low statistics at large $q$. The estimation of $\rho([p_T],V_nV_n(q)^\star)$ is relatively easier but still sufficiently difficult in comparison to the momentum independent correlation coefficient, $\rho([p_T],V_nV_n^\star)$. The main experimental difficulty lies in estimating the  variances in the denominators i.e. $Var\left( V_n(q)V_n(q)^\star \right)$ or $Var\left( V_nV_n(q)^\star \right)$. These quantities require the measurement of a four (or two) particle correlator in the same transverse momentum bin, which turn out challenging for the bins with large $q$ due to limited statistics. Below we propose alternate and approximated choice for the momentum dependent correlation coefficients that could be easily used in experiments.
\begin{figure}[ht!]
\hspace{-0.35 cm}\begin{subfigure}{0.5\textwidth}
\centering
\includegraphics[width= 6.5 cm]{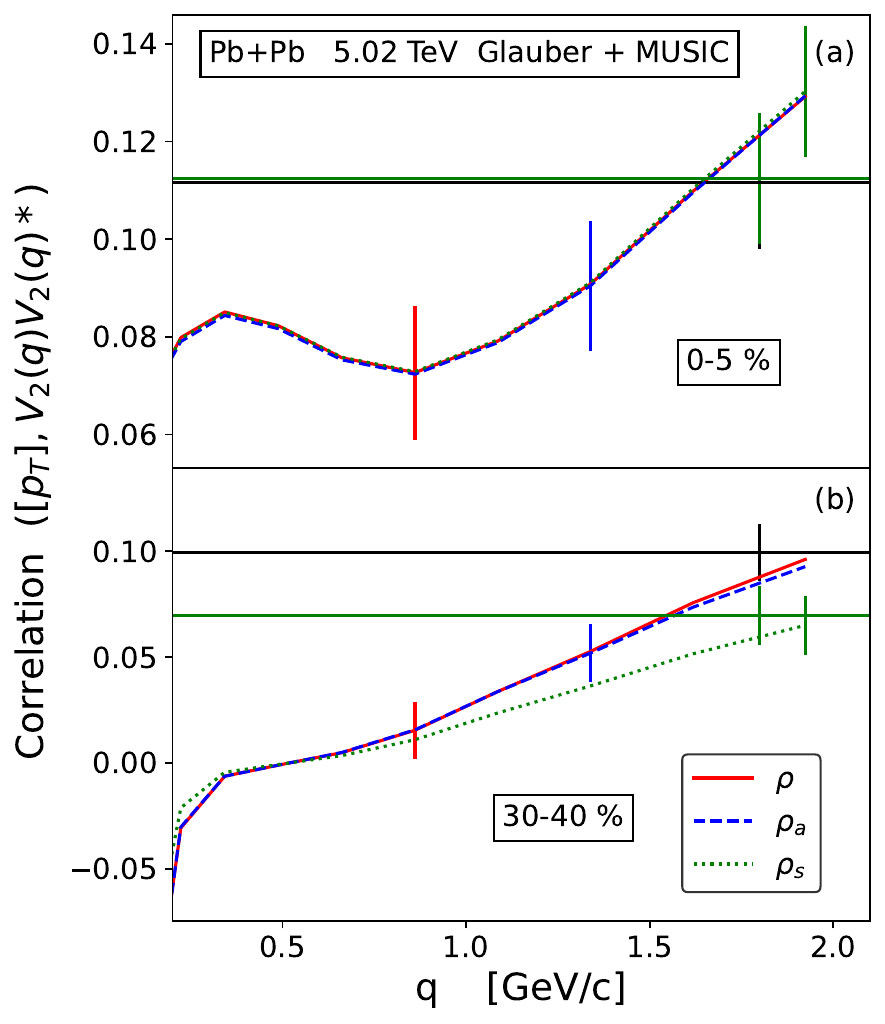}
\end{subfigure}~~~
\begin{subfigure}{0.5\textwidth}
\centering
\includegraphics[width= 6.5 cm]{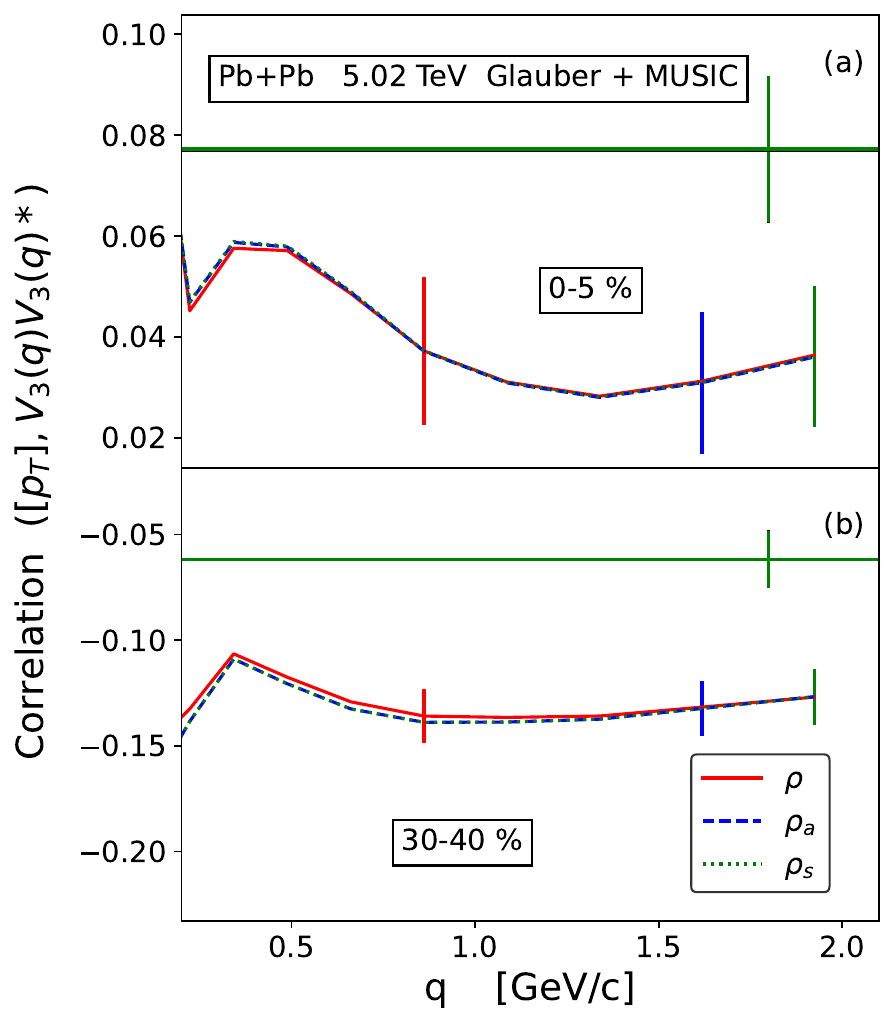}  
\end{subfigure}
\centering
\caption{Momentum dependent correlation coefficient between $[p_T]$ and $V_n(q)V_n(q)^\star$ for 0-5 \% (panel (a) ) and 30-40 \% (panel (b)) centrality in Pb+Pb collision at 5.02 TeV for the elliptic (left) and triangular (right) flow coefficient. The red lines denote the original definition in Eq.~(\ref{eq: pearcorr pt-v2pt}). The blue dashed lines denote the approximated definition for the correlation in Eq.~(\ref{eq: rho_a pt-v2pt}) and the green dotted line denotes the results obtained with the scaled correlation coefficient in Eq.~(\ref{eq: rho_s pt-v2pt}). The figure is from the original publication~\cite{Samanta:2023rbn}, coauthored by the author.}
\label{fig: approximated corr pt-v2pt}
\end{figure}

One of the possibilities would be to use the momentum averaged or integrated variance, $Var\left( v_n^2 \right)$, in the denominator within the square root of the correlation coefficient, because it would be much easier to estimate a four particle correlator in the full transverse momentum acceptance. But then at the same time we need to divide the new construction with $\frac{\langle V_n(q)V_n(q)^\star\rangle}{\langle v_n^2\rangle}$ (or $\frac{\langle V_n(q)V_n^\star\rangle}{\langle v_n^2\rangle}$) in order to properly retain the momentum dependence. With this, the new modified approximate formula for the momentum dependent correlation coefficients would read, 
\begin{equation}
  \rho_a\left([p_T],V_n(q)V_n(q)^\star\right) =
  \frac{Cov\left([p_T],V_n(q)V_n(q)^\star\right) \langle v_n^2\rangle}{\sqrt{Var\left([p_T]\right)Var\left( v_n^2\right)} \langle V_n(q)V_n(q)^\star \rangle }
  \label{eq: rho_a pt-v2pt}
\end{equation}
and
\begin{equation}
  \rho_a\left([p_T],V_nV_n(q)^\star\right) =
  \frac{Cov\left([p_T],V_nV_n(q)^\star\right) \langle v_n^2\rangle}{\sqrt{Var\left([p_T]\right)Var\left( v_n^2\right)} \langle V_nV_n(q)^\star \rangle } \ .
\label{eq: rho_a pt-v2pt expt}
\end{equation}
The approximate formulae in Eqs.~(\ref{eq: rho_a pt-v2pt}) and (\ref{eq: rho_a pt-v2pt expt}) are expected to reproduce very close results as of the original momentum dependent correlation coefficients in Eqs.~(\ref{eq: pearcorr pt-v2pt}) and (\ref{eq: pearcorr pt-v2pt expt}), because if we take the ratios of the new formulae to the original ones then the factors
\begin{equation}
  \frac{ \sqrt{Var\left(V_n(q)V_n(q)^\star\right)}\langle v_n^2\rangle}{\sqrt{Var\left( v_n^2\right)} \langle V_n(q) V_n(q)^\star \rangle}
  \label{eq: factor rho_a}
\end{equation}
and
\begin{equation}
  \frac{\sqrt{Var\left( V_nV_n(q)^\star\right)} \langle v_n^2\rangle}{\sqrt{Var\left( v_n^2\right)} \langle V_n V_n(q)^\star \rangle}
  \label{eq: factor rho_a expt}
\end{equation}
are consistent with $1$ as discussed in Chapter-3~\cite{Bozek:2021mov}.

\begin{figure}[ht!]
\hspace{-0.35 cm}\begin{subfigure}{0.5\textwidth}
\centering
\includegraphics[width= 6.5 cm]{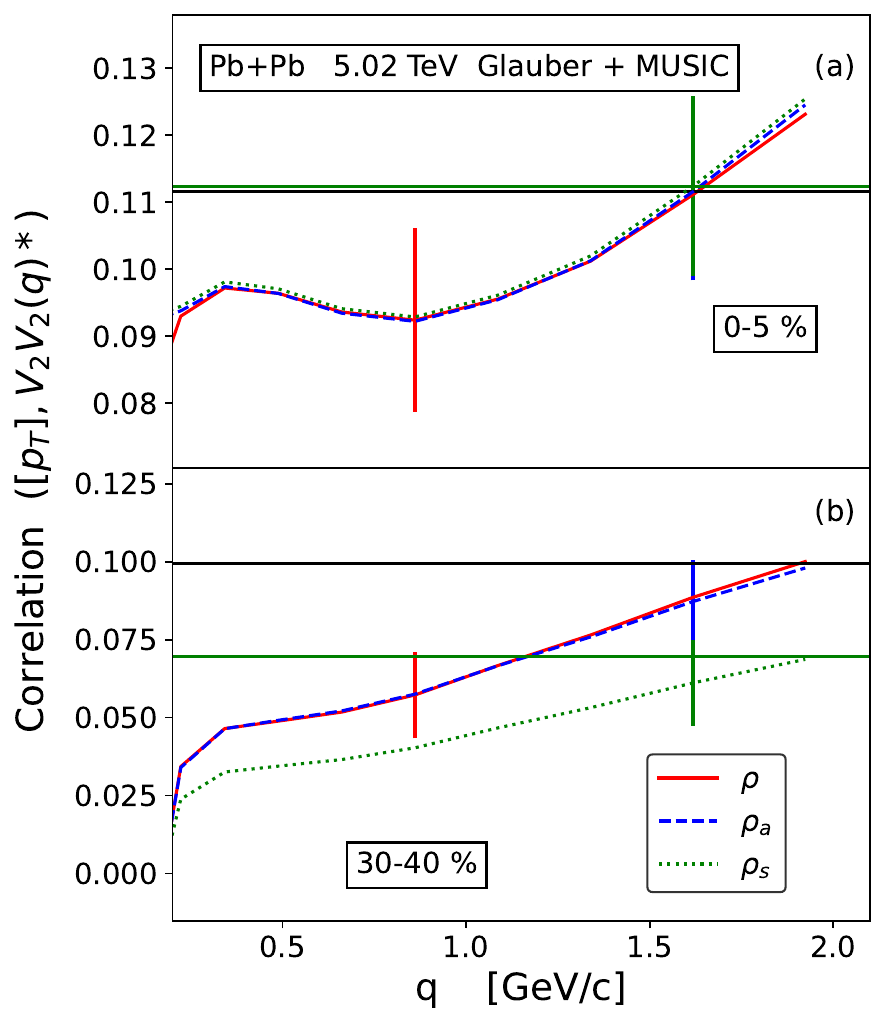}
\end{subfigure}~~~
\begin{subfigure}{0.5\textwidth}
\centering
\includegraphics[width= 6.5 cm]{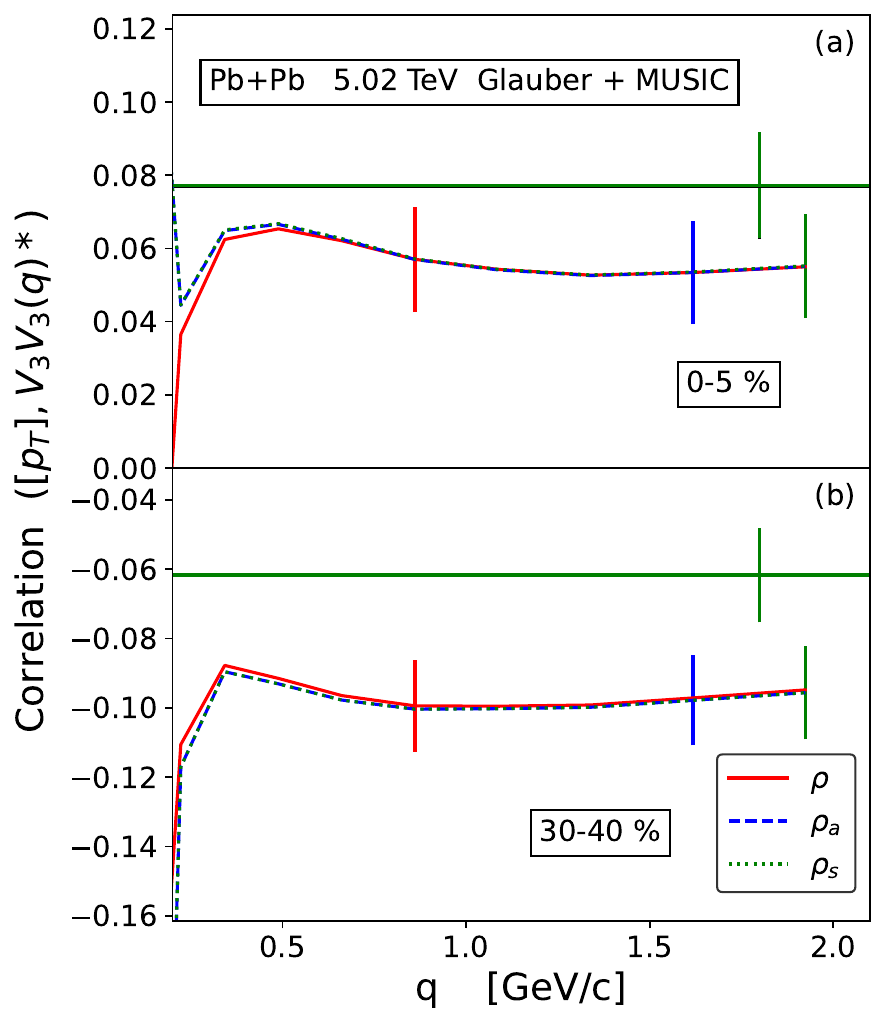}  
\end{subfigure}
\centering
\caption{Same as Fig.~\ref{fig: approximated corr pt-v2pt} but for the correlation coefficient between $[p_T]$ and $V_nV_n(q)^\star$. The figure is from the original publication~\cite{Samanta:2023rbn}, coauthored by the author.}
\label{fig: approximated corr pt-v2pt expt}
\end{figure}
Another possibility to re-construct the correlation coefficients would be to scale the covariance between the mean transverse momentum and the harmonic flow sitting at the numerator of those expressions, by the average of momentum dependent harmonic flow squared instead of the standard deviation. Such modifications are expected to rescale the magnitudes of the correlation coefficient without changing its specific momentum dependence. The formulae for the scaled correlation coefficients are given by,
\begin{equation}
   \rho_s\left([p_T],V_n(q)V_n(q)^\star\right) = 
   \frac{Cov\left([p_T],V_n(q)V_n(q)^\star\right) }{\sqrt{Var\left([p_T]\right)}\langle V_n(q)V_n(q)^\star\rangle)} 
   \label{eq: rho_s pt-v2pt}
\end{equation}
and
\begin{equation}
   \rho_s\left([p_T],V_nV_n(q)^\star\right) = 
   \frac{Cov\left([p_T],V_nV_n(q)^\star\right) }{\sqrt{Var\left([p_T]\right)} \langle V_nV_n(q)^\star\rangle }
    \ .
 \label{eq: rho_s pt-v2pt expt}
\end{equation}
The scaled correlation coefficients, $\rho_s$, are expected to appear as a good approximation of the original correlation coefficients $\rho$ when the harmonic flow is dominated by fluctuations i.e when $ \sqrt{Var\left(V_n(q)V_n(q)^\star\right)} \simeq \langle V_n(q) V_n(q)^\star \rangle$  and  $ \sqrt{Var\left(V_nV_n(q)^\star\right)} \simeq \langle V_n V_n(q)^\star \rangle$. According to the discussions in Chapter-3, this happens in case of elliptic flow in central collision and triangular flow in all centralities. For completeness, the momentum independent version of the scaled correlation coefficient would be,
\begin{equation}
   \rho_s\left([p_T],v_2^2\right) = 
   \frac{Cov\left([p_T],v_2^2\right) }{\sqrt{Var\left([p_T]\right)} \langle v_2^2\rangle }
    \ ,
\label{eq: rho_s pt-v2}
\end{equation}
which is almost same as the normalized symmetric cumulant between the mean transverse momentum $[p_T]$ and the harmonic flow $v_n^2$ as discussed earlier. Moreover, STAR collaboration has used Eq.~(\ref{eq: rho_s pt-v2}) for studying the nuclear deformation in relativistic Au+Au and U+U collisions systems~\cite{JJcph}.

Figs.~\ref{fig: approximated corr pt-v2pt} and \ref{fig: approximated corr pt-v2pt expt} show the comparison between the original momentum dependent correlation coefficient and the approximated definitions for the momentum dependent correlation coefficient $\rho\left([p_T],V_n(q)V_n(q)^\star\right)$ and 
$\rho\left([p_T],V_nV_n(q)^\star\right)$ respectively for $0-5\%$ and $30-40 \%$ centralities and for the elliptic and triangular flow. It could be seen that almost in every cases, within the range $q<2$ GeV, the results with the approximated  expressions coincide with the original formula for the correlation coefficient. In particular, the correlation coefficient $\rho_a$ is very close to the correlation coefficient $\rho$ in all cases and therefore could serve as a good experimental estimate of the momentum dependent correlation between mean transverse momentum and harmonic flow. The scaled correlation coefficient $\rho_s$ also provides equally good approximation except for the elliptic flow in peripheral collisions where fluctuations have less importance than the geometry. Therefore, $\rho_s$ can be used in experiment for the triangular flow and for the elliptic flow in central collisions where fluctuations dominated harmonic flow exists. It should be noted that all of these approximated formulae represent well-defined measures for the momentum dependent correlation coefficients between $[p_T]$ and momentum dependent flow harmonics, and they could be measured in the experiments and compared to the models, although they are not exactly the Pearson correlation coefficients.

\subsection{Scaled covariance}
\label{scaled cov}

As an extension of our study, we also study the momentum dependence of the covariance between the mean transverse momentum and momentum dependent harmonic flow. The covariance appears at the numerator of the momentum dependent correlation coefficients, $\rho\left([p_T],V_n(q)V_n(q)^\star\right)$ or  $\rho\left([p_T],V_nV_n(q)^\star\right)$, and significantly contribute to their strong dependence on the transverse momentum $q$. 
\begin{figure}[ht!]
\includegraphics[width=7.5 cm]{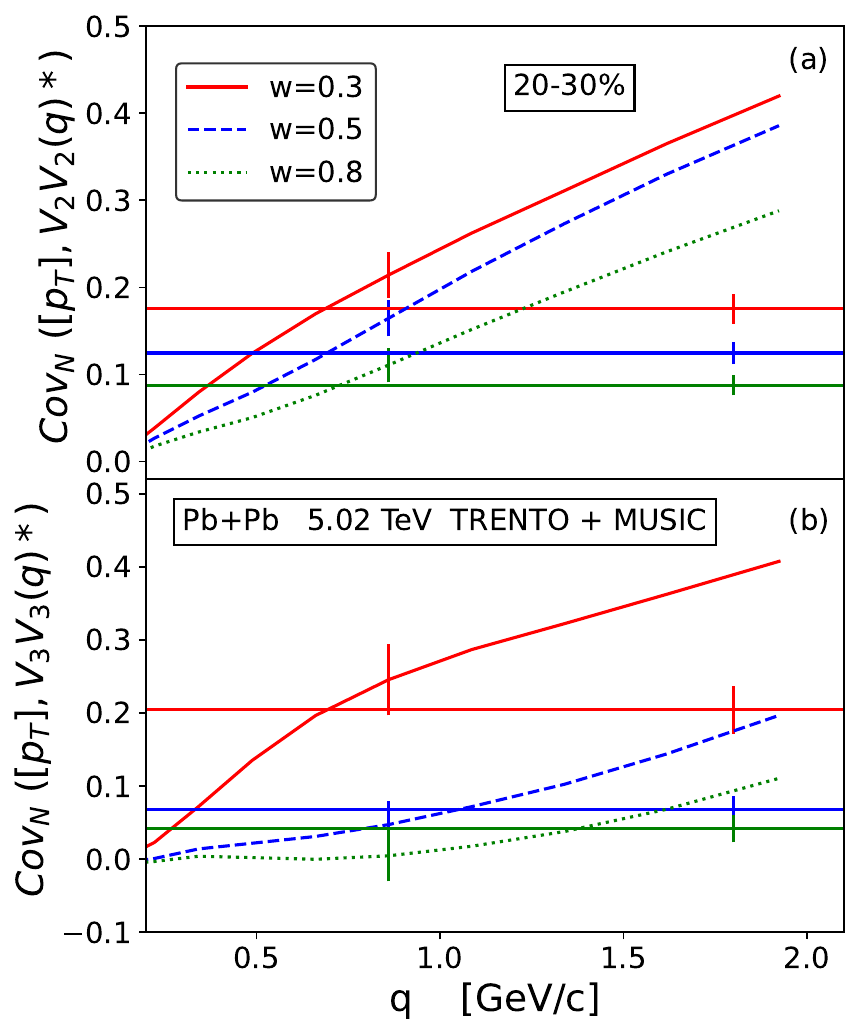}  
\centering
\caption{Normalized covariance between mean transverse momentum $[p_T]$ and momentum dependent harmonic flow $V_nV_n(q)^\star$ for 20-30 \% centrality in Pb+Pb collision at 5.02 TeV with TRENTO initial condition for three different values of $w$: 0.3, 0.5 and 0.8 fm denoted by red, blue and green colors respectively. Panel (a) and (b) represent the results for the elliptic and triangular flow respectively. The figure is from the original publication~\cite{Samanta:2023rbn}, coauthored by the author.}
\label{fig: norm cov}
\end{figure}
Therefore, it would be interesting to look directly at the momentum dependent covariances between the mean transverse momentum and the harmonic flow, which itself is an as good observable as the correlation coefficient and remove the momentum dependence of the denominator. But of course we have to normalize the covariance properly. In particular, momentum dependent normalized covariance can be defined as,
  \begin{equation}Cov_N\left([p_T],V_n(q)V_n(q)^\star\right)=\frac{Cov\left([p_T],V_n(q)V_n(q)^\star\right) }{\sqrt{Var([p_T])} \langle v_n^2 \rangle}  \ .
  \label{eq: norm cov pt-v2pt}
  \end{equation}
  and 
 \begin{equation}
  Cov_N\left([p_T],V_nV_n(q)^\star\right)=\frac{Cov\left([p_T],V_nV_n(q)^\star\right) }{\sqrt{Var([p_T])} \langle v_n^2 \rangle}  \ ,
  \label{eq: norm cov pt-v2pt expt}
  \end{equation}
where the normalization is chosen in a similar way as for the normalized symmetric cumulants between the mean transverse momentum and the harmonic flow coefficients discussed earlier. It is interesting to note that the baselines for both $Cov_N\left([p_T],V_n(q)V_n(q)^\star\right)$ and $Cov_N\left([p_T],V_nV_n(q)^\star\right)$ are given by $\rho_s\left([p_T],v_2^2\right)$ given in Eq~(\ref{eq: rho_s pt-v2}). Such normalized covariances can be useful to constrain the granularity of the initial state in  a better way, in the sense that it is relatively easier to measure in experiment in comparison to the full correlation coefficient. It can also pick up the robust sensitivity of such momentum dependent observable between $[p_T]$ and harmonic flow, which does not get partially washed out by the similar dependence in the denominator.  

Fig.~\ref{fig: norm cov} shows the normalized covariance in Eqs.~(\ref{eq: norm cov pt-v2pt}) and (\ref{eq: norm cov pt-v2pt expt}) for the elliptic and triangular flow in $20$-$30$\% centrality. It is clearly visible that the normalized covariance between the mean transverse momentum and the momentum dependent harmonic flow shows a strong dependence on the transverse momentum $q$ and a remarkable sensitivity to the granularity of initial state or the nucleon width $w$, showing a significant separation between three cases. This occur due to the absence of momentum dependence in the denominator. The effect is particularly pronounced for the triangular flow, where a steepest dependence on $q$ is observed for $w=0.3$ fm in the range of transverse momentum $q<1$ GeV, along with a striking difference from the other two cases. Therefore, the normalized covariance in Eqs.~(\ref{eq: norm cov pt-v2pt}) and (\ref{eq: norm cov pt-v2pt expt}) could serve as an ideal candidate to constrain the granularity in the initial state of heavy-ion collision upon successful verification of our results in the experiments. 

\begin{figure}[ht!]
\hspace{-0.35 cm}\begin{subfigure}{0.5\textwidth}
\centering
\includegraphics[width= 6.5 cm]{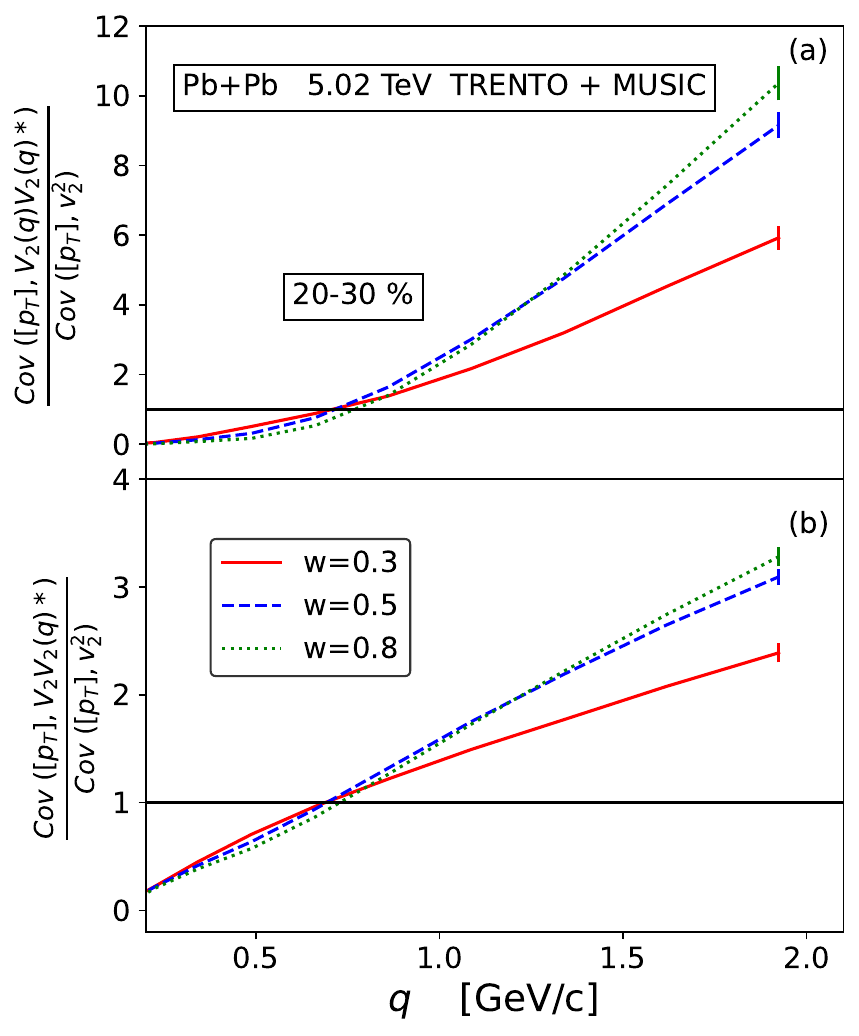}
\end{subfigure}~~~
\begin{subfigure}{0.5\textwidth}
\centering
\includegraphics[width= 6.5 cm]{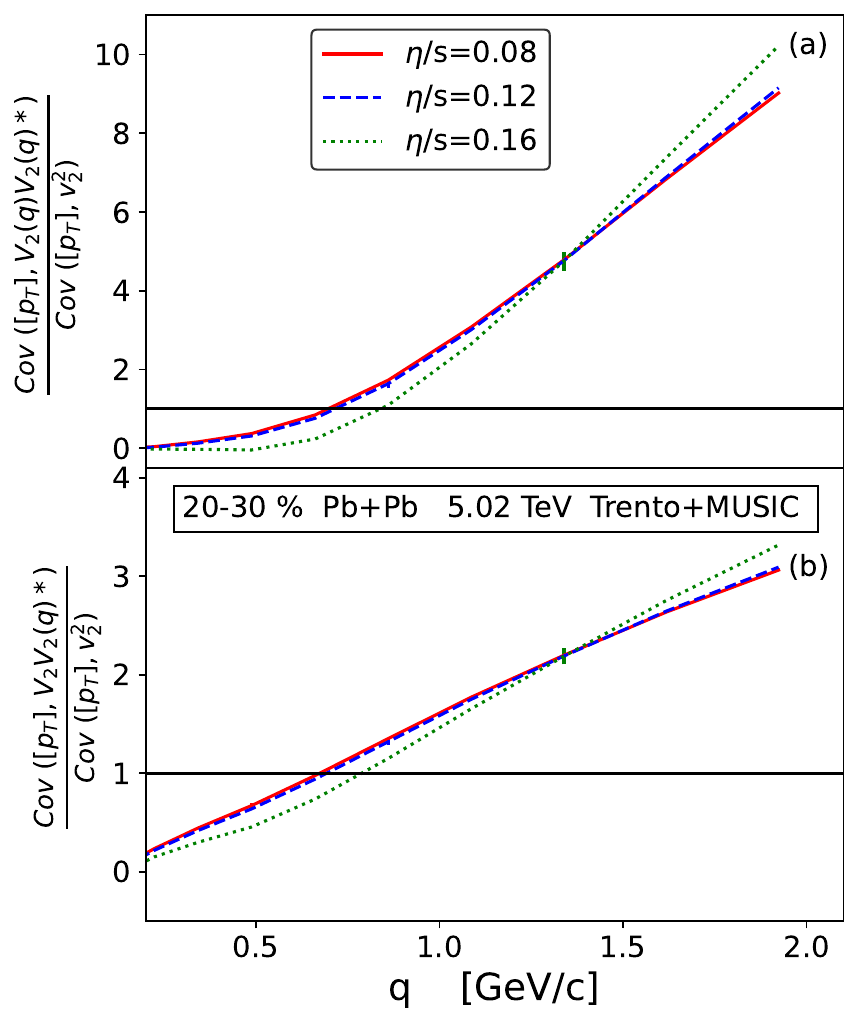}  
\end{subfigure}
\centering
\caption{Left: Ratio of the momentum dependent and momentum independent covariance between mean transverse momentum per particle and harmonic flow coefficients for 20-30 \% centrality in Pb+Pb collision at 5.02 TeV for three different values for granularity in the initial state : $w$= 0.3 (red), 0.5 (blue) and 0.8 (green) fm. In the upper and lower panel, the numerators of the ratios are with momentum dependent flow $V_n(q)V_n(q)^\star$ and $V_nV_n(q)^\star$ respectively. Right: Same as left plot but for three different values of shear viscosity to entropy ratio $\eta/s$: 0.08 (red), 0.12 (blue) and 0.16 (green). The figure is from the original publication~\cite{Samanta:2023rbn}, coauthored by the author.}
\label{fig: cov ratio}
\end{figure}

An alternative way to look into the momentum dependence of the covariance would be directly studying the ratio of momentum dependent and momentum averaged (independent) covariance, given by,
\begin{equation}
  \frac{Cov\left([p_T],V_n(q)V_n(q)^\star\right)}{Cov\left([p_T],V_nV_n^\star\right)} 
  \label{eq: cov ratio pt-v2pt}
\end{equation}
and
\begin{equation}
  \frac{Cov\left([p_T],V_nV_n(q)^\star\right)}{Cov\left([p_T],V_nV_n^\star\right)} \ .
  \label{eq: cov ratio pt-v2pt expt}
 \end{equation}
 The construction of such observables is possible whenever the denominator is not close to zero otherwise it will make the ratio to diverge. In Fig.~\ref{fig: cov ratio} (left), the  simulation results for the covariance ratios in  Eqs.~(\ref{eq: cov ratio pt-v2pt}) and (\ref{eq: cov ratio pt-v2pt expt}) are shown  for three different values of granularity parameter $w$ in collisions with 20-30 \% centrality. The covariance ratios show even more spectacular transverse momentum dependence. For both of the covariance ratios, all lines cross the baseline of $1$, at the average transverse momentum $q\simeq \langle [p_T] \rangle$. Then the lines split at higher momenta depending on the granularity of the initial state given by $w$ and especially showing a remarkable difference for $q \simeq 1$-$2$ GeV. Thus the covariance ratio also shows outstanding sensitivity on the granularity and could be measured in experiment to constrain $w$ with great precision.

To complete the picture, the right panel of Fig.~\ref{fig: cov ratio} shows the shear viscosity dependence of the covariance ratios for different values of $\eta/s$ in the same centrality. The results suggest that the momentum dependence of the covariance ratio has a weak dependence on the value of shear viscosity. The particular shapes of the momentum dependence of such observable could be used as an additional constraint on the value of shear viscosity in Bayesian analysis of model simulation and experimental data \cite{Bernhard:2016tnd,JETSCAPE:2020mzn,Nijs:2020roc}.

%*******************************************************************************
%****************************** Sixth Chapter **********************************
%*******************************************************************************
\chapter{Nuclear deformation through heavy-ion collisions}

% **************************** Define Graphics Path **************************
\ifpdf
    \graphicspath{{Chapter6/Figs/Raster/}{Chapter6/Figs/PDF/}{Chapter6/Figs/}}
\else
    \graphicspath{{Chapter6/Figs/Vector/}{Chapter6/Figs/}}
\fi

Nuclear structure is primarily studied in low energy theoretical and experimental nuclear physics. As we have seen that the information on the nuclear density profile and the Woods-Saxon parameters are particularly important to sample nucleon positions in a nucleus. Such information on nuclear structure can be obtained from low energy electron scattering experiments. So far in our discussions, we were limited to the collision of spherical nuclei i.e. $^{208}$Pb+$^{208}$Pb. However, the majority of atomic nuclei in their ground state are not spherical and posses an intrinsic deformation having in the leading order an axially quadrupole or ellipsoidal structure. Such deformations result in non-zero electric multipole moments of the nucleus with respect to the nuclear wave function. The deformation is quantified by a dimensionless deformation parameter $\beta_n$~\cite{Cline:1986ik, Raman:2001nnq, Shou:2014eya}. Examples of such deformed nuclei are uranium ($^{238}$U), xenon ($^{129}$Xe) etc. which posses a significant non-zero quadrupole deformation $\beta_2$. 

In low energy experiments, information on these deformation parameters are obtained by measuring the electric multipole transition probability of the nucleus from the ground state to excited states~\cite{Cline:1986ik, Raman:2001nnq}. In low energy theory and simulations, the structure and deformation parameters are estimated through {\it ab-initio} calculations~\cite{Warburton:1990zza, Force:2010ng, Moller:2015fba, Yuan:2023teo}. On the other hand, high-energy nuclear experiments or relativistic heavy-ion collisions can serve as an excellent platform for nuclear structure and deformation studies~\cite{Shou:2014eya, Giacalone:2019pca, Giacalone:2020awm, Jia:2021tzt, STAR:2015mki, ATLAS:2019dct}. In particular, over the last few years, there have been numerous theoretical~\cite{Filip:2009zz,Shou:2014eya,Giacalone:2019pca,Li:2019kkh,Giacalone:2020awm,Giacalone:2021udy,Jia:2021wbq,Jia:2021tzt,Bally:2021qys,Jia:2021qyu,Xu:2021uar,Zhang:2021kxj,Jia:2021oyt,Jia:2022qgl,Jia:2022qrq,Bally:2023dxi,Ryssens:2023fkv,Xu:2024bdh,Giacalone:2024ixe} and experimental~\cite{STAR:2015mki,ALICE:2018lao,CMS:2019cyz,ATLAS:2019dct,ATLAS:2020sgl,ALICE:2021gxt,ATLAS:2022dov,STAR:2024eky} efforts to study deformed nuclear structure of heavy nuclei through high energy relativistic heavy-ion collisions at RHIC and the LHC energies. At RHIC, most of these studies have focused on the structures of nuclei such as $^{238}$U, $^{197}$Au and isobars such as $^{96}$Ru and $^{96}$Zr. At the LHC, the study of collisions with the deformed xenon nucleus $^{129}$Xe along with its comparison to spherical lead nuclei $^{208}$Pb, has served as a candidate for nuclear deformation studies.   

The deformation of nuclei has a direct impact on the shape and the geometry of the initial state of heavy-ion collisions, leaving a significant effect on fluctuations and correlations of collective observables in the final state. Especially, the quadrupole deformation of the uranium nucleus is found to have a substantial effect on the event-by-event mean transverse momentum per particle $[p_T]$ and harmonic flow coefficients $V_n$~\cite{Giacalone:2019pca,Giacalone:2020awm}. As a result, the correlations, cumulants and the fluctuation-probing factorization breaking coefficients between those observables are found to exhibit a distinctive behavior~\cite{Giacalone:2019pca,Giacalone:2020awm,Bozek:2021zim,Samanta:2023qem}, paving the path to probe the deformation parameter $\beta_2$ through heavy-ion collisions. In this chapter, we will discuss the observables that are discussed in the earlier chapters, for U+U collision at RHIC energy, which could put exclusive constraints on the quadrupole deformation ($\beta_2$) of uranium nucleus. The following sections are, for the most part, presentations from the original publications~\cite{Bozek:2021zim,Samanta:2023qem}, coauthored by the author. We denote the quadupole deformation by $\beta$ for simplicity, as we will limit our discussion to only quadrupole deformation of uranium nuclei.

\section{Collision of deformed nuclei}
\label{deformed nuclei collision}
If two deformed nuclei are collided in a collision, then the geometry and shape of the overlap region is directly impacted by the deformed structure of the nuclei. This effect is translated to the final state observables such as mean transverse momentum per particle $[p_T]$ and harmonic flow $V_n$ through hydrodynamic expansion. Therefore, to understand how the deformed structure affect these quantities, first we need to implement the nuclear deformation in the Woods-Saxon density distribution for such nuclei, which is used for sampling the nucleon positions in those nuclei before collision.

\subsubsection{Woods-Saxon density distribution for a deformed nucleus:} 
The mass density distribution for a deformed nucleus with axial symmetry is given by three parameter Fermi distribution function with Woods-Saxon parametrization~\cite{Shou:2014eya,Loizides:2014vua,Giacalone:2019pca,Jia:2021tzt,Jia:2022qrq},
\begin{equation}
\rho(r, \theta, \phi ) = \frac{\rho_0}{1+\exp[\frac{r-R_0(1+\beta Y_{2,0}(\theta, \phi)}{a}]} ,
\label{eq: woods-saxon deformed}
\end{equation}
where only the leading order {\it quadrupole deformation} is considered, identified by the dimensionless quadrupole deformation parameter $\beta (\equiv \beta_2)$, which is most relevant for the uranium nuclei. The other parameters $R_0$, $\rho_0$ and $a$ carry similar meaning as before (Eq.~(\ref{eq: wood-saxon general})) with $R \equiv R_0$ for a spherical nucleus. The spherical harmonic $Y_{2,0}(\theta, \phi)$ breaks the spherical symmetry of the nucleus and is given by,
\begin{equation}
Y_{2,0}(\theta, \phi) = \sqrt{\frac{5}{16 \pi}}(3 \cos^2\theta - 1) \ ,
\label{eq: spherical harmonic y20}
\end{equation} 
where $\theta$ and $\phi$ are the polar and azimuthal angles in the intrinsic nuclear frame. The quadrupole deformation parameter $\beta$, given by\cite{Kumar:1972zza,Giacalone:2020awm},
\begin{equation}
\beta \simeq \frac{4 \pi}{5}\frac{\int r^2 \rho(r, \theta, \phi) Y_{2,0}(\theta, \phi) d^3\vec{r} }{\int r^2 \rho(r, \theta, \phi) d^3\vec{r}} \ 
\label{eq: quad deform beta}
\end{equation}
leads to an ellipsoidal structure of the nucleus. If $\beta > 0$ ($^{238}$U) then the nuclear structure is {\it prolate} and the nucleus is {\it oblate} when $\beta < 0$ ($^{197}$Au). 
\begin{figure}[ht!]
\includegraphics[height=7 cm]{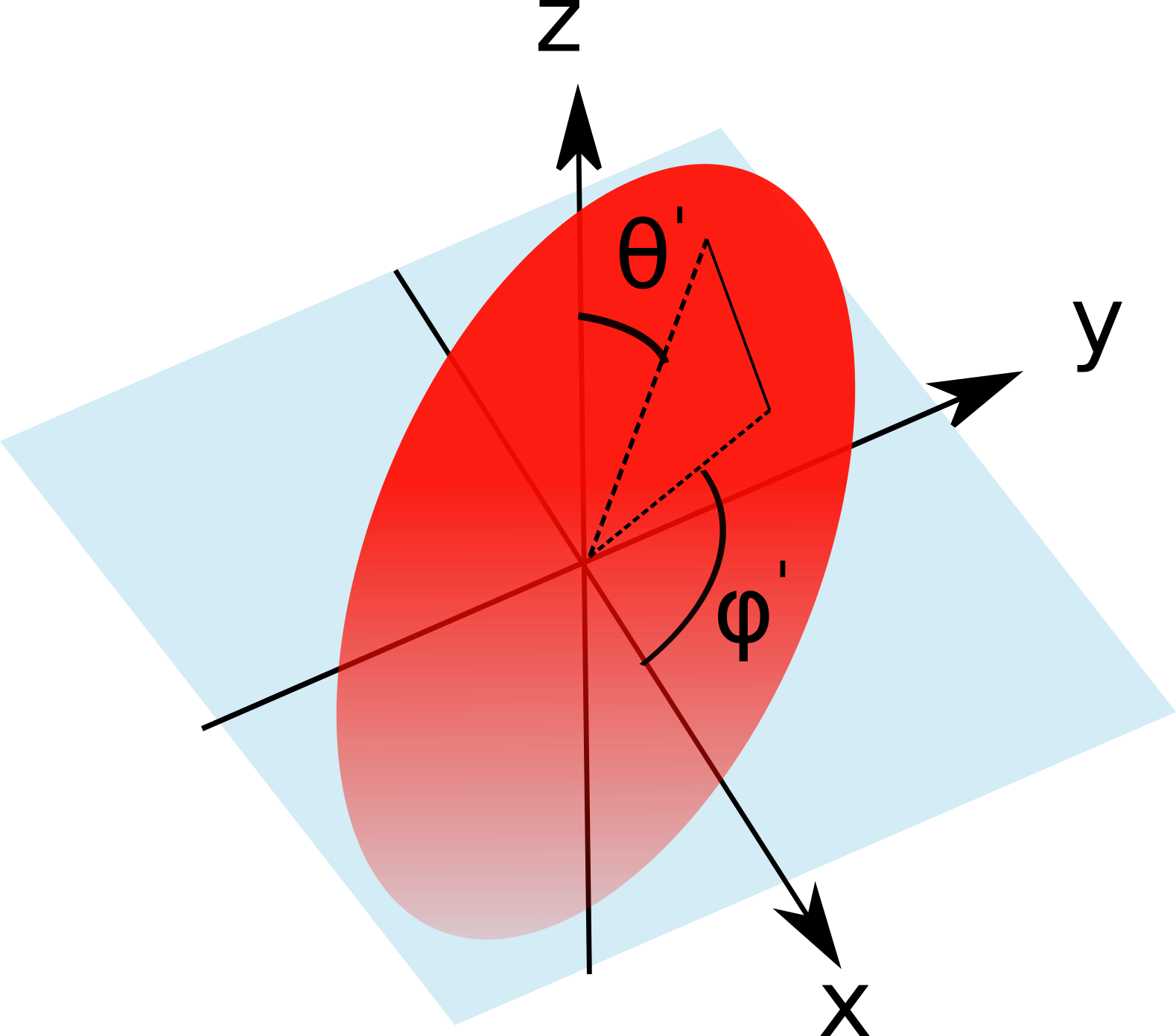}  
\centering
\caption{Deformed uranium nucleus having prolate structure with $\beta > 0$. The nucleus is randomly oriented with respect to the laboratory frame with a polar tilt $\theta'$ and an azimuthal rotation $\phi'$. The z axis represents the beam-axis in laboratory frame and (x,y) plane is the transverse plane. The figure is a modification from~\cite{Giacalone:2020awm}.}
\label{fig: uranium structure}
\end{figure}
\subsection{Geometry of the overlap region in deformed nuclei collisions:}
\label{geometry deformed nuclei}
In heavy-ion collision experiments, the deformed nuclei are injected through the beam pipe of the accelerator, where the colliding nuclei have random orientations in the laboratory frame. This is because the principal axes of the ellipsoids have random angles with the beam-axis. In particular, the principal axis has a polar tilt $\theta'$ and an azimuthal rotation $\phi'$ with respect to the beam-axis $z$ in the laboratory frame, as shown in Fig.~\ref{fig: uranium structure}. As a result, at the time of the collision, the two deformed nuclei collide with two random orientations resulting in different shapes of the overlap area in each collision. This phenomenon has a  great impact on the mean transverse momentum and harmonic flow.  In this chapter, we consider uranium nuclei which have a positive $\beta$ and hence have a prolate structure. Additionally, we restrict ourselves to central collisions where the overlap area of the two nuclei is maximum and the effect of the deformation is especially pronounced.

Let us consider central or ultracentral collisions of uranium nuclei with maximal overlap between the two nuclei. Let us also label the two nuclei as A and B with their orientations with respect to the collision axis in the lab frame, at the time of collision, defined by the sets of angles ($\theta'_A$, $\phi'_A$) and ($\theta'_B$, $\phi'_B$) respectively. There are two extreme cases of fully-overlapping scenario which are particularly interesting \cite{Giacalone:2019pca,Giacalone:2020awm}:
\begin{figure}[ht!]
\includegraphics[height= 12 cm]{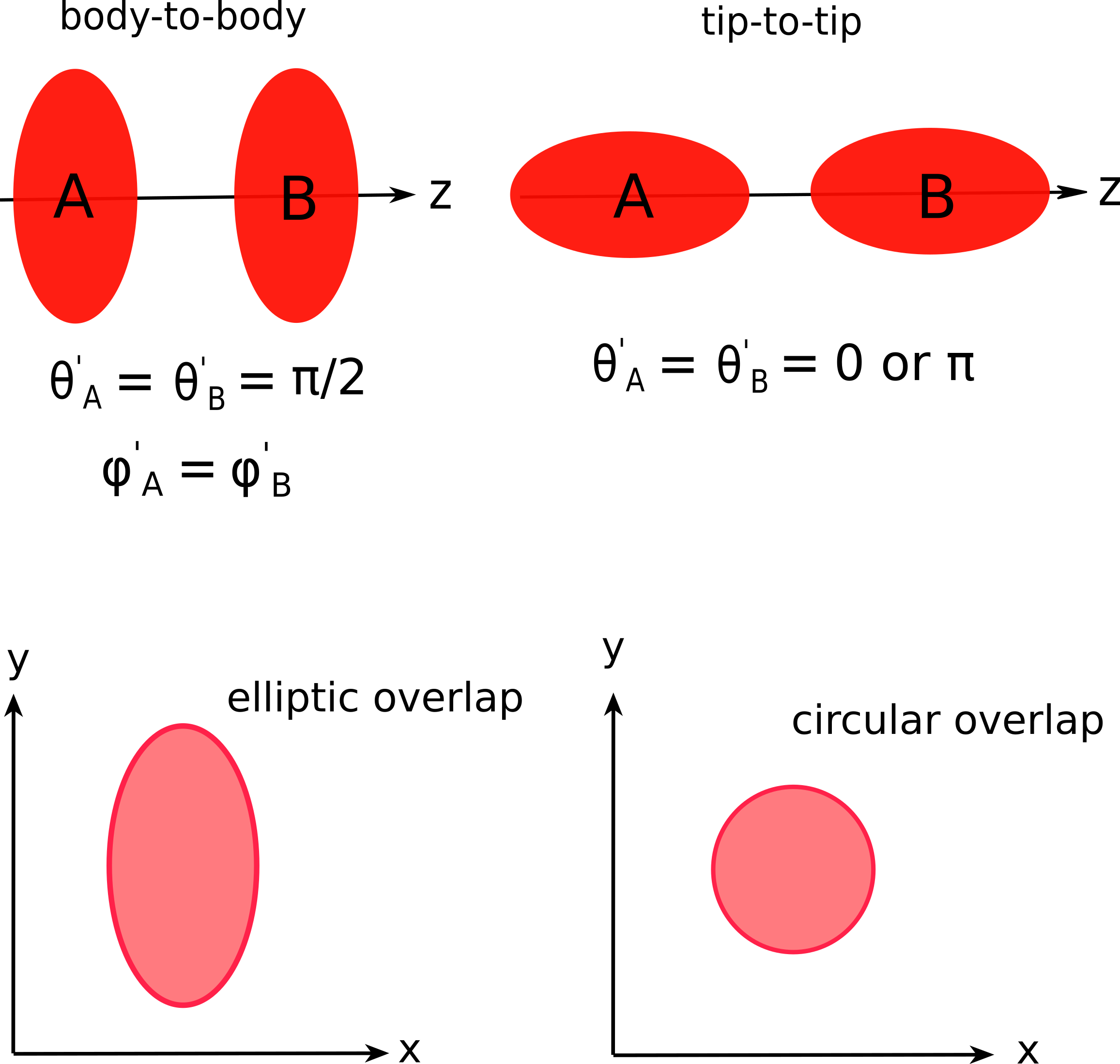}  
\centering
\caption{Pictorial representation of the body-to-body and tip-to-tip collision of deformed uranium nuclei. For body-to-body collision (left) the polar angles of the nuclei with the collision axis assume $\theta'_A=\theta'_B = \pi/2$ and azimuthal angles $\phi'_A = \phi'_B$, resulting in an elliptic overlap area after collision. On the other hand a tip-to-tip collision (right) occurs when $\theta'_A=\theta'_B = 0$ or $\pi$, producing a circular overlap area after collision. The figure is motivated from~\cite{Giacalone:2020awm}.}
\label{fig: overlap U+U}
\end{figure}
\begin{itemize}
\item A {\it body-to-body} collision. In this case two nuclei are essentially perpendicular to the collision axis with $\theta'_A=\theta'_B = \pi/2$ and both nuclei are rotated by same azimuthal angle $\phi'_A=\phi'_B$ as shown in Fig.~\ref{fig: overlap U+U} (left). Then the overlap area on the transverse plane takes an elliptic shape, which originates from the shape of the colliding nuclei, enhancing the ellipticity of the initial state even in the {\it central collisions.}  

\item A {\it tip-to-tip} collision. In such configuration, the principal axis of each colliding nucleus coincides with the beam-axis ($z$) i.e. $\theta'_A=\theta'_B = 0$ or $\pi$. This results in a circular overlap region on the transverse plane, as shown in Fig.~\ref{fig: overlap U+U} (right), having smaller transverse area than the body-to-body collisions. 
\end{itemize}

The other configurations fall in between these two cases. 

\subsection{Deformation effect on the collective observables}
\label{deform effect on pt, vn}
In central collisions of deformed nuclei, the different configurations of the initial states directly influence the collective observables such as mean transverse momentum per particle $[p_T]$ and harmonic flow coefficients $V_n$ in the final state.

\subsubsection{Transverse momentum}
Let us consider the above mentioned configurations for events with full overlap of the nuclei at fixed multiplicity i.e. fixed entropy in the initial state. In the case of a tip-to-tip collision, the transverse size of the overlap area (circular) or collision volume is smaller corresponding to a larger density at fixed multiplicity~\cite{Giacalone:2019pca,Giacalone:2020awm}. According to relativistic thermodynamics, this results in larger temperature, hence larger pressure and larger energy per particle, which eventually give rise to larger transverse momentum per particle $[p_T]$. The opposite situation occurs when the collision is body-to-body, which leads to larger transverse size and hence a smaller $[p_T]$. Thus in central collisions of deformed nuclei, the deformation has a direct impact on $[p_T]$. Nuclear deformation contributes significantly to the event-by-event fluctuations of mean transverse momentum per particle. In general the average transverse momentum in heavy ion collision can be related to the temperature of QGP medium as~\cite{Gardim:2019xjs}, 
\begin{equation}
\langle p_T \rangle \simeq 3 T \ ,
\label{eq: mean pt vs temperature}
\end{equation}
where $T$ is the effective temperature of the medium. The dependence of $[p_T]$ on the transverse size of the overlap region $R$ can be written as~\cite{Giacalone:2020awm},
\begin{equation}
\frac{\delta p_T}{\langle p_T \rangle} \simeq -c_s^2\frac{\delta R}{\langle R \rangle} \ ,
\label{eq: pt fluctuation vs R fluctuation}
\end{equation}
where $\delta p_T = [p_T] - \langle p_T \rangle$, $\delta R = R-\langle R \rangle$, $c_s$ is the speed of sound in the QGP and $\langle \dots \rangle$ denote the event-averaged quantities. The negative sign denotes the anti-correlation between mean transverse momentum per particle and transverse size, as seen in Fig.~\ref{fig: scatter plot pt-R and pt-S}. 

\subsubsection{Harmonic flow coefficients}
\label{deform effect on pt}
The nuclear deformation largely affects the harmonic flow coefficients~\cite{Giacalone:2020awm}, specifically the elliptic flow, in central collisions. In a central collision of spherical nuclei, when two nuclei fully overlap, the harmonic flow is mostly driven by the event-by-event fluctuations. This is also the case for elliptic flow, unlike a non central collision where it is driven by the elliptic geometry of the initial state.  

However, in case of deformed nuclei collision this is no longer true. Let us consider the two extreme cases of fully overlapping nuclei in central collision, as discussed earlier. In body-to-body collision, the overlap region takes the shape of an ellipse resulting in a significantly large ellipticity $\Epsilon_2$ even in central collision. This leads to an enhanced contribution to the elliptic flow $v_2$ in the final state which gets dominating contribution from geometry than fluctuations. On the other hand, in tip-to-tip collision, the overlap area is circular, which does not contribute to the geometrical component of $v_2$ but solely to the component arising from fluctuations. Thus deformation plays a significant role in enhancing the eccentricity and therefore the elliptic flow in central collision. This fascinating phenomena creates an unique opportunity to probe the deformation parameters by studying the correlations and fluctuations of flow and related observables.  

The contribution of deformation to the initial eccentricity can be written as~\cite{Jia:2021tzt},
\begin{equation}
 \Epsilon_2\{2\}^2 = a'_2 + b'_2 \beta^2 \ ,
\label{eq: eps2 vs beta2}
\end{equation}
where, $ \Epsilon_2\{2\} = \sqrt{\langle \Epsilon_2^2 \rangle}$ is the event averaged eccentricity (ellipticity) in analogy to $v_2\{2\}$. As discussed in Chapter-3, the elliptic flow can be related to the initial eccentricity as the hydrodynamic response of the initial state $v_2\{2\}^2=k_2 \Epsilon_2\{2\}^2$, with $k_2$ being the hydro-response coefficient. Therefore, the elliptic flow can be related to the deformation parameter of the nucleus through similar parametric dependence,
\begin{equation}
 v_2\{2\}^2 = a_2 + b_2 \beta^2 \ .
\label{eq: eps2 for deform}
\end{equation}
It should be noted that the coefficients $a'_2$ and $a_2$ quantify the contributions to the participant eccentricity and elliptic flow, that do not arise from the deformation of the nucleus, while $b'_2$ and $b_2$ pick up contributions to the eccentricity and elliptic flow driven by deformation of the colliding nuclei. 

\section{Fluctuations and correlations for deformed nuclei collision}
\label{deform effect on fluct and corr}
Due to nuclear deformation, event-by-event fluctuations of harmonic flow which we studied in terms of the momentum dependent factorization-breaking coefficients, and the correlation coefficients or the cumulants between mean transverse momentum and harmonic flow get impacted, showing peculiar characteristics in central collisions of such deformed nuclei, which can be potentially used to put constraints on the deformation parameter $\beta$.  

For the results presented in this section, we simulate U+U collisions at RHIC energy $\sqrt{s_{NN}}=193$ GeV with TRENTO initial conditions and then relativistic hydrodynamic evolution through MUSIC. The Woods-Saxon and deformation parameters used for uranium-238 nucleus is presented in Table~\ref{tab: Wood-Saxon params deformed}. 

\begin{table}[ht!]
\centering
\begin{tabular}{ c c c c }
   \toprule 
   Nucleus  & $R_0$ (fm)  &  $a$ (fm) & $\beta$ \\ 
    \midrule
    $^{238}$U & 6.86  & 0.42  & 0.265  \\
     \bottomrule
\end{tabular}
\vskip 2mm
\caption{Woods-Saxon parameters for nuclear density distribution (Eq.~\ref{eq: woods-saxon deformed}) in deformed uranium nucleus~\cite{Shou:2014eya,Loizides:2014vua}.}
\label{tab: Wood-Saxon params deformed}
\end{table}

\subsection{Factorization-breaking coefficients}
\label{fact-break for deformed nuclei}

Let us first discuss the factorization-breaking coefficients which measure the decorrelation between flow vectors in different transverse momentum bins as discussed in Chapter-3. Here we want to explore if nuclear deformation plays any role in modifying these factorization-breaking coefficients. For this, we consider three different scenarios: a) U+U collision with deformation ($\beta = 0.265$), which could be considered as the default case. b) Collision of deformed U+U but without any fluctuations in entropy deposition in the initial state. This is achieved by setting the fluctuation-parameter $k \gg 1$. and c) Collision of spherical U+U with $\beta=0$ while keeping all other parameters same. This is not realized in reality but we consider this situation in order to show the impact of deformation more prominently by making a direct comparison with the deformed case.    
\begin{figure}[ht!]
\includegraphics[height=6 cm]{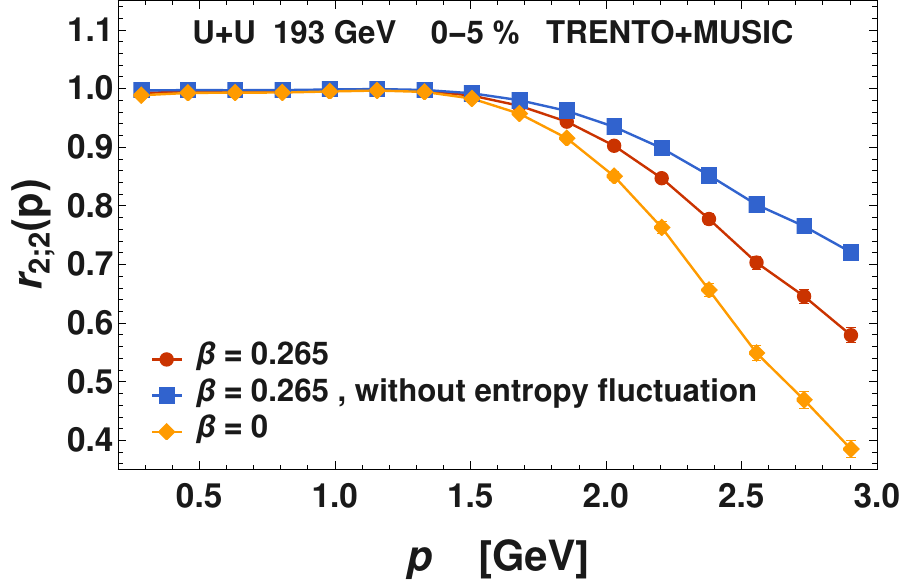}  
\centering
\caption{Factorization-breaking coefficients of elliptic flow vector squared $r_{2;2}(p)$ between momentum averaged flow and momentum dependent flow for 0-5 \% centrality in U+U collision at $\sqrt{s_{NN}}=193$ GeV. The momentum dependent coefficients obtained from hydrodynamic simulation with TRENTO initial conditions for the spherical ($\beta =0$) and deformed ($\beta=0.265$) nuclei collisions, are denoted by orange diamonds and red circles respectively. The blue squares represent the simulation results obtained without fluctuations in the entropy deposition at the initial state, where the parameter $k$ is set to $ \gg 1$ while obtaining the initial conditions. The figure is from the original publication~\cite{Samanta:2023qem}, coauthored by the author.}
\label{fig: flow vec fact braek UU}
\end{figure}

Fig.~\ref{fig: flow vec fact braek UU} displays the factorization-breaking coefficients between flow vectors, where one of the flow harmonics is global or momentum averaged and the other one is momentum dependent as discussed in Chapter-3, for 0-5 \% centrality in U+U collisions. The correlation coefficient gradually deviates from 1 with increasing transverse momentum (shown as $p \equiv p_T$ ), reflecting a significant decorrelation for both deformed (red curve) and spherical (orange curve) nuclei collisions. Spherical nuclei collisions show remarkably larger decorrelation with increasing transverse momentum as compared to the deformed nuclei collision. The particular difference emerges more profoundly when the fluctuations in the initial entropy deposition are switched off (blue curve), where the decorrelation is further reduced.      

Let us explain the above results. The decorrelation of flow vectors in transverse momentum is governed by event-by-event fluctuations of flow, which means that the decorrelation would be larger where fluctuations dominate. In Fig.~\ref{fig: flow vec fact braek UU}, as we consider 0-5 \% centrality, for spherical nuclei collision the elliptic flow is mostly due to fluctuations and the overall magnitude is smaller. Therefore, it is easier to decorrelate and hence we observe a larger decorrelation between the flow vectors in transverse momentum. In case of deformed nuclei collisions, on the other hand, the presence of nuclear deformation increases the geometrical component in the total eccentricity. Therefore, even in central collisions there is significant influence of the elliptic geometry on the observed elliptic flow. The contribution of fluctuations to elliptic flow is partly washed out by the geometrical component resulting in a smaller decorrelation in transverse momentum as reflected in the results. This situation mimics the case of a semi-central collision of spherical nuclei (at a relatively large impact parameter)~\cite{Giacalone:2021uhj}, where the overall magnitude of the flow is larger and the flow vectors in different transverse momentum bins are also more correlated with the overall orientation in the transverse plane, resulting in a smaller decorrelation in transverse momentum. 

In the case of collision of deformed nuclei without fluctuations in the initial entropy deposition, event-by-event fluctuations of the elliptic flow are further reduced, resulting in an even smaller decorrelation as shown by the blue lines. This once again portrays the role of fluctuations in the flow decorrelation. The momentum dependent factorization breaking coefficients $r_{2;2}(p)$ put useful constraints on $\beta$ and could be verified in experiments. 
\begin{figure}[ht!]
\hspace{-0.3 cm}\begin{subfigure}{0.5\textwidth}
\centering
\includegraphics[height=5.1 cm]{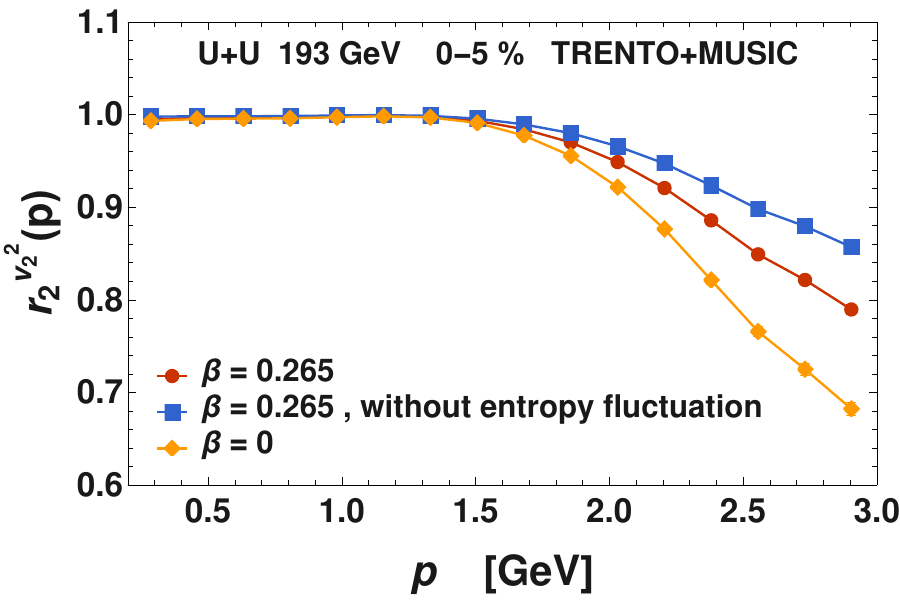}
\end{subfigure}~~
\begin{subfigure}{0.5\textwidth}
\centering
\includegraphics[height=5.1 cm]{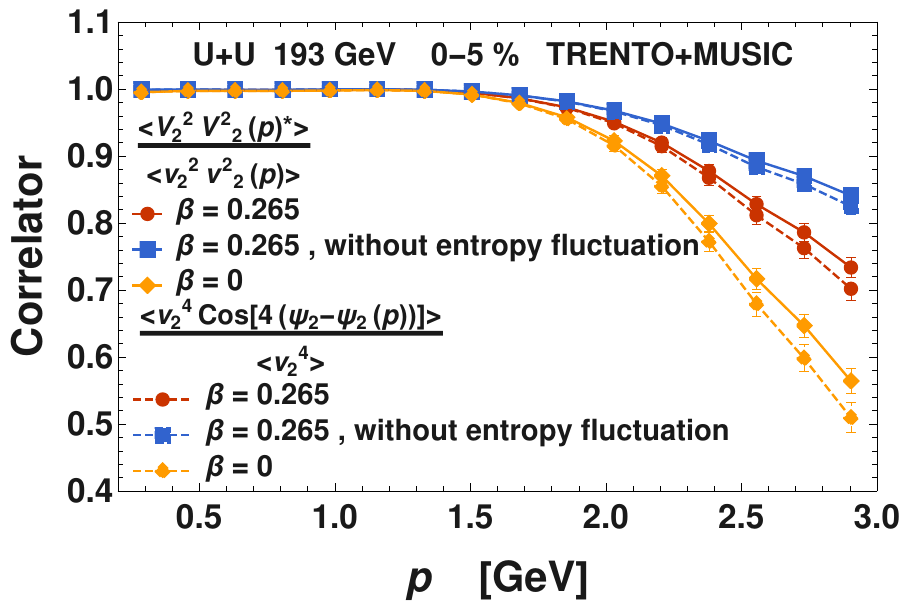}  
\end{subfigure}
\centering
\caption{Left: Flow magnitude squared factorization breaking coefficient for the elliptic flow $r_2^{v_2^2}(p)$ for 0-5 \% centrality in U+U collision at 193 GeV. Right: Flow angle decorrelation between momentum averaged and momentum dependent elliptic flow coefficients.  The symbols with different color have similar meaning as Fig.~\ref{fig: flow vec fact braek UU}. The solid and the dashed lines on the right panel represent the results obtained with experimental estimation and the actual definition of flow angle decorrelation respectively. The figure is from the original publication~\cite{Samanta:2023qem}, coauthored by the author.}
\label{fig: flow mag fact-break and flow ang decorr UU}
\end{figure}

Fig.~\ref{fig: flow mag fact-break and flow ang decorr UU} shows the flow magnitude factorization-breaking coefficients (left) and flow angle decorrelations (right) for the elliptic flow coefficients. Similar trends of the results are observed in both cases; the decorrelation is smaller for the deformed case. Both of these quantities can be measured experimentally and therefore would provide additional constraints on the deformation. Most notably, the relation proposed in Chapter-3, also holds true for the deformed nuclei collision,
\begin{equation}
 \begin{aligned}
[1-r_{n;2}(p)] \simeq 2[1-r_n^{v_n^2(p)}] \ .
 \end{aligned}
\label{eq: relation between flow vec and mag decorr deformed}
\end{equation}
Moreover, the experimental estimate of the angle decorrelation and true angle decorrelation coincide with each other also for deformed nuclei collisions, once again validating this approximation.  
\begin{figure}[ht!]
\hspace{-0.2 cm}\begin{subfigure}{0.5\textwidth}
\centering
\includegraphics[height=4.6 cm]{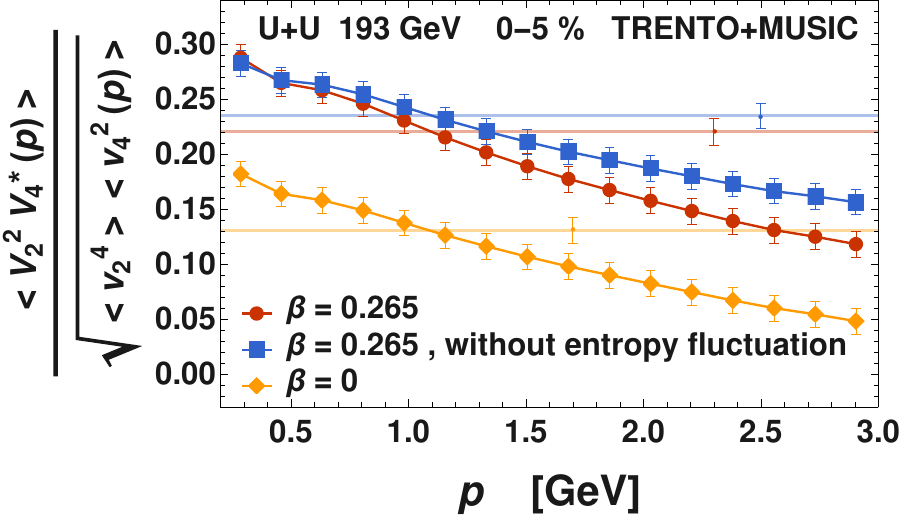}
\end{subfigure}~~~
\begin{subfigure}{0.5\textwidth}
\centering
\includegraphics[height=4.6 cm]{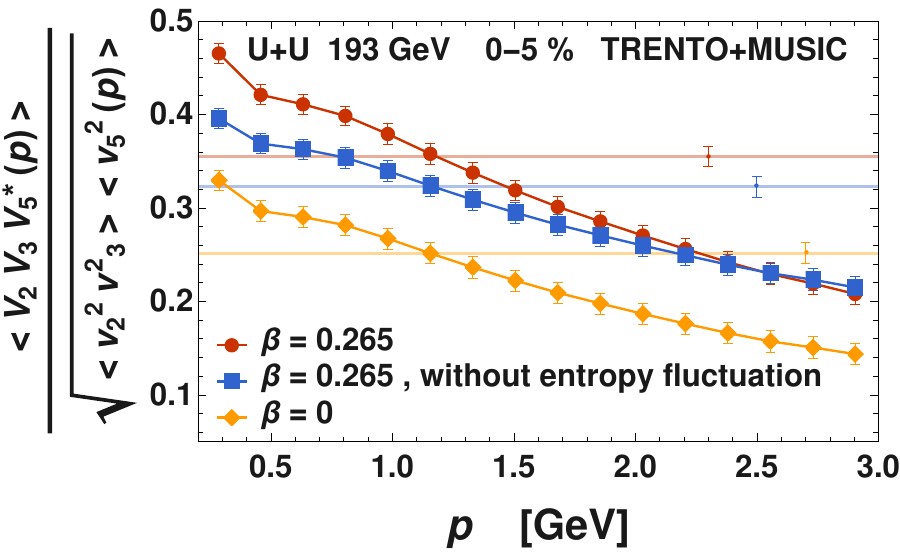}  
\end{subfigure}
\centering
\caption{:Momentum dependent mixed flow correlation between $V_2^2-V_4(p)$ (left) and $V_2V_3-V_5(p)$ (right) for 0-5 \% centrality in U+U collision at 193 GeV. The results obtained for the spherical and deformed nuclei collisions are represented by orange and red lines respectively. The  blue lines denote the results with no entropy fluctuations in the initial state. The horizontal lines with same colors represent the corresponding mixed flow correlations between the momentum averaged flow coefficients. The figure is from the original publication~\cite{Samanta:2023qem}, coauthored by the author.}
\label{fig: mixed flow correlation UU}
\end{figure}

As the triangular flow is largely driven by the fluctuations, nuclear deformation has negligible to no effect on it, and is therefore irrelevant in the present context. To complete the picture, in Fig.~\ref{fig: mixed flow correlation UU} we present the results for the momentum dependent mixed-flow correlation between $V_2^2-V_4(p)$ and $V_2V_3-V_5(p)$. As expected, the correlation is larger for deformed nuclei collisions, also depicted by the corresponding baselines representing the respective correlation between momentum averaged flow. The transverse momentum dependence is similar for the deformed and spherical nuclei with however a slightly larger difference between the two curves at the low momentum. The study of such momentum dependent mixed flow correlations could give further constraints on the deformation and can be tested in experiments.   
\begin{figure}[ht!]
\includegraphics[height=6 cm]{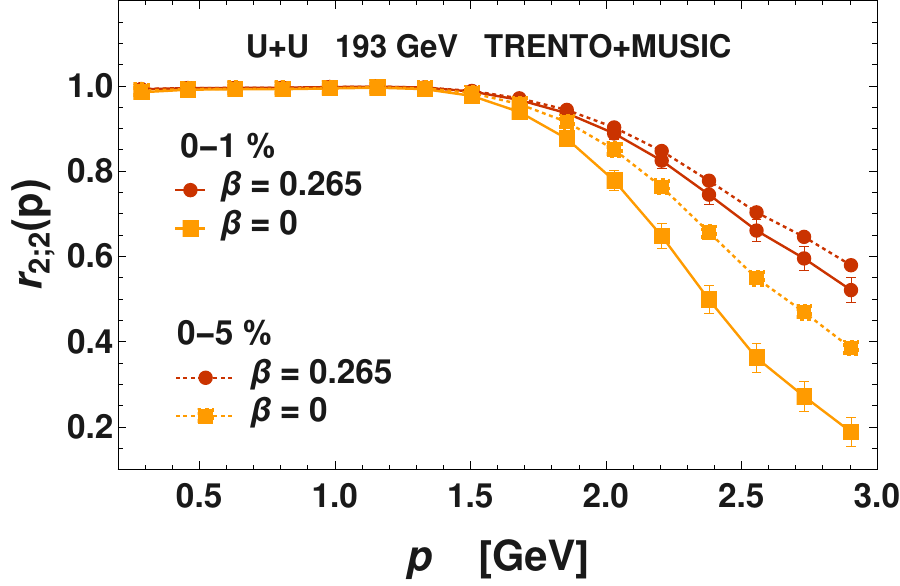}  
\centering
\caption{Comparison of the deformation effect through flow vector factorization breaking coefficient for the elliptic flow in central ($0-5 \%$) and ultracentral ($0-1 \%$) U+U collisions at 193 GeV. The results obtained from hydrodynamic simulation for the spherical and deformed nuclei collisions are represented by the orange and red colors. The solid and the dashed lines represent the results corresponding to $0-1 \%$ and $0-5 \%$ centrality respectively. The figure is from the original publication~\cite{Samanta:2023qem}, coauthored by the author.}
\label{fig: flow vec decorr UU central to ultracentral}
\end{figure}

In order to emphasize the fact that the deformation effect is more pronounced in central collisions or as the collisions are more central, in Fig.~\ref{fig: flow vec decorr UU central to ultracentral} we present results for the factorization-breaking coefficient $r_{2;2}(p)$ for two cases : 0-1 \% centrality (solid lines) which represents the ultracentral collisions and usual 0-5 \% centrality (dashed lines) which represents in general central collisions. It can be seen that the correlation is smaller as we move towards ultracentral collisions for both deformed and spherical nuclei. The difference between the two centralities is larger in spherical case, because of enhanced fluctuations and less prominent for the deformed case due to relatively smaller contributions of fluctuations, reflecting the fact that the eccentricity without deformation is smaller for $0$-$1$\% centrality. On the other hand, the relative difference between the spherical and the deformed case is substantially larger in $0-1 \%$ centrality than $0-5 \%$. Please note that when we say the effect of deformation, we particularly mean this relative difference.

\subsection{Pearson correlation and symmetric cumulants}
\label{rho and SC for deformed nuclei}

The nuclear deformation not only affects event-by-event fluctuations of the mean transverse momentum per particle and harmonic flow coefficients but also significantly impacts the correlation between them. In particular, we see that because of the deformed structure of uranium nuclei, there exist an anti-correlation between $[p_T]$ and $v_2^2$ in central collisions, which reduces the overall correlation between them~\cite{Giacalone:2020awm} as discussed below.    

\subsubsection{Anti-correlation between $[p_T]$ and $v_n^2$}

Let us consider the body-to-body case for fully overlapping nuclei in a central collision. We have seen that because of the larger transverse size or larger collision volume at fixed multiplicity, this leads to a smaller transverse momentum per particle. On the other hand, due to the elliptic shape of the overlap region the elliptic flow is enhanced. Therefore deformation of the nucleus has opposite effect on $[p_T]$ and $v_2$, resulting in a negative or anti-correlation between them, which drastically reduce the correlation and eventually makes it negative as one moves towards ultracentral collisions~\cite{Giacalone:2020awm}.  

In this section we present the results for the Pearson correlation coefficient, normalized and symmetric cumulants as discussed in Chapter-5, for U+U collision at 193 GeV. Our simulation set-up remains same with three cases : deformed ($\beta = 0.265$), spherical ($\beta=0$) and deformed without entropy fluctuations in the initial state ($\beta=0.265, k \gg 1$). 

\begin{figure}[ht!]
\includegraphics[height=6 cm]{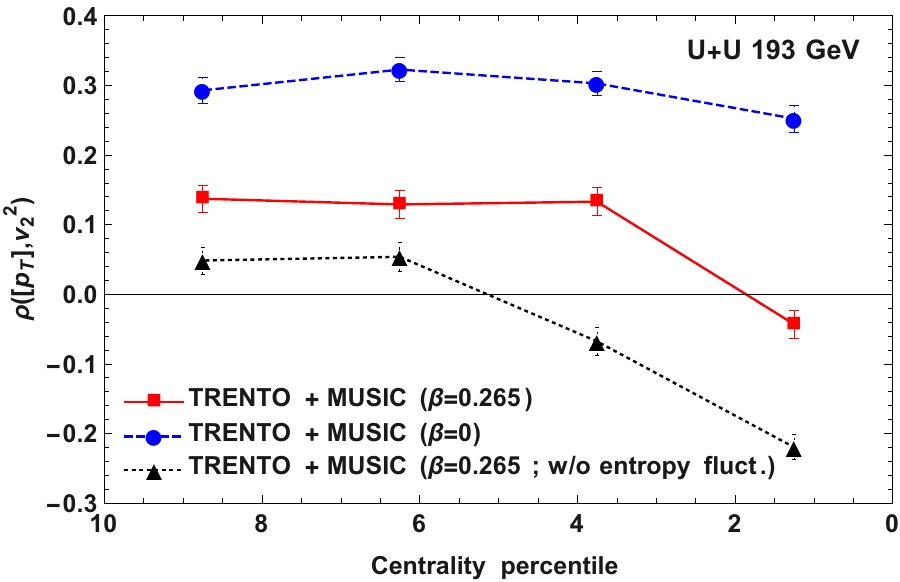}  
\centering
\caption{The Pearson correlation coefficients between mean transverse momentum per particle and elliptic flow coefficient as a function of centrality in $0-10 \%$ central collisions of U+U at 193 GeV. The red squares denote the results obtained from the hydrodynamic simulation with TRENTO initial condition for deformed nuclei ($\beta=0.265$). The blue circles represent the corresponding results for the spherical nuclei ($\beta=0$) collisions. The black triangles depict the results for the deformed nuclei collision without fluctuations in entropy deposition in the initial state. The figure is from the original publication~\cite{Bozek:2021zim}, coauthored by the author.}
\label{fig: rho pt-v2 UU}
\end{figure}

Fig.~\ref{fig: rho pt-v2 UU} displays the results for the Pearson correlation coefficient~\cite{Bozek:2016yoj} $\rho([p_T],v_2^2)$ for the three cases. As expected, the correlation is drastically smaller for deformed nuclei as compared to the spherical one. In the ultracentral region it even goes to negative, as discussed earlier. In the absence of entropy fluctuations in the initial state, the correlation get further reduced, as it should be because it suppresses the fluctuations of both $[p_T]$ and $v_2$ and reduces correlation with multiplicity. In this case, the correlation coefficient is negative for centrality $< 5 \%$. Please not that in this case we do not correct the correlation coefficient or the observables within it for multiplicity fluctuations. Because in central collisions, the covariances of the harmonic flow and mean transverse momentum with multiplicity are small. 
\begin{figure}[ht!]
\hspace{-0.2 cm}\begin{subfigure}{0.5\textwidth}
\centering
\includegraphics[height=4.8 cm]{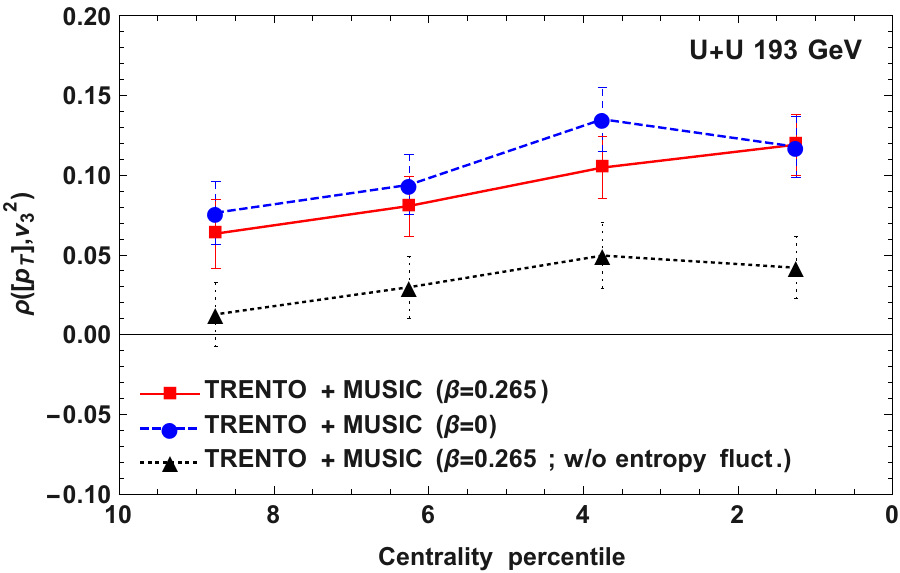}
\end{subfigure}~~
\begin{subfigure}{0.5\textwidth}
\centering
\includegraphics[height=4.8 cm]{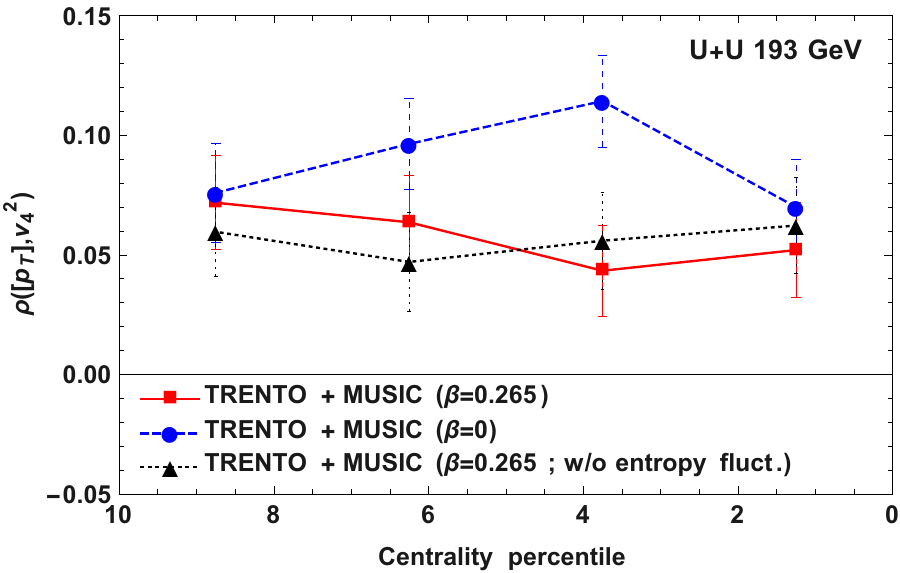}  
\end{subfigure}
\centering
\caption{Left: Pearson correlation coefficient between mean transverse momentum $[p_T]$ and triangular flow $v_3^2$, as a function of centrality in U+U collision. Right: The same for quadrangular flow $v_4^2$. The symbols carry same meaning as Fig.~\ref{fig: rho pt-v2 UU}. The left panel of the figure is from the original publication~\cite{Bozek:2021zim}, coauthored by the author.}
\label{fig: rho pt-v3 and pt-v4 UU}
\end{figure}
Furthermore, here we use (for the Pearson correlation) U+U collision with narrower centrality bins and we have checked that the corrections for multiplicity fluctuations to the correlation coefficients are small. The same holds true for the higher order symmetric cumulants\footnote{Although for the symmetric cumulants we use wider bins for accessing higher statistics}. Moreover, the correction for the multiplicity fluctuations is partially achieved in the third scenario where the fluctuations in entropy deposition in the initial state are switched off, resulting in significantly smaller multiplicity fluctuations. Therefore, in this section we only present results for the uncorrected Pearson correlation coefficients and symmetric cumulants for U+U collisions.

Fig.~\ref{fig: rho pt-v3 and pt-v4 UU} shows the results for the correlation coefficients $\rho([p_T],v_3^2)$ (left) and $\rho([p_T],v_4^2)$ (right). As mentioned earlier, deformation has negligible effect on the triangular flow. Therefore for $\rho([p_T],v_3^2)$ the results are similar (consistent within errors) for the spherical and deformed nuclei collisions. However, the correlations encounter overall decrease in magnitude without initial entropy fluctuations. For the quadrangular flow, $\rho([p_T],v_4^2)$, on the other hand, shows a noticeable difference between the two cases around $2-6 \%$ centrality, although not very large. This could be due to the non-linear mixing of $v_2^2$ with $v_4$ and the difference arises due to the effect of deformation on the elliptic flow. 

\begin{figure}[ht!]
\hspace{-0.2 cm}\begin{subfigure}{0.5\textwidth}
\centering
\includegraphics[height=5.2 cm]{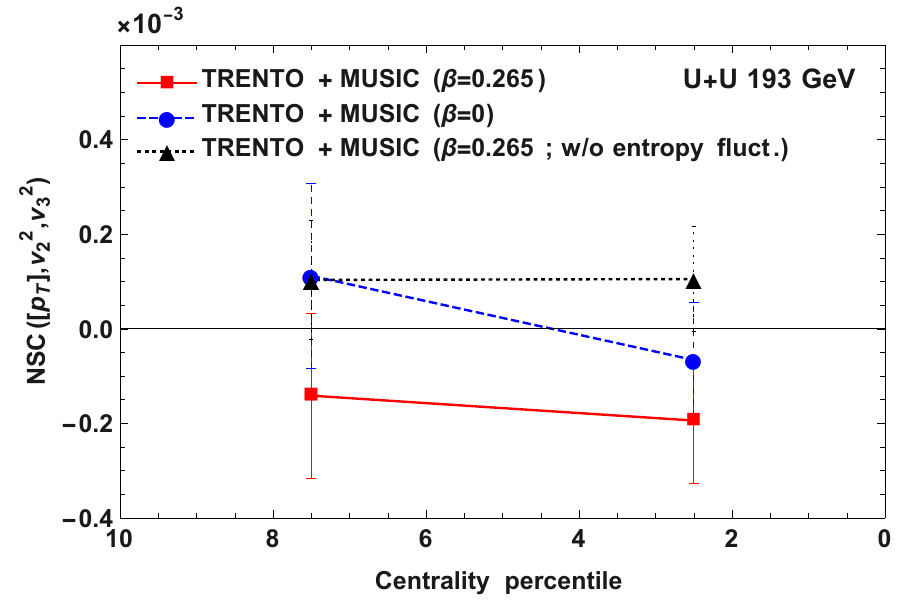}
\end{subfigure}~~
\begin{subfigure}{0.5\textwidth}
\centering
\includegraphics[height=5.2 cm]{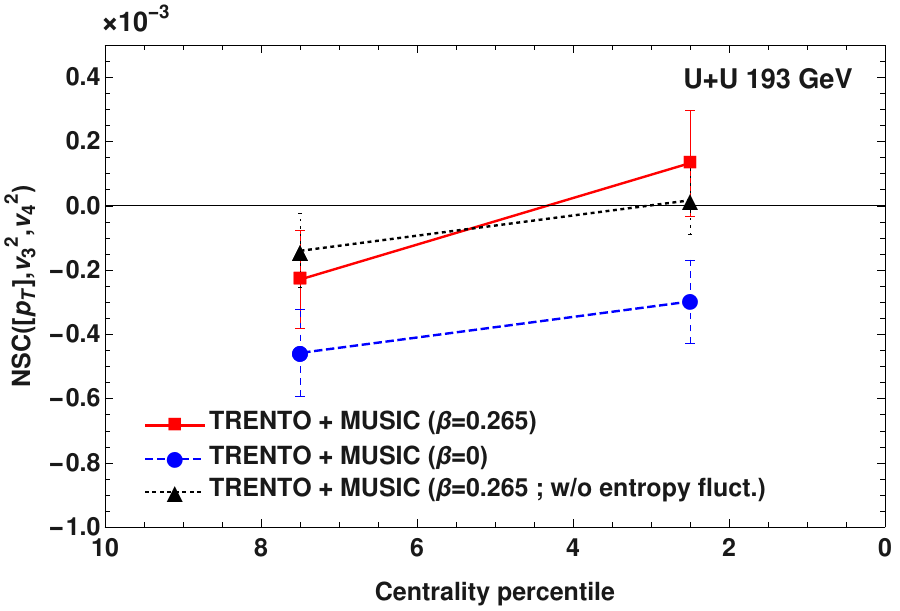}  
\end{subfigure}
\centering
\caption{Third order normalized symmetric cumulant between $[p_T]$, $v_2^2$ and $v_3^2$ (left) and $[p_T]$, $v_3^2$ and $v_4^2$ (right) as a function of centrality with two centrality bins in central collisions of U+U at 193 GeV. The symbols have similar meaning as Fig.~\ref{fig: rho pt-v2 UU}. The left panel of the figure is from the original publication~\cite{Bozek:2021zim}, coauthored by the author.}
\label{fig: NSC pt-v2-v3 and pt-v3-v4 UU}
\end{figure}

Fig.~\ref{fig: NSC pt-v2-v3 and pt-v3-v4 UU} shows the third order normalized symmetric cumulant between $[p_T]$, $v_2^2$ and $v_3^2$ on the left and between $[p_T]$, $v_3^2$ and $v_4^2$ on the right. Both of the symmetric cumulants are very small in comparison to the Pearson correlation coefficients, because they pick genuine higher order correlations between the observables. For $NSC([p_T], v_2^2, v_3^2)$, the correlation is smaller (more negative) for deformed nuclei collisions than for the spherical case and removing the entropy fluctuations increases the correlation between them. However, the difference between the three cases is quite small. On the other hand, for $NSC([p_T], v_3^2, v_4^2)$, the cumulant is mostly negative for all the three cases and the deformed case has larger correlation (less negative) than the spherical nuclei collisions, highlighting the fact that $v_3$ is not influenced by the deformation, with little influence on $v_4$ due to non-linear mode.  

\begin{figure}[ht!]
\hspace{-0.2 cm}\begin{subfigure}{0.5\textwidth}
\centering
\includegraphics[height=5.2 cm]{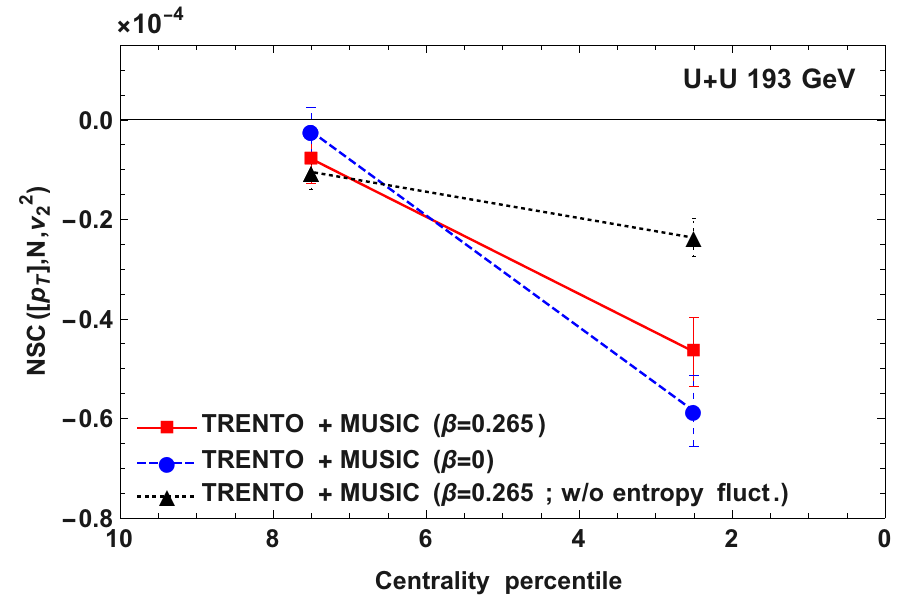}
\end{subfigure}~~
\begin{subfigure}{0.5\textwidth}
\centering
\includegraphics[height=5.2 cm]{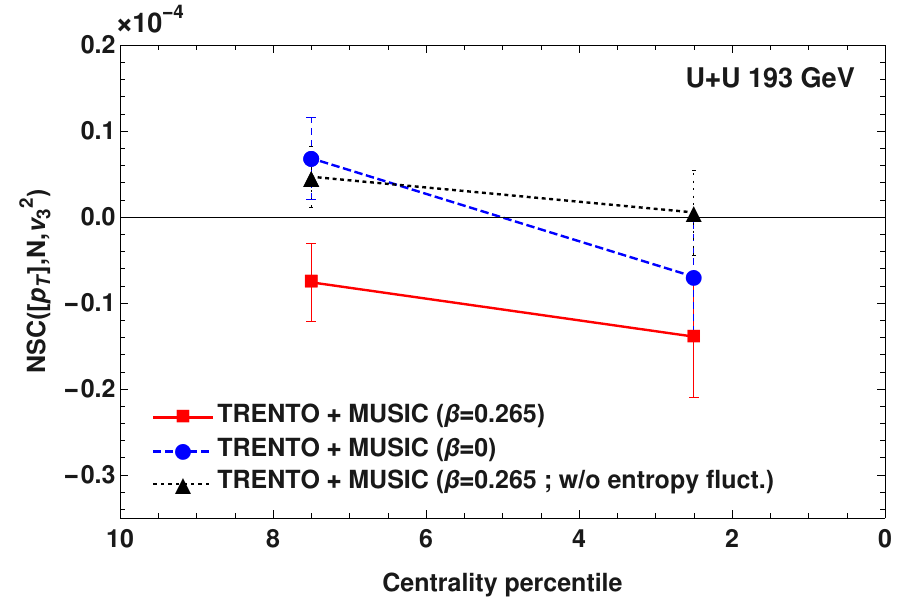}  
\end{subfigure}
\centering
\caption{Left: Third order normalized symmetric cumulant between mean transverse momentum, multiplicity and elliptic flow coefficient  as a function of centrality in central collisions of U+U at 193 GeV (left). Right: Same but for the triangular flow coefficients. The symbols carry similar meaning as Fig.~\ref{fig: NSC pt-v2-v3 and pt-v3-v4 UU}. The left panel of the figure is from the original publication~\cite{Bozek:2021zim}, coauthored by the author.}
\label{fig: NSC pt-n-v2 and pt-n-v3 UU}
\end{figure}

Although the elliptic flow in central collision of deformed nuclei is dominated by the geometry arising from the nuclear deformation, the collective observables in such collisions are in general significantly impacted by the fluctuations in initial entropy and its azimuthal asymmetries, as seen before. Fluctuations of initial entropy dictate the fluctuations of final state multiplicities. Therefore, it would be interesting to study symmetric cumulants involving not only the mean transverse momentum and harmonic flow, but also the multiplicity in the event. Such observables have potential to pick up the effect of initial state entropy fluctuations more prominently. 
\begin{figure}[ht!]
\includegraphics[height=6 cm]{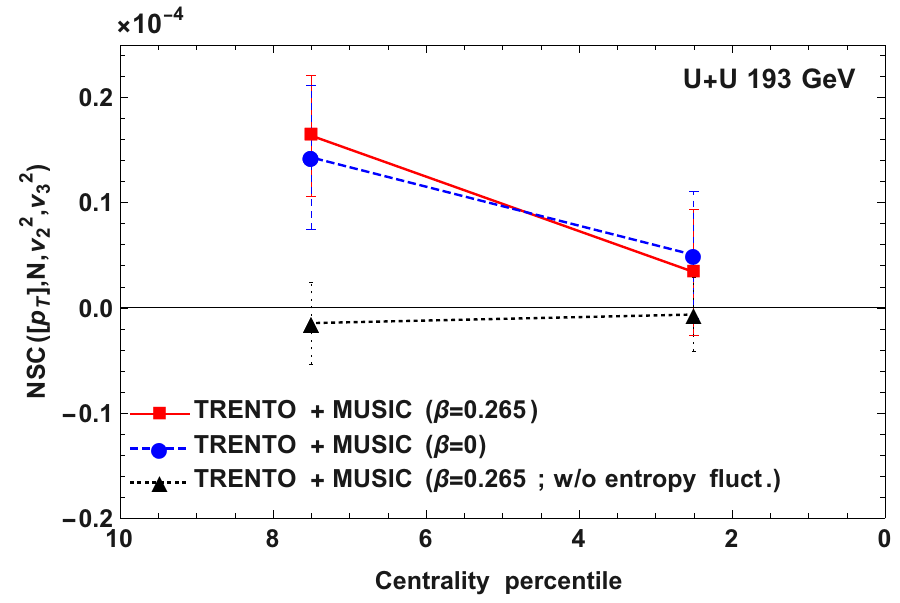}  
\centering
\caption{Fourth order normalized symmetric cumulant between mean transverse momentum, multiplicity, elliptic and triangular flow coefficient as a function of centrality in U+U at 193 GeV. Symbols are similar as Fig.~\ref{fig: NSC pt-n-v2 and pt-n-v3 UU}. The figure is from the original publication~\cite{Bozek:2021zim}, coauthored by the author.}
\label{fig: NSC pt-n-v2-v3 UU}
\end{figure}
It is important to note that the results for the symmetric cumulants involving multiplicity as one of the observables could significantly depend on the definition of the centrality bin. In order to reduce such bias from centrality cuts, these observables could be measured experimentally using centrality bins defined on basis of other observables such as the total transverse energy deposited in the forward calorimeter ($E_T$) as discussed earlier.

The third order normalized symmetric cumulants $NSC(p_T,N,v_2^2)$ and $NSC(p_T,N,v_3^2)$ are shown in Fig.~\ref{fig: NSC pt-n-v2 and pt-n-v3 UU}. As expected, $NSC(p_T,N,v_2^2)$ shows a remarkable sensitivity to the fluctuations of entropy deposition from the participant nucleons as one moves towards more central collision. In particular, the fluctuations in entropy increase the magnitude (more negative) of $NSC(p_T,N,v_2^2)$. The deformation does not have a significant impact, both the spherical and deformed cases are consistent with each other in the centrality range considered. 
\begin{figure}[ht!]
\includegraphics[height=6 cm]{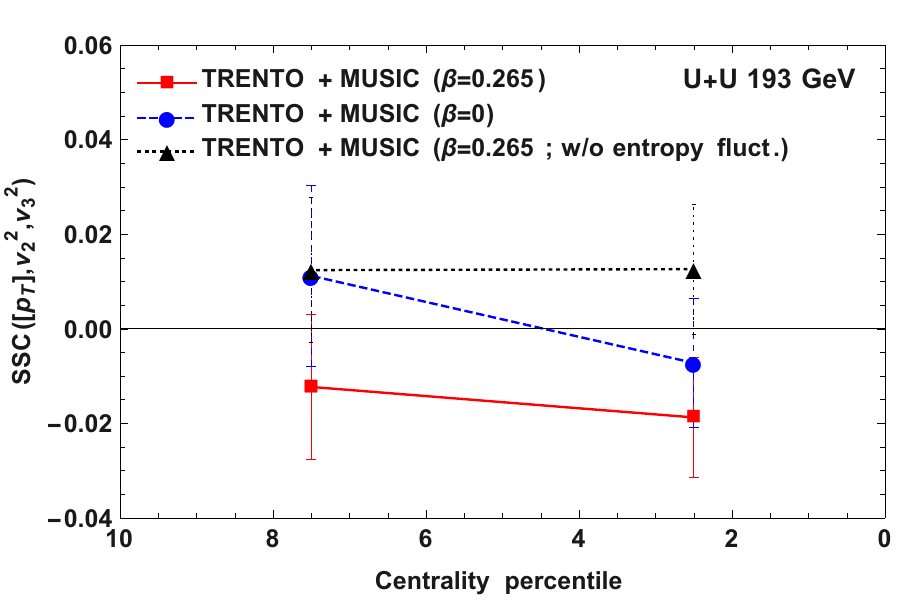}  
\centering
\caption{Third order scaled symmetric cumulant between $[p_T]$, $v_2^2$ and $v_3^2$ in U+U collision at 193 GeV. Symbols carry similar meaning as in Fig.~\ref{fig: NSC pt-v2-v3 and pt-v3-v4 UU}. The figure is from the original publication~\cite{Bozek:2021zim}, coauthored by the author.}
\label{fig: SSC pt-v2-v3 UU}
\end{figure}
A similar effect is observed for $NSC(p_T,N,v_3^2)$ but with mild difference when fluctuations in entropy are turned off, the difference is significant for the centrality $5-10\%$. Such a reverse scenario is observed because $v_3$ is primarily dominated by fluctuations and switching them off affect both $N$ and $v_3$. The fourth order cumulant $NSC(p_T,N,v_2^2,v_3^2)$, shown in Fig.~\ref{fig: NSC pt-n-v2-v3 UU}, is positive for the cases involving entropy fluctuations in the initial states for both spherical and deformed nuclei collisions. With entropy fluctuations switched off, the cumulant decreases\footnote{Please note that the absolute magnitudes of the cumulants involving multiplicity always decrease when there is no fluctuation in entropy deposition in the initial state.} remarkably and is compatible with zero for the centrality range studied. Study of these cumulants showcases the importance of multiplicity fluctuation and its cross-correlation with other collective observables in the central collisions of deformed nuclei.  

In order to complete the picture and maintain consistency with Chapter-5, in Fig.~\ref{fig: SSC pt-v2-v3 UU}, we present the third order scaled symmetric cumulants for mean transverse momentum, elliptic and triangular flow in U+U collisions. The scaled symmetric cumulant $SSC([p_T], v_2^2, v_3^2)$ shows similar behavior as $NSC([p_T], v_2^2, v_3^2)$ and does not provide much new information apart from the fact that it has larger order of magnitude for the correlation, because of the changed normalization in the denominator. Moreover, its study does not require additional information on the average transverse momentum of the particles which is sometimes not known a priori.

%*******************************************************************************
%****************************** Seventh Chapter (Conclusions) **********************************
%*******************************************************************************

\chapter{Summary and outlook}

% **************************** Define Graphics Path **************************
\ifpdf
    \graphicspath{{Chapter6/Figs/Raster/}{Chapter6/Figs/PDF/}{Chapter6/Figs/}}
\else
    \graphicspath{{Chapter6/Figs/Vector/}{Chapter6/Figs/}}
\fi

Finally, it brings us to the last chapter of this thesis, where we summarize the important findings and leave our concluding remarks for the research carried out. We also highlight further research possibilities on the similar topics, which lied beyond the current scope. We briefly discuss our future goals, prospective research directions and new potential problems that might be interesting to investigate in the coming years. 

The main goal of this thesis is to study the properties and the collective dynamics of the QGP medium created at the collision of two heavy nuclei at ultrarelativistic energies. The most distinctive and exotic feature of the heavy-ion collision is the collective anisotropic flow of the final state particles, originating from the geometry of its initial state. One of the most exclusive features of this collective flow is its event-by-event fluctuations mainly stemming from the event-by-event fluctuations in the initial state. Our primary aim is to focus on the fluctuations and correlations of those collective observables such as mean transverse momentum per particle $[p_T]$, harmonic flow coefficients $v_n$ etc. Relativistic hydrodynamics serves as an excellent theoretical tool to study the QGP medium, its initial state properties and unique collective signatures in the final state. This viscous hydrodynamic framework comprising a number of intermediate stages can be utilised in simulations or model, to study those collective observables and make robust predictions. Most interesting features of those observables originate from the initial state of the collision and their event-by-event fluctuations. Below we summarize the most important findings of our research and the conclusions coming out of them :  
\begin{itemize}
    \item The event-by-event fluctuations of the harmonic flow coefficients result in decorrelation between flow vectors in different transverse momentum bins. This decorrelation can be understood by studying the factorization-breaking coefficients between the flow vectors in two bins. In order to address experimental limitations due to low statistics in higher $p_T$-bins, we use one of the flow vectors as momentum averaged~\cite{ALICE:2022dtx}. The flow vector decorrelation is partly due to flow magnitude and partly due to flow angle decorrelation, the experimental measurement of which requires the construction of the factorization-breaking coefficients between the squares of the flow~\cite{Bozek:2021mov}. Our model results for Pb+Pb collision at 5.02 TeV, qualitatively reproduce the ALICE data. The observed mismatch between our model results and the data arise from the potential presence of the non-flow correlations which can be substantially minimized by taking the transverse momentum bins separated by large pseudorapidity gap. We also present predictions for similar momentum dependent mixed-flow correlations which can be used to study non-linearity in the dynamics of the system and can provide useful constraints on the non-linear modes in the initial state, when confirmed in experiments. Factorization-breaking coefficients can serve as robust probes of the fluctuations of flow.  

    \item Fluctuations of mean transverse momentum per particle in ultracentral Pb+Pb collisions display unique peculiar patterns, encoding important physical significance. The sudden steep decrease of $Var(p_T|N_{ch})$ in ATLAS data in the ultracentral region, can be explained by modelling the correlation between $[p_T]$ and $N_{ch}$ at fixed impact parameter through a two dimensional correlated Gaussian distribution~\cite{Samanta:2023amp}. Our model results show that in the ultracentral limit, the contribution of the impact parameter fluctuations or the volume fluctuations gradually goes to zero causing the sharp decline. Our model fit to the data returns a strong correlation between $[p_T]$ and $N_{ch}$ at fixed $b$, which appears to be a natural consequence of thermalization of the QGP system, a fundamental assumption in its hydrodynamic description. Additionally, we present robust predictions for the skewness and kutosis~\cite{Samanta:2023kfk}, characterizing the non-Gaussian features of $[p_T]$-fluctuation, based on a Gaussian-fluctuation model of impact parameter. The skewness and kurtosis show interesting patterns in the ultracentral regime around the knee of the multiplicity distribution, which arise mainly due to impact parameter fluctuations. As hinted by the measurements from the ALICE collaborations, our predictions can be   verified by the upcoming measurement of the ATLAS collaboration in similar bins of centrality estimators. Our results unveil crucial physical aspects of the QGP medium, and highlight the importance of impact parameter and its  fluctuations in heavy-ion collision. 

    \item Correlation between $[p_T]$ and $v_n^2$ can serve as a fine tool to probe the correlation present in the initial state of the collision between eccentricities, transverse size, entropy etc. We present results for the Pearson correlation coefficient between $[p_T]$ and $v_n^2$ by comparing it with the data and the linear predictor from the initial state~\cite{Bozek:2021zim}. To measure higher order correlations, we propose normalized and scaled symmetric cumulants between mean transverse momentum, harmonic flow and multiplicity, measuring genuine correlations between these observables and putting additional layer of constraints on the initial state properties. Similar momentum dependent Pearson correlation coefficients can be defined, which do not depend on the specific $p_T$-cut dependence and show sensitivity to the granularity in the initial state~\cite{Samanta:2023rbn}. We propose alternate constructions of such $p_T$-dependent correlation coefficients, which are experimentally favourable. Covariance between $[p_T]$ and $p_T$-dependent harmonic flow can be studied directly, providing a robust probe to the granularity. Such correlations and covariances between mean transverse momentum and harmonic flow coefficients capture the inter-correlation between the shape and the size of the QGP fireball, while putting additional constraints on the initial state properties.  

    \item Although nuclear structure is primarily studied in low energy nuclear physics, high energy heavy-ion collision can provide an excellent platform to map the deformed nuclear structure through similar collective observables. In particular, we show that deformation has a direct impact on the transverse momentum and harmonic flow coefficients in central collision of uranium nuclei at 193 GeV at RHIC. We show that the factorization breaking coefficients for the elliptic flow~\cite{Samanta:2023qem} and the Pearson correlation coefficient between $[p_T]$ and $v_2^2$ can provide robust probes of the quadrupole deformation parameter $\beta_2$ of the uranium ($^{238}$U) nucleus. We also present momentum dependent mixed flow correlations and propose new different normalized symmetric cumulants constructed with $[p_T]$, $v_n^2$ and $N_{ch}$, which provide additional novel constraints on the deformation of $^{238}$U~\cite{Bozek:2021zim}. Our results show that the collective flow and its fluctuations in central collisions provide a unique opportunity to investigate nuclear deformation through high energy nuclear collisions.          
\end{itemize}

\section*{Future directions}
\label{future plans}
The research carried out in this thesis paves a path for further development and novel research opportunities in future, in each sector. Moreover, the problems encountered in this work have motivated us to formulate new potentially interesting projects which we would like to explore in future. Below we discuss the scopes for further development in the related areas along with our future plans : 

\begin{itemize}
    \item We have studied the deformation of uranium nucleus through correlations and fluctuations-probing observables, which are relevant at RHIC energies. However, there exist other deformed nuclei such as $^{129}$Xe, which are collided at the LHC at 5.44 TeV~\cite{ALICE:2018lao,CMS:2019cyz,ATLAS:2019dct,ATLAS:2020sgl,ALICE:2021gxt,ATLAS:2022dov}. The xenon nucleus would be interesting to explore because, in addition to the quadrupole deformation ($\beta$ or $\beta_2$), it also exhibits axial asymmetry in its structure, which is identified by triaxiality parameter ($\gamma$)~\cite{Bally:2021qys,Jia:2021qyu}. The particular triaxial structure appears when all the three axes of the nucleus are of different length ($0<\gamma<60^\circ$) (Fig.~(\ref{fig: triaxial struct xenon})). Such structures affect the nuclear mass distribution~\cite{Bally:2021qys},
    \begin{equation}
    \begin{aligned}
     \rho(r, \theta, \phi ) &= \frac{\rho_0}{1+\exp[\frac{r-R(\theta, \phi)}{a}]} \ , \\
     \eqsp{with,} R((\theta, \phi) &= R_0 \{1+\beta[\cos \gamma Y_{2,0}(\theta, \phi)+\sin \gamma Y_{2,2}(\theta,\phi)]\} \ .
     \end{aligned}
    \label{eq: woods-saxon xenon}
    \end{equation}
    Due to three possible different structures, the overlap area in central Xe+Xe collisions would assume three possible shapes, inducing the possibility of larger shape fluctuations at fixed multiplicities. 
    \begin{figure}[ht!]
    \includegraphics[height=8 cm]{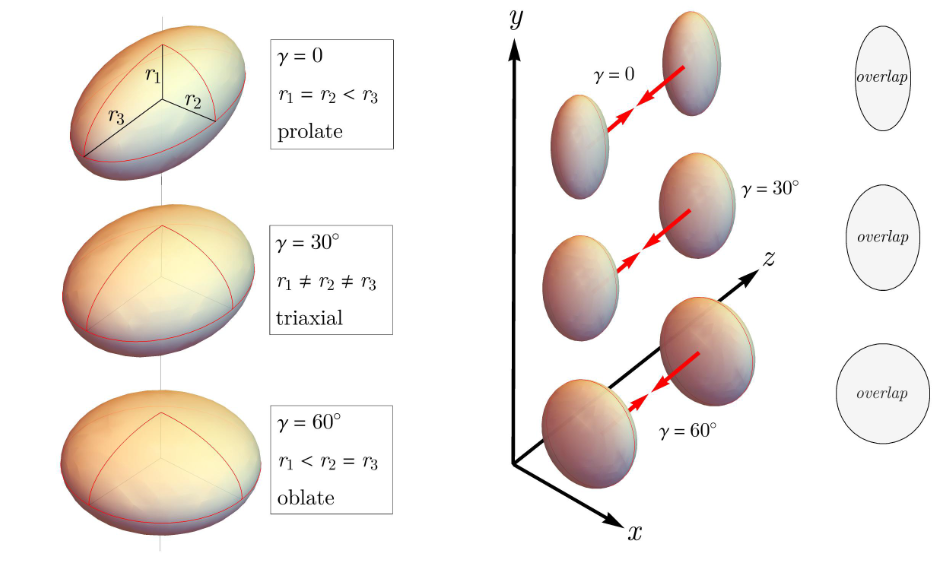}  
    \centering
    \caption{Pictorial representation of the structure of xenon-129 nucleus with three possible configurations shown from top to bottom on the left: prolate, triaxial and oblete. Different configuration results in different shapes of the overlap region in central collisions. Figure taken from~\cite{Bally:2021qys}.}
   \label{fig: triaxial struct xenon}
    \end{figure}
    Therefore it would be interesting to study similar observables such as factorization-breaking coefficients, Pearson correlation coefficients (already has been found to be sensitive to $\gamma$~\cite{Bally:2021qys}), normalized and scaled symmetric cumulants between $[p_T]$, $v_n^2$ and $N_{ch}$ etc. in Xe+Xe collision at 5.44 TeV, and can be compared to the results for Pb+Pb collisions, which are easily accessible at the LHC. Such quantitative and qualitative comparisons can shed light on the structure of Xe nuclei, providing sensitive constraints on its deformation parameter and triaxialiaty through high energy collisions. 

    Moreover, recent studies suggest a significant hexadecapole deformation ($\beta_4$) for the uranium nuclei~\cite{Ryssens:2023fkv,Xu:2024bdh} that can be constrained by the collective observabeles: flow cumulants, non-linear correlations etc. In future, we would be also interested to explore the hexadecapole deformation $\beta_4$ of $^{238}$U, with the correlations, higher order cumulants and factorization breaking coefficients which contain huge potential to put robust constraints on $\beta_4$.

    \item In the present work, we have focused on the transverse momentum fluctuations in ultracentral Pb+Pb collision at the LHC. However, similar study for $[p_T]$-fluctuations can be performed in p+Pb collisions which also show significant collective phenomena at the LHC~\cite{Bozek:2011if,Bozek:2012gr,Bozek:2013df,Bozek:2013uha,Werner:2013ipa,Schenke:2015aqa,Bozek:2017elk,ALICE:2012eyl,CMS:2012qk,ATLAS:2013jmi}. Being a much smaller system as compared to Pb+Pb, multiplicity fluctuations are expected to be significantly large in the ultracentral collisions (highest multiplicity). Therefore the fluctuations of $[p_T]$ at fixed multiplicity are also expected to be large, highlighting the importance of intrinsic or quantum fluctuations. Moreover the role of impact parameter would be interesting above the knee of the multiplicity distribution in p+Pb collision, which covers a large multiplicity range~\cite{Samanta:2024znz}. Therefore, investigation of $[p_T]$ fluctuations in ultracentral p+Pb collision and its direct comparison to Pb+Pb would be very interesting, which could unravel the dynamics of quantum fluctuations and its system-size dependence. A comparison to p+p collision would shed further lights on the issue and can be used as baselines for such studies.
    
    Additionally, we would like to study collectivity in p+Pb collisions using new observables (e.g. factorization-breaking coefficients\footnote{Of course the statistics for such observable in small systems would be a big challenge.}) that have not been studied yet. In particular, the initial state properties and their correlations can be mapped to the final sate by constructing flow cumulants, correlations and symmetric cumulants between mean transverse momentum and harmonic flow coefficients.  

    \item Furthermore, in future, we plan to study small collision systems such as O+O at both RHIC and LHC energies. The STAR collaboration has recently presented measurement on O+O collisions at RHIC energy (200 GeV)~\cite{Huang:2023viw}. O+O collisions at the LHC (7 TeV) are planned for Run3~\cite{Citron:2018lsq,Brewer:2021kiv,ALICE:2022wpn}, with data expected soon. The study of O+O collisions is interesting because it can address several issues such as the structure of $^{16}$O nucleus at the high energy scale, the limit of applicability of hydrodynamics for such small system, a direct comparison of the collective observables to Pb+Pb collision as O+O serves as a similar symmetric collision systems but with different centre of mass energy and much smaller size of the fireball etc. According to the low energy experiments, $^{16}$O nucleus exhibit a tetrahedral structure with alpha-clusters at its edges~\cite{Rybczynski:2019adt,Summerfield:2021oex} shown in Fig.~(\ref{fig: alpha-cluster oxygen}). As a result there exist surging theoretical interest to image the structure of $^{16}$O nucleus and study its unique collective signatures in high-energy O+O collision ~\cite{Sievert:2019zjr,Rybczynski:2019adt,Summerfield:2021oex,Behera:2023oxe,YuanyuanWang:2024sgp,Zhang:2024vkh,Zhao:2024feh}. The measurements from the STAR collaboration highlight significant contributions from sub-nucleonic fluctuations and nucleon-nucleon correlation at the initial state of O+O collision.  
    \begin{figure}[ht!]
    \includegraphics[height=6.5 cm]{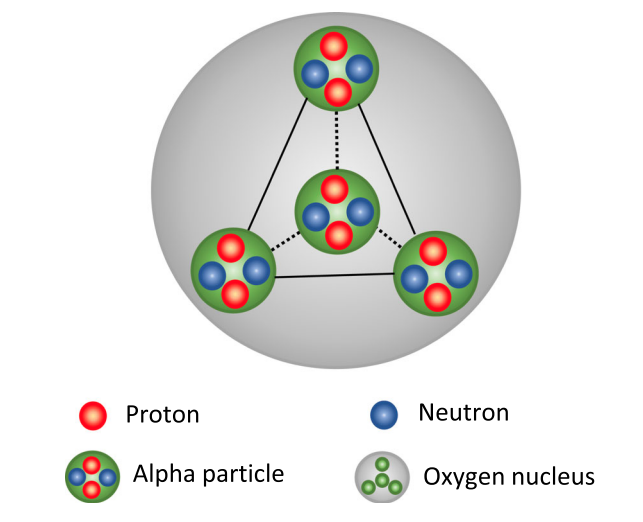}  
    \centering
    \caption{Pictorial depiction of the $\alpha$-clustered structure of oxygen-16 nucleus. Figure taken from~\cite{Behera:2021zhi}}
   \label{fig: alpha-cluster oxygen}
    \end{figure}
    We plan to study O+O collision at both energies by using different approaches for obtaining the initial state density distributions such as fitting Woods-Saxon distribution to the nuclear density obtained from low energy {\it ab-initio} NLEFT, VMC, EQMD or PGCM calculations~\cite{Rybczynski:2019adt,Summerfield:2021oex,YuanyuanWang:2024sgp,Zhang:2024vkh}, applying novel alpha-cluster+sub-nucleonic structure within initial condition model~\cite{Summerfield:2021oex} (e.g. TRENTO), implementing non-zero octupole deformation ($\beta_3$)~\cite{YuanyuanWang:2024sgp} to account for nuclear deformation etc. We would also like to identify the onset time for hydrodynamics in such small systems which is usually larger due to large inverse Reynold's number and Knudsen number~\cite{Summerfield:2021oex}. Investigation of applicability of hydrodynamics in O+O collision would be interesting in general,  which could be applicable for other small system collision. In O+O collision we would also like to map the final state observables to initial state properties through multi-particle correlations and cumulants.  
       
\end{itemize}

In addition to the above mentioned prospective projects, we would also like to explore other small systems such as p+Au, d+Au and He+Au at RHIC energies. In particular, the study of factorization-breaking coefficients would be interesting in such systems probing the fluctuations of harmonic flow at a much smaller scale with comparable system-sizes. This could help us to have a clear insight on the origin of flow-decorrelation in different kinematic bins.

High energy heavy-ion collision has entered its third decade with the current Run3 program at the LHC. ultrarelativistic heavy ion collision has emerged as an established field of research by answering some of the most fundamental questions of physics over the past thirty years. From the first Au+Au collision at RHIC to the latest Run3 measurement of Pb+Pb at the LHC, the study of collectivity has helped us to understand many fundamental properties of the Quark-Gluon-Plasma with hydrodynamic description serving as its fundamental basis. However, in spite of the rapid progress in the field , availability of large data sets and abundance of collision systems, there remain some major fundamental issues which need to be addressed. Some of these include the limitation of hydrodynamic picture and its boundaries in small collision systems, origin of collective flow in small systems, nuclear structure and its accessibility in future collider e.g. at the Electron-Ion-Collider (EIC), Future Circular Collider (FCC) etc. The study of small system collisions at the current and future runs at the LHC (e.g. p+Pb, O+O, p+O, Ne+Ne, Ne+Pb etc.) would help us to remarkably push the limit of our understanding of the QGP medium and hence the quest is on $\dots$

% ********************************** Back Matter *******************************
% Backmatter should be commented out, if you are using appendices after References
%\backmatter

% ********************************** Bibliography ******************************
\begin{spacing}{0.9}

% To use the conventional natbib style referencing
% Bibliography style previews: http://nodonn.tipido.net/bibstyle.php
% Reference styles: http://sites.stat.psu.edu/~surajit/present/bib.htm

% Two possible choices : APS or JHEP
%\bibliographystyle{References/apsrev4-2}
\bibliographystyle{References/JHEP}

\cleardoublepage
\bibliography{References/references} % Path to your References.bib file

% If you would like to use BibLaTeX for your references, pass `custombib' as
% an option in the document class. The location of 'reference.bib' should be
% specified in the preamble.tex file in the custombib section.
% Comment out the lines related to natbib above and uncomment the following line.

%\printbibliography[heading=bibintoc, title={References}]

\end{spacing}

% ********************************** Appendices ********************************

\begin{appendices} % Using appendices environment for more functunality

% ******************************* Thesis Appendix B ********************************

\chapter{Fluctuations of harmonic flow}

\section{Toy model for transverse momentum dependent flow decorrelation}
\label{a: flow decorr toy model}
The relation between flow vector, magnitude and angle decorrelation in Eq.~(\ref{eq: relation between flow vector, magnitude and angle decorrelation}) can be proved based on a simple toy model~\cite{Bozek:2023dwp}. To account for the momentum dependent fluctuations, which are small, let us consider the transverse momentum dependent flow vector $V_n(p)$ as a small deviation from the integrated flow\footnote{The author is sincerely grateful to Jean-Yves Ollitrault and Matthew Luzum for the discussions and independent formulation of this model during his stay at IPhT, Saclay}, 
\begin{equation}
    \vec{V}_n(p) = C(p) \vec{V}_n + \vec{\delta}(p) ,
 \label{eq: pt dependent flow vector in toy model}
\end{equation}
where, $C(p)$ is a scalar and $\vec{\delta}(p)$ is a small vector deviation, both of which depend on the transverse momentum, denoted here by $p$. Then the event-by-event transverse momentum dependent fluctuations of $\vec{V}_n(p)$ is governed by both the scalar $C(p)$ and the vector $\vec{\delta}(p)$ where the latter is solely responsible for the flow angle fluctuations. 

Next we take scalar product of Eq.~(\ref{eq: pt dependent flow vector in toy model}) with $\vec{V}_n$ and then take average over events : 
\begin{equation}
    \langle \vec{V}_n\cdot\vec{V}_n(p)^\star \rangle = C(p) \langle \vec{V}_n \cdot\vec{V}_n^\star \rangle + \langle \vec{V}_n \cdot \vec{\delta}(p)^\star \rangle \ .
 \label{eq: scalar product with integrated flow}
\end{equation}
Now comes our main {\it model} assumption : 
\begin{equation}
    \langle \vec{V}_n \cdot \vec{\delta}(p)^\star \rangle = 0 \ ,
 \label{eq: model assumptio}
\end{equation}
which is based on the fact that transverse momentum dependent fluctuations are small and randomly oriented so that its correlation with the integrated flow can be taken to be zero. With this from Eq.~(\ref{eq: scalar product with integrated flow}), we have ,
\begin{equation}
  C(p)= \frac{\langle \vec{V}_n\cdot\vec{V}_n(p)^\star \rangle}{\langle \vec{V}_n \cdot\vec{V}_n^\star \rangle} = \frac{\langle \vec{V}_n\cdot\vec{V}_n(p)^\star \rangle}{\langle v_n^2 \rangle} \ ,
 \label{eq: expression for Cp}
\end{equation}
where $v_n^2 \equiv |\vec{V}_n|^2$ .

\subsection{Flow vector decorrelation in second order}
\label{flow vec decorr}

The factorization-breaking coefficient between flow vector squared is given by Eq.~(\ref{eq: fact-break flow vector squared one bin}),
\begin{equation}
    r_{n;2}(p)=\frac{\langle \vec{V}_n^2 \cdot \vec{V}_n^*(p)^2 \rangle}{\sqrt{\langle v_n^4 \rangle \langle v_n^4(p) \rangle}} \ .
 \label{eq: flow vec fact-break}
\end{equation}
Using Eq.~(\ref{eq: pt dependent flow vector in toy model}) the numerator becomes up to second order in $\delta (p)$, 
\begin{equation}
 \begin{aligned}
  \langle \vec{V}_n^2 \cdot \vec{V}_n^*(p)^2 \rangle = C(p)^2 \langle v_n^4 \rangle \ ,
  \end{aligned}
 \label{eq: numerator flow vec fact-break}
\end{equation}
where we have used $\langle \vec{V}_n(p) \cdot \vec{\delta}(p)^\star\rangle=0$ and $\langle \vec{V}_n(p)^2 \cdot \vec{\delta}^2(p)^\star\rangle=0$. Similarly, in the denominator,
\begin{equation}
 \begin{aligned}
  \langle v_n^4(p) \rangle = \vec{V}_n^2(p) \cdot \vec{V}_n^\star(p)^2 = C(p)^4 \langle v_n^4 \rangle + 4 C(p)^2 \langle v_n^2 \delta(p)^2 \rangle
  \end{aligned}
 \label{eq: denominator flow vec fact-break}
\end{equation}
where $\delta(p)^2 = \vec{\delta}(p)\vec{\delta}(p)^\star$. With this Eq.~(\ref{eq: flow vec fact-break}) becomes, 
\begin{equation}
     r_{n;2}(p) = \frac{1}{\bigg( 1+4\frac{\langle v_n^2 \delta(p)^2 \rangle}{C(p)^2 \langle v_n^4 \rangle}\bigg )^{1/2}} \simeq 1 - 2 \Delta_n(p)
 \label{eq: flow vec decorr toy model}
\end{equation}
where, 
\begin{equation}
     \Delta_n(p) = \frac{\langle v_n^2 \delta(p)^2 \rangle}{C(p)^2 \langle v_n^4 \rangle }
 \label{eq: decorr factor}
\end{equation}
is the decorrelation factor and it is small, quantifying the amount of decorrelation between transverse momentum dependent flow vector and momentum averaged flow vector. 

\subsection{Flow magnitude decorrelation}
\label{flow mag decorr}

Factorization-breaking coefficient between flow magnitude squared is given by, 
\begin{equation}
    r^{v_n^2}(p)=\frac{\langle |\vec{V}_n|^2 |\vec{V}_n^*(p)|^2 \rangle}{\sqrt{\langle v_n^4 \rangle \langle v_n^4(p) \rangle}}  \ .
 \label{eq: flow mag fact-break}
\end{equation}
The denominator remain same as before. The numerator can be expanded as, 
\begin{equation}
    \langle |\vec{V}_n|^2 |\vec{V}_n^*(p)|^2 \rangle = \langle v_n^4 C(p)^2 \rangle + \langle v_n^2 \delta(p)^2 \rangle = C(p)^2 \langle v_n^4 \rangle \bigg( 1 + \Delta_n(p)\bigg) .
 \label{eq: numerator flow mag fact-break}
\end{equation}
Then Eq.~(\ref{eq: flow mag fact-break}) becomes,
\begin{equation}
     r^{v_n^2}(p)\simeq (1+\Delta_n(p)) (1-2 \Delta_n(p)) \simeq 1-\Delta_n(p)
 \label{eq: flow mag decorr toy model}
\end{equation}
Thus Eqs.~(\ref{eq: flow vec decorr toy model}) and (\ref{eq: flow mag decorr toy model}) satisfies the relation,
\begin{equation}
    1-r_{n;2}(p) = 2 [1-r^{v_n^2}(p)] \ ,
 \label{eq: relation between flow vec and flow mag decorr toy model}
\end{equation}
as given in Eq.~(\ref{eq: relation between flow vector and magnitude decorrelation}). 

\subsection{Flow angle decorrelation}
\label{flow ang decorr}
The experimental estimate of the flow angle decorrelation is obtained by taking ratio of the flow vector and flow magnitude factorization-breaking coefficients and provides a measure of the actual angle decorrelation, 
\begin{equation}
  F_n(p)= \frac{\langle \vec{V}_n^2 \cdot \vec{V}_n^*(p)^2 \rangle}{\langle |\vec{V}_n|^2 |\vec{V}_n^*(p)|^2 \rangle} \simeq \frac{\langle v_n^4 \cos[2n(\Psi_n-\Psi_n(p))]\rangle}{\langle v_n^4 \rangle} .
 \label{eq: flow ang decorr}
\end{equation}
Then using the previous equations one gets,
\begin{equation}
    F_n(p)= \frac{1}{1+ \Delta_n(p)}\simeq 1- \Delta_n(p) . 
 \label{eq: flow ang decorr first order toy model}
\end{equation}

Thus from a simple toy model of transverse momentum dependent flow vectors with random small fluctuations, one can prove the relation presented in Eq.~(\ref{eq: relation between flow vector, magnitude and angle decorrelation}), 
\begin{equation}
 \begin{aligned}
[1-r_{n;2}(p)] \simeq [1-r_n^{v_n^2(p)}] + [1- F_n(p)] \,
 \end{aligned}
\label{eq: relation between flow vec, mag and ang decorr toy model}
\end{equation}
which is also reflected in the results presented in Chapter-3. 

% ******************************* Thesis Appendix B ********************************

\chapter{Transverse momentum fluctuations}

\section{Simulations with hydrodynamics and HIJING}
\label{a: hydro vs hijing}

The setup of our hydrodynamic calculation in Chapter-4 is identical to that of Chapter-3. 
We use a boost-invariant version of the hydrodynamic code MUSIC~\cite{Schenke:2010nt} with the default freeze-out temperature $T_f=135$~MeV. We assume a constant shear viscosity to entropy density ratio $\eta/s=0.12$, and the bulk viscosity is set to zero.
The initial entropy distributions are taken from the TRENTO model~\cite{Moreland:2014oya}, where the parameters are fixed as follows. The most important parameter is the parameter $p$ which defines the dependence of the density on the thickness functions of incoming nuclei, which is set to $p=0$, corresponding to a geometric mean, which is the default choice. The parameter defining the strength of multiplicity fluctuations is set to $k=2.5$ (the default being $k=1$). With this choice, the relative multiplicity fluctuations is compatible (within statistical errors) with ATLAS data in Table~\ref{tab: fit parameters Nch and Et dist}. The nucleon-nucleon cross section is set to $\sigma_{NN}=7.0$~fm$^2$ (instead of the default $\sigma_{NN}=6.4$~fm$^2$).

The normalization of the entropy density from the TRENTO model is adjusted so as to reproduce the charged multiplicity measured by ALICE in Pb+Pb collisions at $5.02$~TeV~\cite{ALICE:2015juo}. 
Despite this normalization, the average mulplicity is $\overline{N_{ch}}=6660\pm 30$, much larger than that seen by ATLAS (Table~\ref{tab: fit parameters Nch and Et dist}). 
The main reason is that some of the particles escape detection, even within the specified angular and $p_T$ range, and the data are not corrected for the reconstruction efficiency. 
In addition, we expect deviations between the model and data for two reasons. 
First, hydrodynamic models typically underestimate the pion yield at low $p_T$~\cite{Grossi:2021gqi,Guillen:2020nul}. 
Since the calculation is adjusted to reproduce the total charged multiplicity, which is dominated by pions, this implies in turn that it should overestimate the yield for $p_T>0.5$~GeV/c, which is the range where it is measured by ATLAS. 
Second, our hydrodynamic calculation assumes that the momentum distribution is independent of rapidity. 
In reality, it is maximum near mid-rapidity, in the region covered by the ALICE acceptance. 
This should also lead to slightly overestimating the multiplicity seen by ATLAS, whose inner detector covers a broader range in rapidity. 

The width of $p_T$ fluctuations from our hydrodynamic calculation is $\sigma_{p_T}(0)=13\pm 1$~MeV$/c$.
Note that they are dynamical fluctuations only. 
The reason is that we do not sample particles according to a Monte Carlo algorithm, but simply calculate the expectation value of $[p_T]$ at freeze-out.
Therefore, the width of $[p_T]$ fluctuations from the hydrodynamic calculation can in principle be compared directly with that measured experimentally. 
Our value is somewhat higher than the value $\sigma_{p_T}(0)=9.357$~MeV$/c$ inferred from ATLAS data (see Fig.~\ref{fig: varpt}). The fact that hydrodynamics overestimates $[p_T]$ fluctuations is an old problem~\cite{Bozek:2012fw}, which can be remedied by carefully tuning the fluctuations of the initial density profile~\cite{Bozek:2017elk,Bernhard:2019bmu,JETSCAPE:2020mzn}.  
It is the reason why we choose to fit the magnitude of $[p_T]$ fluctuations to data, rather than obtain it from a hydrodynamic calculation. 

The Pearson correlation coefficient between $N_{ch}$ and $[p_T]$ from our hydrodynamic calculation is $r_{N_{ch}}\sim0.674$ which is in excellent agreement with the value $r_{N_{ch}}=0.676$ returned by the fit to ATLAS data (Fig.~2 (c) of the paper). 
Simulations with HIJING shown in Fig.~(\ref{fig: b-fluct and pt-Nch scatter plot}) of the paper follow the same setup as in Ref.~\cite{Bhatta:2021qfk}. The average multiplicity is $\overline{N_{ch}}=5149$, somewhat lower than in the hydrodynamic calculation, and the average value of $p_T$, denoted by $\overline{p_T}$, is $941$~MeV$/c$, also lower than in the hydrodynamic simulation ($\overline{p_T}=1074$~MeV$/c$).

\section{Fitting the variance of $[p_T]$ fluctuations}
\label{a: fitting var data}

ATLAS provides us with two data sets for the centrality dependence of the variance, depending on whether centrality is determined with $N_{ch}$ or $E_T$.
We first carry out a standard $\chi^2$ fit for each of these sets, where the error is the quadratic sum of the statistical and systematic errors on the data points. 
The three fit parameters are $\sigma_{ p_T}(0)$ (the standard deviation of $[p_T]$ for $b=0$), $\alpha$ (which defines the decrease of the variance as a function of impact parameter), and the Pearson correlation coefficient $r$ between $[p_T]$ and the centrality estimator for fixed $b$. 
Consistency of our model requires that $\sigma_{ p_T}(0)$ and $\alpha$, whose definition does not involve the centrality estimator, are identical for $N_{ch}$ and $E_T$ based data for a given $p_T$ selection. 
Values of $\sigma_{p_T}(0)$ are identical within less than $1\%$, but values of $\alpha$ differ by $6\%$, with $E_T$-based data favoring a larger $\alpha$. 
We then fix the values of $\sigma_{p_T}(0)$ and $\alpha$ to the average values of $N_{ch}$ and $E_T$-based results, and redo the fits by fitting solely the Pearson correlation coefficient $r$ for each of the two data sets. 
Due to the small tension between the values of $\alpha$,  our fit slightly overestimates the variance for the lowest values of $N_{ch}$, and slightly underestimates it for the lowest values of $E_T$. 
This effect is of little relevance to our study which focuses on ultra-central collisions, and we have not investigated its origin. 

The values of $\alpha$ are close to $1.2$, which implies that the variation of dynamical fluctuations with impact parameter is faster than that of statistical fluctuations, for which $\alpha=1$.  
The parameter $\sigma_{p_T}(0)$ is close to $10$~MeV$/c$, while the average value of $p_T$ is close to $1$~GeV$/c$. 
This corresponds to a relative dynamical fluctuations of order $1\%$ in central collisions. 
The values of the Pearson correlation coefficient end up being similar, between $0.6$ and $0.7$, for both data sets. 

The results shown are obtained by assuming that the variance of the charged multiplicity is proportional to the mean, that is,  ${\rm Var}(N_{ch}|c_b)/\overline{N_{ch}}(c_b)$ is constant. As explained in Sec.~\ref{correlated Gaussian}, we have also tested two alternative scenarios, assuming either that ${\rm Var}(N_{ch}|c_b)$ is constant or that the ratio  ${\rm Var}(N_{ch}|c_b)/\overline{N_{ch}}(c_b)^2$ is constant. 
We have checked that the fit to the data is as good.  
The values of fit parameters vary only by 3\% for $\alpha$, and even less for $\sigma_{p_T}(0)$ and  $r$.

\section{Centrality dependence of multiplicity fluctuations}
\label{a: duke vs jetscape}

\begin{figure}[ht!]
\begin{subfigure}{0.5\textwidth}
\centering
\includegraphics[height=5.5 cm]{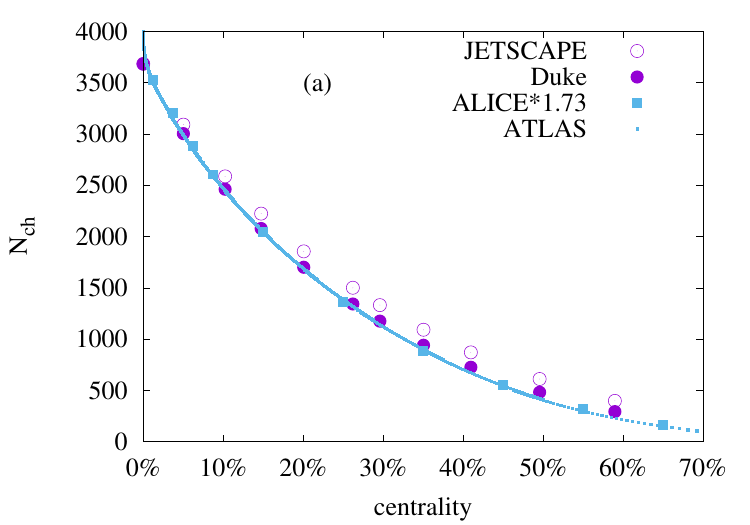}
\end{subfigure}~~
\begin{subfigure}{0.5\textwidth}
\centering
\includegraphics[height=5.5 cm]{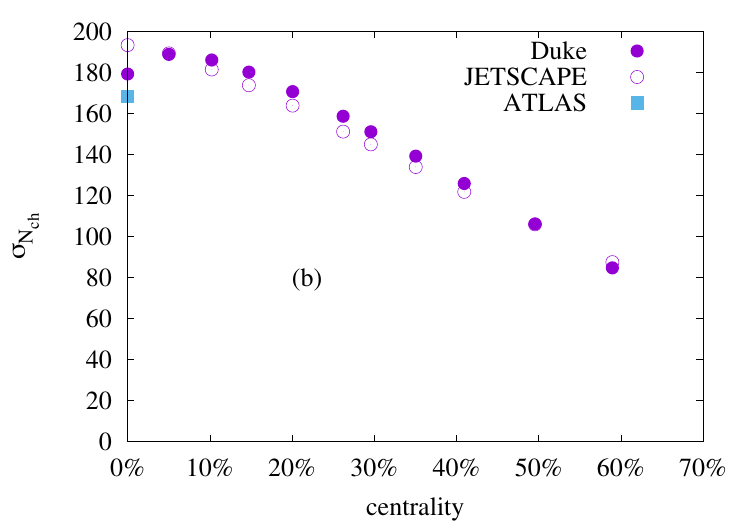}  
\end{subfigure}
\centering
\caption{Left: Variation of charged multiplicity $N_{ch}$ with centrality in Pb+Pb collisions at $\sqrt{s_{NN}}=5.02$~TeV measured by ATLAS~\cite{ATLAS:2022dov} and ALICE~\cite{ALICE:2015juo}. For ATLAS, the centrality is defined from the cumulative distribution of $N_{ch}$ and then divided by a calibration factor $1.153$~\cite{ATLAS:2018ezv}, which corrects for the fact that for the largest centrality fractions, some of the recorded events are fake. The ALICE results have been re-scaled by a factor $1.73$ to correct for the different acceptance and efficiency of the detector. The circles display the centrality dependence of the mean initial energy for the T$_{\text{R}}$ENTo parametrizations used by the Duke~\cite{Moreland:2018gsh} and JETSCAPE analyses~\cite{JETSCAPE:2020mzn}. The centrality is defined as $\pi b^2/\sigma_{\rm Pb}$, where $\sigma_{\rm Pb}=767$~fm$^2$ is the total inelastic cross section. Right: Variation of the standard deviation of $N_{ch}$ with centrality. 
}
\label{fig: mean and std dev of Nch}
\end{figure}

The probability distribution of the multiplicity at fixed impact parameter $b$ is expected to be approximately Gaussian~\cite{Das:2017ned} and can be characterized by its mean $\overline{N_{ch}}(c_b)$ and standard deviation $\sigma_{N_{ch}}(c_b)$, which both depend on $b$. 
The mean can be reconstructed using the simple following rule. 
If a fraction $c_b$ of events have a multiplicity larger than $N$, then $N\simeq \overline{N_{ch}} (c_b)$~\cite{Broniowski:2001ei}. 
This simple rule, which is applied to ATLAS data in Fig.~\ref{fig: mean and std dev of Nch} (a), works well except for multiplicities around and above the knee. 
On the other hand, the centrality dependence of $\sigma_{N_{ch}}$ is not known, and we use state-of-the-art hydrodynamic calculations by the Duke group~\cite{Moreland:2018gsh} and by the JETSCAPE collaboration~\cite{JETSCAPE:2020mzn} to evaluate it. 
However, we want to avoid running massive hydrodynamic calculations, and we therefore  estimate the multiplicity fluctuations from the initial conditions of these calculations. We assume that for every collision event, the multiplicity is proportional to the initial energy. Both Duke and JETSCAPE analyses employ the T$_{\text{R}}$ENTo parametrization~\cite{Moreland:2014oya} for the initial energy density, but with slightly different values of the parameters. 
We run these T$_{\text{R}}$ENTo initial conditions for several fixed values of $b$ (specifically, $b=0$, $3.5$, $5$, $6$, $7$, $8$, $8.5$, $9.25$, $10$, $11$, $12$~fm). 
For each $b$, we generate $10^5$ events with both Duke and JETSCAPE parameters, and we compute the initial energy of each event. 
We rescale this energy by a constant factor so that it matches the ATLAS result for the charged multiplicity at $b=0$~\cite{Samanta:2023amp}.
The variation of the mean $N_{ch}$ with centrality is displayed in Fig.~\ref{fig: mean and std dev of Nch} (a).
Experimental data are also shown. 
One sees that ALICE and ATLAS data are in excellent agreement once properly rescaled.
The calculation using the Duke parametrization agrees very well with experiment. 
Agreement is not quite as good, but still reasonable, for the JETSCAPE parametrization. 

\begin{figure}[ht!]
\includegraphics[height=7 cm]{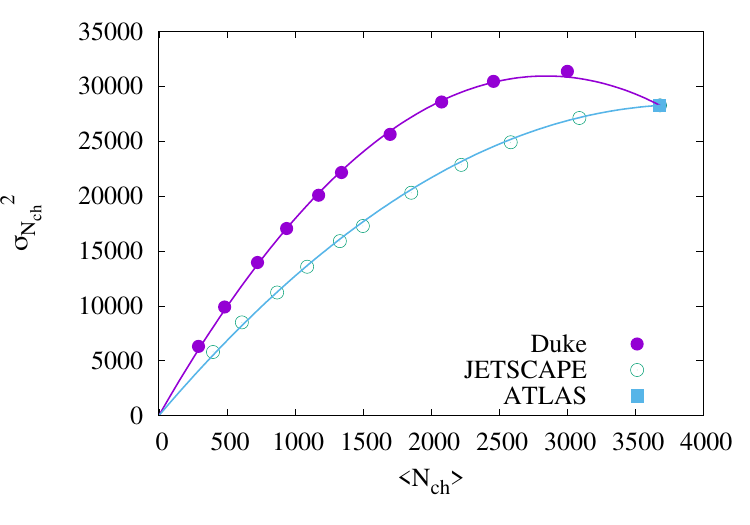}
\centering
\caption{Parametric plot of the mean and variance of $N_{ch}$ as a function of $b$.
Both models have been calibrated in such a way that they match data at $b=0$.
Solid lines are fits using $y=\gamma x+(1-\gamma) x^2$, where $y\equiv\sigma^2_{N_{ch}}(b)/\sigma^2_{N_{ch}}(0)$ and $x\equiv \overline{ N_{ch}}(c_b)/\overline {N_{ch}}(0)$, with $\gamma=2.83$ (Duke) and $\gamma=1.90$ (JETSCAPE). The calculation in Sec.~\ref{Variance} was done with $\gamma=1$ (variance proportional to mean). 
}
\label{fig: variance vs mean Nch}
\end{figure}
We then calculate the standard deviation of $N_{ch}$, $\sigma_{N_{ch}}$, for each value of $b$. 
Results are displayed in Fig.~\ref{fig: mean and std dev of Nch} (b). 
The standard deviation can only be measured at $b=0$~\cite{Yousefnia:2021cup} from the tail of the distribution of $N_{ch}$, therefore, there is only one data point on this plot.
One sees that both model calculations are in reasonable agreement with this data point, but slightly overestimate it. 
We use model calculations only to predict the $b$-dependence of $\sigma_{N_{ch}}$, not the value at $b=0$ which is measured precisely. 
We therefore rescale $\sigma_{N_{ch}}$ from the model calculation by a constant factor so that it matches the experimental value at $b=0$. 
The resulting predictions for $b>0$ are displayed in Fig.~\ref{fig: variance vs mean Nch}.
We plot the variance $\sigma_{N_{ch}}^2$ as a function of the mean.
If $N_{ch}$ is the sum of $k$ identical, uncorrelated distributions, where $k$ depends on $b$, both the mean and the variance are proportional to $k$, therefore, they are proportional to one another.
This behavior is only observed for large values of $b$.
Both model calculations predict that the variance increases more slowly as $b$ decreases. 
The Duke calculation even predicts that it decreases for the smallest value of $b$. 
The two solid lines, which are polynomial fits to our calculations, are used as two limiting cases which define the error bands in Fig.~\ref{fig: skew pt} and Fig.~\ref{fig: kurt pt}.

\end{appendices}

% *************************************** Index ********************************
\printthesisindex % If index is present

\end{document}